This thesis should be referenced as:

Kirby, G.N.C. “Reflection and Hyper-Programming in Persistent Programming Systems”.
Ph.D. Thesis, University of St Andrews (1992).

# Reflection and Hyper-Programming in Persistent Programming Systems

Graham N. C. Kirby

Department of Mathematical and Computational Sciences

University of St Andrews

St Andrews

Fife KY16 9SS

Scotland

// Abstract

In an orthogonally persistent programming system, data is treated in a manner independent of its persistence. This gives simpler semantics, allows the programmer to ignore details of long-term data storage and enables type checking protection mechanisms to operate over the entire lifetime of the data.

The ultimate goal of persistent programming language research is to reduce the costs of producing software. The work presented in this thesis seeks to improve programmer productivity in the following ways:

- by reducing the amount of code that has to be written to construct an application;
- by increasing the reliability of the code written; and
- by improving the programmer's understanding of the persistent environment in which applications are constructed.

Two programming techniques that may be used to pursue these goals in a persistent environment are type-safe linguistic reflection and hyper-programming. The first provides a mechanism by which the programmer can write generators that, when executed, produce new program representations. This allows the specification of programs that are highly generic yet depend in non-trivial ways on the types of the data on which they operate. Genericity promotes software reuse which in turn reduces the amount of new code that has to be written.

Hyper-programming allows a source program to contain links to data items in the persistent store. This improves program reliability by allowing certain program checking to be performed earlier than is otherwise possible. It also reduces the amount of code written by permitting direct links to data in the place of textual descriptions.

Both techniques contribute to the understanding of the persistent environment through supporting the implementation of store browsing tools and allowing source representations to be associated with all executable programs in the persistent store.

This thesis describes in detail the structure of type-safe linguistic reflection and hyper-programming, their benefits in the persistent context, and a suite of programming tools that support reflective programming and hyper-programming. These tools may be used in conjunction to allow reflection over hyper-program representations. The implementation of the tools is described.

# Acknowledgements

I thank my supervisor Ron Morrison for his support, guidance and enthusiasm. He has succeeded in providing a superb research environment.

Ron Morrison, Richard Connor, Quintin Cutts, Al Dearle and Dave Stemple have all been directly involved in the research described in this thesis. I thank them and the other members of the PISA group at St Andrews, Fred Brown, Dave Munro and Craig Baker, for the benefits of numerous discussions. I have also been helped by talking with Alex Farkas, Tim Sheard, John Rosenberg and Malcolm Atkinson.

Ron Morrison, Dave Stemple and Richard Connor provided invaluable constructive criticism during the writing of this thesis.

Thanks to Al for luring me into computing research in the first place with his diving trip tales of browser writing. Finally I thank Charlotte for always cheering me up.

# Contents

# 1 Introduction

## 1.1 Persistence and Software Costs

The persistence of data in a computer system is the length of time for which the data exists; this may range from the very short to the very long. Some examples at the extremes of the spectrum are the temporary data created during the evaluation of an expression in a program and the long-term data stored in a company's customer database. In the first case the data persists for a brief fraction of a single program execution, while in the second the data may outlive the programs that operate on it.

In an orthogonally persistent programming system, the manner in which data is manipulated is independent of its persistence. The same mechanisms operate on both short-term and long-term data, avoiding the traditional need for separate systems to control access to data of different degrees of longevity. Thus data may remain under the control of a single persistent programming system for its entire lifetime. The benefits of orthogonal persistence have been described extensively in the literature [ACC82, ABC+83, ABC+84, AM85, AMP86, AB87, Dea87, MBC+87, Wai87, AM88, Dea88, Bro89, MBC+89, Con90, MBC+90]; only a brief outline will be presented here.

The principal gains provided by orthogonal persistence are presented in the FIDE Course on Database Programming Languages and Persistent Systems [AAC+91] as:

- Improved programming productivity from simpler semantics.
- Without persistence, ad hoc arrangements for long term data storage, and data translations are necessary.
- Type checking protection mechanisms operate over the whole environment.
- Referential integrity is automatically supported.

In a non-persistent system, long-term data is stored in a database such as a file system or relational database. Programs operate on the data by reading it from the database, manipulating it in some way and writing it back to the database. Typically, the format of the data is translated to and from a richer programming language format as it is transferred between the database and programming language domains. Figure 1.1 illustrates the conceptual mappings that must be maintained and understood by the programmer.

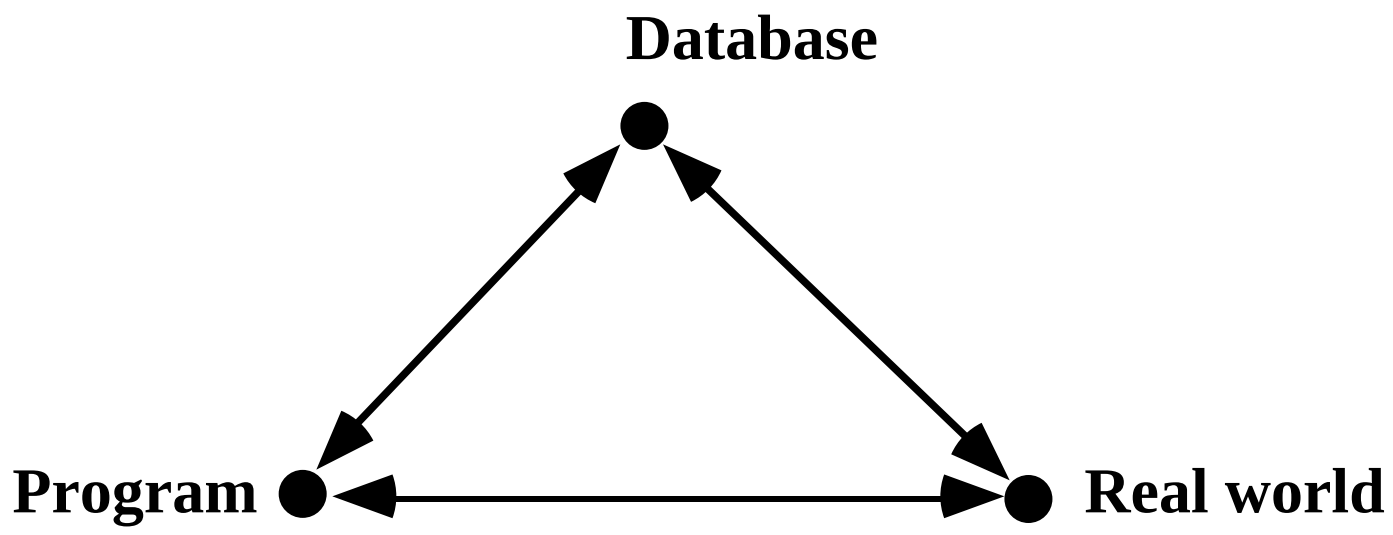

**Figure 1.1: Conceptual mappings in a non-persistent system**

The provision of a persistent language eliminates two of the three mappings since data retains its programming language format over its entire lifetime. This simplification is illustrated in Figure 1.2:

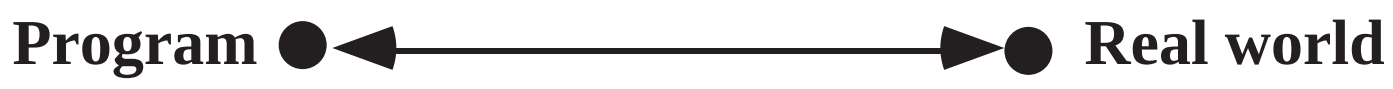

**Figure 1.2: Conceptual mappings in a persistent system**

Further benefits of orthogonal persistence, as described in [AAC+91], are:

- Programs such as procedures and modules can be represented by first class values which reside in the persistent store.
- The persistence mechanism provides incremental loading.
- The persistence mechanism provides linking via incremental data loading.
- The persistence mechanism verifies type correctness of such linking.
- The persistence mechanism supports incremental program construction and replacement.
- The persistence mechanism provides program and data library management.

The ultimate goal of persistent programming language research, as in many other branches of computer science, is to reduce the costs of producing software. These costs are borne during a number of phases in the software development process which include software design, coding, debugging and software evolution. The aim of the work described in this thesis is to investigate and to support certain mechanisms for reducing software costs in persistent systems. Gains in productivity are sought in three ways:

- by reducing the amount of code that has to be written to construct an application;
- by increasing the reliability of the code written; and
- by improving the programmer's understanding of the persistent environment in which applications are constructed.

Details of how these sub-goals are tackled will be elaborated later in this chapter.

## 1.2 The FIDE View of Software Production

The context of this work is the ESPRIT funded FIDE project [FID90], a collaboration in which a number of groups worked towards the integration of programming languages and database systems. The research areas within FIDE included type systems, object stores, compiler technology and integrated programming environments. In [Con90], Connor summarises the goals of the project as follows:

> "An ideal model for building applications in an integrated data-intensive system has been described by the FIDE project, and may be likened to the diagram in Figure 1.3. In such a model as much as possible is factored out of the application programs into a central repository. Using such a model it should be possible to accelerate software production, improve system reliability and gain economies from code re-use.
>
> Currently, however, information systems are usually constructed within a set of loosely connected support systems, such as database systems, programming languages, programming environments and operating systems. Each of these components is often designed independently and built using a separate technology. The resulting inconsistencies between these technologies make programming of data-intensive applications difficult, expensive, and error-prone.

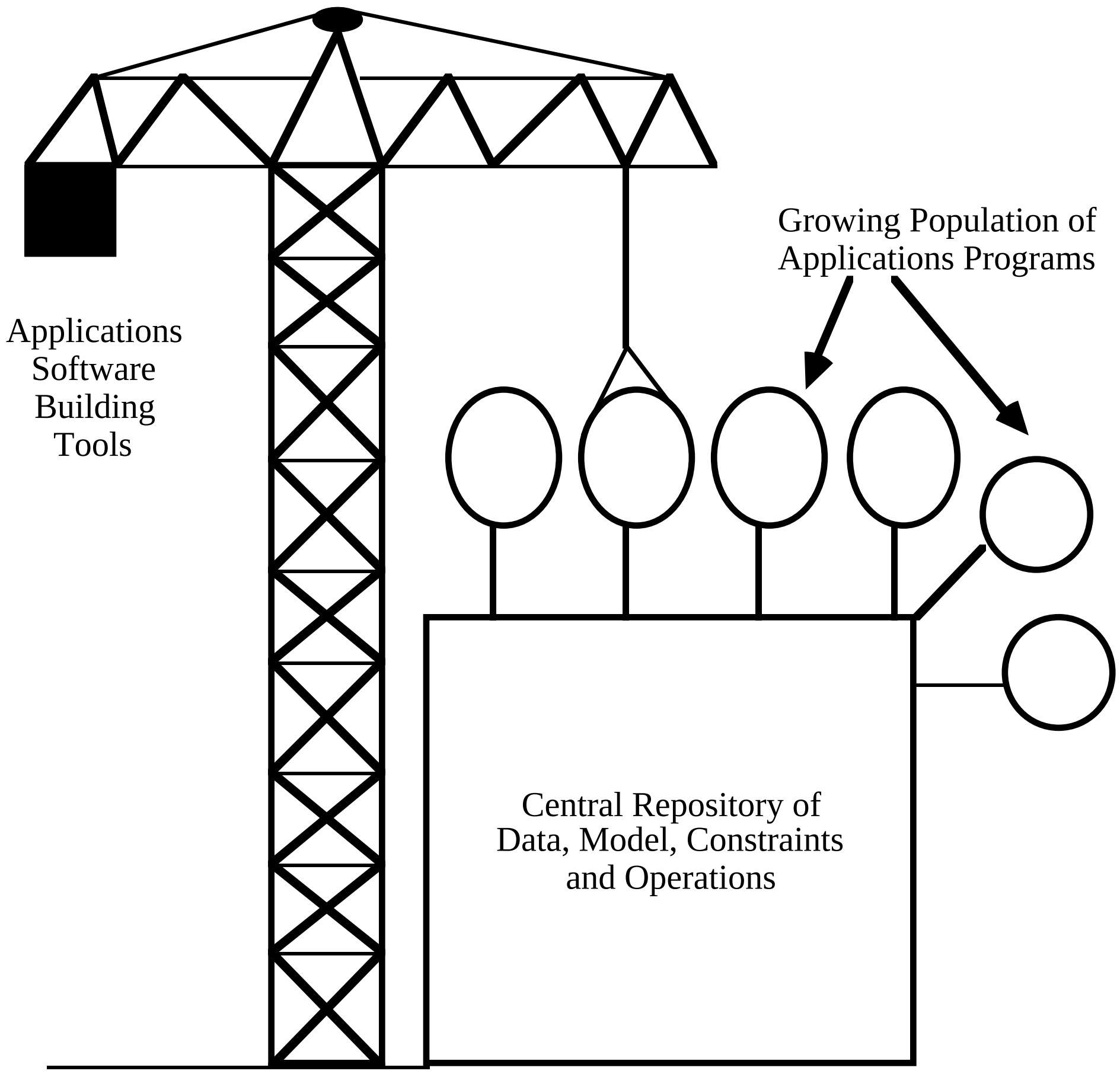

**Figure 1.3: The FIDE model of applications building**

> The programming of systems which deal with large amounts of long-lived data is intrinsically difficult. However, the major failure of current technology as outlined above is that the lack of integration leads to avoidable complexity.
>
> Persistent programming systems are one area in which to experiment with the removal of this avoidable complexity. By merging the distinction between long-term and short-term data, it becomes possible to model all of the activities required for a body of data within a single system."

## 1.3 Persistence as a Platform

A number of persistent programming systems have been proposed, including PS-algol [ACC82], Amber [Car85], Galileo [ACO85], Leibniz [Eve85], Persistent Prolog [GMD85], Poly [Mat85], DBPL [MS89], E [RC90], $\chi$ [HS90], Napier88 [MBC+89] and STAPLE [DM90]. To various extents these systems provide the benefits described earlier. The work presented in this thesis attempts to advance the technology by taking the basic persistent system as given and using it as a platform on which to build further programming support. The aim is to reduce software costs in a number of ways.

### 1.3.1 Writing Less Code

Writing program code is time consuming and error-prone. One way to reduce the costs associated with constructing a software application is to write less code in the process. This can be achieved using a number of methods including:

- using a higher-level programming language;

- reusing existing fragments of code; or
- automating parts of the code writing process.

Work in this thesis addresses the second and third points. One area involves the specification of highly generic program forms. A generic program is more generally applicable than a non-generic equivalent and the range of situations in which it can be reused is thus increased. The thesis describes research on the specification of these forms and also on the structure needed to support their reuse. This structure includes tools to aid the programmer in locating candidate programs for reuse and in composing those selected to form new programs.

Another research area to be described involves allowing the programmer to specify certain parts of an application without having to write code explicitly. In particular the need to write code describing the positions of programs and data in the persistent store is reduced.

### 1.3.2 Writing More Reliable Code

Debugging incorrect programs is costly. In general the earlier in the application development process that errors are detected, the cheaper it is to correct them. The considerations may not be solely financial: consider, for example, the differing opportunities available for correction when an error in a jet engine controller is detected during development as against during operation. To increase the number of errors caught early, many systems subject programs to a variety of static checks before allowing them to be executed. This enables many errors to be corrected before they have an opportunity to cause harm. In addition to static checking, programming systems may employ dynamic checking, performed during program execution in an attempt to detect and prevent harmful operations before they occur.

Most orthogonally persistent systems enforce strong typing. This means that, through a combination of static and dynamic checking, they ensure that no data is operated on in a manner forbidden by its type. This increases program reliability by preventing a particular class of errors defined by the type system. It is widely believed that at least some of the type checking in persistent systems must be performed dynamically if the systems are to be useful. However, it is advantageous for as much of the type checking as possible to be static, as this allows type errors to be detected earlier and thus at less cost. The thesis describes research to increase the proportion of static type checking while retaining the required flexibility. The techniques established also extend to another form of checking, verifying the presence of data in the persistent store. It is also believed that they may be used to allow forms of checking that have been regarded as intrinsically dynamic to be performed statically [Con92].

### 1.3.3 Understanding the Persistent Environment

To make use of existing programs in composing applications, the programmer must be able to obtain descriptions of those programs available. That is, there is a need to explore the persistent environment in which application construction takes place. The thesis describes research to provide tools for this exploration; they assist the programmer to discover what software components are available and whether a given component is type compatible with an intended use. The tools are also able to provide source representations for any executable programs located in the persistent store. This is achieved by enforcing the retention, in the persistent store, of source code at the time it is compiled.

### 1.3.4 Research Topics

The research in this thesis follows the directions described in the previous sections in two main research areas:

**Linguistic reflection** is a technique that allows programs to be treated as data and vice-versa. Using reflection the programmer can write programs that produce new programs. The research involves a study of the nature of reflection and the development of tools to support the writing of reflective programs.

**Hyper-programming** is a technique that allows source programs to contain direct links to values in the persistent store. As with reflection, the research involves a study of the nature of hyper-programming and the development of support tools. A further topic is the investigation of how reflection and hyper-programming may be used in conjunction.

## 1.4 Linguistic Reflection

With linguistic reflection, programs can create new program representations and transform them into executable programs. Type-safe linguistic reflection is a kind of linguistic reflection in which the program representations are checked for type correctness. The technique may be used to specify highly generic program generators that create programs tailored to the types of the data to be operated on. This genericity extends beyond that available in current polymorphic systems and thus provides greater opportunities for software reuse.

A related use is in accommodation to system evolution: using linguistic reflection, applications may adapt to changes in the structure of the data on which they operate, while retaining a high degree of static type checking. Support for adaptable applications assists the goal of writing less code, by reducing the need to re-implement applications as data evolves.

Several systems that support type-safe linguistic reflection have been implemented in the past; these include PS-algol [PS88], Napier88 [MBC+89]. and TRPL [She90]. Various applications of the technique are described in [Coo90a, Phi90, She91, HKS92, Kir92, SSF92, SSS+92]. This thesis describes research that focuses on the following aspects:

- classification and analysis of the anatomies of reflective systems;
- identification of issues affecting the useability of linguistic reflective systems; and
- investigation of the interaction between linguistic reflection and persistence.

## 1.5 Hyper-Programming

A hyper-program is a source program that contains links embedded in the text, in the same way that a fragment of hyper-text contains links to other fragments. The difference is that hyper-program links may to refer to data of any type in the persistent store, rather than being restricted to textual data [FDK+92, KCC+92b].

The provision of hyper-programming facilities assists the three goals of writing less code, writing more reliable code and understanding the persistent environment. The writing of less code is achieved by allowing more succinct programs, as a textual description of how to access a data item may be replaced by a link to the data. Code reliability is improved by enabling certain program checking to be performed statically rather than dynamically. Finally, the use of hyper-programs enables source representations to be supplied for certain programs that may exist in the persistent store but admit no purely textual representation. This assists the programmer in understanding the nature of the software available for reuse.

This thesis describes the first known implementation of hyper-programming. A related research area is the interaction of hyper-programming with type-safe linguistic reflection.

## 1.6 Software Products

The software produced during the work described in this thesis comprises the following:

- a set of reflective programming tools;
- a programming environment in which the programmer may browse the contents of the persistent store and compose hyper-programs linked to data found there;
- the support technology on which these tools are based:
  - a graphical user interface tool-kit;
  - an interactive persistent store browser;
  - a hyper-text editor which is used to support editing of hyper-programs.

## 1.7 Thesis Structure

Chapter 2 analyses the components of reflective systems, identifying two main types, behavioural reflection and linguistic reflection. The category of linguistic reflection is further classified into type-safe and non-type-safe varieties. Examples of type-safe linguistic reflection in different languages are given. The chapter analyses the nature of the program generators used in linguistic reflective systems and identifies some aspects that lead to their being difficult to write.

Chapter 3 describes the concept of hyper-programming and explains the benefits obtained. Chapter 4 gives details of the user interface of the hyper-programming environment. To illustrate this it shows how a simple application may be constructed by combining new code with reused existing programs.

Chapter 5 presents some proposals for improving the ease with which reflective program generators may be written. A generator model which supports the implementation of these proposals and allows hyper-program representations to be manipulated is described. The chapter then illustrates the user interface to the programming tools with which these generators are constructed.

Chapter 6 describes the principal features of the implementations of the hyper-programming and reflective programming tools. This includes the implementation of the underlying window management facilities, the hyper-text editor, the persistent store browser, and modifications to the Napier88 compiler to support compilation of hyper-programs.

# 2 Reflection

## 2.1 Introduction

Reflection involves programs being able to modify their own behaviour in the course of their evaluation. They can achieve this in two ways: by changing the way that programs are evaluated in the system, or by changing their own structures. Existing reflective languages can be placed in one of two categories according to the variety of reflection that they support. These varieties are *behavioural reflection* and *linguistic reflection*. One particular form of reflection in the latter category, *type-safe linguistic reflection*, allows all reflective operations to be type-checked. This form is suitable for use in strongly typed persistent systems in which the integrity of large amounts of data depends on the prevention of operations that contravene type rules.

This chapter describes some of the forms of reflection used in existing systems, and analyses two forms of type-safe linguistic reflection, run-time reflection and compile-time reflection. Applications of these techniques are described. These are: attaining high levels of genericity, accommodating evolution in persistent systems, implementing data models, optimising implementations and validating specifications. Some of the factors that affect the useability of type-safe linguistic reflective systems from the point of view of the programmer are identified. The nature of the generators, the functions that produce new program representations, is described in detail.

Although its use is not confined to persistent systems, type-safe linguistic reflection is particularly useful in such systems. As well as facilitating the evolution of persistent data, it provides the basis for implementing hyper-program capabilities, as will be described in Chapter 3. Conversely, the provision of a persistent store may enhance the usefulness of reflective techniques by allowing both generic program forms and specialised versions to persist.

### 2.1.1 Behavioural Reflection

With behavioural reflection a program can alter its own meaning. It does this by manipulating its evaluator. One way to achieve this is for an interpreted language to allow access to the internal structures of the interpreter at run-time. Thus a program may change the behaviour of the interpreter as the program is being interpreted. This results in the interpreter performing different actions in the process of interpretation, effectively altering the meaning of the program. Another possible technique is for a program to change its own compiler during its compilation, although no implementations of this are known.

Another mechanism, used in object oriented languages, is to provide a meta-object for every object in the system. A meta-object controls some aspects of the behaviour of its associated object, for example, how it inherits from super-classes, how its methods are invoked or how to make a copy of it. In effect the meta-object is a mini-interpreter for its object. By sending messages that change the way these aspects are handled to its meta-object, an object can indirectly modify its own behaviour.

A number of existing systems that support behavioural reflection are listed below:

**Brown** is a variant of Lisp [MAE+62]. It allows the definition of reflective functions that manipulate interpreter data structures [FW84].

**Meta-Prolog** is a logic based language based on Prolog [Kow79, CM84]. Meta-theories that describe the deduction process itself can be defined [Bow86].

**SOAR** is a rule-based reflective language. The programmer may specify meta-rules that control the operation of the inference system itself [LRN86].

**3-KRS** is an object oriented language. Every object in the language has a meta-object. Meta-objects are themselves objects so they have their own meta-meta-objects and so on. Meta-objects are created lazily. Each meta-object contains information about the implementation and interpretation of its corresponding object. By sending messages to its meta-object an object can alter itself indirectly [Mae87].

**Modulex** is a strongly-typed language based on Modula-2 [Wir83] and supports both object oriented and relational models. All objects have meta-objects. The programmer can control the degree to which the representations of those meta-objects are accessible to other users. Making the representation of a meta-object visible allows users to run queries against it [Ala90].

## 2.1.2 Linguistic Reflection

### 2.1.2.1 Characterisation

In linguistic reflection systems, programs can change themselves directly, in contrast to the indirect manner supported by behavioural reflection. A linguistic reflective program manipulates data structures that represent itself in some way, and any changes are integrated into the current computation. The program may, for example,

- in an interpreted language, alter the data structure that represents the program;
- in a compiled language, alter the compiled code that is being executed; or
- generate new data structures to be interpreted or new code to be executed.

These options offer different trade-offs between flexibility, execution efficiency and assurance of program correctness. For example the first two options give a high degree of flexibility, but without constraints on the nature of allowable updates to the program there is little opportunity for program verification.

The systems described below all support the third method of achieving reflection, where programs generate new code fragments that are incorporated into the programs. There are no known examples of systems supporting the first two methods; some possible reasons for this are described in Section 2.2. The third method of reflection may be used to cut out a level of interpretation and provide more succinct notations. This method also facilitates implementation of checks on the new code before it is executed. Some of the systems described perform type checking on the new code, while others simply attempt to run whatever code is generated.

This informal definition of reflection will be supplemented by a more detailed description later.

### 2.1.2.2 Lisp

Lisp supports linguistic reflection through the *eval* function which allows an S-expression to be constructed and evaluated at run-time. This enables a program to change itself by creating new program fragments and integrating them into the current execution. Figure 2.1 shows an example:

```
(setq code-rep '(defun id-fun (x) x) )
(eval code-rep)
(id-fun 3)

=> 3
```

**Figure 2.1: Example of reflection in Lisp**

The first line of the program defines *code-rep* to be an expression that represents the source code of a function definition, *defun id-fun (x) x*. In the second line this expression is evaluated with the result that the function *id-fun* comes into scope. The function is then applied to the value 3 to give the result, also 3.

Lisp and its variants support both behavioural and linguistic reflection, the former through allowing programs to alter the interpreter, as in Brown, and the latter through the ability to create new expressions that are interpreted. Another Lisp variant, Scheme [RC86], supports reflection at compile-time through macros. The reflection here is of the same form as that of POP-2 macros, described below.

#### 2.1.2.3 POP-2

POP-2 [BCP71] is an untyped language that supports linguistic reflection at both compile-time and run-time. The former is achieved using macros that are executed during compilation to produce lists of lexical items. A macro is executed when its identifier is encountered in the compiler input stream. The macro body may read further lexical items from the input stream. After execution of the body the list of items produced is substituted for the macro identifier in the input stream and compilation resumes. Thus a program containing macro calls may alter itself by creating new program fragments that are integrated with it during compilation.

Figure 2.2 shows the definition of a macro, due to Atkinson [Atk91], that replicates the items in the input stream, up to **end**, a specified number of times.

```
macro repeat;
            vars count, res, rep, next;
            itemread() → count;                      !! get number of repetitions
            nil → rep;                               !! sequence of items to repeat
collecting: itemread() → next;                       !! next item in repeated list
            if next = "end" then goto collected;     !! terminated by end
            rep <> [% next %] → rep;                 !! append next item
            goto collecting;                         !! repeatedly join this to res
collected:  nil → res;                               !! res holds repeated groups
expanding:  if count <= 0 then goto expanded;        !! how many times done?
            rep <> res → res;                        !! join repeated element on
            count - 1 → count;
            goto expanding;
expanded:   macresults( res )
end;
```

**Figure 2.2: Example of compile-time reflection in POP-2**

The macro first declares variables *count*, *res*, *rep* and *next*, reads a value for *count* from the compiler input stream, and initialises *rep* with the empty list. The next four lines perform a loop, reading the compiler input stream and building up a list of lexical items in *rep* until the item **end** is encountered. The syntax *rep <> [% next %]* denotes a list obtained by appending the list *rep* with a list containing the single item *next*. Following the label *expanding* the macro builds up a new list in *res* by appending *rep* to it *count* times. Finally the list *res* is returned as the macro result and the lexical items in it are inserted into the compiler input stream.

The macro *repeat* may then be used as follows:

```
[% repeat, 5, a, b, c, end d %]
```

This program is equivalent to:

```
[% a, b, c, a, b, c, a, b, c, a, b, c, a, b, c, d %]
```

POP-2 also supports reflection at run-time, with the pre-defined function *popval*. This function takes as argument a list of lexical items and treats it as a program representation which is compiled and executed. An example of a call to *popval* is shown below:

```
popval([ x + y => goon ]);
```

This compiles the program representation up to the reserved word **goon** and executes it, resulting in the value of $x + y$ being written out. Here linguistic reflection is achieved by allowing programs to create new program fragments that are integrated into the current execution.

#### 2.1.2.4 TRPL

TRPL [She90] is a statically typed language that supports compile-time linguistic reflection. New code is generated and integrated with the original program during compilation. To do this the programmer writes *context sensitive macros* which are functions that produce code representations. When the compiler encounters a call to a macro it is executed immediately and the resulting code replaces the original call. Within a macro definition the programmer can access the information accumulated by the compiler during compilation up to the macro call. In particular this allows the types of identifiers introduced earlier in the program to be discovered, allowing the action of the macro to depend on the types of its arguments. The code representations manipulated by the macros are in a parsed, abstract syntax form, expressed as TRPL values.

The abstract syntax produced by the execution of a macro is fed back into the compiler and compiled as normal. The new code may in turn contain macro calls, leading to further macro executions as they are compiled. So long as this terminates eventually, at the end of compilation the compiled program contains no reflective language constructs. Since all the new code is processed by the compiler in the normal way it is guaranteed to be type-correct once compilation succeeds. Linguistic reflection is achieved through the ability of a program with macro calls to create new program fragments, during compilation, that are integrated into the original program.

Figure 2.3 shows a TRPL program that defines a macro to expand any integer identifier into an expression that increments it by one:

```
macro INC (x := make_id (?, ?));
env e;
let x_type := type_of (x, e) in      @ get the type of x
case x_type
      {     TYPE (integer)      → EREP (a := a + 1, a := x)
            others              →  x}

variable i : integer := 3;
INC (i)
```

**Figure 2.3: Example of reflection in TRPL**

The macro header in the first line specifies that the macro takes one parameter called *x*. All macro parameters are code representations, i.e., instances of abstract syntax. The header also contains a pattern, *make_id(?, ?)*, that the parameter must match. This ensures that the parameter passed to the macro is the representation of an identifier. The components of the *make_id* construct are not important here so the 'don't care' symbol *?* is used. If the macro is passed a code representation that does not match the pattern an error is reported and compilation fails.

On the second line the variable *e* is bound to the compilation environment at the point of the macro call. It is then used in the following line in the call to the pre-defined function *type_of*. This returns a representation of the type of the given code expression in that environment. It is the ability to discover this information that allows the output of a macro to depend on the types of its parameters. In this example the type representation is used to check that the identifier passed to the macro is of type *integer*. If so the result of the macro is an expression that increments the variable by 1. The pre-defined macro *EREP* is used to avoid having to write down a messy abstract syntax expression. *EREP* produces a parsed version of the expression passed to it, with optional substitutions. If the identifier is not of the correct type the *others* branch of the case clause is executed and the result of the macro is the input expression unchanged.

After the macro definition a variable *i* is declared, followed by a call to the macro. When the program is compiled the resulting executable code will be equivalent to that produced by compiling the following program:

```
variable i : integer := 3;
i := i + 1
```

The checking of the type of the parameter in this example was shown to illustrate how macros have access to type information. Even if no checking was performed, incorrect uses of the macro, such as passing it a string identifier, would be detected at compilation time by the type-checking of the macro's output.

The reflective facilities in TRPL do not enlarge the class of programs that can be written. For any TRPL program with macro calls there is also an equivalent non-reflective program that contains the code produced by the macros. The power of the reflection is that it allows the programmer to write highly generic, yet statically type-checked, functions that vary their behaviour depending on the types of their parameters. This can give major savings in the total amount of code that the programmer writes [SSF92].

#### 2.1.2.5 PS-algol

In contrast to the compile-time reflection of TRPL, reflection in PS-algol [PS88] occurs at run-time, by allowing running programs to access the PS-algol compiler which is a procedure in the persistent store. Linguistic reflection is achieved through programs altering themselves by creating new program fragments, at run-time, which are compiled and integrated into the current execution.

The compiler operates on string representations of programs rather than abstract syntax forms. The programmer passes a representation of the expected type of the result to the compiler along with a string to be compiled. To compile successfully the string must represent a procedure value. The compiler returns an instance of type *pntr* which is the infinite union of all structure types. If the compilation was successful and the type of the result matched that specified, the *pntr* is a structure containing the result. Otherwise an error is reported and the *pntr* is a null value. In this scheme the subsequent use of the compiled result can be type-checked statically. A dynamic check is required to determine whether the compilation succeeded. Figure 2.4 shows an example of run-time reflection in PS-algol:

```
let compilerDb = open.database( "compiler", "friend", "read" )
let compileHolder = s.lookup( "compile", compilerDb )

structure compilerPack( proc( string, pntr → pntr ) compileFn )
let compile = compileHolder( compileFn )

let mkFun = proc( → string )
begin
      write "enter real expression over x"
      let expr = reads()

      "proc( real x → real ) ; " ++ expr
end

let newFunCode = mkFun()

structure procHolder( proc( real → real ) fun )
let dummyProcHolder = procHolder( proc( real a → real ) ; 0.0 )
let resultHolder = compile( newFunCode, dummyProcHolder )

if resultHolder is procHolder then
begin
      let newFun = resultHolder( fun )
      let res = newFun( 1.3 )
end
else writeString "compilation failed"
```

**Figure 2.4: Example of reflection in PS-algol**

The program begins by linking to the compiler procedure in the persistent store. It calls the pre-defined procedure *open.database* to access one of the roots of persistence, an associative table. The parameters specify the name of the table, its password and ‘read only’ mode. Line 2 uses another pre-defined procedure, *s.lookup*, to look up the pointer indexed by the key "compile". Next, the form of the structure that the pointer is expected to point to, *compilerPack*, is defined and the pointer dereferenced to give the compiler procedure. Following this is the definition and call of the procedure *mkFun* which takes no parameters and returns a string result. When called it generates the string representation of a new procedure, a function that maps reals to reals. The body of the function is obtained, in string form, from the user via a call to the pre-defined input procedure *reads*. There is no guarantee at this stage that the result of *mkFun*, *newFunCode*, is a valid representation of a procedure. The program then defines a new structure, *procHolder*, with one field of the expected type of the result of compilation, which is a procedure that takes a real and produces another as its result. A pointer, *dummyProcHolder*, to an instance of the structure containing a dummy procedure value is also defined. This pointer is passed to the compiler procedure along with the new code representation. The next line tests whether the result produced is a pointer to an instance of *procHolder*. If so the compilation has succeeded and the structure is dereferenced and the new function called. Otherwise an error is reported.

#### 2.1.2.6 Napier88

Napier88 supports run-time linguistic reflection of a similar style to that of PS-algol. Linguistic reflection is achieved through creating new program fragments in string form to be compiled at run-time, and incorporating the resulting values into the running program. The main difference is in the interface to the compiler. Napier88 allows code representing values of any type to be compiled, rather than only procedures. The result returned from the compiler is an instance of the infinite union type *any*. If any errors have occurred during compilation the *any* contains a string describing them, otherwise the *any* contains a procedure that will execute the compiled code when called. Figure 2.5 shows the same example in Napier88:

```
use PS() with compilerEnv, IO : env in
use compilerEnv with compile : proc( string → any ) in
use IO with  readString : proc( → string );
             writeString : proc( string ) in
begin
     let mkFun = proc( → string )
     begin
          writeString( "enter real expression over x" )
          let expr = readString()

          "proc( x : real → real ) ; " ++ expr
     end

     let newFunCode = mkFun()
     let compiledCode = compile( newFunCode )

     project compiledCode as result onto
     proc( → proc( real → real ) )
     begin
          let newFun = result()
          let res = newFun( 1.3 )
     end
     string : writeString( "compilation failed because " ++ result )
     default : {}
end
```

**Figure 2.5: Example of reflection in Napier88**

The program starts by linking to the environments *compilerEnv* and *IO* in the persistent root environment obtained by calling the pre-defined procedure *PS*. Environments are extensible collections of bindings, used to structure the persistent store and support incremental evolution [Dea89]. The program then links to the compiler procedure and procedures to read and write strings. The expected types of the procedures in the persistent store are declared. The program then defines and calls the generator procedure as before and the resulting string is passed to the compiler procedure. A projection clause matches the type of the value in the *any* returned. If compilation succeeds the value is a procedure that will execute the compiled code and return the result of type *proc( real → real )*. This is called in the corresponding branch. In the case of compilation failure the *any* is projected onto a string describing the error.  The default branch is required by Napier88 rules but should never be executed in this program as one of the first two branches should always match.

Current work on the Napier88 compilation system is investigating further refinements to the compiler interface [Cut92].  These involve a separation of the processes of syntax checking, code generation and linking existing values into compiled code.  This could allow a program to be compiled in the context of an existing environment, or to be partially compiled and then have different copies linked into different environments.

## 2.2 Type-Safe Linguistic Reflection

In general linguistic reflection involves a program modifying some form of itself in a way that affects its continued execution.  The previous section described languages in which this modification takes place during the compilation of the program (Scheme, POP-2, TRPL), and languages in which it takes place during execution (Lisp, POP-2, PS-algol, Napier88).  TRPL, PS-algol and Napier88 are strongly typed, that is, all computations are checked for type correctness before they are executed.  Some restrictions on the forms of linguistic reflection allowable are necessary if strong typing is to be enforced efficiently.

Program compilation involves the translation of a source program into another form which is interpreted at execution time.  There is a trade-off between compiler and interpreter complexity.  Where a compiler performs a large amount of processing the compiled form may be low-level, requiring only a simple interpreter.  At one end of the spectrum the compiled form is native machine code, and interpretation reduces to direct execution by the CPU.  Where the compiler is simple, however, a higher-level compiled form is produced and an interpreter of greater complexity is required.  Thus at the other extreme the compilation phase is omitted and the source code is interpreted directly.

Where reflection takes place at compile-time a program contains distinguished sections that are interpreted during the compilation process and affect the resulting compiled form:

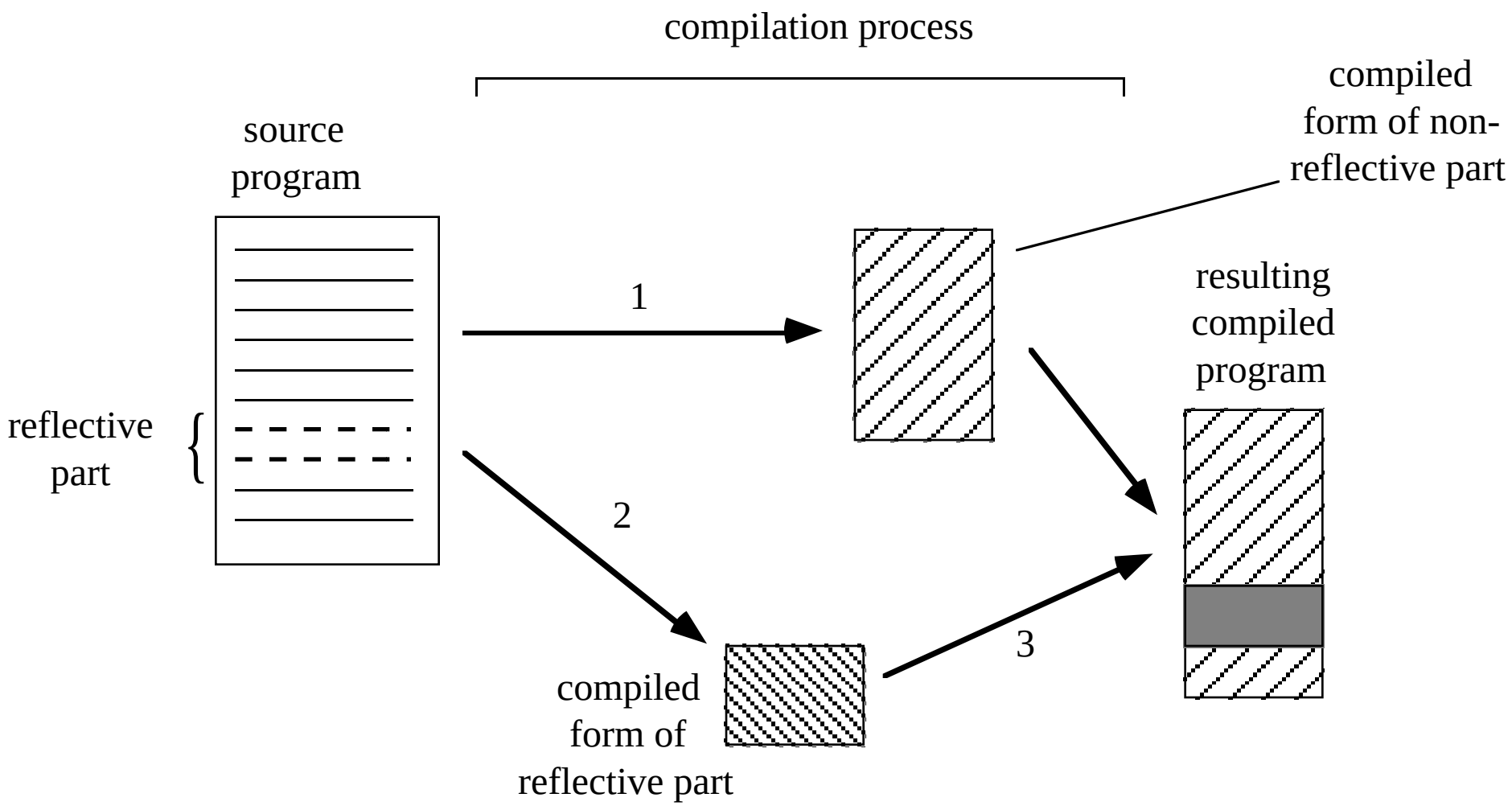

1 source program translated into compiled form
2 reflective part of source program translated into compiled form
3 compiled form of reflective part interpreted, affecting final compiled form

**Figure 2.6: Compile-time reflection**

With reflection at run-time, the action of interpreting the reflective parts of the compiled form causes parts of the compiled form itself to be modified as shown in Figure 2.7:

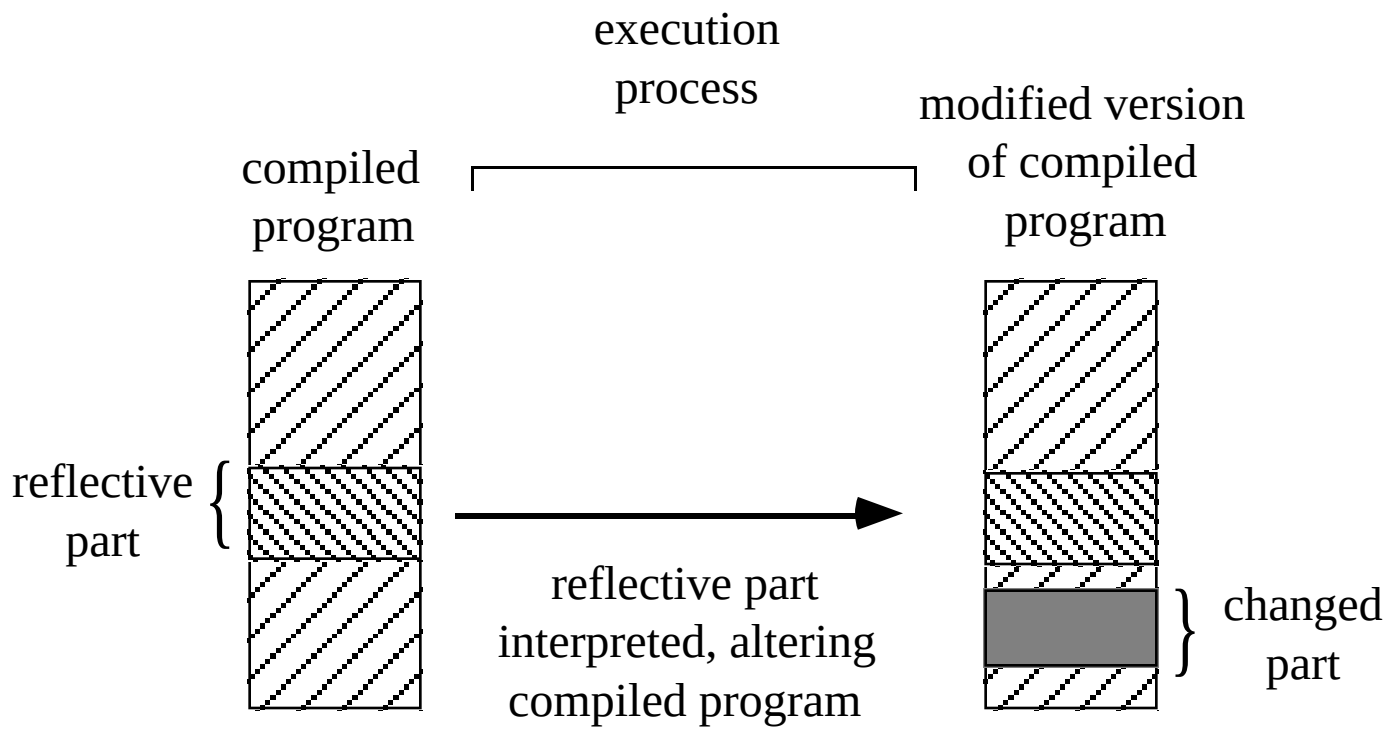

**Figure 2.7: Run-time reflection**

In strongly typed languages every action performed by a program is type-checked before it takes place, in order to enforce the modelling and protection roles of the type system. To attain type-safe linguistic reflection, the subject of the remainder of this chapter, all modifications made by a reflective program to itself must be checked for type correctness before the modified sections are executed. In the general case a modification to a program involves both the removal or overwriting of some existing code, and the addition of some new code. Strong typing thus requires checking that the program remains type correct after the removals and additions.

A simple way to achieve this is to impose the following two restrictions on the forms of reflection:

- that modifications be allowed to source programs only; and
- that the only modifications allowed be additions.

The first restriction allows the type checking to be performed using existing type-checking technology which operates on source programs or parse trees. The second restriction, to incremental modifications, means that only the newly added code needs to be checked, rather than the entire modified program.

Future research may address the issues of validating program modifications that do not adhere to these restrictions. In the first case this would involve either type checking low-level compiled forms, from which much type information had been ‘compiled away’, or constraining the modifications to those that preserved type correctness. In the second case it would involve re-checking the entire program or verifying that the removal of a fragment of code from a correct program left the program still correct.

With type-safe compile-time reflection, as in TRPL, newly generated sections of source program are incorporated into the source program being compiled and they are type-checked in the normal course of compilation. This is illustrated in Figure 2.8:

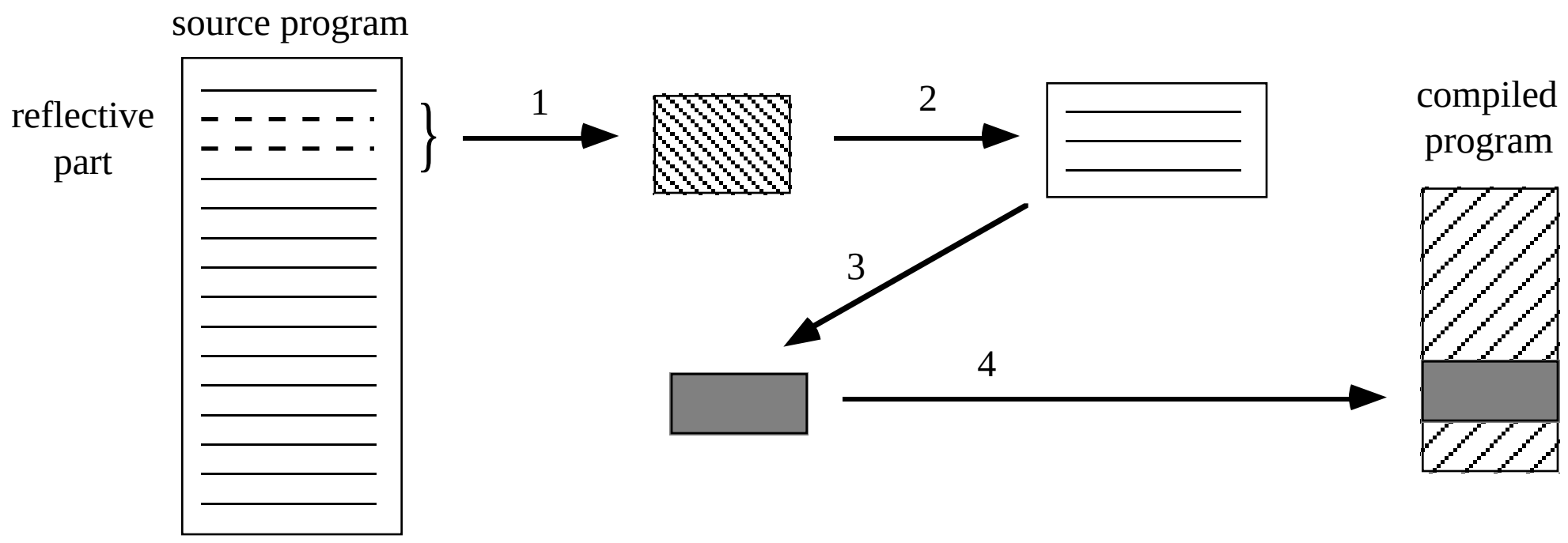

compile time
- 1 reflective part translated into compiled form
- 2 compiled form of reflective part interpreted to generate new source code
- 3 new source code translated into compiled form
- 4 compiled form integrated with rest of compiled program

**Figure 2.8: Compile-time type safe linguistic reflection**

With type-safe run-time reflection, as in PS-algol and Napier88, the newly generated sections are compiled and the resulting compiled forms incorporated into the compiled form being interpreted. This is illustrated in Figure 2.9. This style of reflection requires a dynamic linking mechanism: in PS-algol and Napier88 it involves dereference of pointers and projection of *any*s respectively. As the new sections are compiled in isolation from the original source program they cannot refer to identifiers in scope in the original program. It will be described later how this restriction may be relaxed to allow the incrementally added new code to communicate with the existing code.

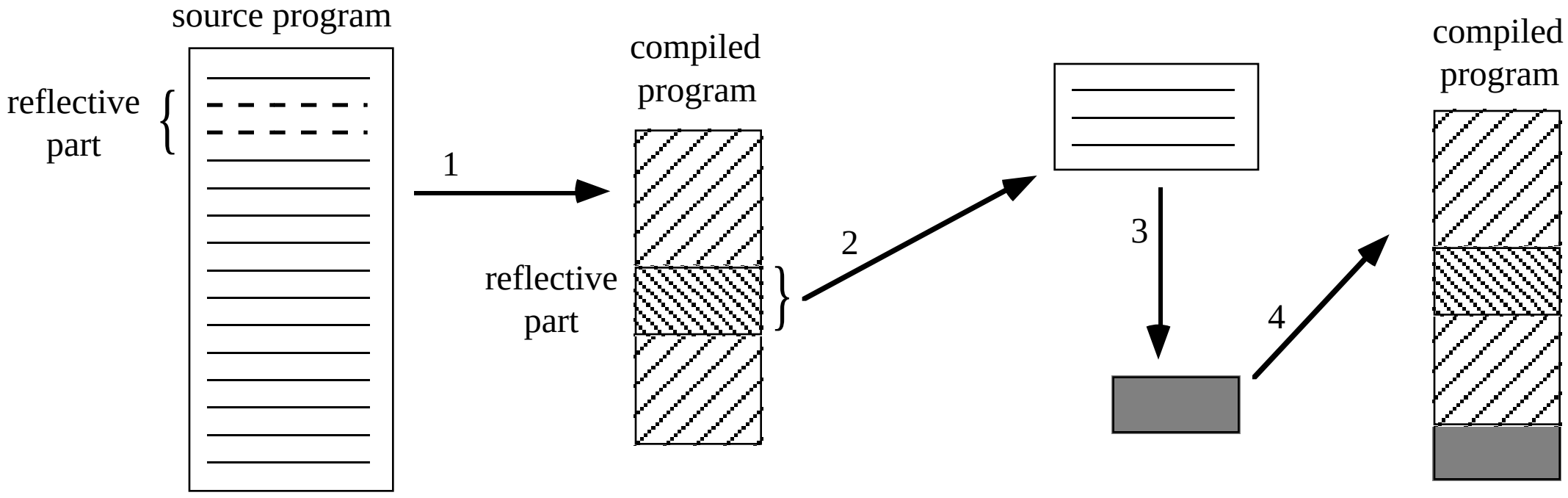

compile-time { 1 source program translated into compiled form

run-time
- 2 compiled form of reflective part interpreted to generate new source code
- 3 new source code translated into compiled form
- 4 compiled form incorporated with original compiled program

**Figure 2.9: Run-time type-safe linguistic reflection**

## 2.3 Anatomy of Type-Safe Linguistic Reflection

### 2.3.1 Reflection in General

This section will attempt to define the nature of type-safe linguistic reflection more precisely. Given a language, *L*, and a domain of values, *Val*, the nature of execution of a program in *L* will be discussed. The function *eval* is the evaluation function:

$$eval : L \rightarrow Val$$

The domain of values, *Val*, differs for different languages. Examples of *Val* include numbers, character strings, final machine states, the state of a persistent object store, and the set of bindings of variables produced by the end of a program's execution.

For linguistic reflection to occur, there must be a subset of *Val*, called $Val_L$, that can be mapped into $L$. For example, $Val_L$ could be the set of character strings containing syntactically correct $L$ expressions. Since $Val_L$ is a subset of *Val* that may be translated into the language $L$ it may be thought of as a representation of $L$.

A subset of $L$ consisting of those language constructs that cause reflective computation is denoted by $L_R$. $L_R$ is called the reflective sub-language and $Val_{L_R}$ stands for its representation. An evaluation of an expression in $L_R$ invokes a generator. In linguistic reflection the generators, the programs that produce other programs, are written in a subset of the language $L$ which will be denoted by $L_{Gen}$. $L_{Gen}$ may include all of $L$ but the programs written in $L_{Gen}$ must produce results in $Val_L$. The major relationships among the language and value sets are:

$$L_R \subset L$$
$$Val_{L_R} \subset Val_L \subseteq Val$$
$$L_{Gen} \subseteq L$$

The significance of the proper subset relationships is explained later. Two functions are required for a full description of linguistic reflection. The first, *drop*, takes a construct in $L_R$ and produces a generator in $L_{Gen}$. The second, *raise*, takes a value in $Val_L$ and produces an expression in $L$:

$$drop : L_R \rightarrow L_{Gen}$$

$$raise : Val_L \rightarrow L$$

Linguistic reflection is defined as the occurrence of the following pattern of computation, within the *eval* function, in the evaluation of a program in $L$:

```
procedure eval( e : L ) → Val      ! This types e as L and eval as L → Val.
      case
      ...
      inL_R (e) => eval( raise( eval( drop( e ) ) ) )
      ...
```

**Figure 2.10: The linguistic reflective nature of *eval***

where the ellipses cover the evaluations of non-reflective constructs. The construct *eval( raise( eval( drop( e) ) ) )* represents the intuition that during the evaluation of a reflective expression the result of the evaluation is itself evaluated as an expression in the language.

The expression produced by *drop* is a generator that is evaluated by the inner *eval*. The type of a generator $g$ in $L_{Gen}$ is:

$g : Val \rightarrow Val_L$

The generator takes some arguments and returns as its result an expression in $Val_L$ which is then translated into *L* by *raise*. The result is finally evaluated by the outer *eval*. This is illustrated in Figure 2.11. The whole diagram represents the *eval* function, as does the box containing *eval* within the diagram. This nested structure is a consequence of representing the recursive function *eval* by a flow diagram and will be a feature of other diagrams.

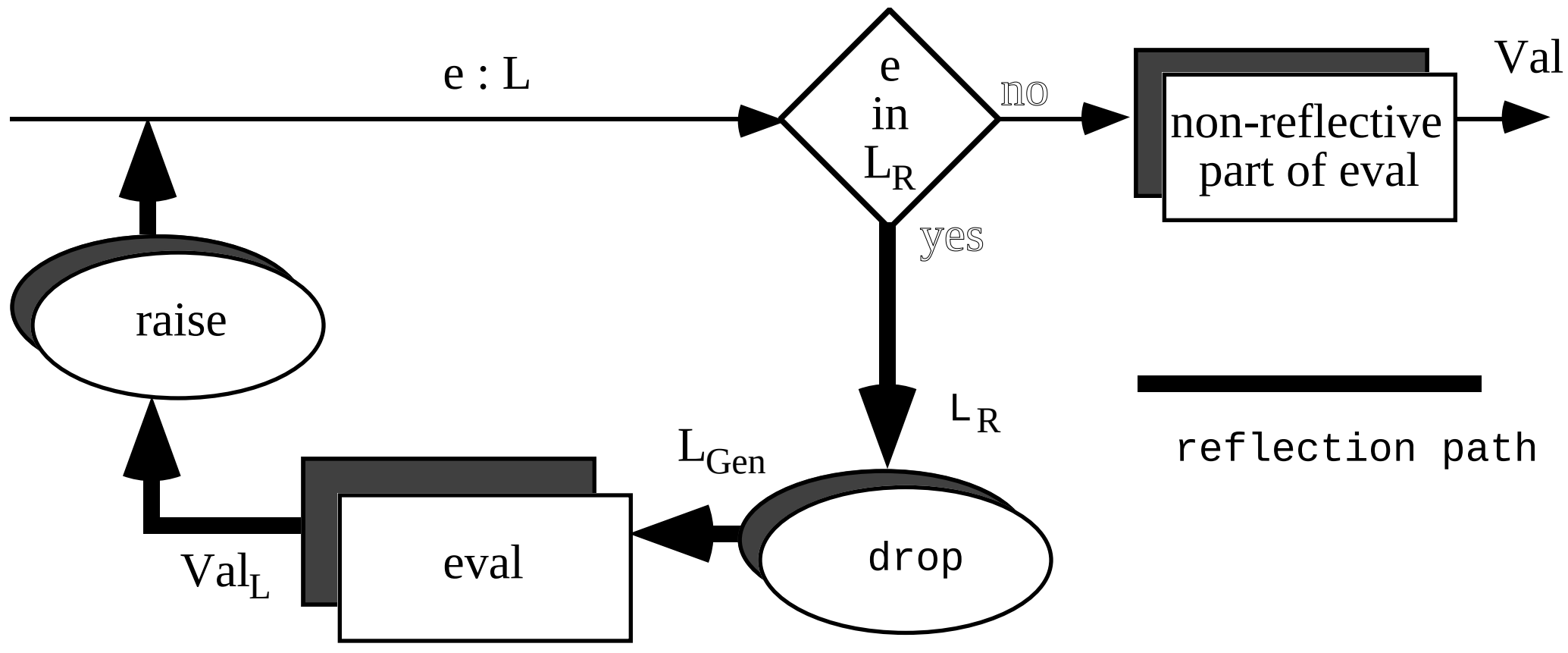

**Figure 2.11: *eval* in linguistic reflection**

In order to make these concepts more concrete an example language is introduced. In this language:

- $Val_L$ is the set of character strings that represent sentences in *L*;
- $L_R$ contains the single verb **execute** which initiates reflection;
- *drop* presents a string expression to the inner evaluation function;
- *raise* maps a string representation to the corresponding sentence in *L*;
- $L_{Gen}$ is the set of expressions in *L* that result in character strings in $Val_L$.

In this example language, *drop* and *raise* are defined by:

drop( **execute**( stringExpression ) ) = stringExpression

raise( "expression" ) = expression

Figure 2.12 gives an example of linguistic reflection in this language. The symbol ++ denotes string concatenation. The inner *eval* concatenates the strings to produce the string *"2+3"*, while the outer *eval* evaluates the expression ***execute**( "2+" ++ "3" )* by the following sequence:

```
eval( execute( "2+" ++ "3" ) )

      ! the reflection is recognised
=> eval( raise( eval( drop( execute( "2+" ++ "3" ) ) ) ) )
=> eval( raise( eval( "2+" ++ "3" ) ) )
=> eval( raise( "2+3" ) )
=> eval( 2+3 )
=> 5
```

**Figure 2.12: An example of linguistic reflection**

The above example shows that some reflective expressions may be evaluated statically, at compile-time, since here all the information to perform the inner *eval* may be found statically. Uses of this style of reflection are described later. This is not always the case however and some reflective computation may have to be delayed until run-time. Figure 2.13 shows such a computation in which run-time input is solicited by the *readString* procedure. The inner *eval* is thus constrained to execute at run-time.

```
execute( "2+" ++ readString() )
```

**Figure 2.13: Example of run-time linguistic reflection**

A reflective computation is well formed if it terminates and the output of each inner *eval* is syntactically correct and typed correctly. Termination requires that the inner *eval* must eventually result in a value in $Val_{L\text{-}L_R}$, the set of values that represent non-reflective program constructs. Syntactic correctness requires that the result of *eval( drop( e ) )* is in $Val_L$ for all reflective expressions. A generated expression must be internally type consistent as well as typed correctly for its context.

In general, type correctness must be checked for each individual generated expression. Type checking generators for the types of all their possible outputs is a topic for further research, and is undecidable in general.

## 2.3.2 Compilation

This section is concerned with the mechanisms for linguistic reflection in compiled languages and the anatomy given so far must be further refined to describe these. Figure 2.14 shows the structure of *eval* as a composition of two functions: *compile* and *eval'*. The function *compile* takes an expression in language *L* and produces another in a target language *L'*. The function *eval'* is the evaluation function for *L'*. The types of the functions are defined by:

$$compile : L \rightarrow L'$$

$$eval' : L' \rightarrow Val$$

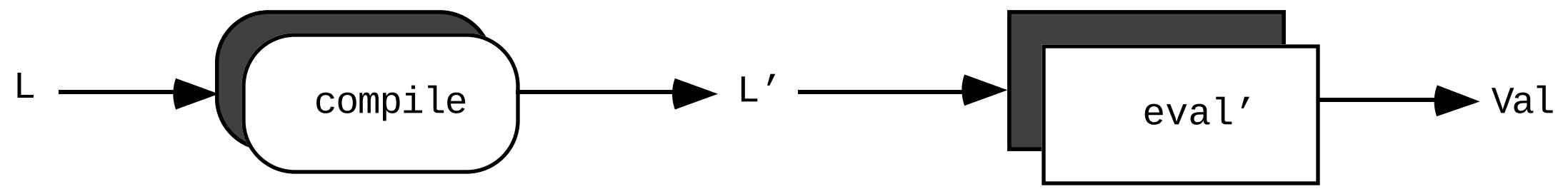

**Figure 2.14: *eval* as function composition**

### 2.3.3 Compile-Time Linguistic Reflection

One way in which linguistic reflection can be accomplished in a compilation environment is for reflective constructs to be compiled and executed during the compilation of a program containing them. This is limited to cases where the reflection is over compile-time information, that is static, and cannot be used for reflection that depends on values that are only available at run-time.

In such a system, generators are used to express computations over the syntactic elements of a program. As in any form of linguistic reflection, the computations are expressed in the subset $L_{Gen}$ of the language $L$. The reflective sub-language $L_R$ contains the calls to the generators. That is, the pattern of evaluation that defines $L_R$ is only initiated by these reflective calls. A possible *drop* function in this architecture is a function that takes a reflective call, finds its generator definition and uses the definition and the call arguments to form a call to the generator. The inner *eval* executes the call at compile-time to produce a new expression in $Val_L$. This in turn is transformed to an *L* expression by *raise* and presented to the outer *eval*. Figure 2.15 illustrates this model. Such a pattern of reflection is called *compile-time linguistic reflection* since the reflection is performed at compile-time even though the evaluator, *eval'*, is called. The type checking of the generated expressions is performed by the compiler. The pattern of *eval* is given by:

```
procedure eval( e : L ) → Val      ! This types e as L and eval as L → Val.
      eval'( compile( e ) )

procedure compile( e : L ) → L'
      if inL_R(e)
      then compile( raise( eval'( compile( drop( e ) ) ) ) )
      else translate( e )
```

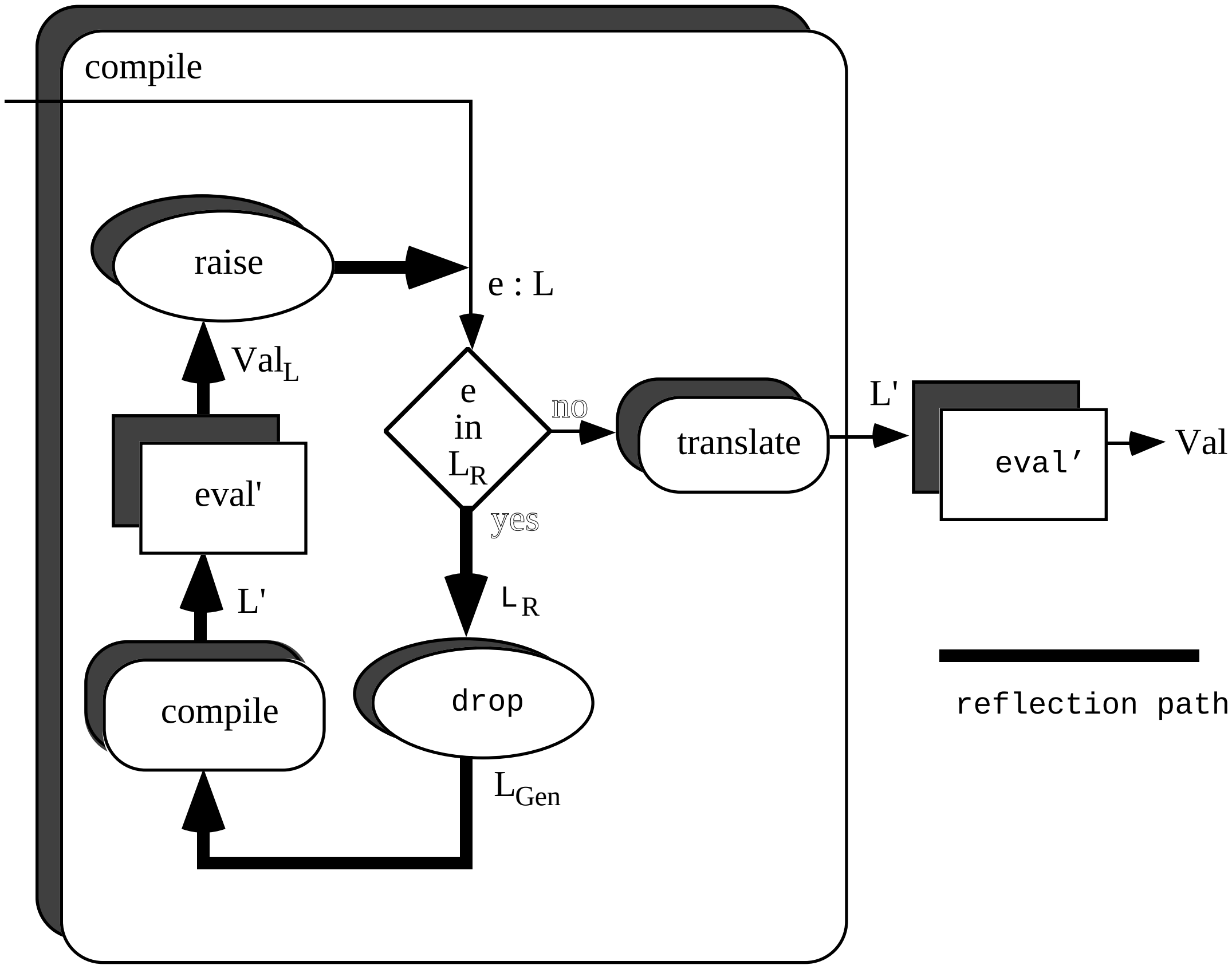

**Figure 2.15: *eval* in compile-time linguistic reflection**

The macro facilities in POP-2 and Lisp contain this style of compile-time reflection without the type checking.

### 2.3.4 Optimised Compile-Time Linguistic Reflection

An optimised variant of the previous architecture can be produced by having the parser generate abstract syntax as values in *Val*. This choice of $Val_L$ allows the result of the inner *eval* to be passed directly to the post-parse compiler called *postParseCompile*. The *raise* function reduces to the identity function in this optimisation. The *drop* function here produces a compiled version of the generator in the target language generator subset $L_{Gen}$. This optimised *drop* function is denoted by $drop_{Opt}$. The structure of *eval* in this case is shown in Figure 2.16. Here $e_v$ denotes the parsed form of *e* expressed in $Val_L$ and $L_{R_v}$ the parsed forms of $L_R$. The pattern of *eval* is given by:

```
procedure parse( e : L ) → Val_L
      ...

procedure eval( e : L ) → Val
      eval'( compile( e ) )

procedure compile( e : L ) → L'
      postParseCompile( parse( e ) )
```

**procedure** postParseCompile( $e_v$ : $Val_L$ ) $\rightarrow$ L'
  **if** $inL_{R_v}$( $e_v$ )
  **then** postParseCompile( $raise_{Opt}$( eval'( $drop_{Opt}$( $e_v$) ) ) )
  **else** translate( $e_v$ )

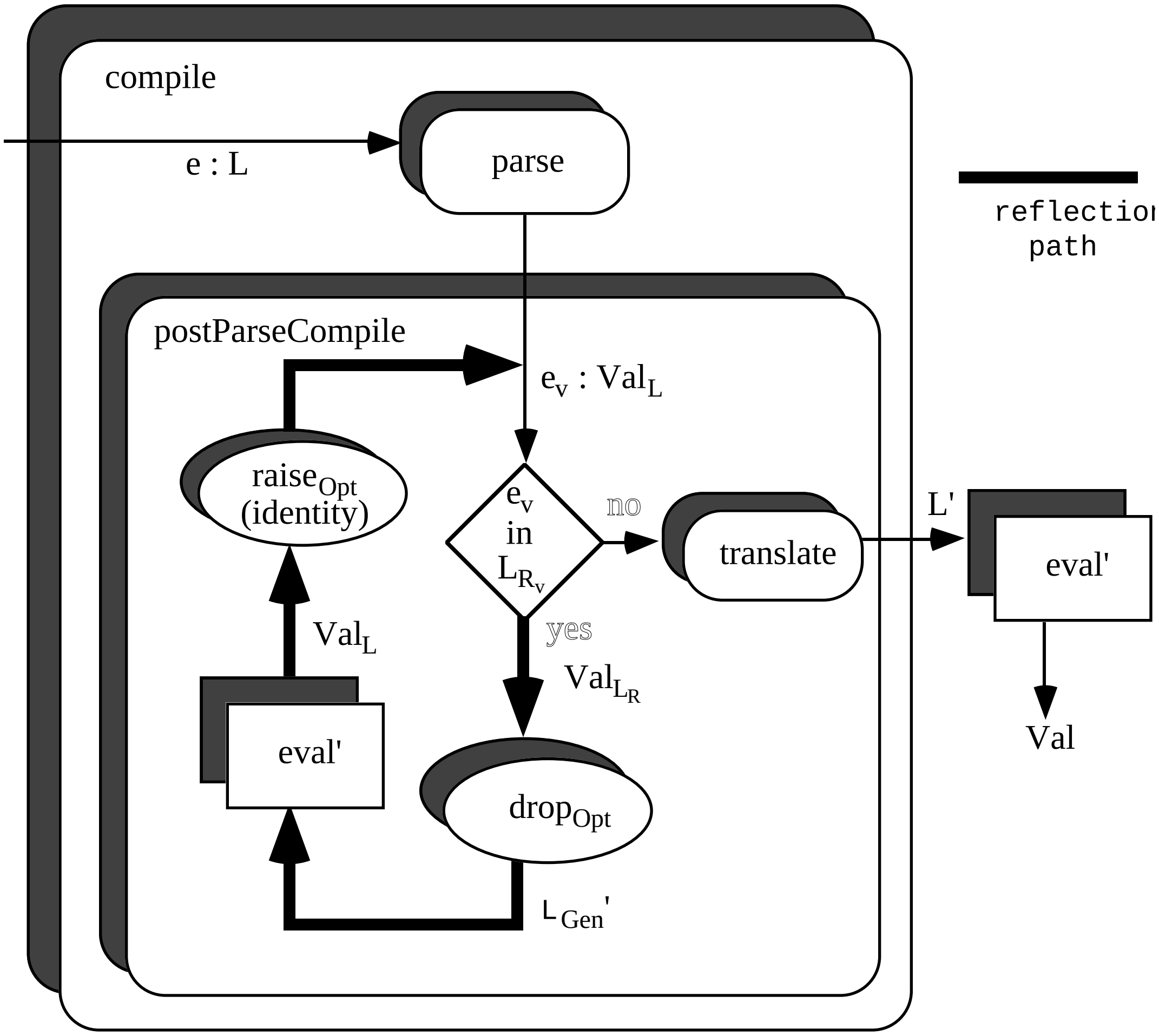

**Figure 2.16: *eval* in optimised compile-time linguistic reflection**

The two versions of the same function *eval'* in both Figure 2.15 and Figure 2.16 highlight some implementation choices. Although both *eval'* functions are semantically the same, they may be implemented differently. For example, the *eval'* within the compiler could be an interpreter function and the right hand *eval'* could be the machine executing machine code. The details of these implementations are not germane to this description.

An example of such an architecture is the implementation of TRPL. The TRPL reflective constructs are TRPL context sensitive macro calls, the elements of $L_R$. The $drop_{Opt}$ function takes the parsed arguments of a macro call and passes them to the macro definitions which have been compiled into target language functions (generators) ready for *eval'*. Thus a call of the compiler is avoided in the reflective *eval*. The result of executing the compiled macro definitions is to produce new TRPL code expressed in the parsed form $Val_L$. This code can contain new function, type and even macro definitions. This new code is presented to the post-parse compiler for compilation and evaluated using *eval'*. Type checking is performed after each inner *eval'*.

Figure 2.17 gives an example of optimised compile-time reflection as it occurs in TRPL. $e_v$ denotes the $Val_L$ form of *e*, while $\underline{e}$ denotes its compiled form.

eval( **execute**( "2+" ++ "3" ) )
=> eval'( compile( **execute**( "2+" ++ "3" ) ) )
=> eval'( postParseCompile( parse( **execute**( "2+" ++ "3" ) ) ) )
=> eval'( postParseCompile( (**execute**( "2+" ++ "3" ))$_v$ ) )

! the reflection is recognised
=> eval'( postParseCompile( raise$_{Opt}$( eval'( drop$_{Opt}$( (**execute**( "2+" ++ "3" ))$_v$ ) ) ) ) )

! *drop$_{Opt}$* produces *"2+" ++ "3"$_{Gen}$* which denotes compiled generator of *(2+3)$_v$*
=> eval'( postParseCompile( raise$_{Opt}$( eval'( "2+" ++ "3"$_{Gen}$ ) ) ) )
=> eval'( postParseCompile( raise$_{Opt}$( (2+3)$_v$ ) ) )

! *raise$_{Opt}$* is the identity function
=> eval'( postParseCompile( (2+3)$_v$ ) )
=> eval'(2+3)
=> 5

**Figure 2.17: Optimised compile-time linguistic reflection in TRPL**

The original expression, ***execute**( "2+" ++ "3" )*, is parsed and then examined by the post-parse compiler which recognises that it is a parsed form of a reflective construct. A generator previously compiled into its *L'* form from a definition of ***execute*** is produced by *drop$_{Opt}$* using the parsed form of ***execute***'s input. This generator, *"2+" ++ "3"$_{Gen}$*, evaluates to the *$Val_L$* form of *2+3*. The inner *eval'* executes the generator and the parsed form *(2+3)$_v$* is produced. This is passed to the post-parse compiler, which completes its compilation. It is eventually evaluated in its compiled form by *eval'* as the completion of the original *eval*.

## 2.3.5 Run-Time Linguistic Reflection

Where reflection occurs at run-time the expression in $L_R$, which causes the reflection, has already been compiled. That is, it is the *eval'* function that recognises the expression in $L_R'$, the compiled form of $L_R$, to initiate reflection. The original expression *e* is in the process of being evaluated by:

eval( e )
=> eval'( compile( e ) )
=> eval'( e ) ! where e is the compiled form of *e*

The pattern of *eval'* in this case is shown by

**procedure** eval'( e : L' ) → Val ! This types e as *L'* and *eval'* as *L'* → *Val*.
  **if** inL$_R$'( e )
  **then** eval( raise$_{Run}$( eval'( drop$_{Run}$( e ) ) ) )
  **else** eval''( e )

Notice that the outer evaluation function is *eval* whereas the inner one is *eval'*. The outer *eval* encompasses the compiler since it expands to *eval'( compile( … ) )*. The *drop$_{Run}$* function has the type $L_R' \rightarrow L_{Gen}'$. This is illustrated in Figure 2.18, where *eval''* denotes the non-reflective part of *eval*.

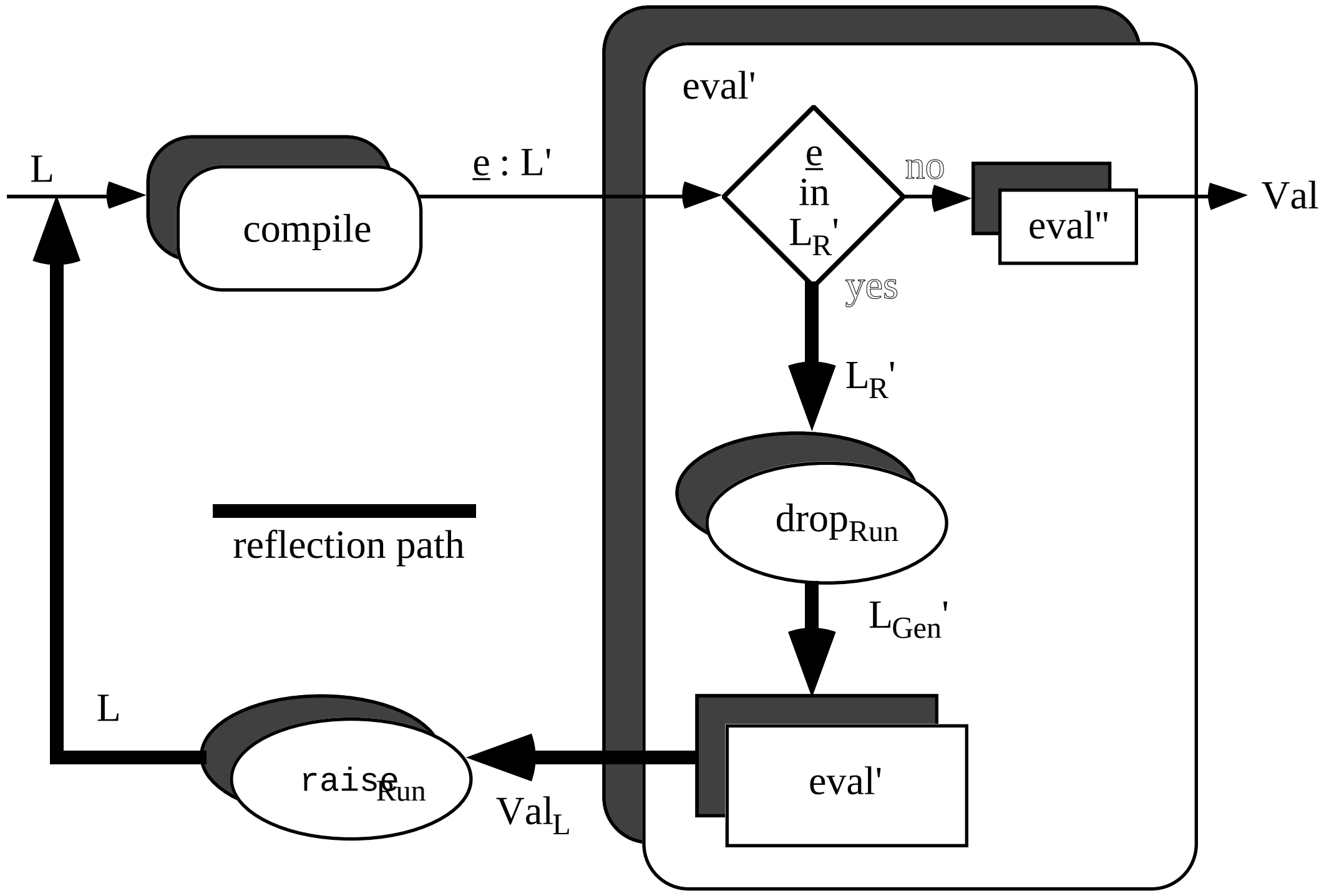

**Figure 2.18: *eval* in run-time linguistic reflection**

An example of this form of reflection is the use of a run-time callable compiler together with the ability to bind and execute newly compiled program fragments within the running program. PS-algol and Napier88 with their callable compilers and incremental loaders are examples of languages that provide run-time linguistic reflection. The function *eval* in Lisp and the function *popval* in POP-2 are early examples of untyped run-time reflection.

Figure 2.19 gives an example of run-time reflection as it occurs in Napier88.

eval( **execute**( "2+" ++ readString() ) )
=> eval'( compile( **execute**( "2+" ++ readString() ) ) )
=> eval'( **execute**( "2+" ++ readString() ) )

! now the reflection is recognised
=> eval( raise$_{Run}$( eval'( drop$_{Run}$( **execute**( "2+" ++ readString() ) ) ) ) )
=> eval( raise$_{Run}$( eval'( "2+" ++ readString() ) ) )

! if "3" is input for the call of *readString*
=> eval( raise$_{Run}$( "2+3" ) )

! applying *raise$_{Run}$* and expanding *eval*
=> eval'( compile( 2+3 ) )
=> eval'( 2+3 )
=> 5

**Figure 2.19: Run-time linguistic reflection in Napier88**

The original expression is first compiled and is in the process of being evaluated by *eval'* when the reflection is discovered. The compiled form ***execute( "2+" ++ readString() )*** is presented to *drop$_{Run}$* which removes the **execute** verb. The inner *eval'* reads in the string and concatenates it with *"2+"*. If the string read in is *"3"* then the result of the concatenation is *"2+3"*. This expression is in $Val_L$ and is transformed into $L$ by $raise_{Run}$. Finally the expression *2+3* is compiled and evaluated by *compile* and *eval'*.

### 2.3.6 Combined Compile-Time and Run-Time Linguistic Reflection

The integration of compile-time and run-time linguistic reflection is a topic for further research. Figure 2.20 shows a possible structure for *eval* in a combined system:

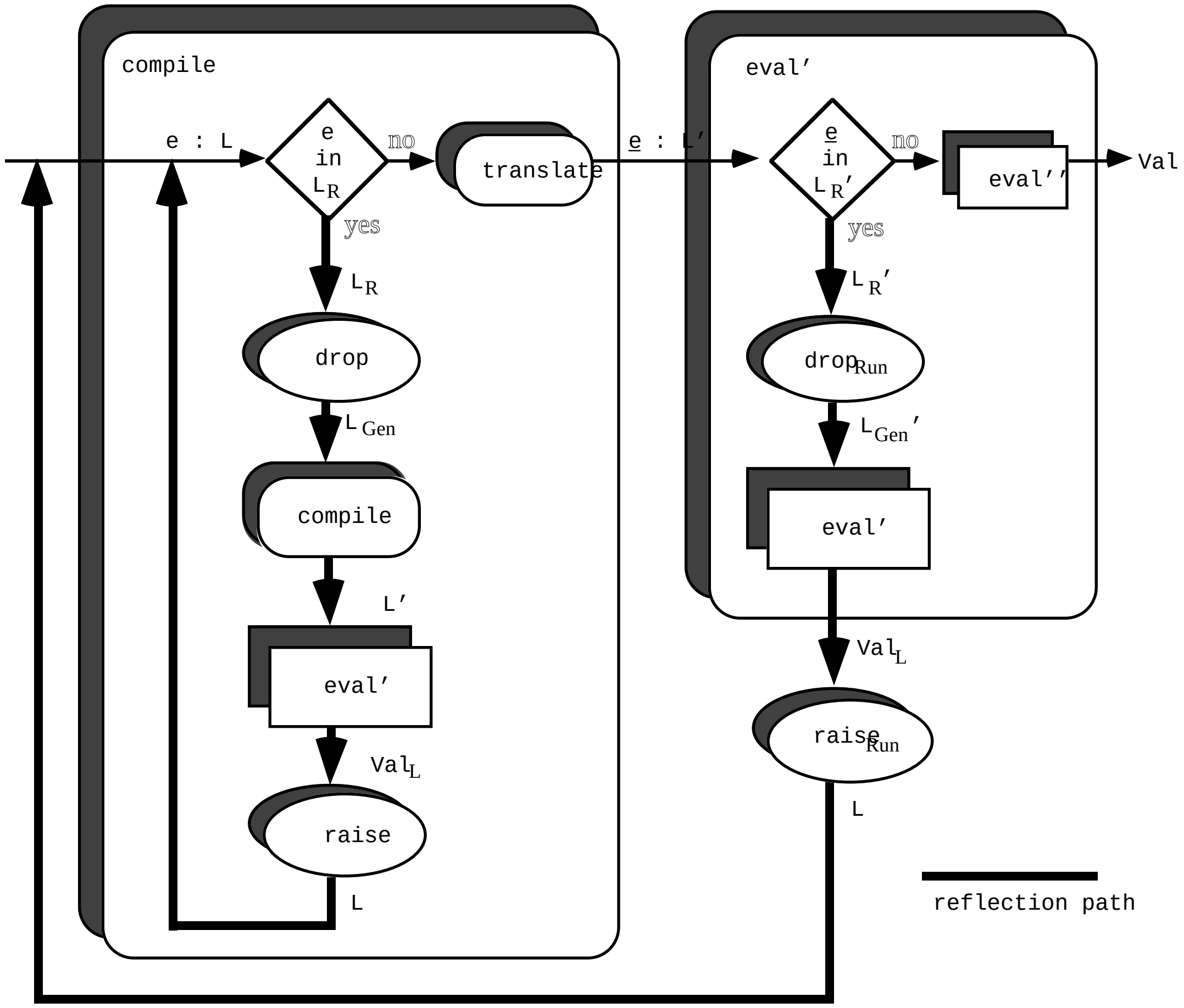

**Figure 2.20: *eval* in combined compile-time and run-time linguistic reflection**

## 2.4 Dimensions of Type-Safe Linguistic Reflection

The previous sections have described some particular mechanisms for type-safe linguistic reflection. The reflective process has a number of dimensions, choices for each of which must be made when designing a reflective system. These are:

- the way that the reflection process is initiated;
- the nature of the generators;
- the time at which the generators are executed;
- the environment in which the generators are executed;
- the time at which generated code is type checked;
- the environment in which the generated code is executed; and
- the way that generated code is linked into the original program.

### 2.4.1 Initiation of Reflection

Linguistic reflection is initiated by an expression in the reflective sub-language, $L_R$, being evaluated. Some examples of the form of $L_R$ expressions are:

- a simple verb applied to a program representation, such as ***execute*** *"3+4"*;
- a macro call, such as *INC (i)*;
- a call to a compiler function, such as *compile( newFunCode )*.

### 2.4.2 Time of Generator Execution

In existing implementations the generators are executed at fixed points in the evaluation process, either during program compilation or at run-time. A goal of current research is to unify compile-time and run-time reflective technology in a single system, providing a consistent notation for specifying both styles of reflection. One proposal for such a notation, due to Connor [Con91], suggests the provision of two $L_R$ constructs. Both perform the same function but they differ temporally and therefore in the environments in which they operate. They are:

**force** which forces the reflective evaluation on the first encounter and replaces the *force* construct with the generated result. It therefore performs the inner *eval* and the *drop*.

**delay** which delays reflective evaluation. That is the inner *eval* and *drop* are not performed until the program is executing after the initial compilation phase.

Compile-time linguistic reflection uses *force* implicitly whereas run-time linguistic reflection uses *delay*. The two constructs could also be used in combination to give finer control over the time at which reflection occurs.

### 2.4.3 Nature of Generators

Since the generators are all written in the language subset $L_{Gen}$, it is the nature of the language forms that they manipulate that distinguishes different linguistic reflective languages. The generators compute over and produce expressions in $Val_L$. In some systems this is simply the set of strings.

Where some processing of the expressions has already taken place, there is a possibility of using more structured forms for $Val_L$. In optimised compile-time linguistic reflection the generators operate over parsed forms of $L$. Thus $Val_L$ can be the abstract syntax trees constructed by the parser. The generators compute over these abstract syntax trees and form new ones.

Readability of the generators is an important issue with either kind of program representation, and is discussed in Section 2.7. When available, pre-defined abstractions operating over program representations aid the task of defining new program representations through analysis and composition of existing ones.

### 2.4.4 Execution Environment of Generators

The time of reflective evaluation affects the environment that is available to a generator. There are two environmental issues here. First of all the generators may need access to the details of the compilation such as the symbol table which provides type, scoping and identifier definitions. This is trivially available in compile-time linguistic reflection but it is also possible to parameterise the generators, with an environment, and to arrange that the compiler environment is preserved and available at run-time for run-time linguistic reflection.

The second issue is that generators may link to existing values. This linking may be to R-values, by copy, or L-values, by reference, and may be resolved at compile-time or delayed until run-time.

### 2.4.5 Time of Type Checking of Generated Code

In optimised compile-time linguistic reflection the result of the generation is integrated into the program being compiled. The internal type consistency of the new program fragment and its type compatibility with the environment into which it is placed are both checked by the post-parse compiler before execution.

In run-time linguistic reflection the result of the generation is type checked when it is presented to the compiler as part of the outer *eval*. This checks for the fragment's internal consistency. The type compatibility of the fragment with its environment is checked when it is linked into the original running program.

### 2.4.6 Execution Environment of Generated Code

In compile-time linguistic reflection the generated code fragments are incorporated into the main program and are thus executed in the environment created by that program. In existing run-time reflective systems the generated fragments are compiled and executed in isolation, and the results then linked into the running program. It is possible for a generated program to access values in scope in the original program that initiated the reflection, but only indirectly through the persistent store: the main program can place a value in the store where it can be accessed by the generated program.

More flexibility could be provided by parameterising generated code fragments with an environment in which to execute, together with a mechanism to allow the environment at any given point to be captured. Another mechanism that allows generated code to refer directly to values available in the generator will be described in Chapter 5.

### 2.4.7 Linking Generated Code into the Original Program

In compile-time linguistic reflection the code produced by executing a generator is incorporated into the program being compiled. The new code is compiled just as if it had been part of the original program, thus no special linking mechanism is needed.

In run-time linguistic reflection the generated code is compiled and executed independently of the main program and the results linked into the running program. This requires a dynamic linking and type-checking mechanism.

### 2.4.8 Characterisation of TRPL, PS-algol and Napier88

The type-safe reflective languages TRPL, PS-algol and Napier88 represent three sets of choices among the dimensions described above. The languages can be characterised as follows:

#### 2.4.8.1 TRPL

Linguistic reflection in TRPL is initiated when the post-parse compiler encounters the parsed form of a macro call. The macro arguments, which are instances of TRPL abstract syntax, i.e., elements of $Val_L$, are passed to the compiled form of the corresponding generator. The generator is executed to produce new instances of abstract syntax representing function and type definitions and an in-line code expansion. Pre-defined abstractions are provided for manipulating type and code representations.

Each generator has access to the compiler environment at the point of the macro call, allowing the types of expressions to be queried. The type representations obtained have a structured form which facilitates analysis to obtain type representations of sub-components, such as a field of a record type. Existing data in the file system or from the user may be accessed in a generator body using the standard IO facilities.

Abstract syntax produced by the execution of a macro is fed back into the post-parse compiler and type-checked in the course of its compilation. This includes both checking the internal consistency of the new program fragment and its compatibility with the surrounding program. At run-time the code produced by a generator is executed in the context of the preceding program, thus values introduced earlier in the program may be in scope in the generated code.

#### 2.4.8.2 PS-algol

Linguistic reflection in PS-algol is initiated at run-time when the evaluator encounters the compiled form of a call to the following procedure:

compile : **proc**( **string**, **pntr** → **pntr** )

The procedure takes a program representation and a pointer to a structure of the appropriate form to contain the expected result, and returns a pointer to a new structure containing the compiled result. Each generator is a procedure that produces a string as its result. Thus the inner *eval'* involves the evaluation of a procedure body. In the cases that a generator executes without errors the string is an element of $Val_L$, as it represents a PS-algol procedure. The only pre-defined operations on program representations are string concatenation and sub-string copying. The generators are executed at run-time and may access values in the persistent store but have no direct access to compilation information. As a generator can take arbitrary parameters it may be passed type representations in the same way as any other values. Type representations can be obtained using the following pre-defined procedure:

class.identifier : **proc**( **pntr** → **string** )

This returns a string representation of the form of the structure denoted by the pointer. To obtain a type representation for a non-structure value the programmer must create a structure with the value as one of the fields, pass it to *class.identifier*, and process the resulting string to extract the appropriate part.

Type checking of the program fragment produced by a generator occurs in two stages. The first occurs during compilation of the main program, when the compatibility of the expected type of the generated code with the rest of the program is checked. Secondly, at run-time the compiler checks that the fragment is internally consistent and that its type matches the expected type. If this fails the pointer returned by the compiler denotes an error structure that contains information about the reasons for failure.

Generated program fragments are executed in isolation from the main program. If required the programmer can arrange for values in scope in the main program to be accessible by generated fragments, by placing them in the persistent store. Note that the result of successfully executing a generated fragment is always a new procedure value, and that the new procedure is itself used in the context of the main program, thus values in scope there may be passed to it as parameters when it is called. The new procedure is linked into the original program by dereferencing the structure returned by the compiler procedure.

### 2.4.8.3 Napier88

The provision of linguistic reflection in Napier88 is similar to that in PS-algol. The principal differences are the use of the run-time compiler, the form of the type representations, and the way in which generated code is linked into the original program. The compiler procedure, *compile*, has the following type:

```
compile : proc( string → any )
```

The procedure *compile* is not limited to procedure representations but may be applied to a representation of any Napier88 value or sequence of commands. The value injected into the resulting *any* is a parameter-less procedure that will execute the compiled code in the case of successful compilation, or a string explaining the fault in the case of a compilation error. The programmer does not have to supply a representation of the expected type to the compiler procedure. Some examples are shown below:

```
compile( "3 + 4" ) => any( proc( → int ) ; 3 + 4 ) => 7

compile( "xor screen onto screen" ) => any( proc() ; xor screen onto screen )
      => screen cleared

compile( "proc( i : int → int ) ; i + 1" )
      => any( proc( → proc( int → int ) ) ; proc( i : int → int ) ; i + 1 )
      => proc( i : int → int ) ; i + 1

compile( "abc" ) => any( "error at line 1: …" )
```

**Figure 2.21: Examples of use of Napier88 *compile* procedure**

A type representation for a value injected into an *any* may be obtained using the pre-defined procedure:

```
getTypeRep : proc( any → TypeRep )
```

This returns a structured type representation, the form of which is described in [Cut92] and defined in Appendix B. Generated code is linked into the original program with a run-time type check of the *any* as illustrated below:

```
let result = compile( "…" )

project result as exec onto
proc( → int ) :  { let seven = exec() }
proc() :         exec()
string :         writeString( "compilation failed: " ++ exec )
default :        writeString( "result of unknown type" )
```

**Figure 2.22: Run-time checking of *compile* result**

## 2.5 Applications of Type-Safe Linguistic Reflection

### 2.5.1 Genericity and Efficiency

There are a number of application areas for the styles of strongly typed linguistic reflection described earlier. One of these is in supporting highly generic programs efficiently. The advantage of genericity is that it may promote software reuse, with associated economic benefits, by making the programs that are written more generally applicable than their non-generic counterparts. Thus for a given application it is more likely that existing software will be available, reducing the amount of new code that needs to be written.

A number of languages support polymorphic functions [Mat85, Tur85, Per87, MTH89, MBC+89, DM90, HWA+90, She90]. These achieve genericity by allowing the programmer to abstract over details of types. For example, a single function that counts the lengths of homogeneous lists of any element type may be defined. This is possible because the type of the list elements does not affect the way in which the length of the list is calculated. This variety of polymorphism is known as parametric polymorphism. Another variety, inclusion polymorphism, allows types to be partly abstracted over. For example, a function that expects a record parameter with a single field *name* may also be passed a record with two fields, *name* and *address*. The extra information is ignored by the function.

While these forms of polymorphism allow generic functions to be defined, their use is confined to cases where the generic computation does not depend on the types of the operands. There also exist application areas where a generic operation may sensibly be defined over many different types, but where the type of the data does affect the computation. Some examples are: natural join, deep equality testing and pretty-printing. In these cases the 'same' operation may be performed on instances of many different types, with details of each computation being determined by the particular type. For example with natural join the types of the input relations determine both the type of the result relation and the algorithm to produce the result. This constitutes a form of *ad hoc* polymorphism [Str67].

Generic programs whose behaviour depends on the types of their data can be written using type-safe linguistic reflection. The technique involves defining generators that, supplied with the types for a particular call, produce source representations of code to perform the operation for those types. The generators are used differently in compile-time and run-time reflection. With compile-time reflection the following actions are performed:

- During compilation, generator definitions are compiled.
- Also during compilation, calls to generators are executed. The generators produce new source code that is specialised to operate on particular types. They have access to type information accumulated by the compiler during compilation up to the point of the generator calls.
- The new source representations are incorporated into the original source program, replacing the calls to the generators.
- After each generator call, compilation continues from the point where the call was encountered. The new code produced by the generator is compiled and type-checked as if it had been part of the original program.
- When compilation has been completed all the reflective constructs have been 'compiled away' from the resulting compiled code.

With run-time reflection the following actions are performed:

- Generator function definitions are compiled along with the rest of the source program.

- When a generic operation is required to be applied to some data during execution, a generator is called and the data passed to it. The generator is able to discover the types of the data.

- The generator produces new source code to operate on those particular types.

- The new source code is compiled. If compilation succeeds the resulting compiled code is applied to the data.

In both compile-time and run-time reflection the code produced by the generators may be executed many times after the process of reflection has taken place, with no further overheads due to the genericity. This contrasts with the interpretive scheme that would be required to provide the same genericity if reflection were not used. In such a scheme the costs of specialisation would be borne every time a generic operation was performed.

To illustrate this difference, consider both reflective and non-reflective implementations of a generic operation to perform natural join in a language without built-in support for relations. The reflective implementation produces a specialised version of natural join whenever it is required. This version is specialised to the types of the input relations, specifying the names and types of their attributes, and type of the result relation. It is possible to verify before any call to the specialised function that it is supplied with relations of the correct types, thus the body of the function itself need not contain any checking for the well-formed-ness of the input relations. In addition the computation of the result type and the algorithm for producing the result tuples can be performed in the generator rather than the specialised function, which may be executed many times for each execution of the generator.

Without reflection, interpretation is required to provide the genericity. This solution requires a more loosely typed representation of relations, where all relations have the same type, for example a list of attribute names together with a two dimensional array of values. A single natural join function can then be defined for all relations. The disadvantage is that more computation is required at run-time: the compatibility of the input relations must be checked and the algorithm to produce the result tuples determined from examination of the input relations.

In the reflective solution to the natural join problem, the type dependent details of instances of a family of functions are generated. Thus the generator can be thought of as a highly generic abstraction over the functions. Another example of this approach is a set of four traversal functions over recursive data types [She91]. These functions generalise the list *map* and *fold* functions allowing them to be applied to any recursive data type. Sheard has also used the technique to define a deep equality test for any type [She90]. Similarly, forms systems for data entry and access can be automatically generated from type definitions. Cooper has used such a technique to provide a rich repertoire of interaction modes over any structures that may be defined in a range of data models [Coo90b].

The genericity achievable via linguistic reflection depends on the ability of a generator to access type details and generate program fragments that are tailored to the types given when the generator is executed. This constitutes a form of *ad hoc* polymorphism, but the genericity attained in these examples exceeds the capabilities of current polymorphic type systems [SFS+90]. In most polymorphic systems, the behaviour of polymorphic functions must be essentially invariant over the range of input types. The examples listed above have behaviour that varies too much to be accommodated by current polymorphic systems.

In conclusion, linguistic reflection supports the definition of generic programs whose behaviour depends on the types of their inputs, and that are more efficient at run-time than the equivalent interpretive versions. Efficiency is gained by allowing the input data to be represented in a more specialised form while still supporting generic abstractions over the data. This allows validity checking and algorithm construction to be performed earlier.

### 2.5.2 Software Evolution in Persistent Systems

Type-safe linguistic reflection may also be used in accommodating the evolution of strongly typed persistent object stores. Characteristics of such stores are that the type system is infinite and that the set of types of existing values in the store evolves independently from any one program. This means that when a program is written or generated some of the values that it may have to manipulate may not yet exist, and their types may not yet be known for inclusion in the program text. For strong typing these values must have a most general type but in some applications their specific types can only be found once they have been created.

An example of such a program is a persistent object store browser [DB88, DCK90] which displays a graphical representation of any value presented to it. The browser may encounter values in the persistent store for which it does not have a static type description. This may occur, for example, for values which are added to the store after the time of definition of the browser program. For the program to be able to denote such values, they must belong to an infinite union type, such as Amber's *dynamic* [Car85], PS-algol's *pntr* or Napier88's *any*.

Before any operations may be performed on a value of an infinite union type it must be projected onto another type with more type information. This projection typically takes the form of a dynamic check of the value's type against a static type assertion made in the program that uses it. A projection of a Napier88 *any* value was illustrated earlier, in Section 2.4.8.3.

The browser program takes as parameter an infinite union type to allow it to deal with values whose types were not predicted at the time of implementation. However the program cannot contain static type assertions for all the types that may be encountered as their number is unbounded. There are two possibilities for the construction of such a program: it may either be written in a lower-level technology [KD90] or else be written using linguistic reflection.

The linguistic reflective implementation of the browser program in Napier88 has a number of components. First of all the value of the union type passed to the program is interrogated to yield a representation of its specific type. If it is one of the base types such as *string*, *int*, etc., a method built into the browser is used to display the value. Otherwise the type representation is used to construct a representation of a Napier88 program. The compiler is called dynamically with this code representation as its argument, and returns some executable code which is capable of performing the appropriate projection of the union type, along with the required operations to browse the value. This new code is type-safe since it has been checked by the compiler. A different program is generated for each different type of value which is encountered during the browsing of the persistent store.

Figure 2.23 illustrates these actions. The types of the procedures used by the browser are shown in the rectangles. Instances of the type *TypeRep* are used to represent Napier88 types. Instances of the union type *any* are used as inputs to the browser and are also produced by the compiler procedure. In this case the values produced by the compiler are themselves procedures containing the code to display the values being browsed.

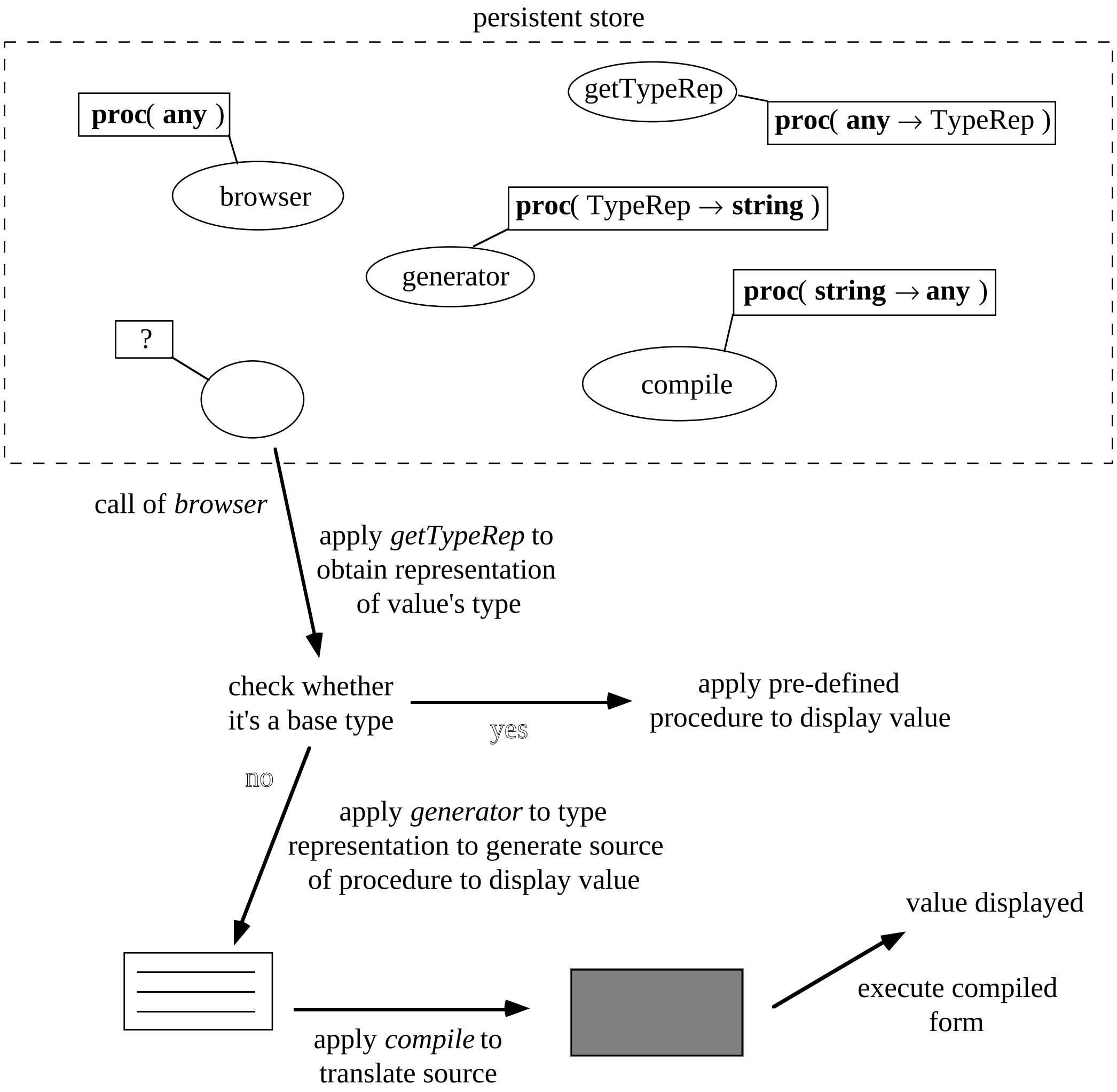

**Figure 2.23: Actions of reflective store browser**

An outline of the browser code is shown in Figure 2.24:

```
let browser = proc( val : any )
begin
    let valTypeRep = getTypeRep( val )

    if valTypeRep denotes a base type then use built-in method else
    begin
        case true of
        valTypeRep denotes a structure type :
        begin
            let new = compile( structureDisplayerGenerator( valTypeRep ) )

            ! new is of type any.
            ! structureDisplayerGenerator builds up a string program
            ! representation through analysis of valTypeRep.

            project new as compiledCode onto
            proc( → proc( any ) ) : compiledCode()( val )
            default :               writeString( "error in compilation" )
        end
```

```
        other cases : use similar methods for other type constructors
    end
end
```

**Figure 2.24: Browsing using run-time linguistic reflection**

For brevity the definitions of the procedures *getTypeRep*, *generator*, *menu* and *writeString* have not been shown. Assume that a value of the following type, injected into *any*, is passed to the browser:

```
type Person is structure( name : string ; age : int )
```

To display the value the browser needs to be able to construct and display a menu window with an entry for each field. It must also be able to extract the field values for further browsing should the user select one of the menu entries. The string produced by the generator is shown in Figure 2.25. The single quote character is used in Napier88 as an escape character to allow double quotes to be included in strings.

```
"type T is structure( name : string ; age : int )

proc( x : any )
    project x as specificX onto
        T : menu( '"name : string'", '"age : int'",
                proc() ; browser( any( specificX( name ) ) ),
                proc() ; browser( any( specificX( age ) ) ) )
        default : writeString( '"error'" )"
```

**Figure 2.25: String produced by generator in browser**

Note that the program produced by the generator itself contains a call to the browser procedure. This is achieved by linking the browser procedure into a location in the persistent store where it can be accessed by the generated program. The details of this access have also been omitted.

The algorithm shown is potentially inefficient as it requires reflection to be performed on every encounter with a structure type. This can be improved by using the persistent store as a cache for the results of reflection so that the generator call and compilation need not occur for types encountered previously. To achieve this the browser maintains a persistent table of display procedures, keyed by type representations. Each time the browser is called it checks whether the table contains a procedure to display values of the same type as the value passed to it, and if so that procedure is used. If there is no such procedure a new one is generated using reflection and entered in the table before it is called. In this way the use of reflection is only necessary on the first encounter with a particular type.

This example illustrates the use of linguistic reflection to define programs that operate over values whose type is not known in advance. These programs potentially perform different operations according to the type of their operands but without endangering the type security of the system. The requirement for such programs is typical of an evolving system where new values and types must be incrementally created without the necessity to re-define or re-compile existing programs.

Linguistic reflection can be used to accommodate a wide range of system changes. For example the schema changes of typical database applications become type changes in database programming languages, and reflective programs that are based on type details can regenerate code whenever a schema changes. If algorithms such as joins or form generation are systematically derived from the type information these derivations will be re-computed. With run-time reflection this happens lazily which may save computation since many systems undergo a sequence of changes between runs of many of their applications. In contrast the hand crafted method of providing the same functionality requires that a programmer locate all the places where changes are necessary, perform all the changes correctly and then re-validate the software. The reflective method gains particularly well in this case as it may avoid the need for re-validation as is discussed below.

### 2.5.3 Implementing Data Models

A data model is typically defined by a data description language and by one or more data manipulation languages (including query languages). Linguistic reflection allows these languages to be implemented efficiently, avoiding any additional levels of interpretation. Sentences in the data description language introduce new model constructs. A reflective generator translates these sentences into type declarations and declarations of associated procedures and introduces these into the computational context. Sentences in the data manipulation language are then translated into corresponding algorithms against these representational types and executed via reflection. In a persistent language this provides a very rapid means of prototyping and evaluating a data model [CAD+87, Coo90a, CQ92]. With the optimisation strategies discussed below this can be developed into a reasonable quality implementation of a DBMS for the data model.

This use of reflection to implement languages is not confined to data models. The technique is applicable to any language and has been used in a commercial system to develop a set of requirements analysis tools based on process modelling [War89, Bru91, BPR91]. Philbrow has used the same technique to provide polymorphic indexing mechanisms over arbitrary collections [Phi90].

### 2.5.4 Optimising Implementations

A form of optimisation has used linguistic reflection to directly declare data structures and to manipulate them directly avoiding a level of interpretation. In addition to this optimisation, a generator that develops concrete code for high level abstractions can choose from implementation strategies in order to minimise costs [CAD+87]. Relational query optimisation, for example, can be integrated directly into the compilation process via linguistic reflection. Run-time reflection allows re-compilation and new optimisation as the statistics of the database change [Cut92]. More general transformations of high level specifications into implementations can also be accomplished using linguistic reflection [FS91].

### 2.5.5 Validating Specifications

There are various ways linguistic reflection can be used to support validation of programs. The first derives from the fact that generated program fragments are stereotyped in their form. This stereotyping can be aimed toward producing forms that facilitate verification efforts [FSS92, SSF92]. Generators themselves can be analysed in order to verify properties of all generated expressions. Though this is a second order problem, there is the possibility of stereotyping the generator programs themselves to produce sub-languages that support the second order reasoning. Validating generators would be especially useful since it would mean that programs that were regenerated as a result of system evolution such as changes to types would not need to be re-validated.

Theorem proving itself can be integrated with compilation using linguistic reflective capabilities. A version of the Boyer-Moore theorem prover kernel has been implemented in TRPL working over the parsed form of TRPL's functional core language. Using this kernel, validation of properties of TRPL functional programs can be performed as a part of the compilation process. For example, the problem of verifying that database integrity constraints are invariants of transactions can be addressed by this approach [SS89].

## 2.6 Anatomy of Generators

A number of uses for type-safe linguistic reflection have been described, giving significant benefits in the areas of software reuse and system evolution, implementation and optimisation. Currently the main constraint on the wider use of the technique is the difficulty of writing and understanding generators. This section will examine the various components found in generator bodies, describe the existing type-safe reflective languages in terms of the framework developed, and identify factors that affect the ease with which generators may be understood.

### 2.6.1 Generator Components

Each generator contains a result expression that when evaluated produces the generated program fragment. The code in this expression itself represents code, thus it belongs to the subset of $L$ containing sentences that, when evaluated, produce values in $Val_L$. This subset will be denoted by $L_L$. The set $L_L$ can be partitioned into two subsets, $L_{L_{Const}}$ and $L_{L_{Var}}$. The former, $L_{L_{Const}}$, contains those sentences that produce the same values in $Val_L$ for all executions, while the latter, $L_{L_{Var}}$, contains sentences that may evaluate to different values on different executions. This is illustrated in Figure 2.26:

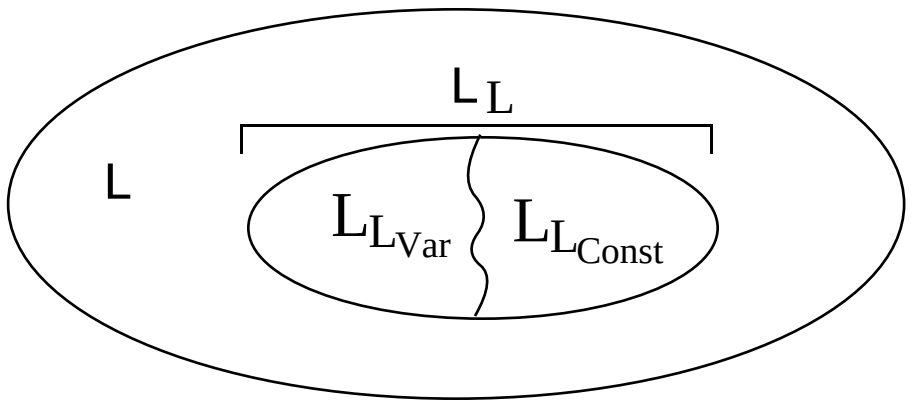

$L_L \subset L$

$L_{L_{Const}} \subset L_L$

$L_{L_{Var}} \subset L_L$

$L_{L_{Const}} \cap L_{L_{Var}} = \varnothing$

$L_{L_{Const}} \cup L_{L_{Var}} = L_L$

**Figure 2.26 Subset relationships between code categories**

Some examples of sentences in these sets are shown below for a language in which members of $Val_L$ are strings and ++ denotes concatenation:

| | | |
|---|---|---|
| a := a + 1 | $\in$ | $L$ |
| "a := a + 1" | $\in$ | $L_{L_{Const}}$ |
| "a := " ++ makeCode() | $\in$ | $L_{L_{Var}}$ |

Note that the last example is itself a composition of two code fragments, one a member of $L_{L_{Const}}$ and the other a member of $L_{L_{Var}}$. In general a generator body contains a section of code in $L$, here termed the *prelude*, followed by a section in $L_{L_{Var}}$ that defines the resulting generated fragment, here termed the *result definition*. This is illustrated in Figure 2.27. The purpose of the prelude is to set up an environment in which the result definition is evaluated.

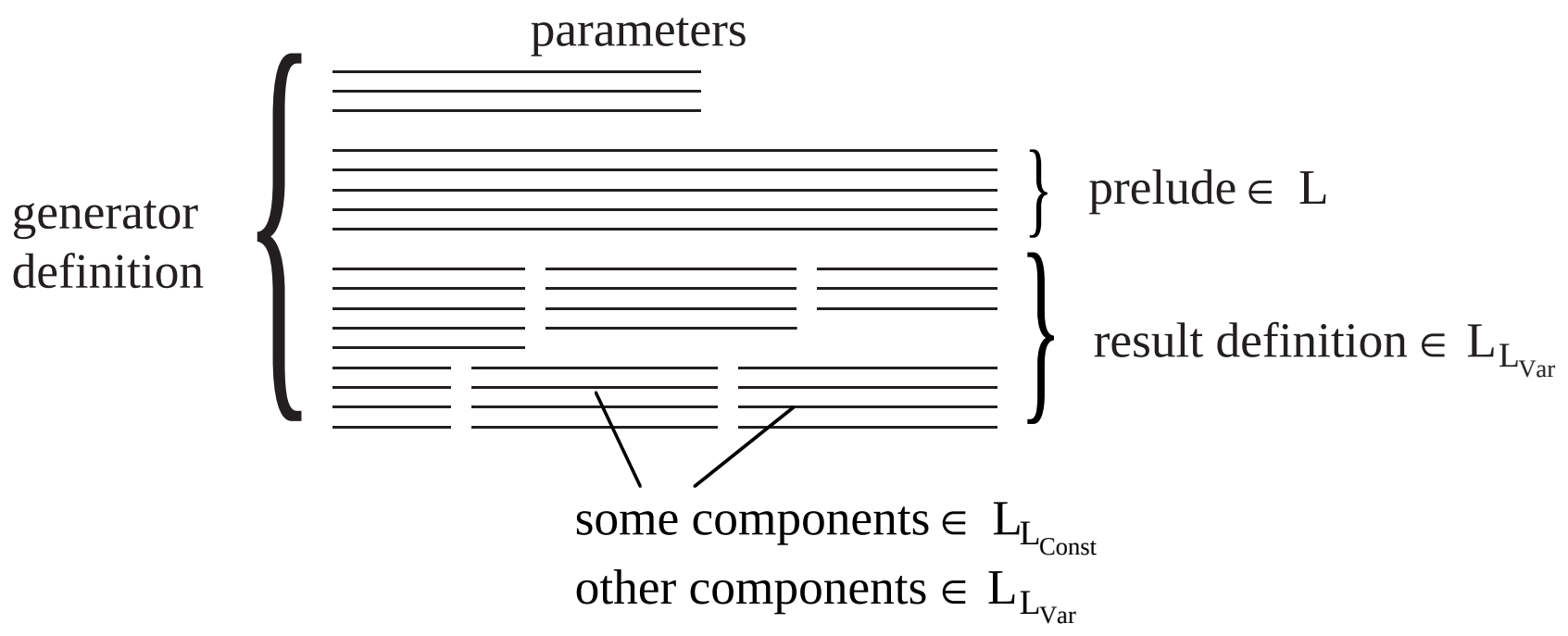

**Figure 2.27: Structure of a typical generator**

In simple cases the generator body may contain only the result definition, and that code may lie in $L_{L_{Const}}$ rather than $L_{L_{Var}}$. In the general case the execution of a generator involves the evaluation of the prelude and those parts of the result definition that lie in $L_{L_{Var}}$, i.e., the variable parts. The parts in $L_{L_{Const}}$ do not need to be evaluated as they are constant over all executions of the generator. Typically the evaluation of the prelude affects the program fragments produced in the result definition. The result of the generator is obtained by composing the newly created fragments with the constant parts of the result definition.

### 2.6.2 Components in TRPL Generators

The generators in reflective TRPL programs are context sensitive macros. The example from Figure 2.3 is reproduced in Figure 2.28, with the constant parts of the result definition shown in outline text, and the variable parts in italic text. To reduce confusion the reserved words have not been emboldened.

```
macro INC (x := make_id (?, ?));          @ Plain text is in L.
env e;                                    @ Outline text is in  L_L_Const.
let x_type := type_of (x, e) in           @ Italic text is in L_L_Var.
case x_type
    {    TYPE (integer)    → EREP (a := a + 1, a :=  x)
         others            → x}
```

**Figure 2.28: Code categories in a TRPL generator**

The body of the macro contains a call to the pre-defined macro *EREP*, which expands to an abstract syntax representation of the code passed to it. That code is *a := a + 1*, where *a* is substituted by whatever identifier has been passed to *INC*. Although a parsed form of code representation is used this is disguised by *EREP* which allows the code in $L_{L_{Const}}$ to be written textually. The substitution written after the main code *a := a + 1* provides the means for composing the constant and variable parts of the result definition. Further examples of reflection in TRPL are given in [SS91].

### 2.6.3 Components in PS-algol Generators

The generators in reflective PS-algol programs are procedures that return strings. The example from Figure 2.4 is reproduced in Figure 2.29 below, with the same formatting as the TRPL example above:

```
let mkFun = proc( → string )                 ! Plain text is in L.
begin                                        ! Outline text is in L_LConst.
      write "enter real expression over x"   ! Italic text is in L_LVar.
      let expr = reads()

      "proc( real x → real ) ; " ++     expr
end
```

**Figure 2.29: Code categories in a PS-algol generator**

Here the code in $L_{L_{Const}}$ is enclosed by quotes. It is composed with the variable parts of the result definition by string concatenation.

In [Coo90a], Cooper describes a variation on this notation in which the main result definition is a single string with embedded place-holders, of the form *#IDENTIFIER*, as shown below:

```
let mkFun = proc( → string )                 ! Plain text is in L.
begin                                        ! Outline text is in L_LConst.
      write "enter real expression over x"   ! Italic text is in L_LVar.
      let expr = reads()

      let program := "proc( real x → real ) ; #EXPRESSION"
      replace( program, "#EXPRESSION", expr )
      program
end
```

**Figure 2.30: Code categories in a PS-algol generator with place-holders**

Each place-holder corresponds to a variable part of the result definition. Following the definition of the string the programmer specifies substitutions for the place-holders using the procedure *replace*. In fact *replace* is not a real procedure since the textual form of the entire generator is pre-processed before compilation to give a form equivalent to that in Figure 2.29. In more complex generators this scheme improves readability by reducing the syntactic noise of string concatenation in the result definition, providing meaningful names for each section of $L_{L_{Var}}$ code, and making those sections easy to pick out.

### 2.6.4 Components in Napier88 Generators

The generators in reflective Napier88 programs have the same form as in PS-algol. For completeness the example from Figure 2.5 is shown below. It is very similar to that in Figure 2.30 apart from minor syntactic differences.

```
let mkFun = proc( → string )                         ! Plain text is in L.
begin                                                ! Outline text is in L_LConst.
      writeString( "enter real expression over x" )  ! Italic text is in L_LVar.
      let expr = readString()

      "proc( x : real     → real ) ;     " ++ expr
end
```

**Figure 2.31: Code categories in a Napier88 generator**

### 2.6.5 Factors in Understanding Generators

Programmers writing generators in various languages have reported that generators are considerably more difficult to write and understand than conventional programs. Some possible reasons for this are:

- A generator may describe a large class of programs rather than a single one. Although a conventional program may have many different possible execution paths, its structure is fixed. The structure of different programs produced by a single generator may differ widely. To understand a generator the reader needs to be able to determine the features common to all programs produced by it, and to understand how the parts that vary among the resulting programs relate to the input parameters to the generator.

- The constant and variable parts of the result definition appear different even though they both represent parts of the resulting program fragment. By the end of the generator execution they are integrated seamlessly but this is not apparent from inspection of the generator source code.

- Code in different parts of the generator are evaluated at different times. During the execution of the generator, the prelude and those parts of the result definition in $L_{L_{Var}}$ are evaluated. Later during the reflection process the new code produced by the generator, comprising the $L_{L_{Const}}$ parts of the result definition composed with the fragments produced by the evaluation of the $L_{L_{Var}}$ parts, is evaluated. Thus adjacent parts of the result definition may be evaluated at different times and in different environments.

- The programmer must understand several mappings:

  — between sentences in $L$ and their representations in $L_{L_{Const}}$,
  e.g. a := a + 1 → "a := a + 1"

  — between sentences in $L$ and their representations in $L_{L_{Const}}$ or $L_{L_{Var}}$ used in code manipulation functions,
  e.g. a := a + 1 → "a := a + 1"
  or a := a + 1 → assign( "a","a + 1" )

  — between sentences in $L_{L_{Const}}$ or $L_{L_{Var}}$ and the sentences in $L$ which they represent,
  e.g. "a := a + 1" → a := a + 1
  or assign( "a","a + 1" ) → a := a + 1

- In languages where $L_L$ comprises string expressions, manipulation of program representations is unwieldy. One example of such a manipulation is determining the result type of a procedure from its representation in $Val_L$. This is non-trivial when the representation is a string, as it involves parsing the string. A more structured representation might contain a representation of the result type as a component that could be accessed directly.

## 2.7 Research Areas

The useability factors identified above suggest several areas for research in reflective programming. One is to improve support for writing generators: some desirable features are listed below.

- When reading a generator definition, it should be easy to identify which parts of the result definition are constant, in $L_{L_{Const}}$, and which parts are variable, in $L_{L_{Var}}$.

- It should be possible to use different code representation forms in the constant and variable parts of the result definition. A textual form, such as strings, is easy to read in the constant parts as it gives the simplest mapping between $L_{L_{Const}}$ and $L$. An abstract syntax form may be more suitable for the variable parts as it facilitates the expression of code representation manipulations.

- Programming tools could aid understanding of generators. For example, a tool could be provided to display the resulting code produced by a generator for any given inputs. This could help in understanding the relationships between generator parameters and the code fragments produced by $L_{L_{Var}}$ code.

Another research area is the provision of flexible and general linking mechanisms to support linking between generators, generated program fragments, the compilation environment and a persistent store.

Chapter 5 describes an interactive system that is designed to assist in constructing reflective Napier88 programs. A window-based generator editor reduces syntactic noise in generator result definitions, and allows the programmer to view the resulting code for any particular generator execution. Hyper-program linking facilities, to be explained in Chapter 3, allow both generators and generated program fragments to contain direct links to values in the persistent store.

## 2.8 Conclusions

Behavioural and linguistic reflection allow a programming system to affect its own behaviour. In behavioural reflection this involves a program altering the way it is interpreted, while in linguistic reflection a program can change itself. A style of linguistic reflection appearing in strongly typed programming languages has been identified, defined and described. This style, termed type-safe linguistic reflection, can extend the class of algorithms that can be written in a type-safe manner. Linguistic reflection is characterised by the ability of a program to generate code in its language that is to be integrated into its own execution. This ability provides a base for generator technology that can be integrated with a programming language in a uniform and type-safe manner. While this capability has been a feature of many interpreter based languages with weak type systems, it is relatively new in compiler based, strongly typed systems. Two styles of linguistic reflection have arisen in database programming languages, compile-time and run-time. Both have been described in detail, allowing a comparison of the mechanisms as currently implemented.

Many uses have been found for linguistic reflection in the database programming area. These uses are characterised by a need for a high level of genericity in specifying data and procedures, a requirement that has proved problematical to meet using programming language type systems alone. Two such uses have been detailed and several more discussed.

Type safety has been achieved in PS-algol, Napier88 and TRPL by type checking each generated program segment, which is necessary when the complete programming language can be used to write generators. Limiting the language subset available for writing generators may allow the generators to be type checked for the type of all output at one time. This is a topic for future research. Other work to be done includes combining the two styles of reflection presented here, and exploring the relationship of linguistic reflection with other kinds of reflection.

The structure of the generators in existing systems has been analysed, three categories of generator code identified, and proposals made toward making generator definitions easier to understand. Ease of use is a significant problem in existing reflective language systems. Although once a generator is written it can be used to effect by any programmer, only a minority of the programming community are likely to write their own generators. This situation might be improved with better generator definition notations and programming tools.

# 3 Hyper-Programming

## 3.1 Introduction

Most programs written in persistent languages access data in a persistent store. Programs contain denotations for this data, which may include values, store locations that contain values, and types. At some stage during the software development process the denotations in a program are resolved to the actual data. The terms defined below will be used in describing this resolution:

**data item**: a value, or a location containing a value, in the persistent store;

**access path**: a description of the position of a data item in the persistent store;

**access specification**: the access path of a data item together with a description of its expected type.

Each program contains an access specification for each of the data items that the program links to. In PS-algol and Napier88 this code is executed at run-time and any related errors occur at that time. Errors can arise due to the data not being present at the specified position in the store, or its type being different from that specified. In addition to accessing data items, programs can also access types in the persistent store but the nature of this access will be discussed later.

Often the programmer knows that some of the data items linked into a program exist in the persistent store at the time the program is written. In such cases a language system could allow the programmer to indicate the required data items by interactive gesture with a mouse, instead of writing an access specification for each data item. Graphical representations of the appropriate data items of the store could be pointed to by the programmer. No access specification would be required as the access paths and type descriptions could be obtained from the system's internal representation of the persistent store.

The linking of the program to the data items specified in this manner could be handled by the system in several different ways. It could perform the same actions as if the programmer had written normal access specifications. It would thus verify at run-time that the store contained data items with the specified access paths and types, and then link those items into the running program. Another strategy is to link the data items directly into the program as it is written.

This chapter will explore the possibilities of the latter option. This scheme requires a change in the nature of the source program: the textual description of the computation is augmented with tokens that denote the persistent data items the program accesses. By analogy with a hyper-text graph, in which a piece of text contains embedded links to other pieces of text, this type of source program is termed a hyper-program [FDK+92, KCC+92b]. It consists of a textual source program with embedded links to values, locations and types. To construct a hyper-program the programmer types in program text and then inserts tokens that denote data items identified by pointing to their graphical representations.

Figure 3.1 shows an example of a hyper-program. The hyper-program contains both text and a token that denotes a data item in the persistent store, a procedure to write out strings. The hyper-program contains a link to the procedure itself rather than an access specification for it.

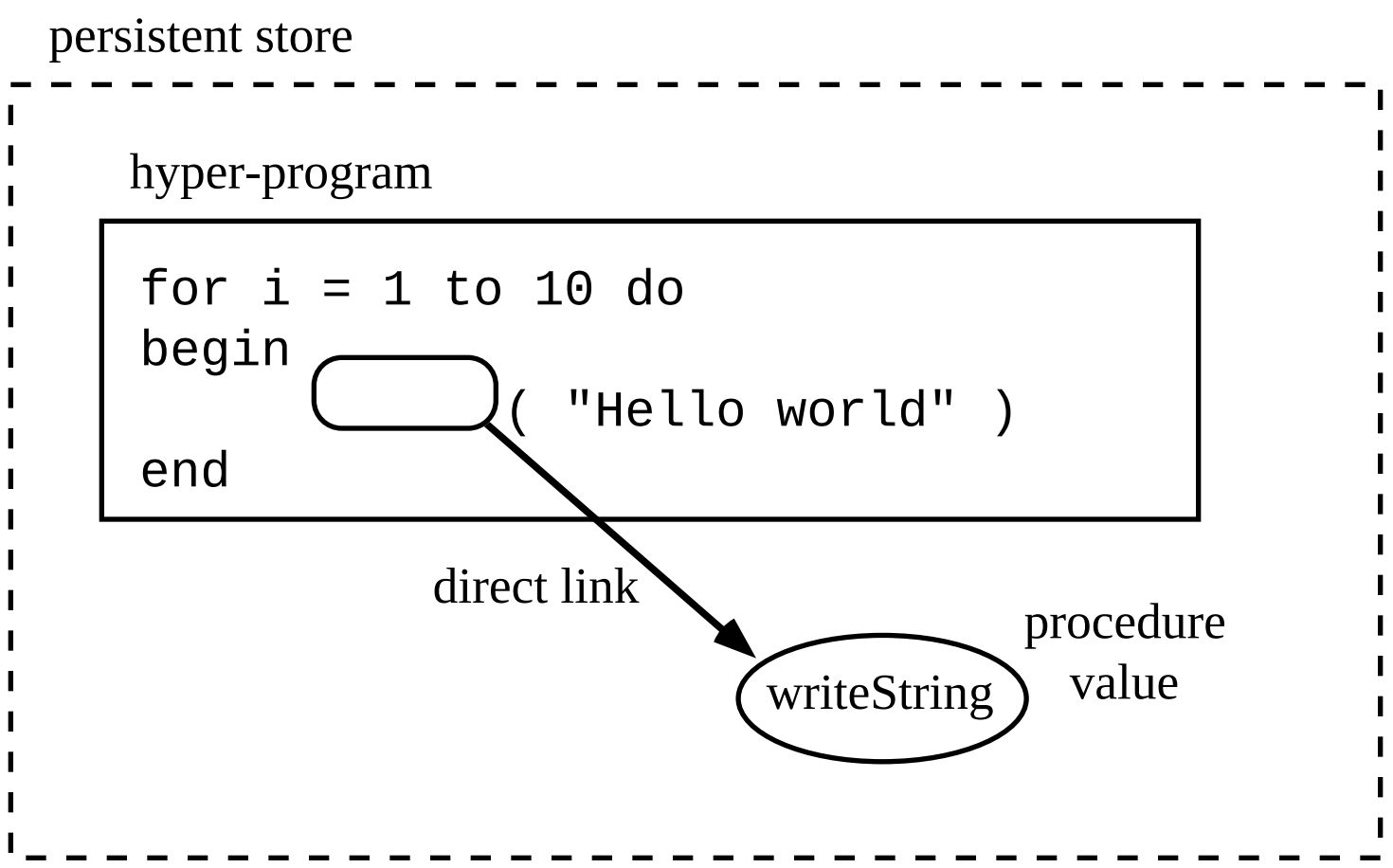

**Figure 3.1: A hyper-program**

The tokens in a hyper-program denote references to, rather than copies of, values and locations in the persistent store. This contrasts with linking mechanisms in file-based languages such as C [KR78], with which a program can be linked to a number of other programs. These involve copying code from other programs into the main program code. In addition, hyper-programs can contain links to any data items rather than just to sections of code.

A number of benefits of using hyper-programs will be described. These are:

- support for program composition and software reuse;
- being able to perform program checking early;
- being able to enforce associations from executable programs to source programs;
- availability of an increased range of linking times;
- reduced program verbosity; and
- support for source representations of procedure closures.

The principal requirement for supporting a hyper-programming system is a persistent store to contain the program representations and the data items corresponding to the tokens in the programs. The assumption is made here that the store is stable and that it supports referential integrity. This means that once a reference to a data item in the store has been established, the data item will remain accessible for as long as the reference exists.

Secondly, the hyper-program source representations must be denotable values in the programming language. Linguistic reflective facilities are required to support the conversion of hyper-program representations into executable programs. Where the executable programs produced by reflecting over the hyper-programs are themselves language values, a suitable representation is required. One possibility is to use procedure closures; these are already supported as first class values in a number of languages.

A third requirement is for tools that provide the programmer with the graphical representation of the persistent store. The representation shows the values, locations and types in the store and the links between them. The programmer can point to the representations of specific data items and obtain tokens for them to be incorporated into hyper-programs.

These implementation requirements are discussed later in Chapter 6. This chapter elaborates on the motivations for building a hyper-programming system and on its benefits. To be useful in practice a hyper-programming system will also have to support additional facilities for ‘programming in the large’, that is, building large applications from smaller components. These include facilities for controlling the sharing of components between applications, for limiting the visibility of some components for protection reasons, and for imposing a degree of partitioning on the persistent store to aid intellectual manageability and execution efficiency. A model to support these facilities, the *hyper-world* model, is proposed.

# 3.2 Motivations and Benefits

## 3.2.1 Program Composition

The primary motivation for providing a hyper-programming system is to allow the programmer to compose programs interactively, navigating the persistent store and selecting data items to be incorporated into the programs. This reduces the need to write access specifications for persistent data items that are accessed by a program.

Existing languages that allow a program to link to persistent data items at any time during its execution, such as PS-algol and Napier88, require it to contain code to specify the access path and type for each data item. The access path defines how the data is found by following a particular route through the persistent store starting from a root of persistence. The type specifies the expected type of the data at that store position. When a program is compiled the compiler checks that subsequent use of the data is compatible with its expected type. When the program is executed the run-time system checks that the data is present at the declared position and that it does have the expected type.

This mechanism gives flexibility because a program can link to data in the store at any time during its execution. However in many cases the programmer knows that a particular data item is present in the store at the time the program is written. Although the programming system could obtain all the information in the access specification by inspecting the data item at that time, the programmer must still write the access specification.

In a hyper-programming system the programmer has the option of linking existing data items into a program by pointing to graphical representations rather than writing access specifications. One example of such a graphical interface is described in Chapter 4. Note that the ability to link to data items at run-time is still required in the cases where data becomes available only after a program is written.

## 3.2.2 Early Checking

Hyper-programming can provide improved safety in several ways. One of these is that it allows some program checks to be performed earlier than normal, subsequently giving increased assurance of program correctness. This is possible because data items accessed by a program may be available for checking before run-time. Referential integrity then ensures that the checked data remains available at run-time.

Checking can be performed at several stages in the program development process in existing systems. The principal opportunities are at compilation-time when a program is translated into an executable program, and at run-time when the executable program is executed. Categories of checking include checking programs for syntactic correctness and type consistency, and checking persistent data access. Usually the program checks are performed at compilation-time, although in some syntax directed programming systems [AHM88] type consistency is verified as a program is constructed. The ability to bring forward the checking of persistent data access in a hyper-programming system is now discussed.

### 3.2.2.1 Checking Persistent Data Access

In conventional strongly typed persistent systems a program contains an access specification for each persistent data item used. These access specifications are checked at run-time: at that time the system verifies that each data item is in fact present in the store, with the previously declared access path and type. This is illustrated in Figure 3.2:

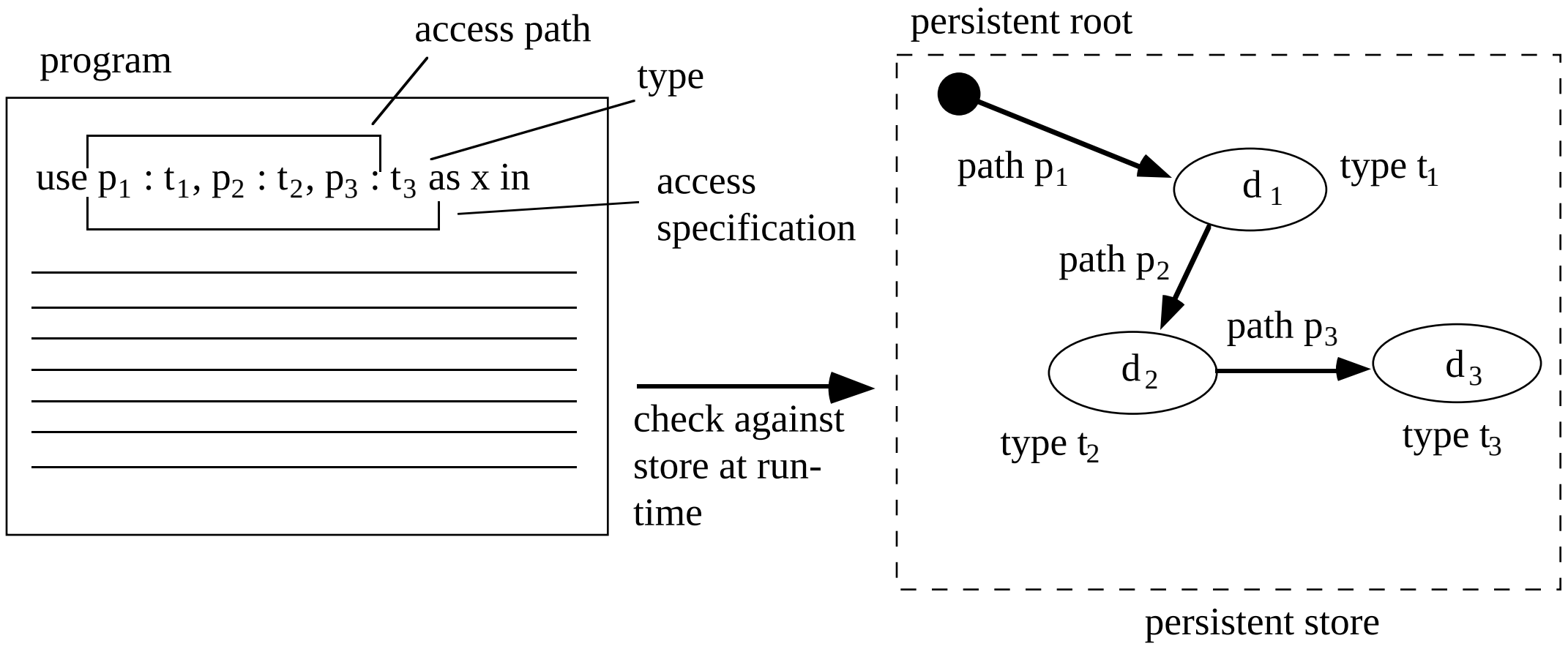

**Figure 3.2: Access specification with run-time checking**

In the program the identifier *x* is introduced to denote the data item obtained by traversing the access path $p_1 : t_1, p_2 : t_2, p_3$. In the diagram this data item is labelled $d_3$. The type of *x* is declared to be $t_3$. Each component of the path—$p_1$, $p_2$ and $p_3$— is a fragment of code that defines a route between two data items. $p_1$ is first applied to the persistent root to give data item $d_1$, then $p_2$ is applied to $d_1$ to give $d_2$, and finally $p_3$ to $d_2$ to give $d_3$. The types of the intermediate data items, $t_1$ and $t_2$, also form part of the access path. Note that there may be other routes to $d_3$ apart from the one shown. At compilation-time the system checks that the access specification is consistent with the rest of the program. At run-time it checks that the access specification is valid with respect to the current state of the store, i.e., that $d_3$ can in fact be accessed along the given path and that it does have the declared type $t_3$.

A program execution will fail if the store does not contain a route to a data item corresponding to the access path specified in the program. Thus even if it is known at the time of writing that a particular program will execute correctly, it cannot be predicted when it may fail on some future execution.

The use of hyper-programs as source representations allows the checking of access specifications to be performed before run-time. Each token embedded in a hyper-program denotes a data item that exists in the store at the time the hyper-program is composed. The process of checking the access path is moved from run-time to program composition time. The access path is established incrementally as the programmer manipulates the graphical representations of the data in the store to locate the required data item. Once the path has been established the data item at the end of it is linked into the hyper-program and the path need not be followed again at execution time. This is illustrated in Figure 3.3. The hyper-program will be unaffected if the access path is then removed. This might occur, for example, due to the link from $d_2$ to $d_3$ being overwritten by a link to some other data item.

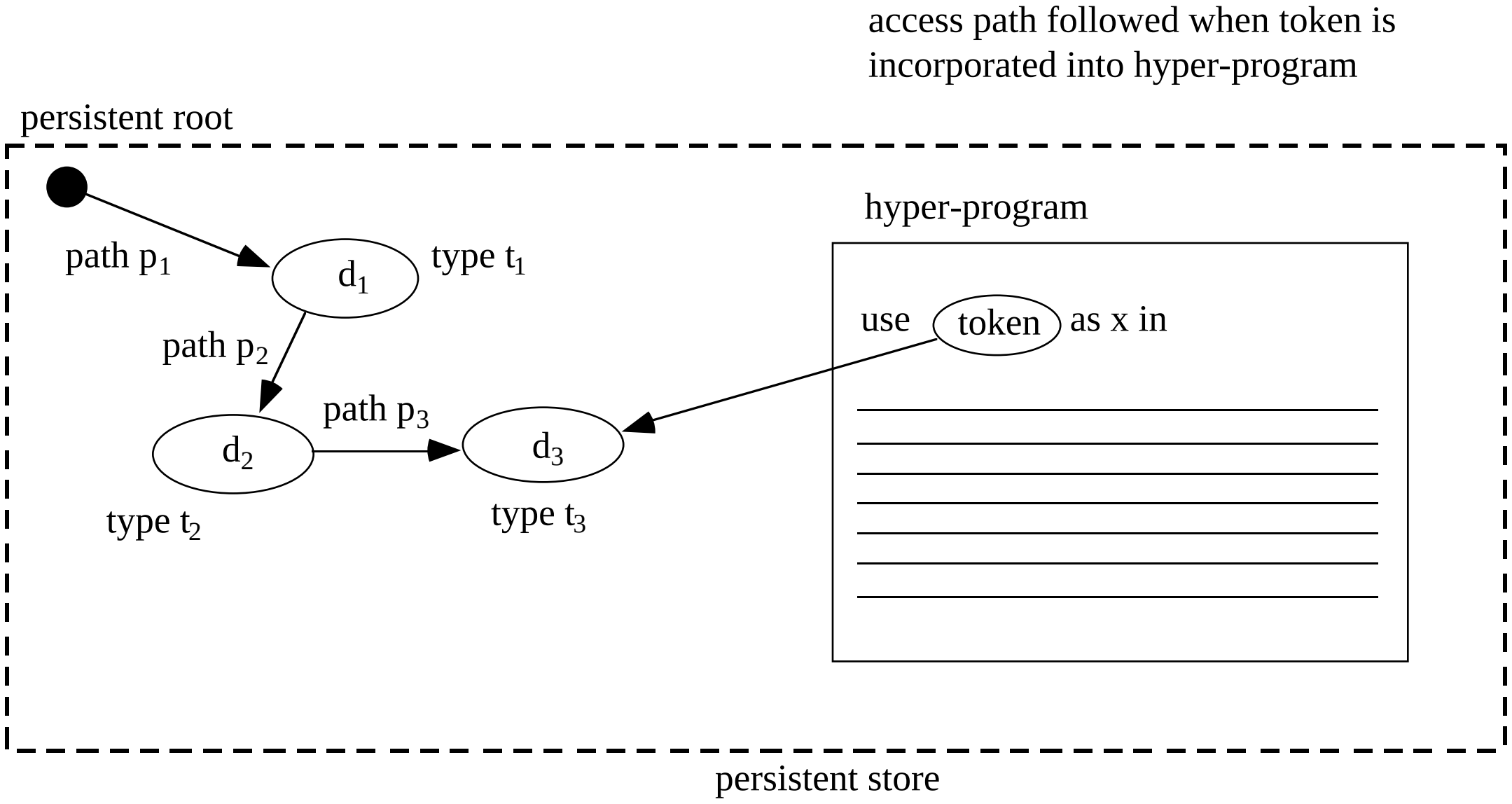

**Figure 3.3: Access path with hyper-program**

The access path part of the access specification is established during hyper-program composition. The other part, the type specification of the data item, is checked when the type consistency of the hyper-program is verified at or before compilation-time. The system checks that the type of the data item denoted by the token is compatible with the use of the token in the program. The various checking phases are summarised in Figure 3.4:

| | **composition-time** | **compilation-time** | **run-time** |
|---|---|---|---|
| **conventional persistent** | | check type consistency of program | check presence and type of data |
| **hyper-programming** | locate data | check consistency of program with data | |

**Figure 3.4: Run-time data checking and hyper-program checking**

Creating direct links from a hyper-program to values in the persistent store, with the associated safety benefits described above, is only applicable where values are present in the store at hyper-program composition time. Added flexibility can be gained by using tokens to denote mutable locations in the store. Linking a location into a hyper-program involves the same processes as for linking a value, with the difference that the value associated with the token changes when the location is updated. Updates to the location may occur at any time after the composition of the hyper-program. Strong typing ensures that the type of any value assigned to a location is compatible with the type of its original contents. This allows the type checking of persistent locations to be performed at compilation-time only. The values in locations associated with the tokens in a hyper-program can vary but their types will always remain the same. Where a token denotes a location, that location is linked directly into the executable program produced from the hyper-program, so that updates to the location also affect the executable program.

#### 3.2.2.2 Other Kinds of Checking

Language systems also perform other kinds of checking at run-time. Some of this checking can also be performed earlier in a hyper-programming system. An example of this is dependent type checking.

A dependent type is a type that depends on a value, requiring run-time type checking in conventional systems. To determine whether two dependent types are compatible, the language's type checker takes account of the associated values as well as their structure. An example of a dependent type is the generic type *map* [ALP+91], instances of which are associations between sets of values. To create a map the programmer calls a generating procedure, passing it as a parameter another procedure that determines whether two given values of the domain type are equal. This equality-testing procedure is used in the implementation of the map. Language rules define that two maps are type equivalent if and only if their respective domain and range types are equivalent, and their equality-testing procedures are identical. Because of this it is not generally possible to type-check at compilation-time a program that contains map operations, as the map values themselves must be tested.

However, in a hyper-programming system the value on which a dependent type depends may be linked directly into a program, and may thus be available for checking at compilation-time. This makes it possible for the system to check operations on dependent types at compilation-time rather than planting code in the executable program to perform the checking at run-time. The system may also provide tools that allow the programmer to verify the type compatibility of selected values before they are linked into the hyper-program. Transmission of the results of such checks to the compilation system is a topic for future research.

More generally the programmer may perform arbitrary checks on data values before linking them into a hyper-program, by writing and executing other programs that compute over them. If the checks succeed, the code that performs the checking can then be omitted from the main hyper-program.

### 3.2.3 Source Code Control

#### 3.2.3.1 Relationships Among Program Forms

Safety can also be improved with respect to the relationships between executable programs and source programs. In a programming system it is often desirable to maintain links between executable programs and their corresponding source code programs, to facilitate debugging and software evolution. These links enable the system to show the source code corresponding to the point where an error occurs in a running program, or to supply the source code for a given executable program so that it can be modified and a new version created.

In existing systems these links operate by conventions and can be corrupted by programmer actions that do not conform to those conventions. Given a language that supports executable programs as first class values—for example, procedures—a hyper-programming system can enforce links from executable code to source code. To illustrate this, the relationships between these different forms of code and other data values will be described, first in general and then with particular reference to file-based systems, persistent systems and finally hyper-programming systems.

Application development involves a number of activities including the following:

- constructing source code programs;
- compiling source code to give intermediate programs;
- linking intermediate programs to give executable programs;
- linking existing data items into executable programs; and
- executing linked programs in a run-time environment.

The software entities involved in these activities are:

- source programs;
- intermediate programs—these are not executable as the code in them makes unresolved references to other programs;
- executable programs—these can be executed directly; and
- data items that are manipulated during execution.

Language systems support several varieties of relationships between the software entities listed above. These are *causations*, *associations* and *direct links*.

Causations are one-way ‘cause and effect’ relationships. A causation from an entity *A* to another entity *B* exists if a change to *A* results in a corresponding but indirect change to *B*. Indirect means that some other process must be performed for the change to propagate. An example of a causation is the relationship between a source program and the corresponding compiled version. A modification to the source program causes a corresponding change in the compiled program but only after the process of compilation.

Associations are general relationships between entities. An example is an association between an executable program and the corresponding source program, maintained by a source level debugging system. This information is not intrinsic to the associated entities themselves but is maintained by an external mechanism. In general the accuracy of associations depends on adherence to conventions: if changes to the entities are made outside the control of the external mechanism the associations may become invalid. In the example the source program could be updated without notifying the debugging system, in which case its association with the executable program would become invalid.

Direct links are references between entities in the run-time environment. A direct link from an entity *A* to another entity *B* exists if a change to *B* results in a corresponding and immediate change to *A*. This could be implemented by storing the address of *B* inside *A*. The language systems considered here support identity, that is, a reference to a given entity is guaranteed to remain valid and to refer to the same entity for as long as the reference exists. Thus a direct link from *A* to *B* always remains valid regardless of the operations performed on *B*. A change to *B* has an immediate effect on *A* without the need for any intermediate process.

#### 3.2.3.2 Languages with External Storage Systems

In languages such as Pascal [Wir71], Ada [DOD83] and C, the persistent data, that which survives for longer than the program execution that creates it, is manipulated differently from the transient data. It is held in a storage system, separate from the run-time environment, with which programs communicate through an interface. An example is the Unix file system [RT78].

The program entities listed earlier, source programs, intermediate programs and executable programs, all reside in the external storage system. Source programs are compiled to produce intermediate programs. Where necessary a linker is then used to link in existing intermediate and executable programs from a program library. This linking involves combining the intermediate program with copies of the library programs to produce a new executable program. At run-time the resulting executable program is itself copied into the data space of a run-time environment and evaluated in that context. The running program may create new data items (values and locations) with direct links between them. It may also access existing data in the external storage system. The run-time environment disappears at the end of execution, along with any new data items created in it.

Figure 3.5 summarises the causations and associations between the various entities:

| causations | | associations | |
|---|---|---|---|
| **from** | **to** | **from** | **to** |
| source program | intermediate program | intermediate program | source program |
| intermediate program | executable program | executable program | intermediate program |
| intermediate library program | executable program | executable program | intermediate library program |
| executable library program | executable program | executable program | executable library program |
| executable program | run-time data item | | |
| file system data | run-time data item | | |
| run-time data item | file system data | | |

**Figure 3.5: Causations and associations**

Associations involving run-time data items are not maintained, as the data items are transient. The relationships are illustrated pictorially in Figure 3.6. Here rectangles represent source programs, rounded rectangles represent intermediate programs, diamonds represent executable programs and ellipses represent data items that can be denoted in the programming language.

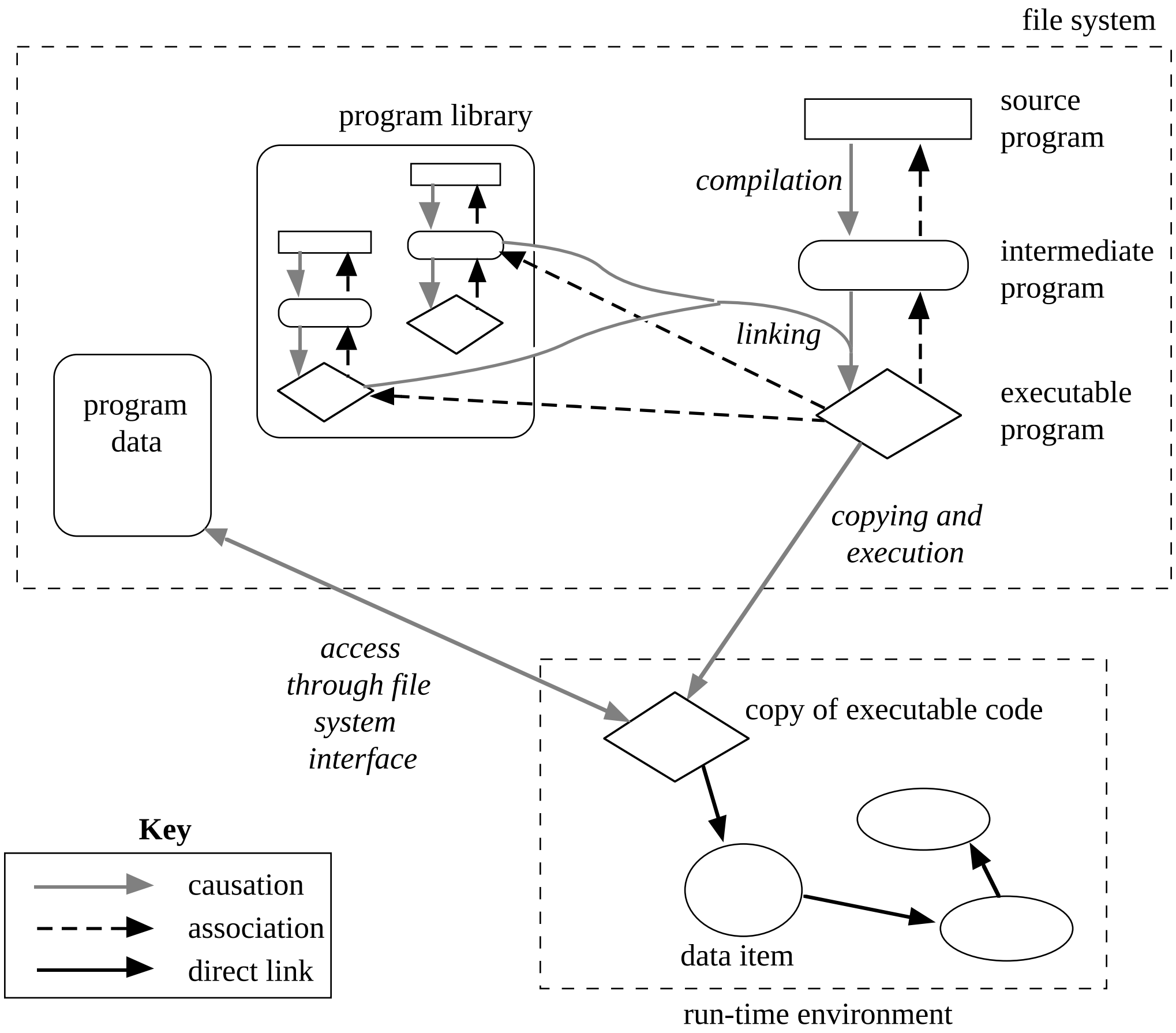

**Figure 3.6: Relationships in a file-based system**

A file-based system may involve linguistic reflection. For example in the Unix environment C programs are used to convert other C source programs into intermediate and executable programs which are executed within the system.

#### 3.2.3.3 Persistent Languages

Persistent languages that support first class procedures are now considered. Examples of these are PS-algol, Napier88, Galileo [ACO85, AGO88], P-Quest [BMM+92, MMS92] and STAPLE [DM90]. The model of persistence in these languages is persistence through reachability [ABC+83]: this means that a data item will persist at the end of a program's execution if and only if it is reachable from one or more persistent roots.

In these languages executable programs can be represented as procedures or functions and can thus be stored in a persistent store rather than a file system. Since each executable program is a language value it can contain direct links to other data items, and other values can contain direct links to it. A separate program library is not necessary as direct links to other executable programs in the store can be incorporated into an executable program when it is formed. Programming techniques to achieve the effects of incremental linking in this way are described in [AM84, AM85, AM86, DCC92]. As executable programs are values, incremental linking of code and incremental loading of data reduce to the same problem and are handled by the run-time system.

Note that although the languages listed above use procedure closures to represent executable programs this is not essential to the schemes described in this section. All that is required is some mechanism to denote executable programs as values in the programming language.

The persistent store may subsume the functions of the file system, or the persistent store and file system may be used together. Figure 3.7 shows the relationships in a hybrid system in which source programs are kept in the file system and executable programs in the store. Here the program library contains only source programs; the corresponding executable programs reside in the store. The combined ellipses and diamonds in the diagram represent these procedure values. As the linking process can be achieved without a separate linker, no intermediate programs are required.

The figure shows causations and associations between source programs and executable programs as before. There is also a causation from the main executable program $e_1$ to the data item $v_1$ which is created by execution of that program. Data item $v_1$ contains a direct link to data item $v_2$, as does $e_1$, which also contains direct links to other executable programs; these direct links replace the associations between executable programs and library programs shown in Figure 3.6.

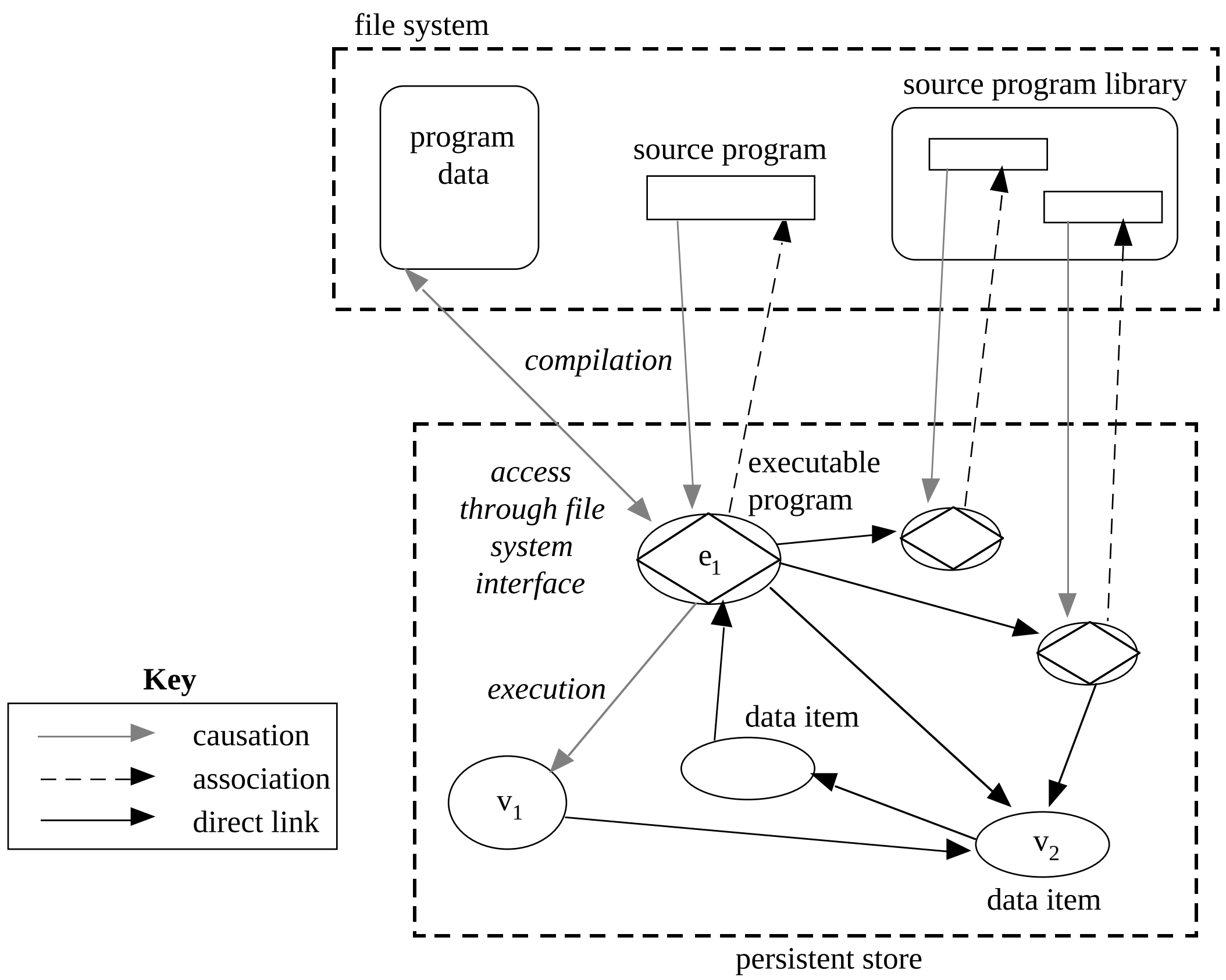

**Figure 3.7: Relationships in a hybrid persistent / file-based system**

Figure 3.8 shows the relationships in a persistent system where all components and data reside in the persistent store. The combined ellipses and rectangles represent source programs that are denotable values in the programming language. These values may be, for example, text strings or abstract syntax trees.

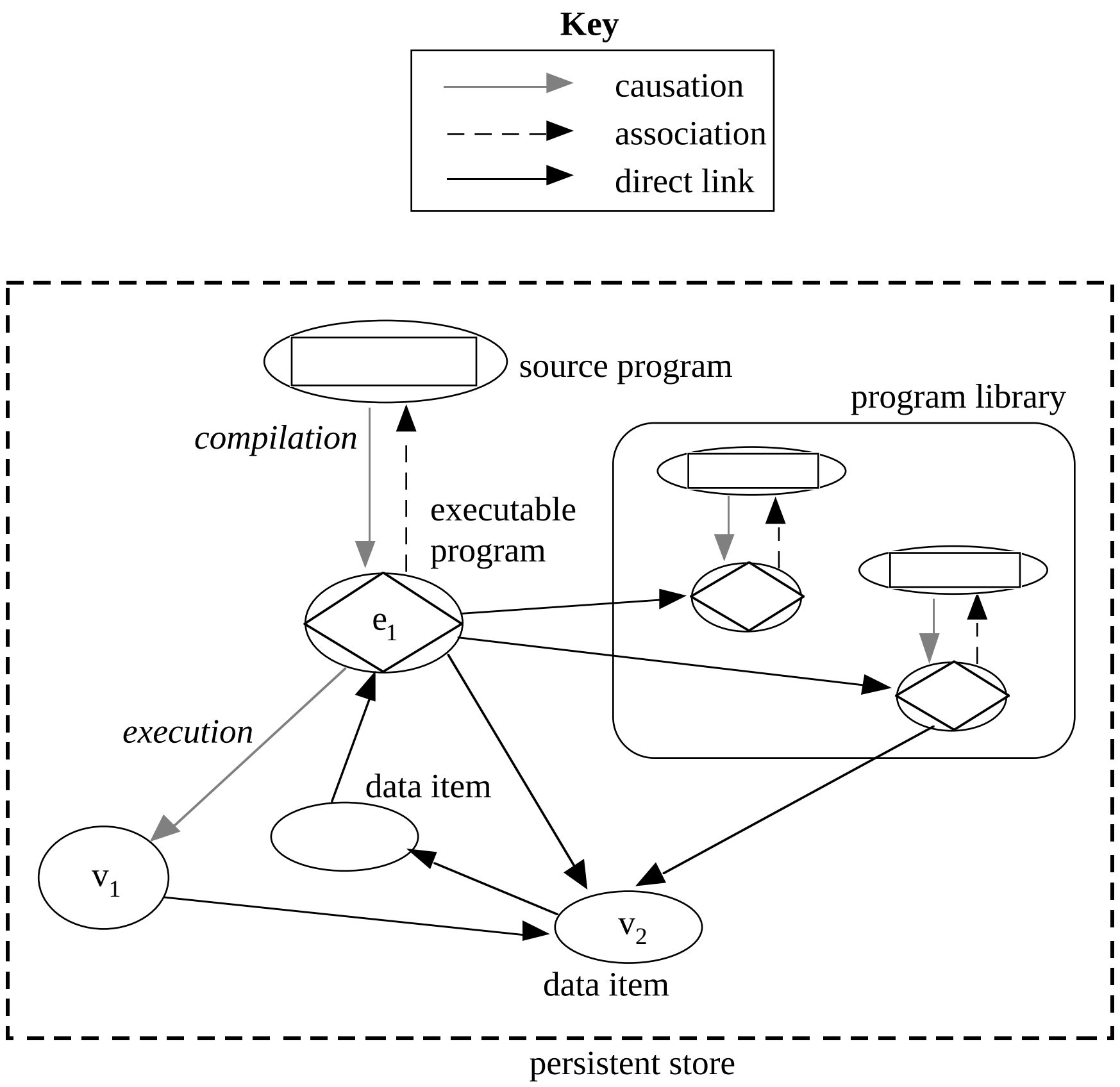

**Figure 3.8: Relationships in a persistent system**

Both schemes shown have the advantage that executable programs are associated with the others that they use by direct links. Once established these links are guaranteed to remain in place. In contrast, the integrity of the associations between executable programs that reside in the external storage system, in a non-persistent system, depends on the programmer following certain conventions. For example the deletion of a source program from the program library might break these conventions.

The scheme shown in Figure 3.8 has the further advantage that the source programs, being in the persistent store, are brought under control of the language. This allows the system to be self-supporting: the environment in which programs are composed, compiled and executed can itself be implemented using the same programming language. Functions that are normally controlled by the operating system can then be integrated with the programming language. These include source code control and versioning, source level debugging, controlling the configuration of applications built from multiple components, documentation, etc. A number of workers are currently addressing the problems of supporting the whole software engineering process within an integrated persistent system [Coo90a, Far91, DCC92, DMD92, KCC+92b]. Type-safe linguistic reflection is needed to implement such a system.

#### 3.2.3.4 Hyper-Programs

Bringing executable programs into the persistent store allows associations between them to be enforced by direct links. It would be beneficial for the associations between executable programs and source programs to be replaced by direct links also, for the same reason, i.e., they could not then be accidentally corrupted. Then each executable program would contain a direct link to its corresponding source program. As an executable program can also contain direct links to other data items in the persistent store, a source program must be able to denote those data items in order to represent the executable program accurately. This requires the use of hyper-programs as source representations.

Figure 3.9 shows the relationships in a hyper-programming system. Each executable program contains a direct link to its source hyper-program. Each of the other direct links contained in an executable program is duplicated in its corresponding hyper-program.

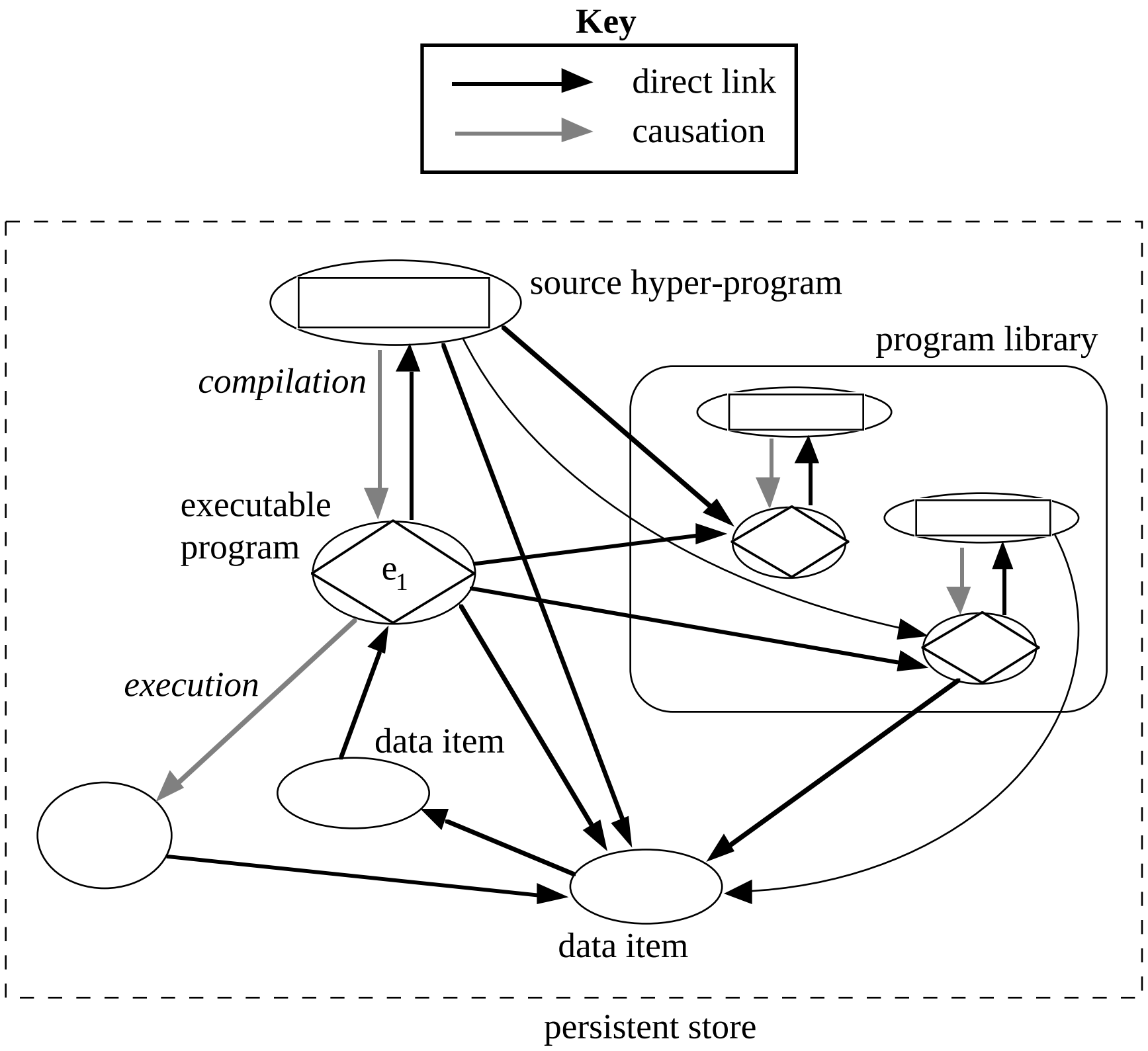

**Figure 3.9: Relationships in a hyper-programming system**

To illustrate the necessity of hyper-programs for providing accurate source representations of executable programs, consider the situation where multiple executable programs have direct links to a store location as illustrated in Figure 3.10:

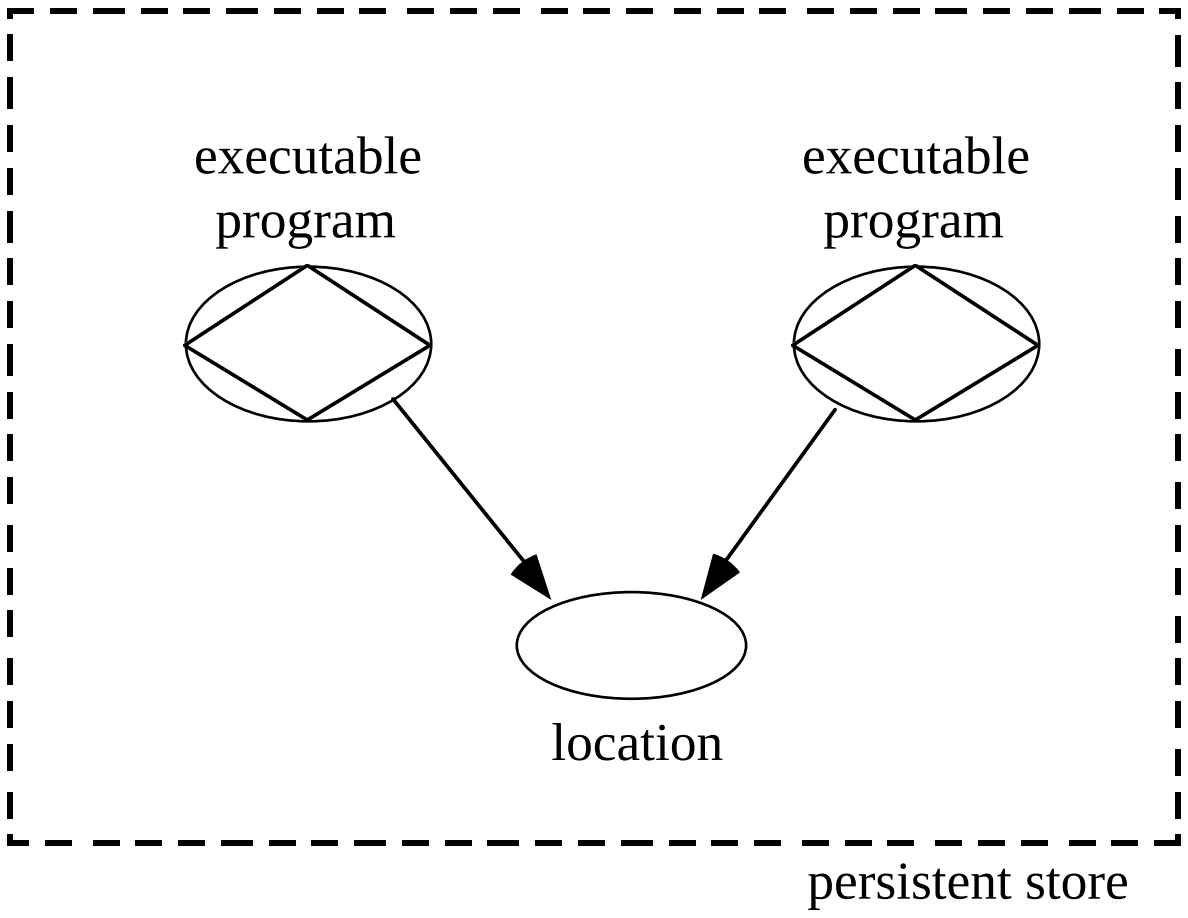

**Figure 3.10: Executable programs sharing a location**

The problem arises in supplying separate source programs for each of the executable programs. Unless there is a direct access path to the location from a persistent root, and in general there does not have to be one, conventional source representations do not provide any notation with which the location can be denoted in a source program.

To illustrate how the situation in Figure 3.10 might arise, Figure 3.11 shows how the executable programs could be created in Napier88:

```
let i := 0

in PS() let inc := proc() ; i := i + 1
in PS() let get := proc( → int ) ; i
```

**Figure 3.11: Creating a shared location in Napier88**

This program first initialises an integer variable *i* with the value 0. It then creates two persistent procedures that operate on *i*, the first incrementing it by 1 and the second returning its current value. The procedures are made persistent by declaring them in the context of the persistent root environment, obtained by calling the pre-defined procedure *PS*. Although the store location corresponding to the variable *i* is not declared in the persistent environment, it will persist because it is reachable from the procedures *inc* and *get* which are themselves persistent. The result of executing this program is that the persistent store contains the two procedures and the shared integer location which is not directly accessible from the persistent root.

The problem in supplying source representations for *inc* and *get* is to denote the same integer location in both source representations. With existing language notations the only way to achieve this is to supply a single source program that represents both procedures, such as that in Figure 3.11. However this is unsatisfactory in general as it forces all executable programs that share locations to be represented in the same source program. This could involve most of the components of a large application, in which case it would nullify one of the benefits of splitting the application into smaller components, that of being able to modify a component independently of the others.

A better solution is to change the program notation by introducing hyper-programs as source representations. It is then possible to denote a shared location in the source program for a single executable program, by including a token for the location within the hyper-program. This makes it feasible for every executable program to contain a direct link to its own source hyper-program. Figure 3.12 illustrates this for the procedure *inc*:

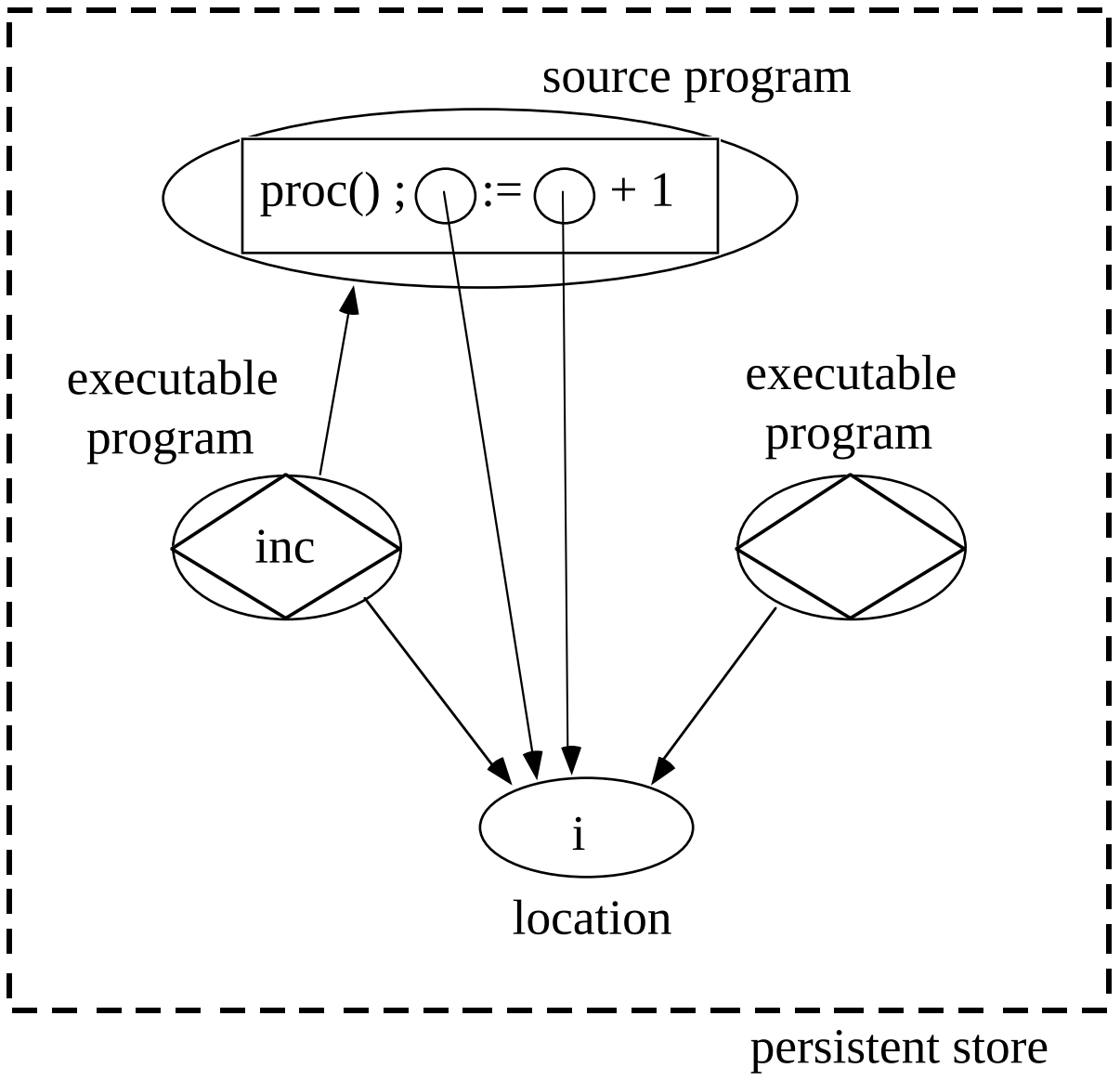

**Figure 3.12: Executable program with direct link to hyper-program**

Thus the use of hyper-programs as source representations allows associations from executable programs to source programs to be replaced by direct links, further improving the robustness of the programming system by eliminating accidental changes to or deletions of source programs.

### 3.2.4 Flexible Linking Mechanisms

Programming languages support a number of different mechanisms for establishing direct links from programs to persistent values, locations and types. The degrees of freedom include constancy or variability, linking to L-values or R-values [Str67], and the time at which the linking takes place. The focus here is on the range of times available. Some possible times are during program composition, during compilation, during a separate linking phase, and during execution.

The principal varieties of programming system identified earlier were file-based, persistent and hyper-programming systems. Another possibility is a compile-time linking system in which the tokens embedded in a program are associated with data items in the persistent store when the program is compiled rather than when it is written. The linking times possible in each of these systems are shown in Figure 3.13. From here on it will be assumed that the hyper-programming systems under consideration incorporate facilities for compile-time linking as well as composition-time linking.

| **System** | **Linking Time** | | | | | | | |
|---|---|---|---|---|---|---|---|---|
| | composition | | compilation | | linking phase | | execution | |
| | program | data | program | data | program | data | program | data |
| file-based | | | | | • | | | • |
| persistent | | | | | • | • | • | • |
| compile-time linking | | | • | • | • | • | • | • |
| hyper-programming | • | • | • | • | • | • | • | • |

**Figure 3.13: Comparison of possible linking times in various systems**

File-based systems allow links to existing data to be formed only at run-time. Links to existing programs are formed during a linking phase by copying library programs into the main program. In persistent systems a linking phase can be simulated using first class functions. As executable programs are a form of data, linking to both programs and data can be performed either at link-time or run-time. Compile-time linking systems support these same linking times and also allow linking to programs and data at compilation-time.

A hyper-programming system supports all the linking times described. The programmer can specify various linking times as appropriate for different components of an application. Deciding when components should be linked into a main program involves trade-offs between program safety, flexibility and execution efficiency.

Run-time linking gives flexibility as the data (*data* will now be used to denote both programs and other kinds of data) accessed does not have to be present in the persistent store, file system or database before run-time. Indeed the access path to the data may not be known until run-time. Program safety is low as the data may not be present when the program is run, causing a run-time failure. Execution overheads are also higher, in strongly typed systems, as the type of the data must be checked dynamically. This kind of linking is possible in many systems, for example, C, Pascal, Ada, Smalltalk-80 [GR83], PS-algol, Napier88.

A distinct linking phase occurs in some file-based systems between compilation and execution, involving the copying of other executable programs into the main executable program. A similar effect can also be achieved in persistent languages with higher-order procedures, where it allows all types of data to be linked into an executable program before run-time. In the latter case it provides improved safety and efficiency over run-time linking, as checks for the data's existence and type are performed before run-time. Flexibility is reduced as its use requires the data to be present earlier.

Linking at compilation-time increases safety and efficiency, bringing checks further forward in time, and reduces flexibility correspondingly. With this mechanism the data linked into an executable program is fixed.

Composition-time linking is the least flexible of the alternatives described as the data bound to must be present at the time that the program is written. It offers the greatest safety as access to the data is always maintained once it is bound into the source code, even if the source code is edited and re-compiled. This is not true of the other linking styles where editing of the source code requires all links to be re-established. Efficiency is slightly increased overall as the access path to the data, whether it is expressed by textual code or by user gesture, must be followed only once, at composition-time, and not on every re-compilation.

Figure 3.14 shows the linking opportunities in several different systems, for a program that accesses persistent data repeatedly throughout its execution. The line above each individual diagram shows the range of times during which the *first* linking to persistent data may be performed, while the line below shows the range during which the *last* linking may take place. Each linking process may cause a failure due to the data not being found or not having the expected type.

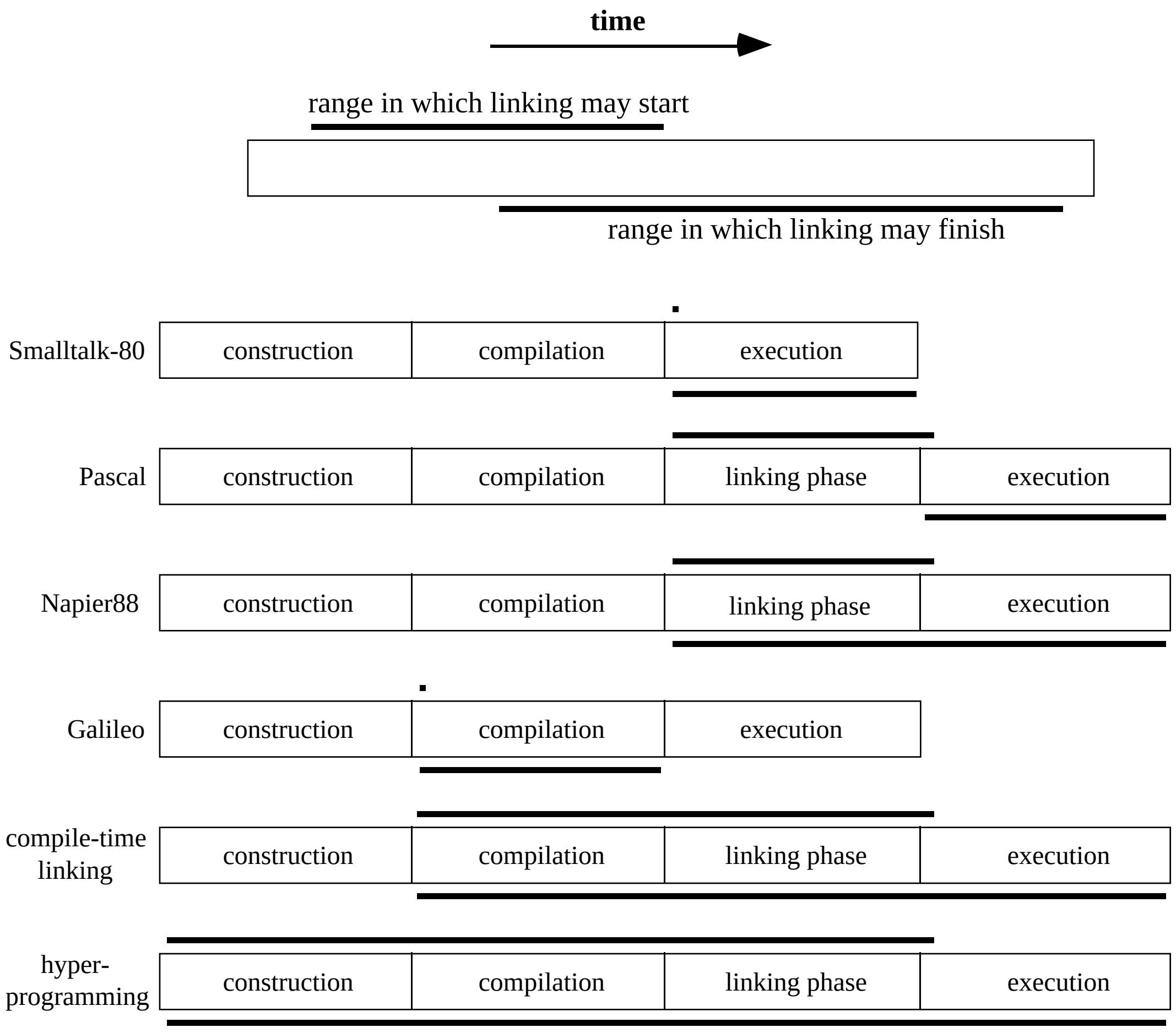

**Figure 3.14: Comparison of linking opportunities in various systems**

The systems shown exhibit a spectrum of possible linking times, from Smalltalk-80 where all linking is performed at run-time, to the hyper-programming system where linking may be performed during any of the phases.

**Smalltalk-80**: All linking is performed at run-time. With the example program that accesses persistent data repeatedly, the first linking occurs near the beginning of execution, and the last linking at any time during execution.

**Pascal**: A distinct linking phase allows other executable programs to be linked into the program before execution. Non-program data, from the file system, is linked to at run-time. Complex data structures must be reconstructed from a flattened form. As Pascal allows direct links between data items in the run-time environment, links to persistent data items need only be established the first time the program accesses them. This means that the establishing of links may finish before the end of execution.

**Napier88**: Persistence allows links to be formed to data structures directly rather than having to reconstruct them. The establishing of links can start during a simulated linking phase or at the beginning of execution, and can finish any time after the beginning of the linking phase.

**Galileo**: Programs are compiled in the context of a persistent environment, thus linking to persistent data starts at the beginning of compilation, and may finish at any stage during compilation. Note that the programmer does not have explicit control over compilation: program fragments are compiled and executed interactively as they are entered.

**compile-time linking**: Linking opportunities are similar to Napier88 except that linking may start or finish as early as the beginning of compilation.

**hyper-programming**: Linking may start or finish even earlier, at the beginning of the program composition process. This gives the widest range of possible linking times and thus the greatest flexibility.

The positions of the left hand ends of the lower lines in each diagram are significant. These show the earliest possible times by which all the linking and checking for the program may be completed. The further a line extends to the left, the earlier it is possible to be confident that suitably written programs in that system will not fail due to linking errors. The exception to this is the hyper-programming system, where although linking to data may be completed during program composition, the type checking of its compatibility with the program is not performed until compilation-time.

### 3.2.5 Program Succinctness

Persistent systems offer significant savings over non-persistent systems regarding the data access code required. One empirical study concluded that 30% of the code in a large set of commercial non-persistent programs was dedicated to transferring data to and from an external storage system [IBM78]. In a persistent system this code is replaced by access specifications. Recent measurements of Napier88 programs have suggested that these access specifications occupy around 13% of program code [Sjø92], a considerable reduction on 30%. The intellectual effort required to write the code is also significant: in writing access specifications in a persistent system the programmer is not concerned with programming transformations between structured and flattened formats.

A hyper-programming system gives a further improvement in conciseness as the access specifications can in some cases be replaced by tokens that denote persistent data items. The information that was specified in the access specifications is provided by the interactive gesturing by which the programmer points out data items to be linked in. The measurements of Napier88 programs found around 20% of identifiers referring to persistent data. Further work is required to measure the proportion of this data that is available for linking at hyper-program composition time.

Figure 3.15 shows the persistent data access code that appears in source programs in the various cases:

| System | Access path code |
|---|---|
| non-persistent | file access + importing + exporting |
| persistent | access path + type description |
| hyper-programming<br>(data present at composition time)<br>(data not present at composition time) | <br>token<br>access path + type description |

**Figure 3.15: Comparison of access path code**

## 3.3 Procedure Representations

As hyper-programs can contain direct links to values and locations in the persistent store they can be used to represent executable programs, including those with links to shared locations. This provides a convenient representation format for procedure values, the benefits of which are now described.

As described earlier, associations between executable programs and source programs can be replaced by direct links. When a procedure value is created, the compilation system can insert a direct link to its hyper-program source program. Given referential integrity, the source code will then remain accessible for as long as the procedure value.

The presence of hyper-program source representations allows browsing tools to display meaningful representations of procedure values, showing both source code and direct links to persistent data items. This may aid software reuse since documentation in the form of the original source code can be made available for every procedure value in the persistent store.

Hyper-programs allow separate procedure source representations since shared locations can be denoted by tokens. A further consequence is that one of a group of procedures that share values or locations can be replaced by a refined version without the need to replace the others. This reduces the cost of modifying applications that are composed of multiple procedures.

Figure 3.16 shows the example program given earlier:

```
let i := 0

in PS() let inc := proc() ; i := i + 1
in PS() let get := proc( → int ) ; i
```

**Figure 3.16: Procedures with a shared location**

After this program has been executed and the procedures *inc* and *get* linked into the persistent store, a hyper-programming system allows one of them to be replaced by a new version that shares the same location, without having to replace the other. For example, *inc* can be replaced by a version that increments by 2 on each call. To achieve this the programmer first obtains the hyper-program source representation for *inc* and makes a copy of it. This new copy has the same store location, containing the value of *i*, linked into it. The copy is then edited to change the increment value to 2, compiled and executed to produce a new procedure value with that store location linked into it. Finally the new procedure is assigned to the store location of *inc*. The sharing between the new procedure and *get* is preserved without any change to *get*. This is illustrated in Figure 3.17:

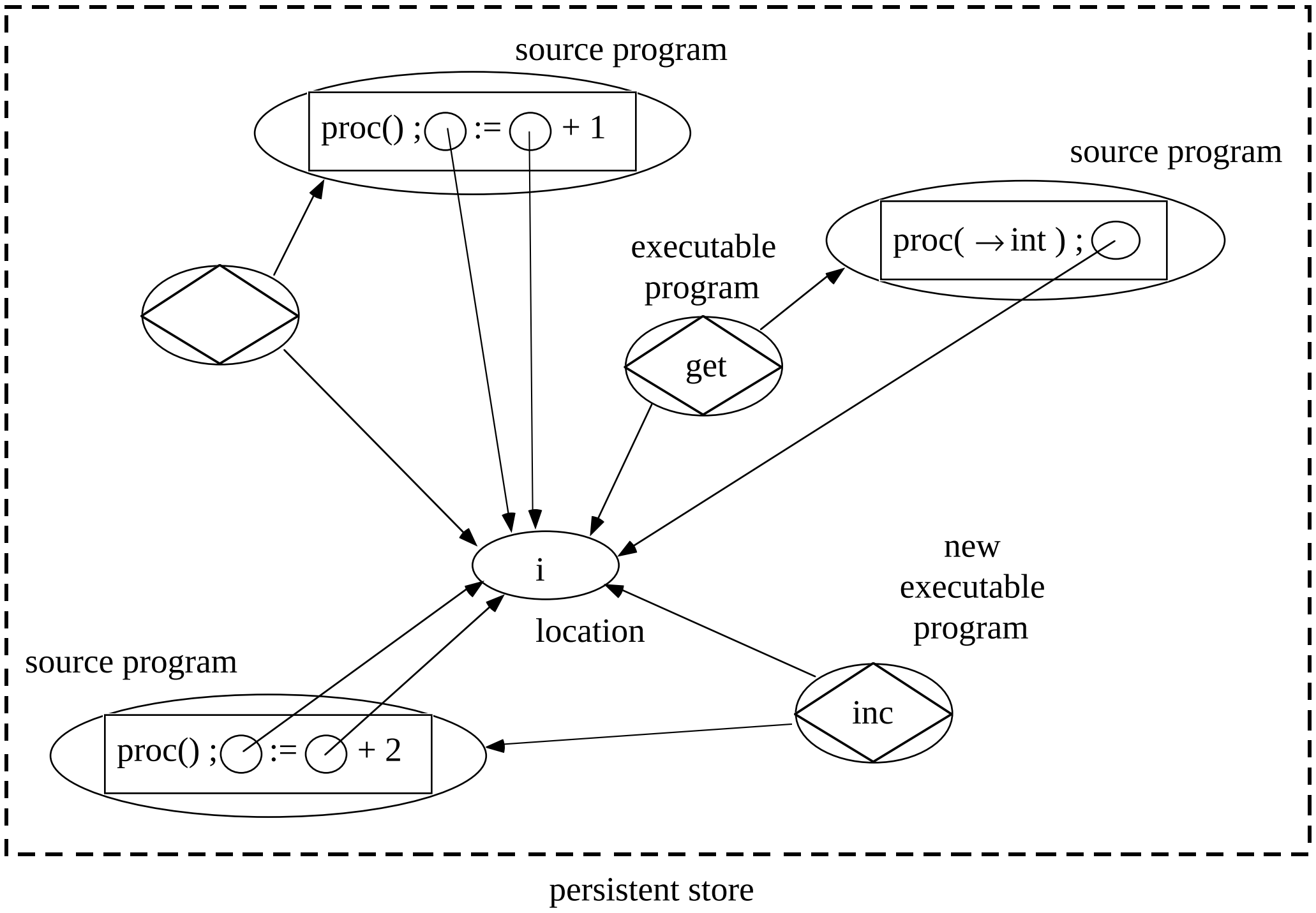

**Figure 3.17: Preserving sharing of a location**

The use of hyper-program source representations for procedures in this way avoids having to replace all procedures that share locations when a single one is changed. Another advantage is that the same shared locations are retained after the replacement of a procedure. Without hyper-program source representations not only do all the procedures have to be replaced in order to preserve sharing, but new shared locations must be created and the values that were previously shared copied into the new locations.

There is some tension between the benefits of being able to inspect procedure closures, described above, and the protection role in which procedures are sometimes used [AM85, MBC+90]. Procedures with encapsulated state may be used to control and limit access to that state. This was illustrated in the example where the location *i* was not directly accessible after the execution of the program, but only through the procedures *inc* and *get*. The implementor of an application may wish to prevent direct access to the internal implementation details, or even for those details to be completely hidden so that users cannot discover how the application is implemented. The ability of a hyper-programming system to support access to the source code of a procedure and the state bound into it may give the user too much freedom. It may be necessary for the system to support different access privileges for different procedures. For some the hyper-program source code could be freely available, while for others access might be restricted to the original implementor by use of a password protocol [CDM+90], or even completely unavailable. Restricting source access to the implementor would allow implementation data structures to be examined or repaired when bugs in the procedures that operate on them were discovered, without unduly compromising protection from users.

## 3.4 Hyper-Worlds

There are a number of components that a persistent programming environment should support if it is to provide for the software engineering process as a whole. These include:

- program composition, compilation and execution;
- storing of source and compiled versions of programs;
- debugging;

- documentation;
- decomposition of large application programs into components, and organisation of those components;
- navigating the persistent store to locate programs and other data with given attributes;
- querying of the types of programs and data in the persistent store.

The model of hyper-programming as described so far allows source programs to contain links to any other data in the persistent store. In large scale systems this generality may lead to several problems. Firstly, the store may become intellectually unmanageable as the number of links increases. Secondly, evolution of application programs by substituting new versions of their components becomes difficult to manage if unrestricted linking to the components is permitted—it may be necessary to locate each data item linked to the component being substituted and determine whether a new version of the data item is required in turn. In addition the model described does not provide a uniform framework for storing meta-data about application components.

One research topic is the provision of additional structure over a basic hyper-programming system to address these needs. The *hyper-world* model offers the programmer a loose coupling mechanism to offset the disadvantages of the tight coupling made possible by hyper-programming. In this model, based in part on that described in [WA86], the persistent store is partitioned into a number of application spaces or hyper-worlds. Each hyper-world contains the program components and data used by an application, and a schema that describes their relationships. Each hyper-world has a single visible component which may be linked to from outside the hyper-world; no other components inside the hyper-world may be linked to from outside.

The schema includes documentation information, a type description and hyper-program source for each component. It also includes a representation of the component linking topology, and a list of type definitions local to the hyper-world. This allows the programmer to perform various queries over the components, and to determine the implications of replacing a component with a changed version.

The partitioning supported by hyper-worlds may reduce problems such as keeping track of inter-component links to a manageable scale, by restricting the region of interest from the entire persistent store to the hyper-world. It may also allow type-checking to be performed more efficiently.

Figure 3.18 shows a representation of a persistent store containing nested hyper-worlds and linked components:

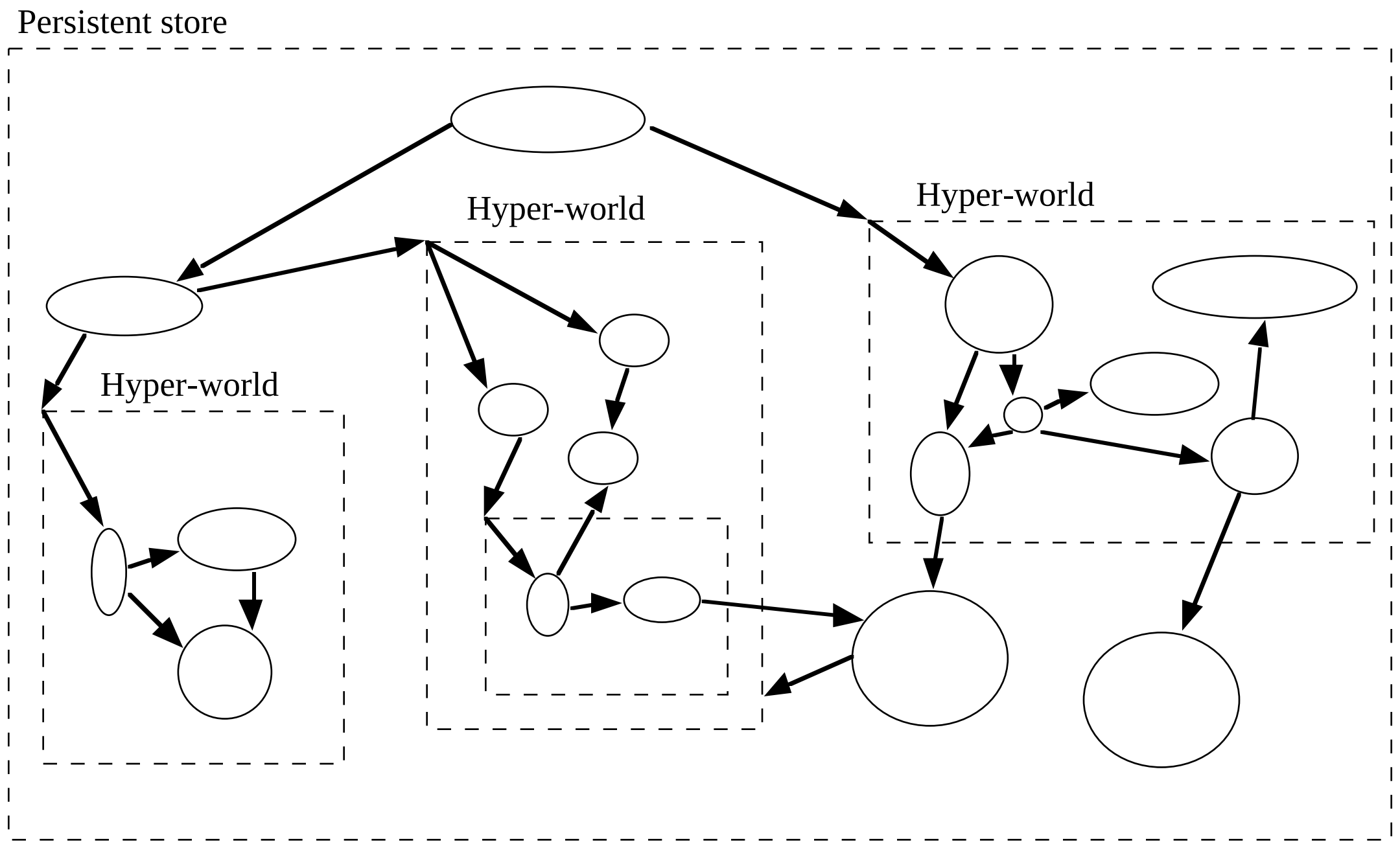

**Figure 3.18: A store with hyper-worlds**

## 3.5 Conclusions

There are many situations when the programmer writes code to access data items in the persistent store, knowing that those data items are present in the store at the time of writing. This chapter has shown how data can be linked directly into a source program as opposed to the program containing instructions on how to link to it at run-time. This gives the benefits provided by interactive languages: greater program safety as there is no danger of losing access to the data during the time between writing and execution, and better efficiency as run-time type and access path checks are factored out, while retaining the flexibility of being able to link to the store dynamically when required.

An analysis has been given of the program entities and their inter-relationships in a hyper-programming system, and compared to those found in file-based and existing persistent systems. A number of benefits of using hyper-programs have been described. These include being able to: perform program checking early; enforce associations from executable programs to source programs with direct links; support an increased range of linking times; reduce program verbosity; and provide source representations for procedure closures.

A framework, hyper-worlds, has been proposed for supporting ‘programming in the large’ in the context of a hyper-programming system. It allows the programmer to impose a degree of partitioning on the persistent store, in order to aid intellectual manageability and improve execution efficiency.

# 4 Hyper-Programming Tools

## 4.1 Introduction

The previous chapter stated that it was desirable for the programmer to be able to write programs that operate on data items in the persistent store, without having to write textual access specifications for the data items in the program. An access specification describes the type of a data item and a path by which it may be reached from a root of the persistent store. Instead of supplying this explicitly the programmer may select a graphical representation of the data item by gesture and have the system incorporate some specification of that data into the program. This specification may take a number of forms, depending on the stage in the software development process at which the mapping from specification to data is resolved.

- When the mapping is resolved during program composition the specification inserted into the program is a token representing a direct link to the data item itself. This occurs in hyper-programming.

- When the mapping is resolved during compilation the specification inserted into the program is a tag identifier. At the time of tagging, at program composition, the tag identifier may be associated with the data item itself, or with its access specification. In the first case a link to the data item currently associated with the tag identifier is inserted into the executable code at compilation-time, while in the second case the data item that currently has the tagged access specification is linked.

- Similarly, when the mapping is resolved at run-time a tag identifier is inserted into the program during composition. At compilation-time the access specification of the tagged data item is inserted into the executable code. The program thus operates on the data item that currently has the tagged access specification at run-time. It may also be necessary to provide a mechanism for the programmer to tag access specifications that do not correspond to any existing data item at the time of composition.

The main part of this chapter describes the programmer's view of a hyper-programming system for Napier88, in order to give an impression of the technique's impact on the programming process. The system also provides some support for compilation-time linking as described above and these facilities are described at the end of the chapter. Support for linking access specifications evaluated at run-time has not been implemented.

## 4.2 Hyper-Programming Tools

The system provides hyper-programming tools that support two main functions:

- Locating data items in the persistent store, either values, locations or types.

- Displaying and editing hyper-programs. This involves being able to link data items from the persistent store into the hyper-programs.

Browsing tools are used to display representations of data items in the persistent store, and to allow the programmer to explore the store by navigating along links between data items. Graphical representations emphasise their linking topology.

The hyper-program editing tool displays hyper-programs as text with embedded light-buttons representing the tokens that denote data items in the persistent store. As well as conventional text editing it allows tokens to be inserted and deleted, and the data items associated with tokens to be examined.

These tools are used in conjunction to support the construction and editing of hyper-programs. The programmer uses the browsing tools to identify and select data items in the persistent store. Tokens for them are then linked into hyper-programs under construction. The browsing tools may be used again to display representations of the data items linked into existing hyper-programs.

The tools implemented represent one particular set of solutions to the requirements of hyper-program construction, and others are possible. They incorporate features based on a number of other systems [DB88, DCK90, KD90, Far91, FDK+92, KCC+92a].

### 4.2.1 Data Representation Display Format

The hyper-programming system uses windows to display hyper-programs, messages and representations of Napier88 data items. The windows are similar to those used in the Open Look [Sun89] and Macintosh [App86] graphical user interfaces.

The form of the data representations varies according to the type of the data. Instances of the scalar types *int*, *real*, *string*, *bool*, *pixel* and *file* are displayed textually in a single output window. Instances of the types *image* and *pic* are displayed graphically in individual output windows. All other types are displayed as menu windows, with an entry for each component of the type. Although there are an infinite number of Napier88 types, the number of type constructors is small and finite and all instances of a given constructor are displayed in the same format. The constructors are: *structure*, *variant*, *proc*, *abstype* and *vector.* Instances of the type *env* are also displayed as menus.

All windows displayed by the system have a title bar at the top, and some have a close box at the left of the title bar and resize handles at the corners. Any window can be moved around the screen by dragging its title bar using mouse button 1. When present the close box can be used to convert a window to its iconic form by clicking on it, and a resize handle can be used to alter the size of a window by dragging it, both with mouse button 1.

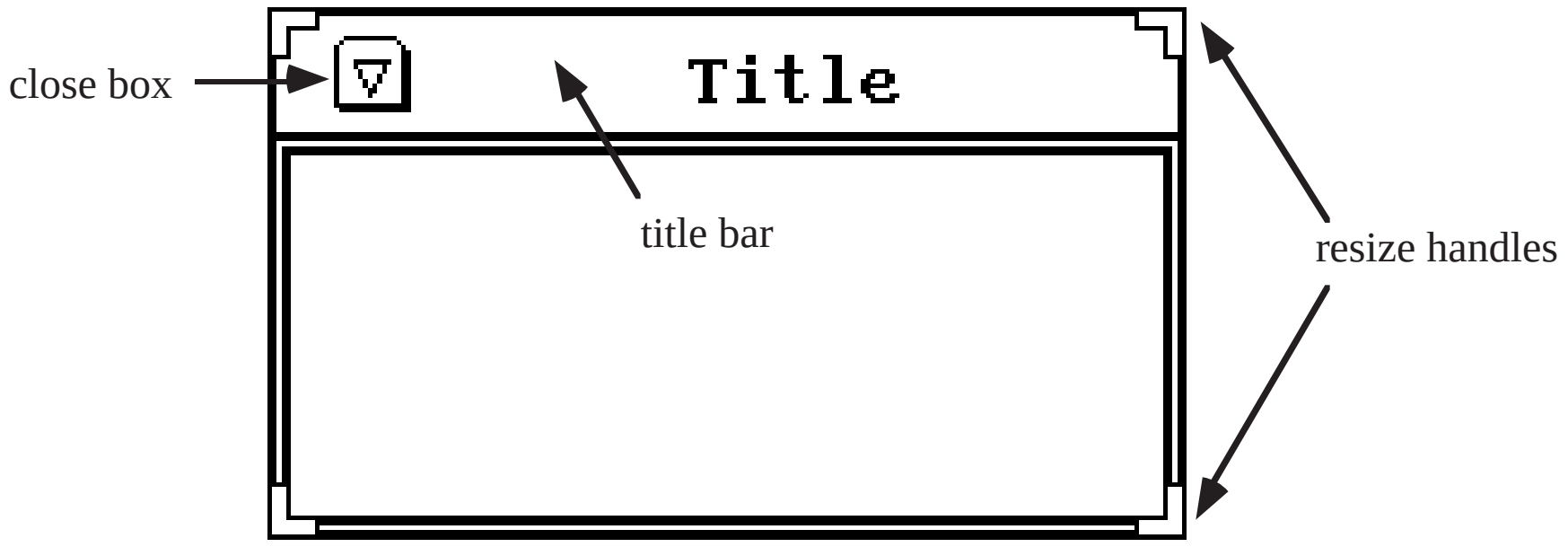

**Figure 4.1: Parts of a window**

The browser displays connecting arrows between menu windows to show direct links between the data items they represent. A menu entry can be selected to cause the browser to display the value of the corresponding data item. Some examples of browser windows are shown in Figure 4.2:

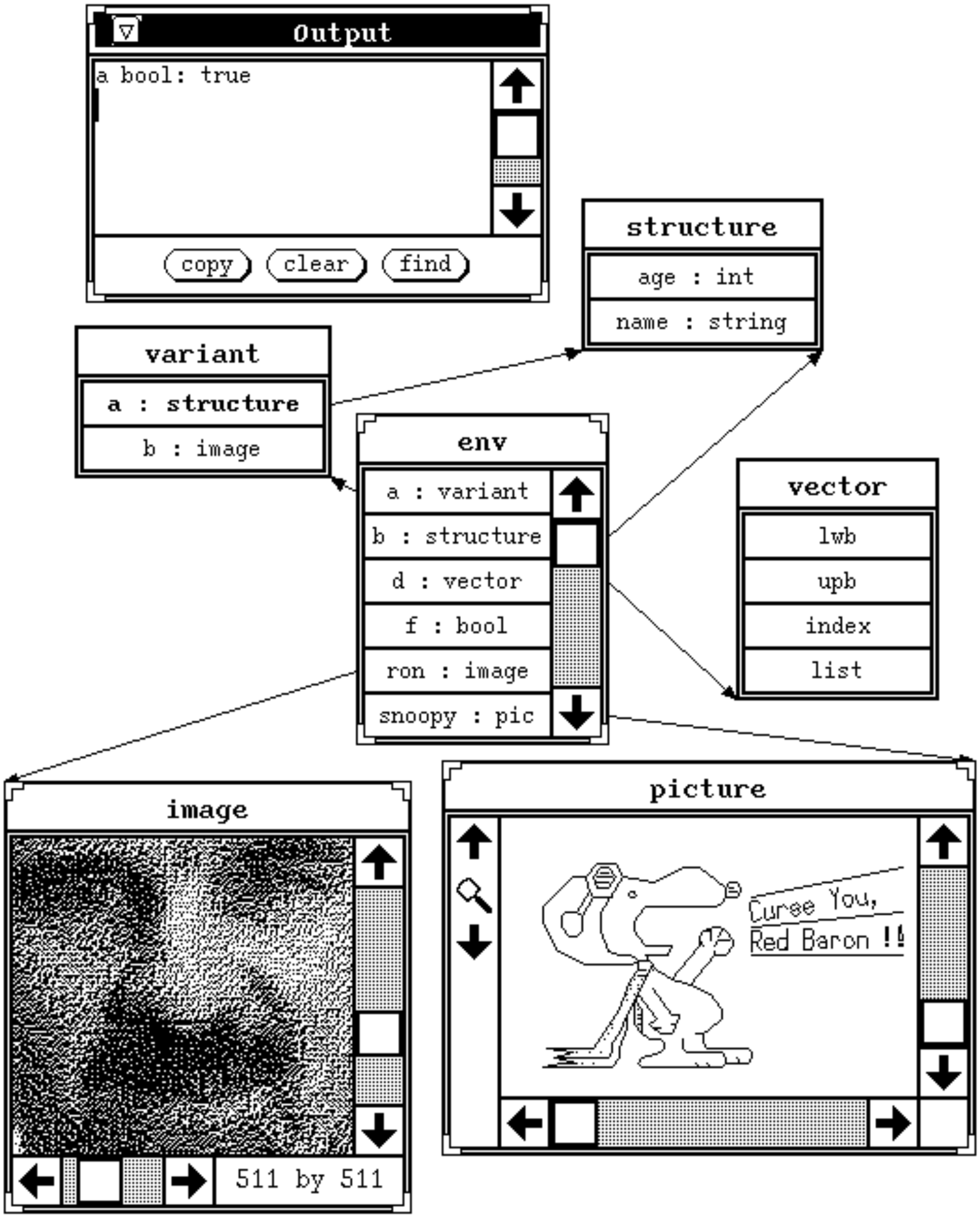

**Figure 4.2: Graphical representations of Napier88 values**

Figure 4.2 shows an environment in the centre and the values of some of its bindings. The arrow between the variant and the structure shows that the structure value present in the variant is also a component of the environment. Structures and environments are displayed as menus, the entries of which can be selected to examine individual fields. Each entry shows the type in the case of a base type, or the type constructor in the case of a constructed type. Variants are also displayed as menus, with the difference that the branch actually present is shown in bold type and this is the only entry that can be selected. Vectors are displayed as fixed menus with the four entries shown. When selected they display the bounds of the vector in the text output window, a particular element, and all the elements respectively. Images, which represent bitmaps, and pictures, which represent line drawings, are displayed in scrollable windows. Picture windows support altering the magnification to allow zooming in and out. Procedures are represented by menus with a single entry to display the source code. The use of procedure menus will be described in more detail later.

At any one time there may be at most one window or menu entry *highlighted*. This is indicated by a highlighted title bar or entry label respectively. Various operations can be performed on the highlighted window or menu entry. To highlight a window the programmer clicks mouse button 1 on the window border. Figure 4.3 shows the highlighting of a window that represents an environment value:

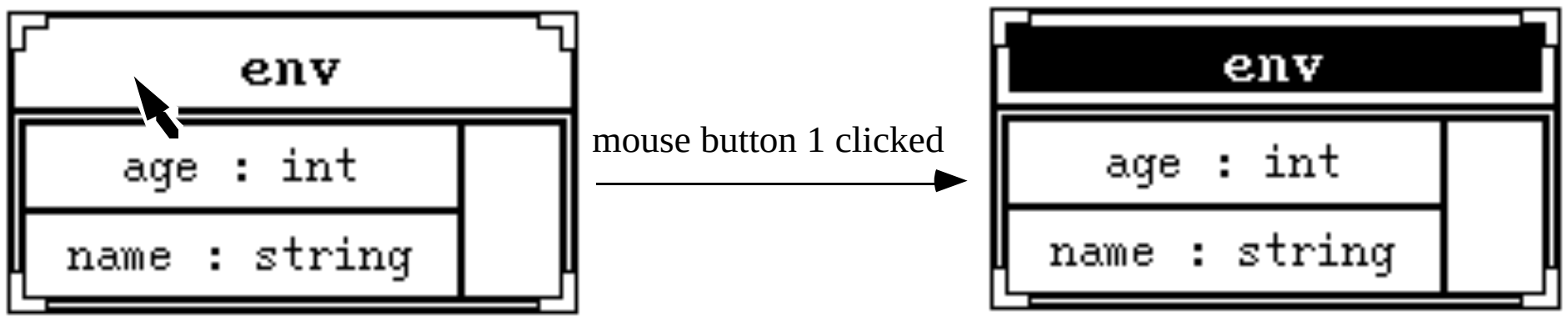

**Figure 4.3: Highlighting a window**

A menu entry can either be highlighted in the way described above, or *selected*, in which case the corresponding data item is displayed by the browser and the menu entry does not remain highlighted. The former is achieved by clicking mouse button 1, and the latter by holding down mouse button 3 until a sub-menu appears and then releasing the button to select the *show* sub-menu entry:

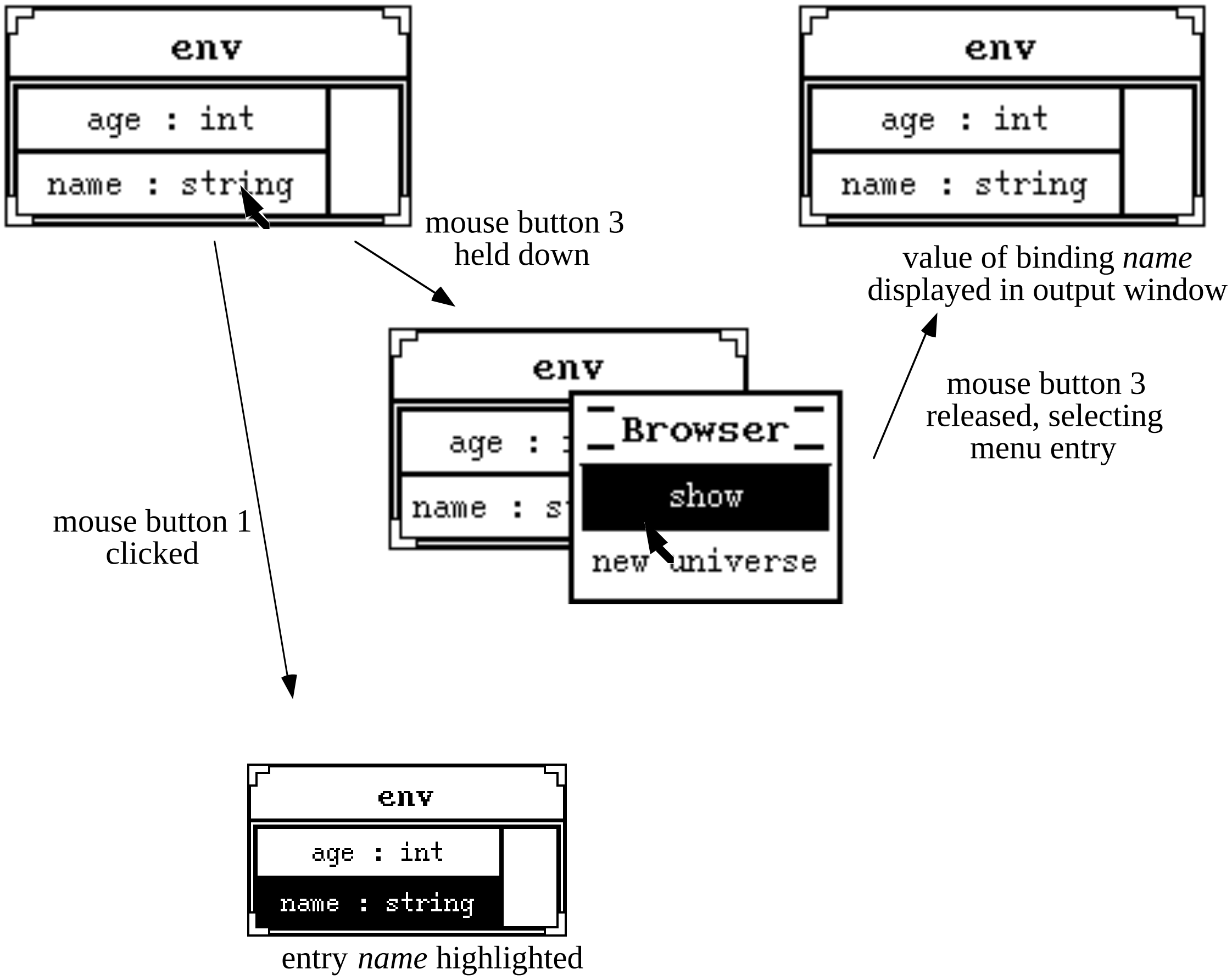

**Figure 4.4: Selecting and highlighting a menu entry**

The screen may become cluttered when the programmer browses a large data structure. *Universes* help the programmer to organise the screen area. Each universe is a window containing a separate invocation of the browser, allowing representations of values to be displayed and moved around independently of other universes. To create a new universe the programmer highlights a menu entry and then selects *display in new universe* from the background menu brought up by holding down mouse button 3 over the background:

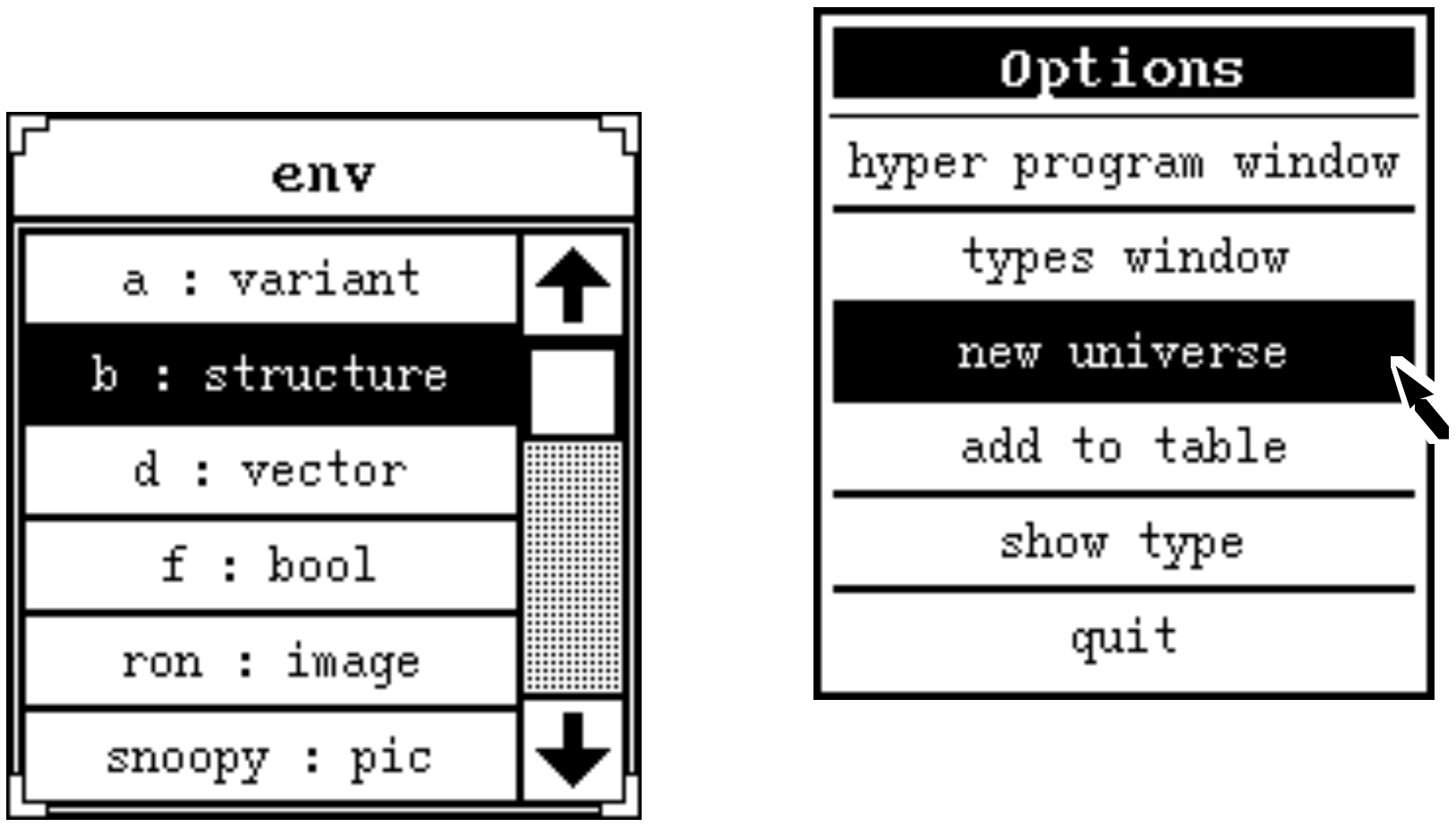

**Figure 4.5: Creating a universe**

A new universe can also be obtained by holding down mouse button 3 over a menu entry, and then selecting *new universe* from the sub-menu that appears. A new window is then created and the value of the data item displayed within it as shown in Figure 4.6:

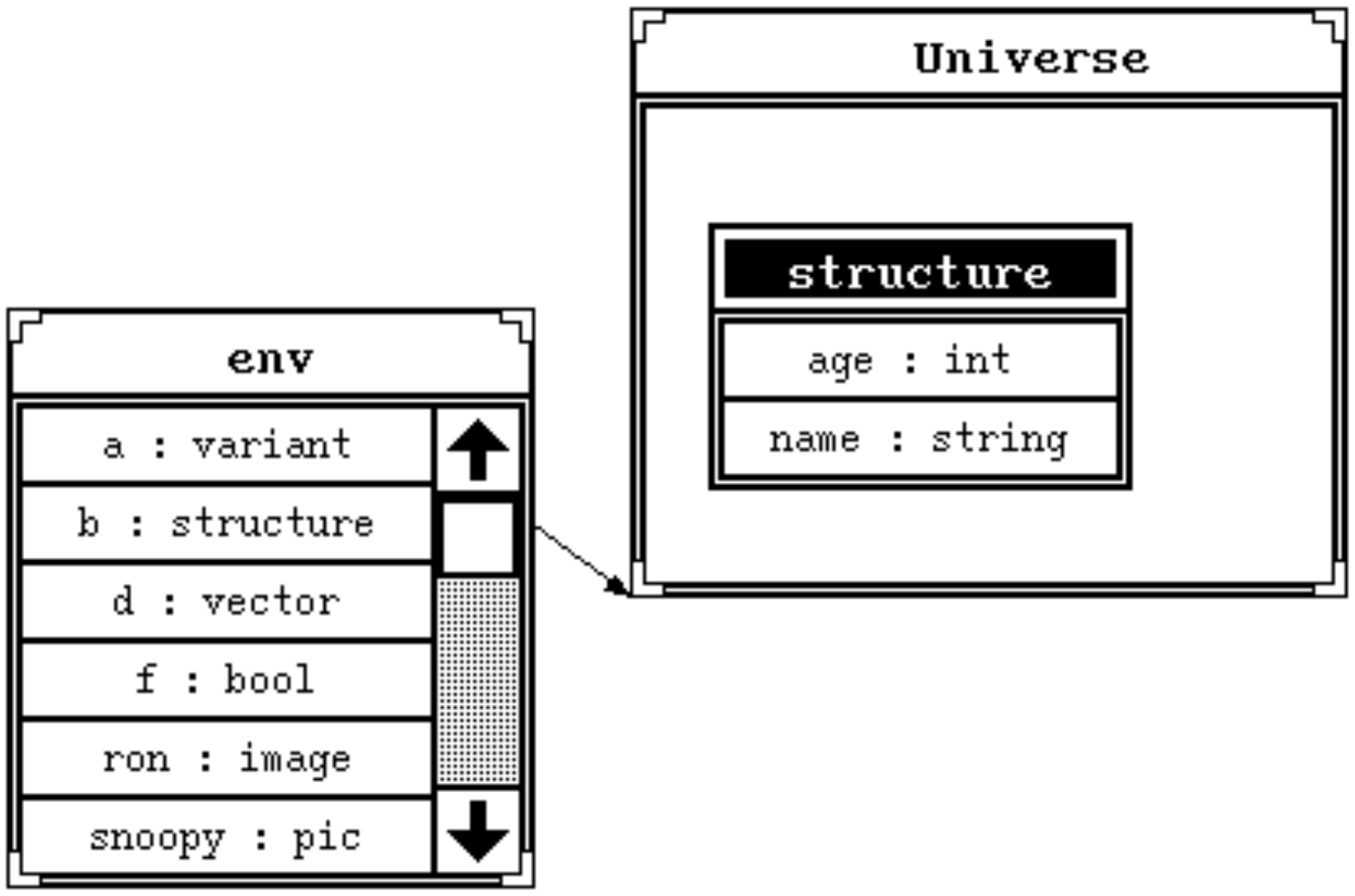

**Figure 4.6: Value displayed in a new universe**

The representations of the new value and any others accessed from it are confined to the universe window, so they are kept separate from the rest of the visible data. Universes provide a grouping mechanism in that all the objects in a universe can be moved or deleted in one action by operating on the window containing them. Any number of universes can be created and they can be nested to any degree.

### 4.2.2 Constructing a Hyper-Program

The use of the hyper-programming system will be illustrated with an example. The example involves constructing a persistent procedure that takes a picture, performs a transformation on it and copies the result repeatedly onto the screen to give a tiling effect. The procedure takes as a parameter a procedure to perform the transformation. The main procedure also has two data items linked into it: the picture itself and a procedure to make a single copy of a picture on the screen.

The following requirement for the main procedure is assumed, that on each execution it operates on the same picture, but on the most up-to-date version of the display procedure. This is achieved by linking the main procedure to the picture value and to the environment location containing the display procedure. Figure 4.7 shows how the example can be programmed in standard Napier88, using access specifications that are evaluated at run-time.

```
use PS() with fishPics : env in
use fishPics with shark : pic ; displayFish : proc( int, int, pic ) in
begin
    let constShark = shark

    let drawShark = proc( transform : proc( pic → pic ) )
    begin
        for x = 1 to 30 do
            for y = 1 to 20 do
                displayFish( x, y, transform( constShark ) )
    end

    in PS() let drawShark := drawShark
end
```

**Figure 4.7: An example Napier88 program**

The first two lines of the program give the access specifications for a picture of type *pic* and a procedure of type *proc( int, int, pic ).* Both are accessed from an environment that is itself accessed from the root environment via the name *fishPics*. Inside the main block of the program a local identifier *constShark* is declared, with the value of the picture. This declaration ensures that the procedure will continue to operate on the same picture even if the environment location originally containing it is updated with a different picture. The program then declares the main procedure *drawShark* which takes as its parameter a procedure that maps pictures to pictures. Two nested loops in the body of *drawShark* draw a transformed version of the picture over the screen. Finally the procedure is made persistent by creating a binding to it in the root environment.

There now follows a description of how an equivalent program may be constructed in the hyper-programming system. Of course the same program could be entered if the programmer required run-time linking to the persistent store. The method to be illustrated shows how composition time linking may be used. To construct the hyper-program the programmer first enters the textual part and then positions the insertion marker at the point where the first direct link is to be inserted. Figure 4.8 shows an editor window containing the textual part of the hyper-program. Missing at this stage are the type of the *transform* parameter, the display procedure and the picture to be displayed.

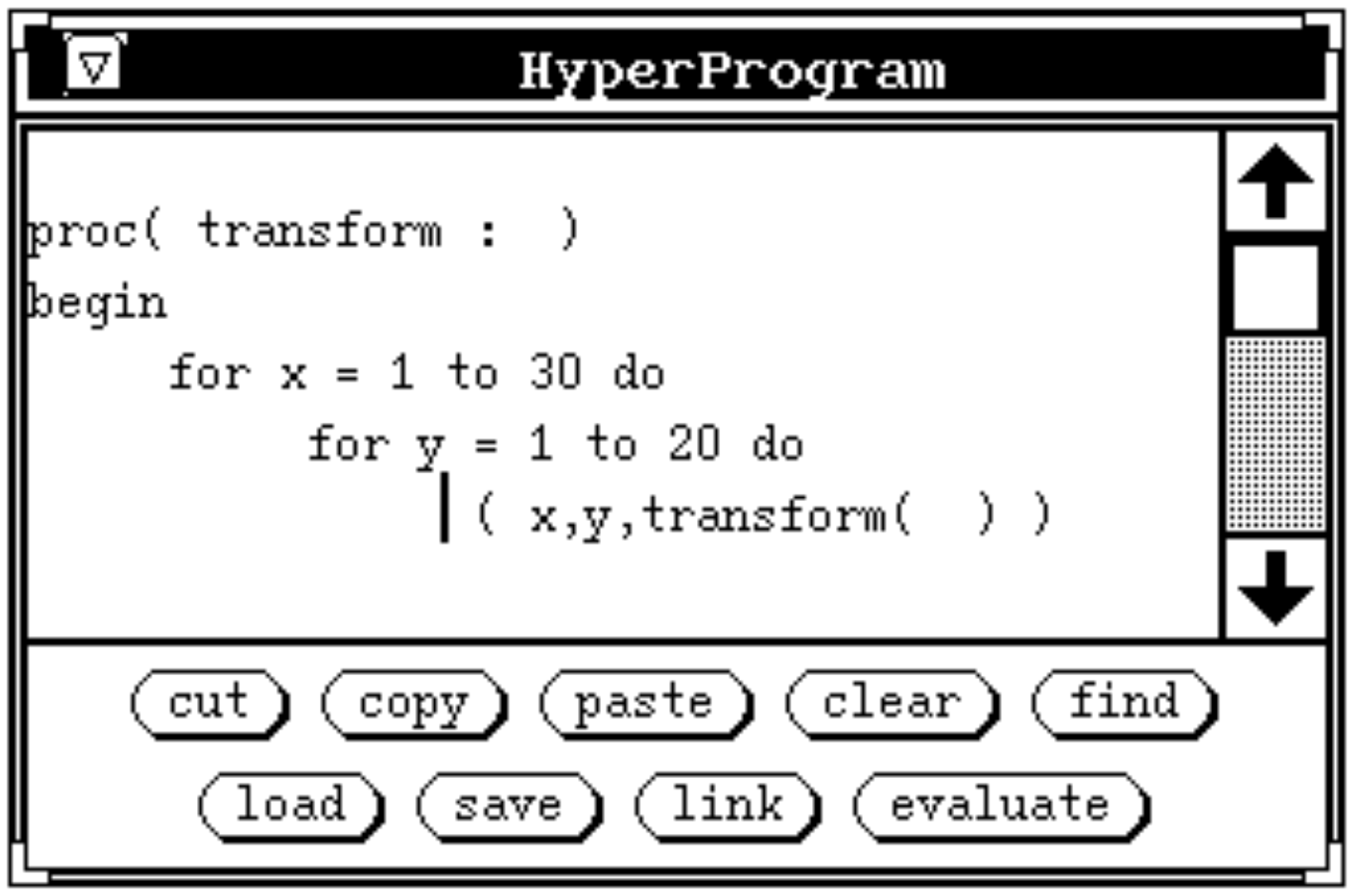

**Figure 4.8: Textual part of a hyper-program**

The location of the display procedure will now be bound in. To do this the programmer navigates through the store from the persistent root with the browser until the procedure is located.

In the example the programmer selects the entry for the environment *fishPics* in the root environment. The procedure required is accessible from *fishPics* through the environment binding with the name *displayFish*. These identifiers will be used in this description to denote the environment and procedure respectively; however, note that the identifiers are really associated with the particular access paths shown rather than the values themselves. Figure 4.9 shows the browser display after the programmer has selected the entry for *displayFish*, resulting in the display of a window representing the procedure.

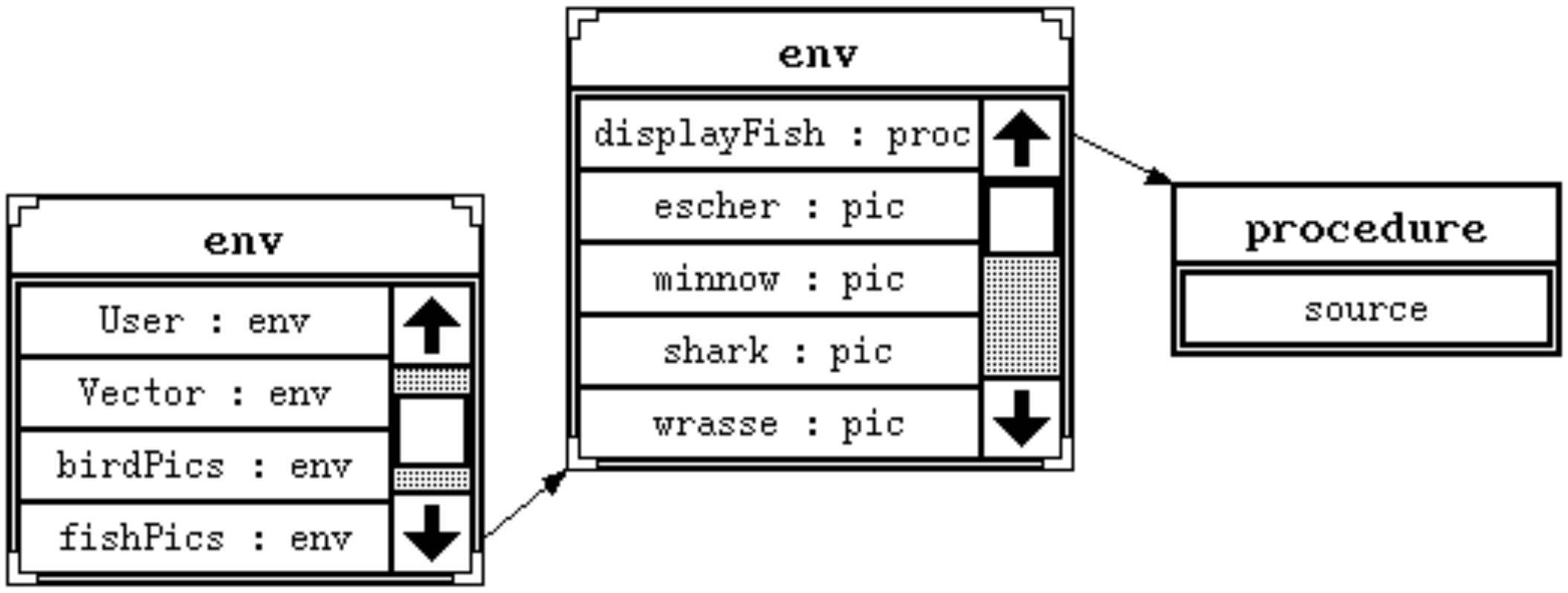

**Figure 4.9: Browser display of a procedure value**

A representation of the type of the procedure is obtained by highlighting the procedure window and then selecting *show type* from the pop-up background menu as shown in Figure 4.10:

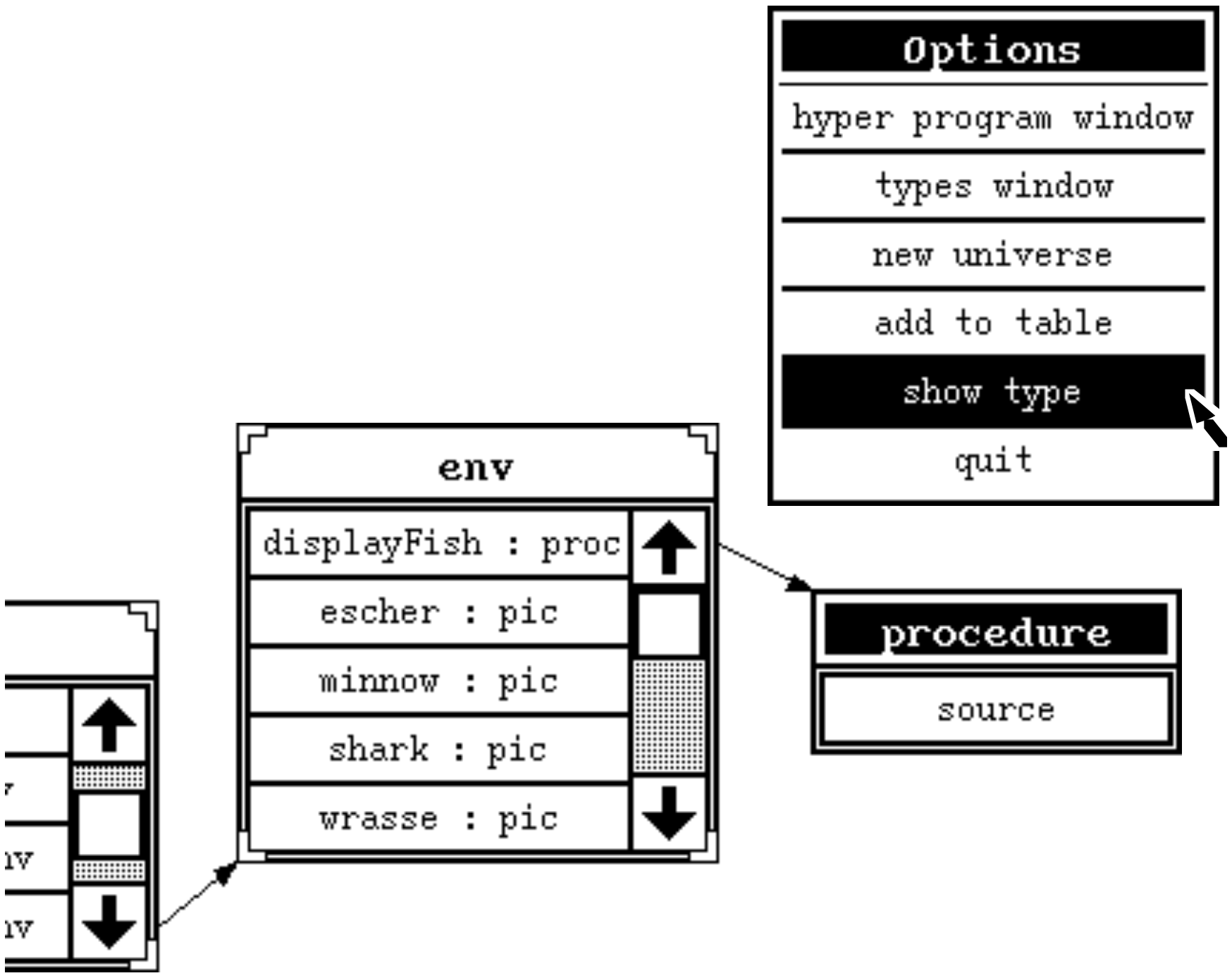

**Figure 4.10: Obtaining the type of a value**

A textual representation of the type of the procedure is then displayed in a window attached to the procedure window as shown in Figure 4.11:

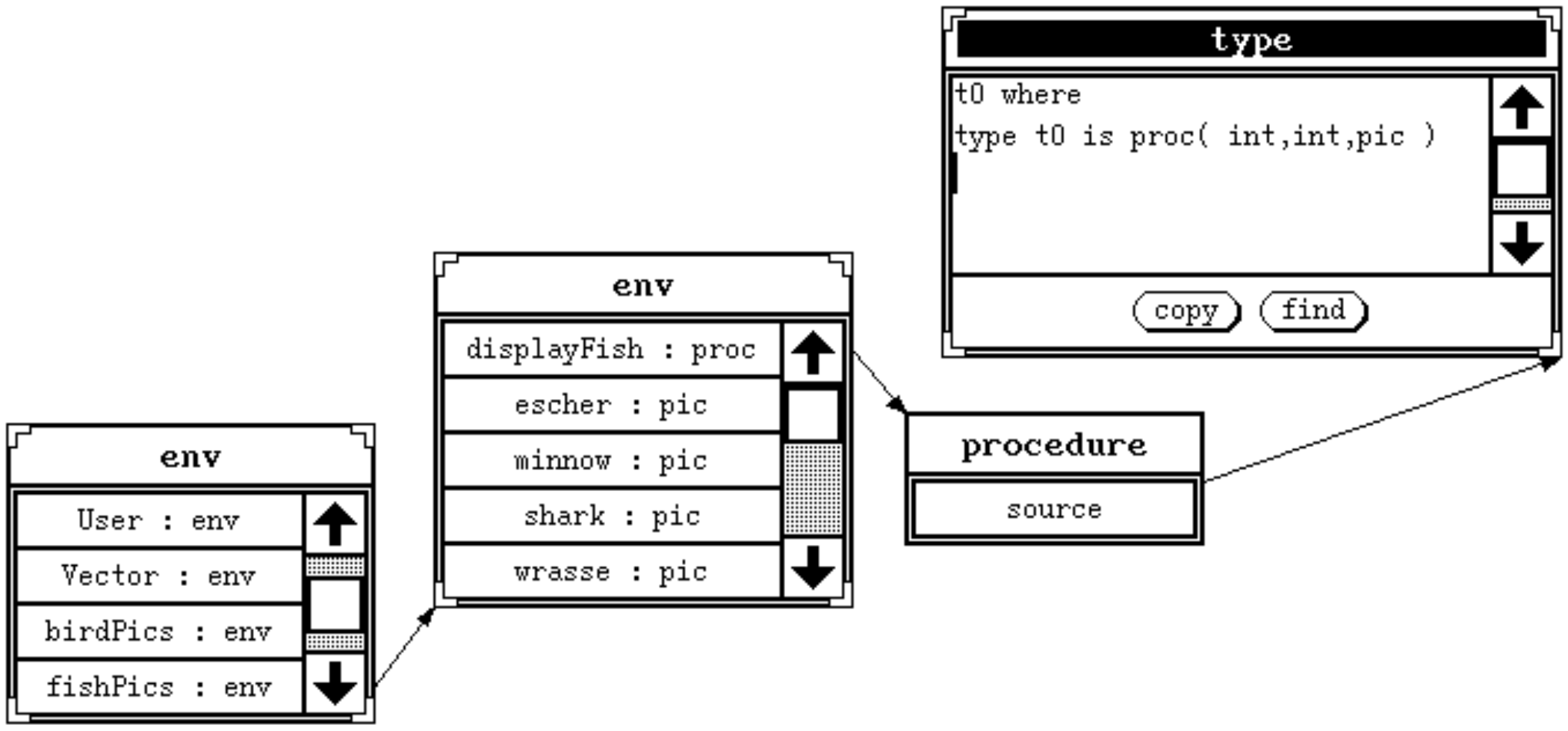

**Figure 4.11: Browser display of a procedure type**

The programmer now selects the location containing the procedure. This is done by highlighting the procedure entry in the environment window using mouse button 1, with the result as shown in Figure 4.12. The root environment and type representation windows have been omitted for brevity.

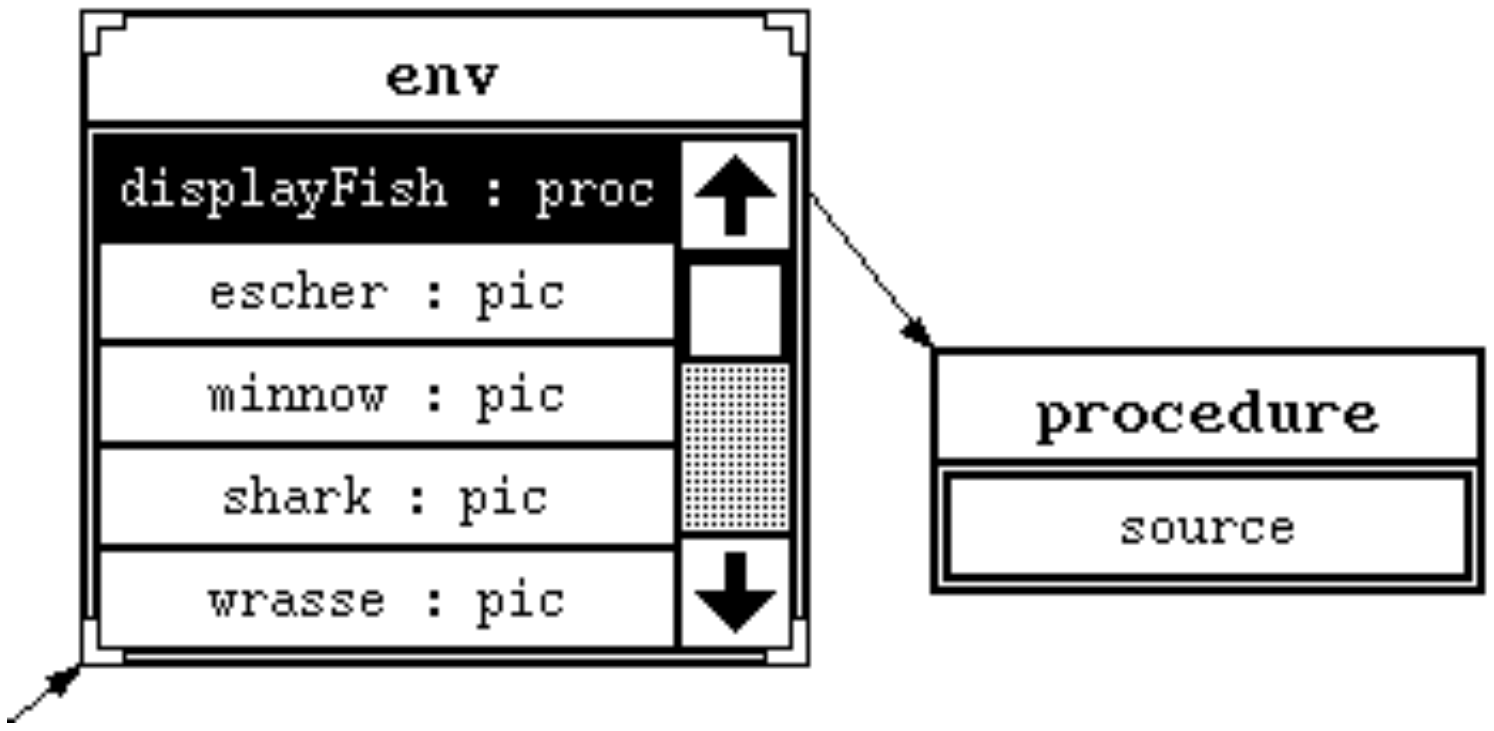

**Figure 4.12: Browser display of an environment location**

Locations in structures, abstract data types and vectors can also be selected in a similar way.

The programmer now presses the *link* button in the editor window to link the selected location into the hyper-program at the position of the insertion point. A button denoting the location appears in the hyper-program as shown in Figure 4.13:

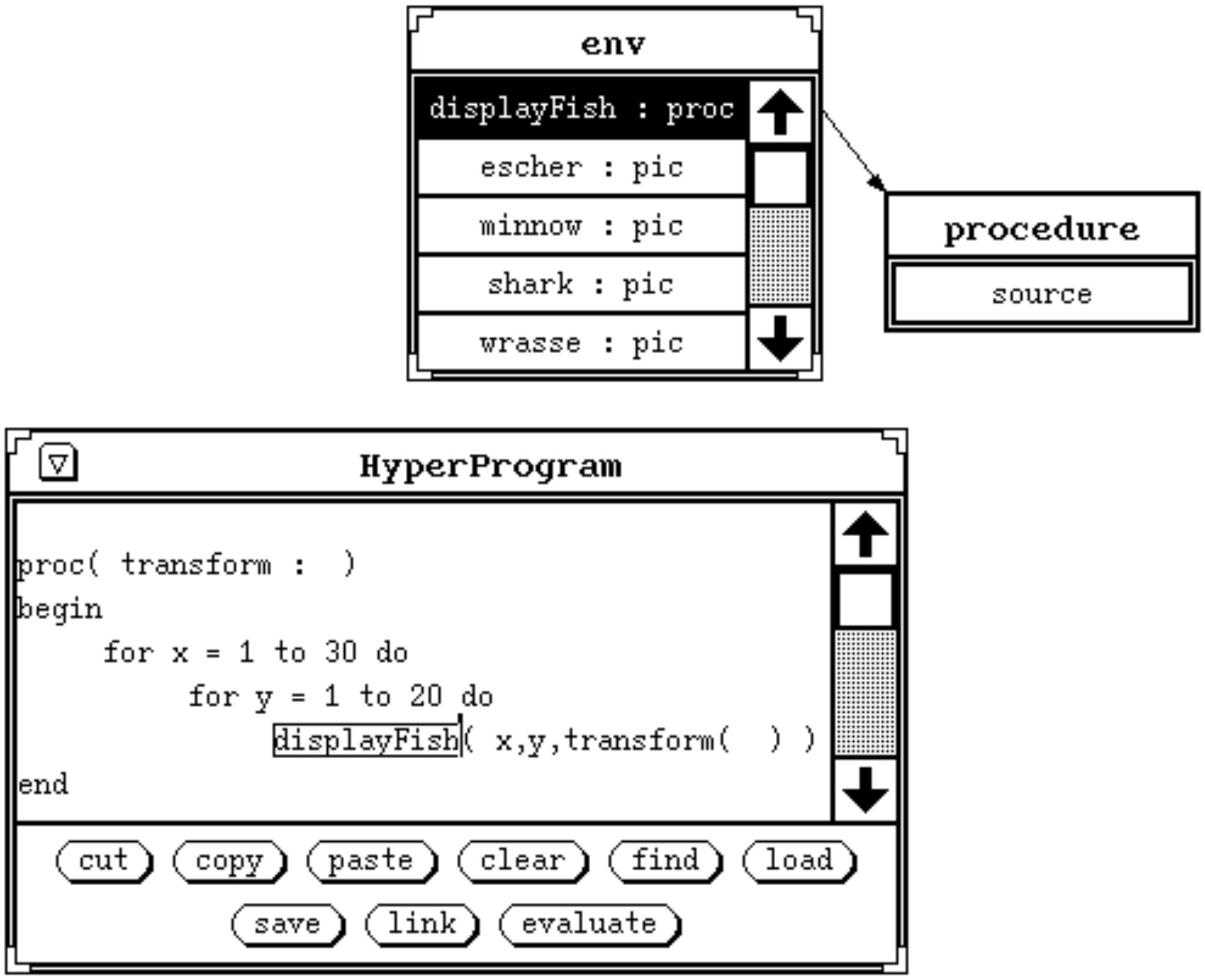

**Figure 4.13: Location linked into a hyper-program**

A similar method is used to link the appropriate picture into the hyper-program. The programmer selects the *shark* entry from the environment to display a representation of the picture. In this case it is the value itself rather than the environment location that is selected. The programmer indicates this by highlighting the picture window rather than the environment location. Figure 4.14 shows the display after the picture has been linked into the hyper-program:

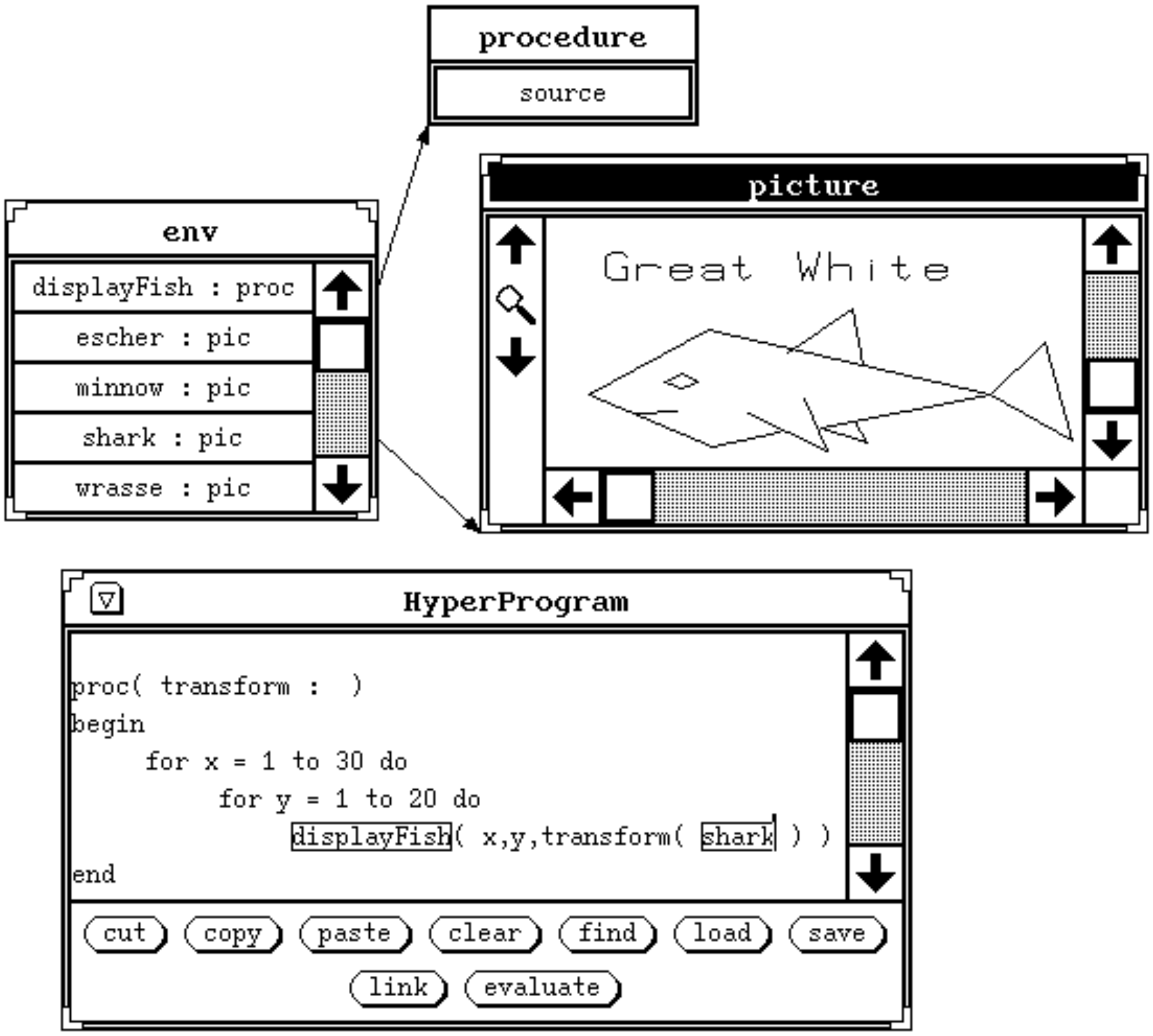

**Figure 4.14: Value linked into a hyper-program**

The buttons embedded in this hyper-program are labelled with the identifiers from their environment entries. The identifiers are not significant to the meaning of the hyper-program and can be changed without affecting the program. They are used only as an aid to legibility. In some cases the system will not be able to find an appropriate label for a button, for example when the environment window pointing to the representation of the value bound in has been removed from the display. In such cases the label will be blank initially. To change the label on an embedded button the programmer presses it using mouse button 2. A dialogue then prompts for the new label.

The final stage in composing the hyper-program is to link in the procedure argument type. To do this the programmer highlights an existing value of the required type, accessed from the *fishPics* environment with the name *fishTransformer*. The programmer then selects *show type* from the background menu, highlights the resulting type window, and presses *link* in the hyper-program window. This sequence of actions results in the insertion of a button to represent the type. Figure 4.15 shows the situation before the *link* button is pressed.

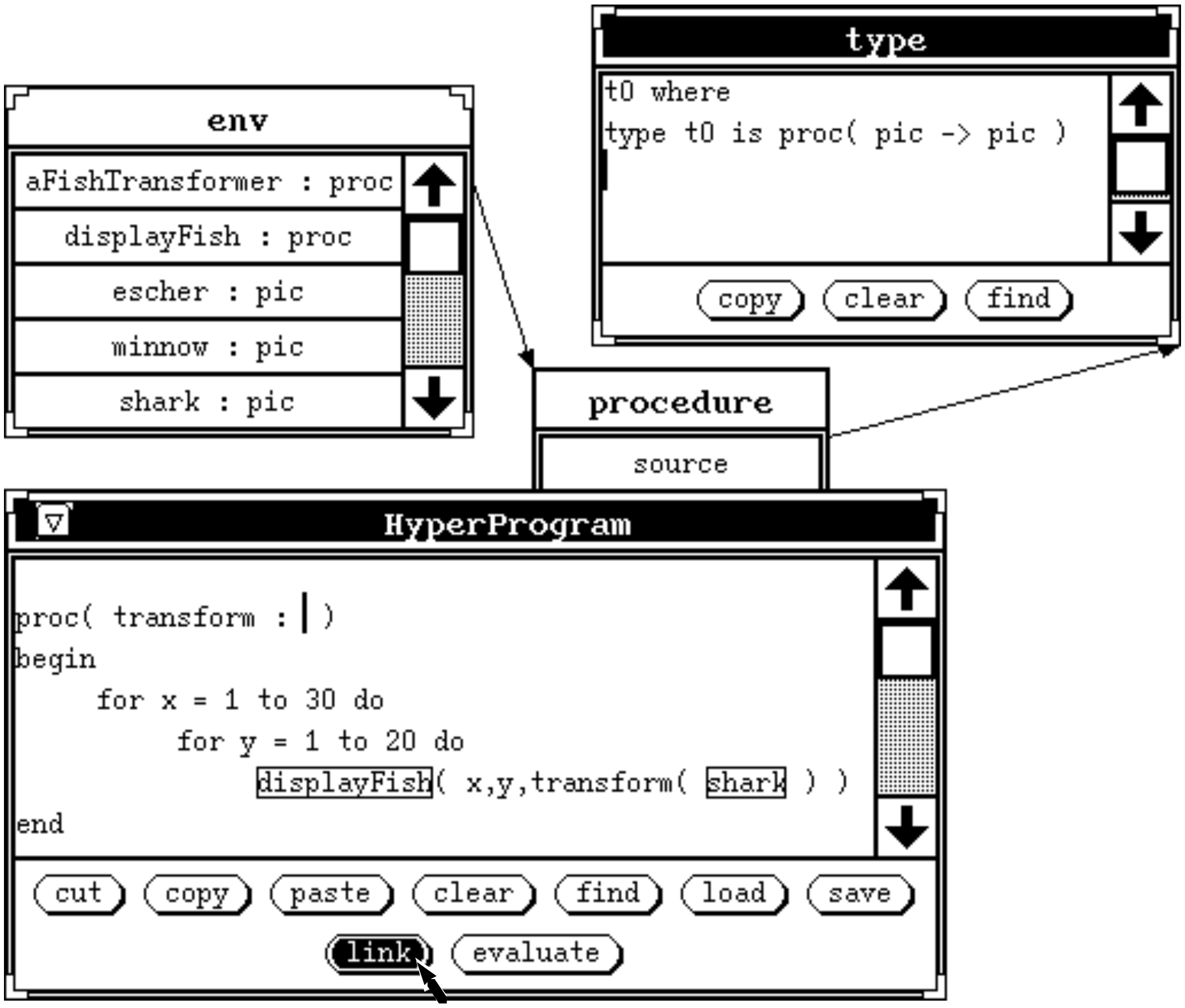

**Figure 4.15: Linking a type into a hyper-program**

The programmer then gives the type button a name *T*, to make the program easier to read, and presses *evaluate* to compile and execute the hyper-program. The result of execution in this case is a new procedure value, a representation of which is automatically displayed by the browser as shown in Figure 4.16:

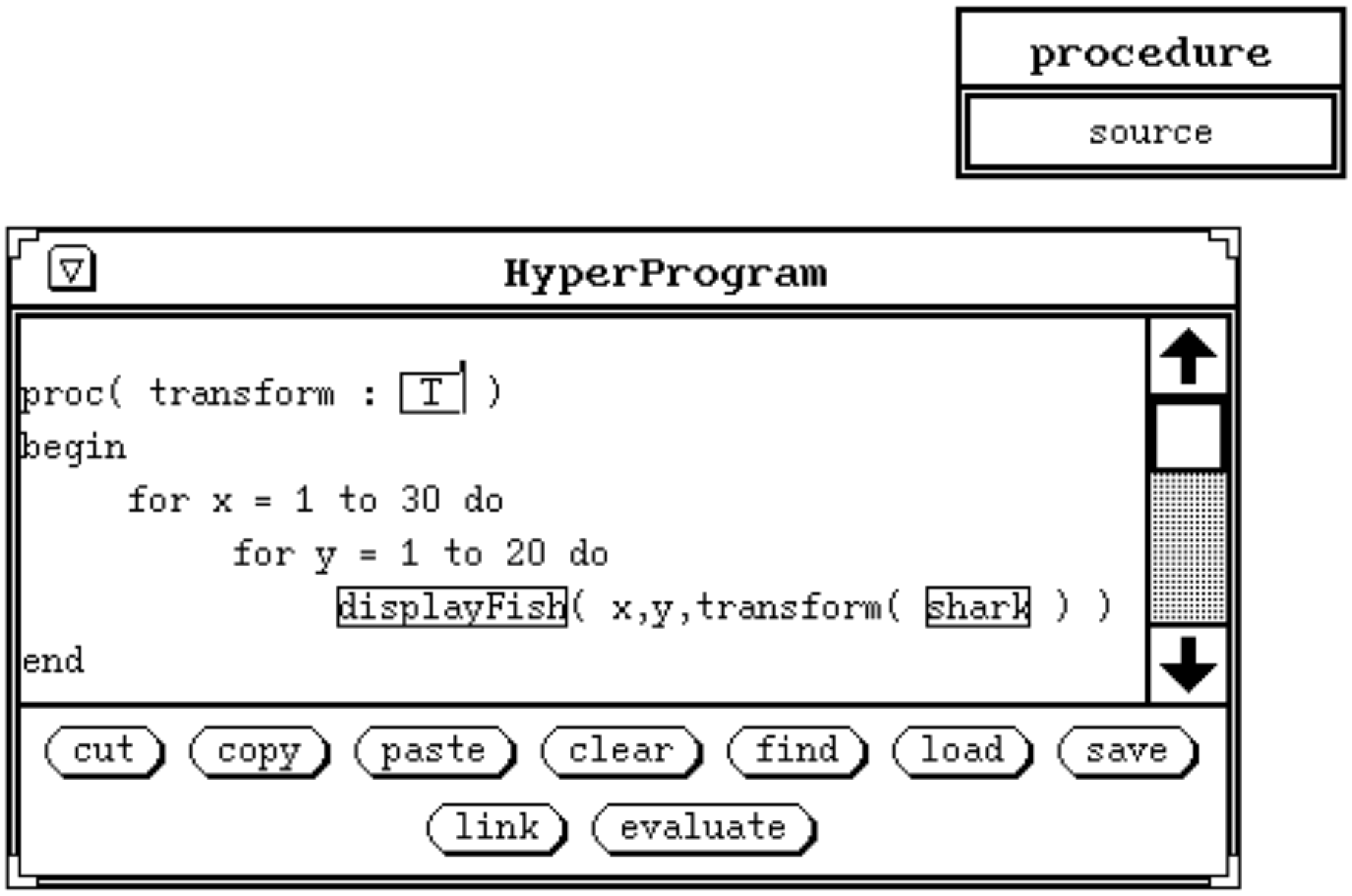

**Figure 4.16: Value resulting from evaluation of a hyper-program**

The new procedure value is now available for linking into other hyper-programs. This is another example of an 'anonymous' value: the system does not supply an initial label for any button denoting the value since there is no identifier associated with it. The procedure value has the hyper-program source code bound into it; this can be recalled later for examination and editing.

The programmer can now construct other hyper-programs that call the procedure. Alternatively the hyper-program might simply be made persistent for later use. Figure 4.17 shows a hyper-program that will link the new procedure into the persistent store. The button in the program denotes the procedure.

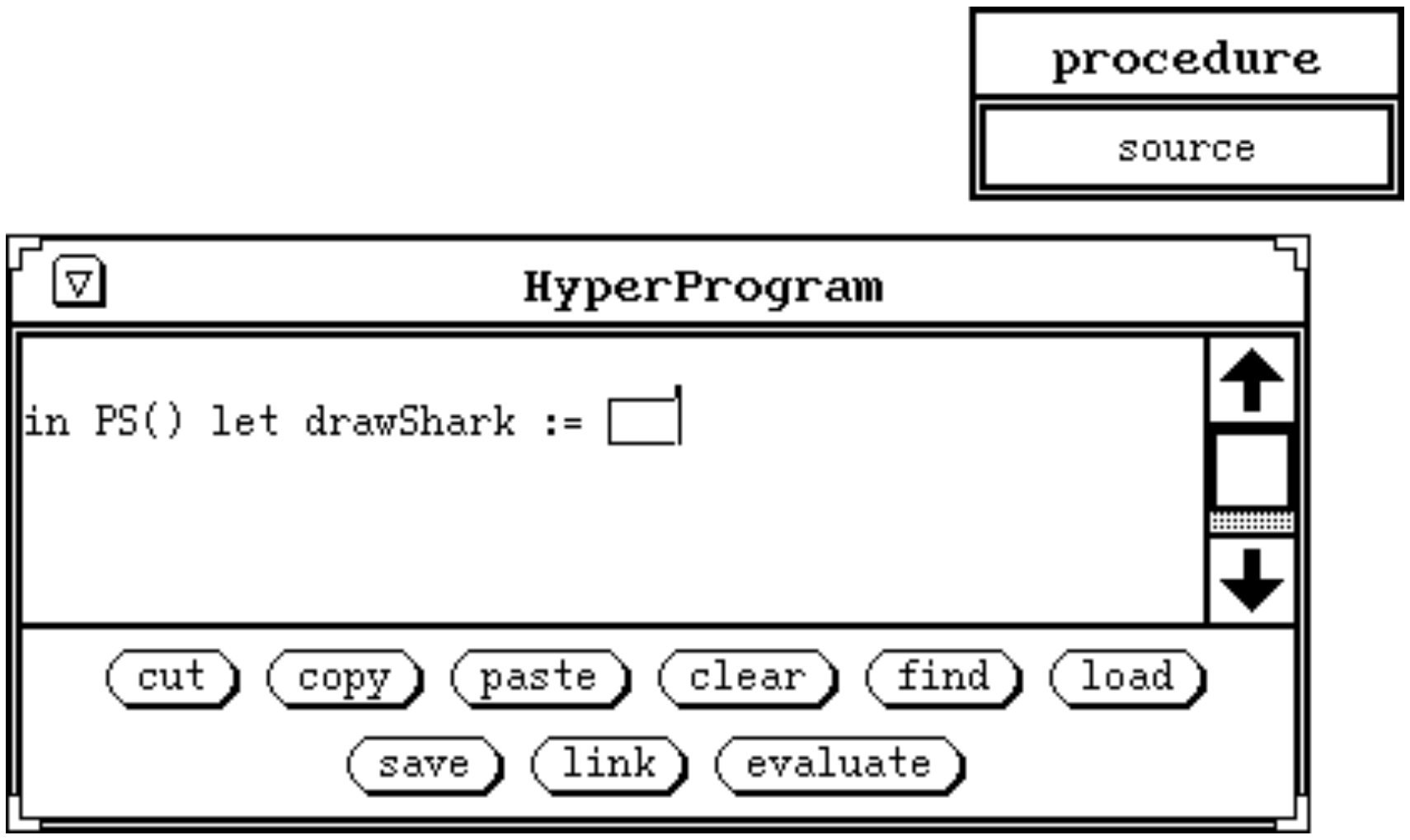

**Figure 4.17: Making a value persistent**

The linking requirements with respect to the procedure and its components have been met. Because the picture value has been linked into the original hyper-program, and consequently into the closure of *drawShark*, the procedure will be unaffected by any subsequent update of the environment location containing the picture, or by the location being dropped from the environment. The location of *displayFish* has been linked, so dropping the location from the environment will not affect *drawShark*, but an update of the location will result in the new value being used inside *drawShark*.

### 4.2.3 Editing a Hyper-Program

Several modes of hyper-program editing are possible in general, including editing text, deleting and inserting tokens, associating existing tokens with different data items, and examining and updating data associated with tokens. The system being described implements a particular selection of these facilities.

The hyper-program editor treats the light-buttons that represent tokens as single characters, so they can be cut, copied and pasted in the same way as text. New tokens are inserted in the manner described in the previous section. The data associated with a token can be inspected by pressing the corresponding light-button: the browsing tool then displays and highlights a representation of the data. That data could then be updated by linking it into another hyper-program that performed some operation on it.

Figure 4.18 shows the display after the programmer has pressed the *displayFish* button in the original program. This highlights the environment location containing the procedure.

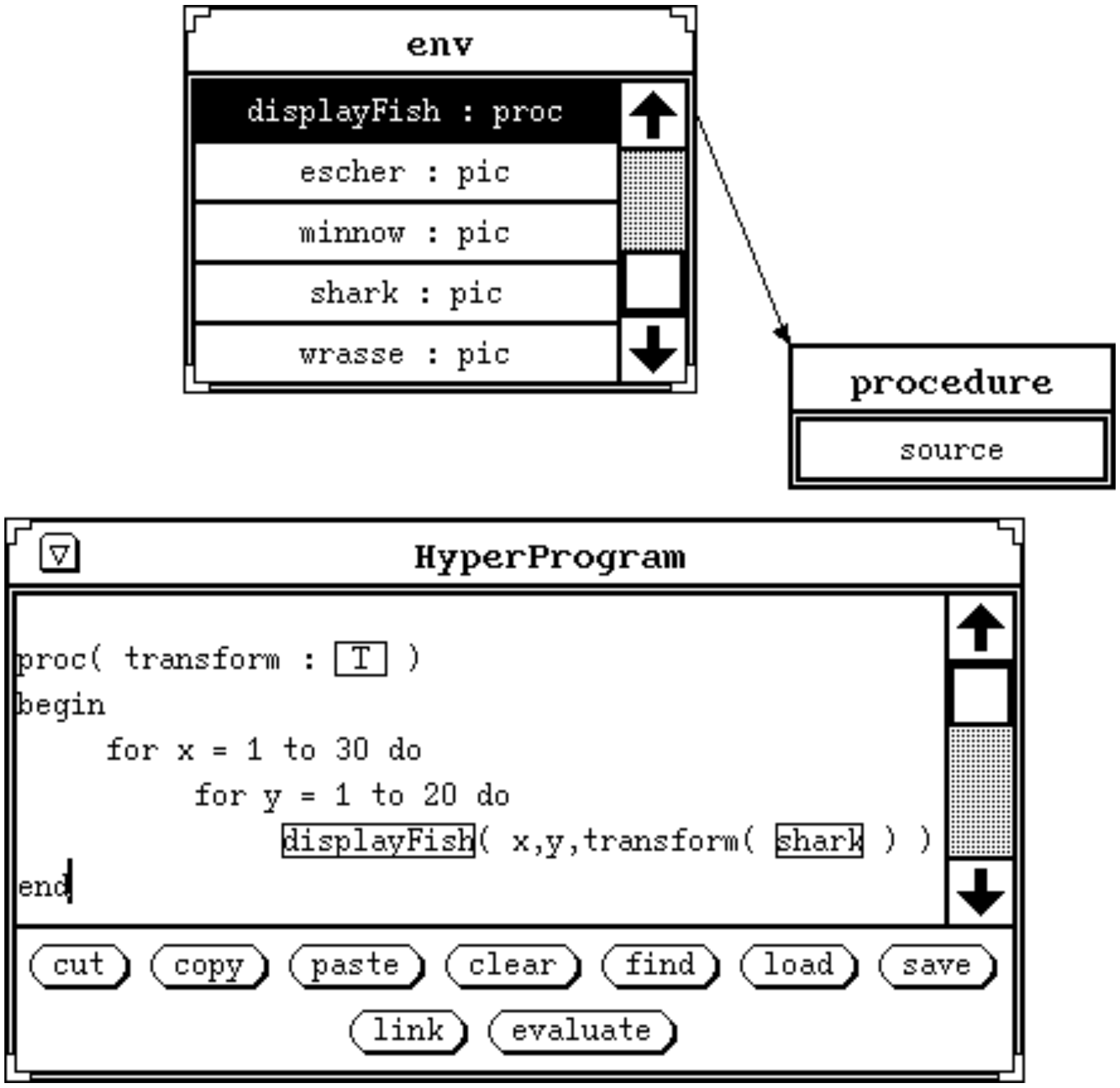

**Figure 4.18: Browsing a link in a hyper-program**

Continuing with the example, Figure 4.19 shows how the environment location could be updated with a refined version of the procedure using standard Napier88.

```
use PS() with fishPics : env;
              drawShark : proc( proc( pic → pic ) ) in
use fishPics with shark : pic ; displayFish : proc( int, int, pic ) in
begin
    let constShark = shark

    drawShark := proc( transform : proc( pic → pic ) )
    begin
        for x = 1 to 30 do
            for y = 1 to 20 do
                if x = 1 or y =1 or x = 30 or y = 20 do
                    displayFish( x, y, transform( shark ) )
    end
end
```

**Figure 4.19: Updating environment location in standard Napier88**

The first three lines of the program give the access specifications for the environment location to be updated and for the other data items as before. A new procedure value of the same type as the original is then assigned to the location *drawShark*. This new procedure draws copies of the transformed picture around the edge of the screen rather than over the whole screen. Note that the picture linked into the new version of the procedure is the picture accessible from the environment *fishPics* at the time that the new version is installed. This will not be the same picture that the original procedure operated on if the picture location has been updated since the original procedure was created. Although this may not be the desired semantics the programmer has no choice given that no special arrangements were made to maintain a reference to the original picture.

It will now be shown how the programmer may achieve a similar update to the environment location using the hyper-programming system. One of the benefits of the system is that links to particular data items may be preserved in a modified version of a procedure if required. In the example this enables the refined procedure to contain a link to the original picture.

The programmer first obtains the source code of the original procedure by selecting the *source* entry from its menu:

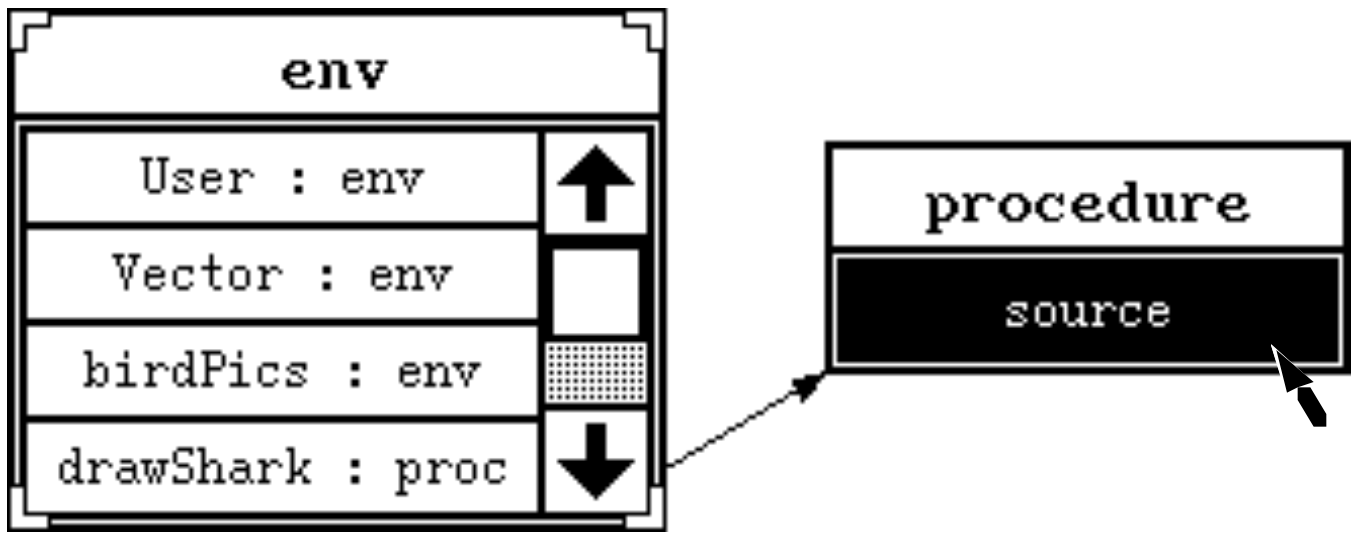

**Figure 4.20: Obtaining the hyper-program source code of a procedure**

This results in the display of an editor window containing the hyper-program source. The system does not allow this source program to be modified, so as to enforce the association from the procedure value to its source. Instead the programmer creates a new editor window by selecting *hyper-program window* from the background menu, and copies the source code into the new window using the *copy* and *paste* buttons. The copied code contains direct links to the same data items as the original, i.e., to the picture, the location containing the display procedure, and the parameter type. The programmer then edits the text of the new hyper-program so that the picture is drawn only around the edges of the screen, and presses the *evaluate* button. If compilation and execution is successful the representation of a new procedure is displayed, as shown in Figure 4.21:

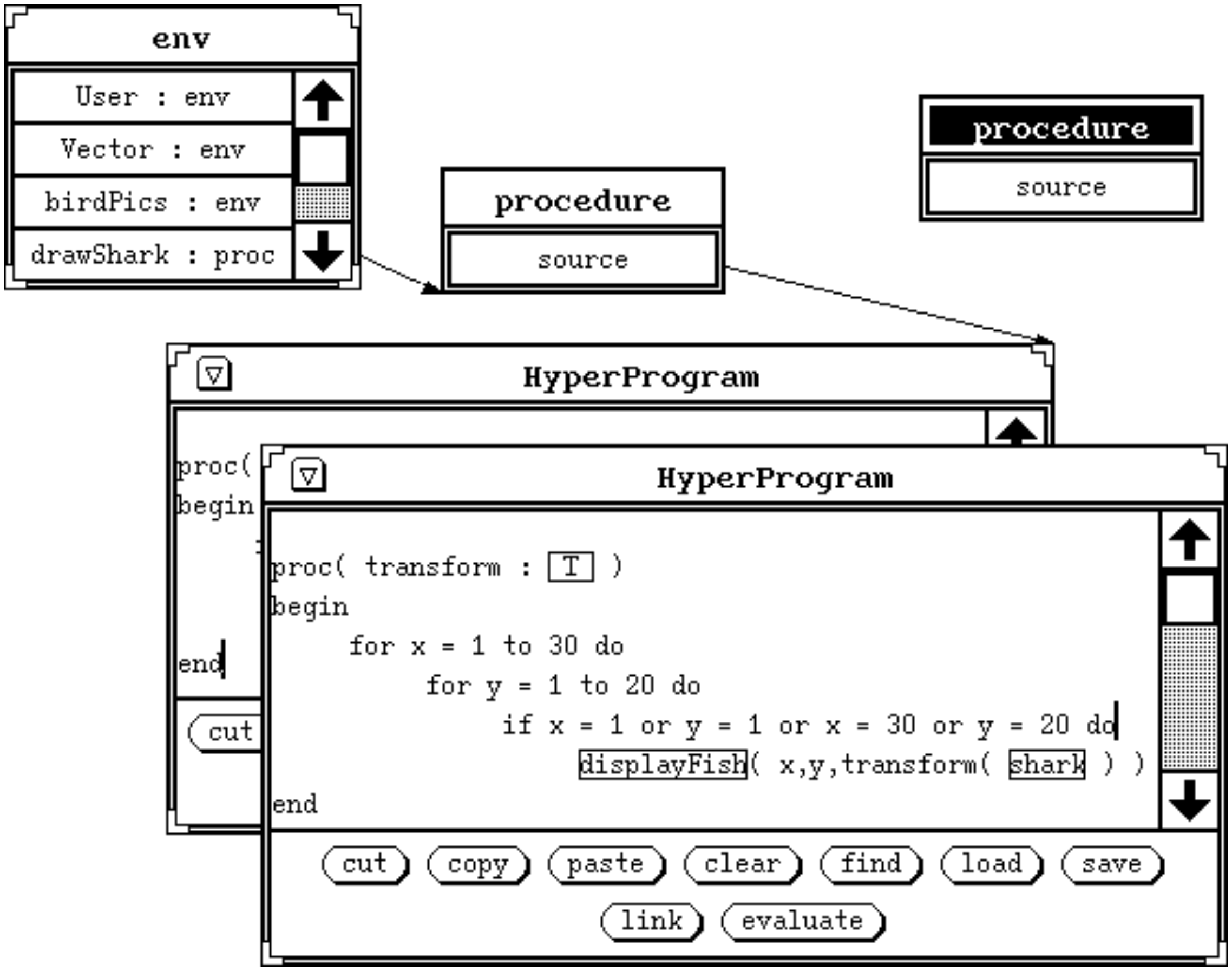

**Figure 4.21: Value resulting from evaluation of modified hyper-program**

This new procedure has different behaviour from the original application but contains the same direct links. The old version can be overwritten with the new one by a program such as that shown in Figure 4.22:

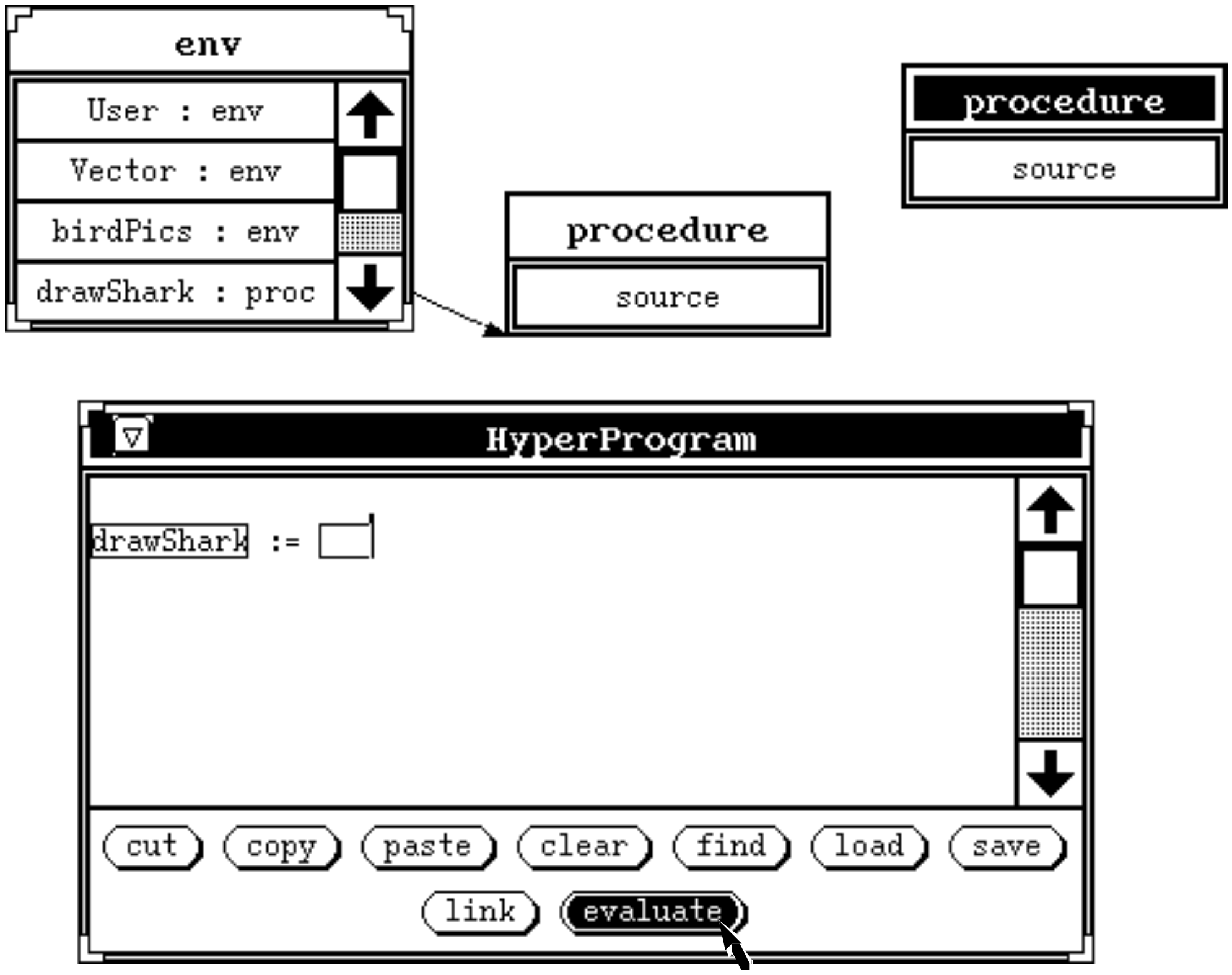

**Figure 4.22: Installing a modified version of the procedure**

### 4.2.4 Compile-Time Linking

The prototype system also supports compile-time linking as described in the introduction. This involves the insertion of tag identifiers into a program. When the program is compiled the tags are resolved into references to data items and these references incorporated in the executable program. This resolution is performed using a shared table that maps identifiers to data items. Any program compiled in the system may contain identifiers from the shared table. The entries in the table appear in a menu window labelled *Shared Table*.

Figure 4.23 shows how the programmer adds an entry to the shared table by highlighting the representation of a data item and selecting *add to table* from the background menu:

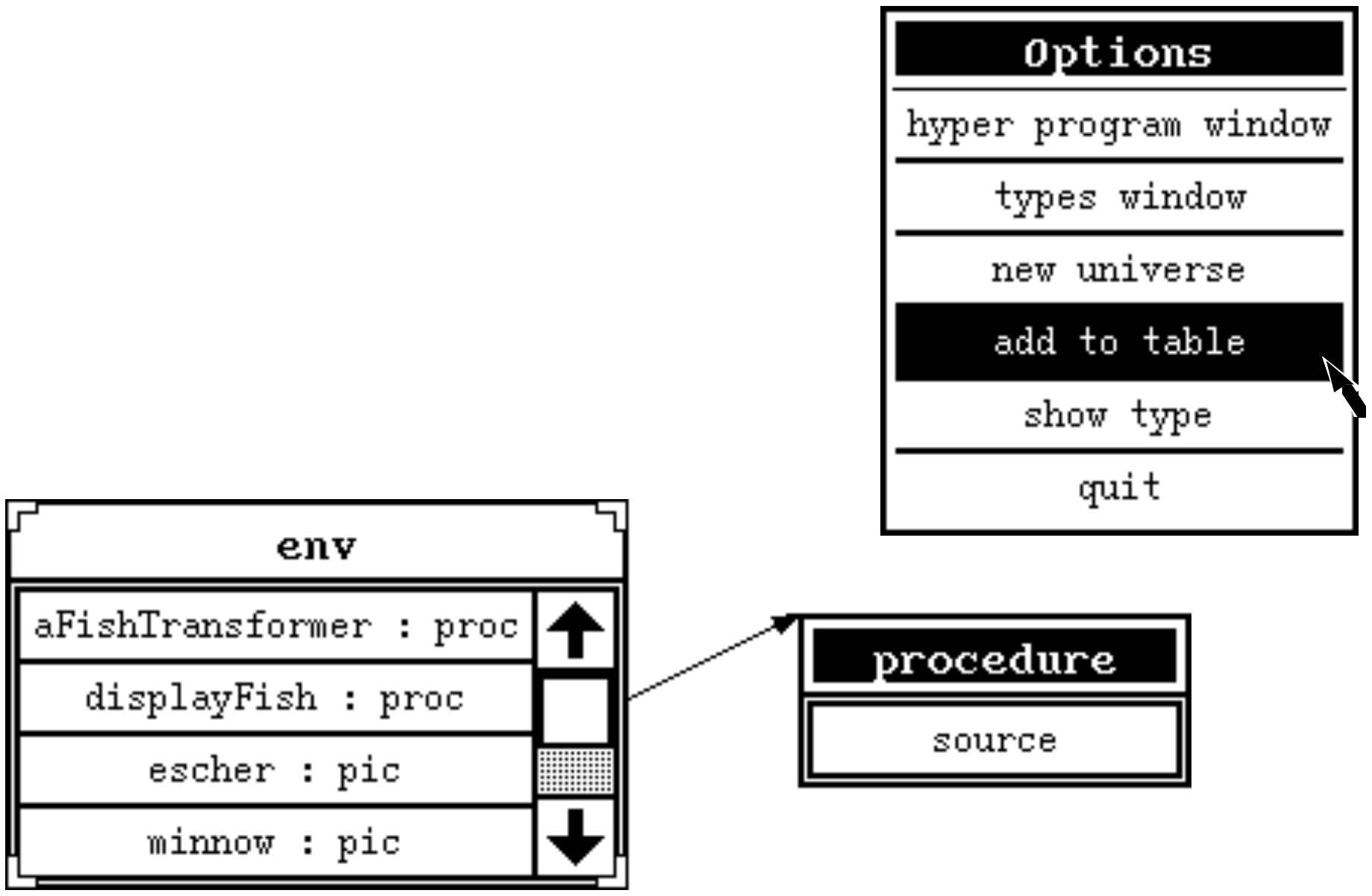

**Figure 4.23: Adding a data item to the shared table**

The system prompts for a name which is then added to the table and appears in the *Shared Table* window. The programmer may then use that name to refer to the data item in programs. Figure 4.24 shows a program that creates a link to the display procedure in the root environment, after the programmer has entered the name *myProc* to denote the procedure in the shared table:

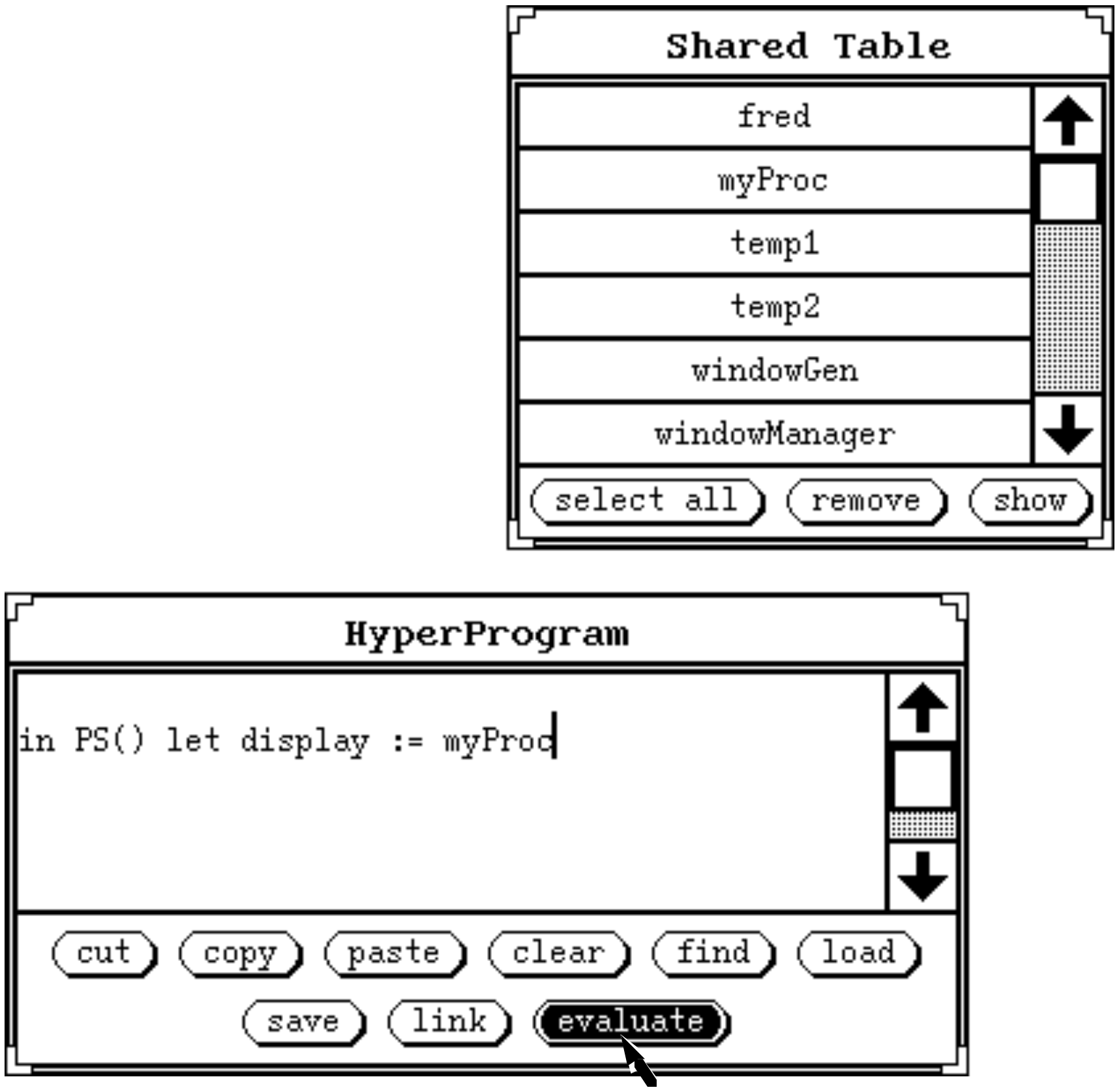

**Figure 4.24: Compile-time linking**

### 4.2.5 Comparison with Other Systems

The hyper-programming system described has evolved from several other strongly typed object browsers, described below.

#### 4.2.5.1 PS-algol Browser

Developed by Dearle and Brown, the PS-algol browser [DB88] was originally designed as an aid to debugging. The system allowed the user to scan the table that is used to structure a PS-algol database, to traverse pointers between structures, and to display the contents of structure fields and vectors. As a linked data structure was traversed the browser maintained a stack of menus, with only the menu showing the current object being visible at one time. The user could pop the menu stack to backtrack along the original route. The restriction to one visible menu made it difficult to visualise complex data structures—for example it was not possible to determine the size of a circular list in which all the data elements were the same.

#### 4.2.5.2 Refined PS-algol Browser

Dearle, Cutts and Kirby produced a prototype of a refined version of the PS-algol browser [DCK90] which could display multiple menus linked by arrows. Although an improvement on the existing PS-algol browser, a full implementation was never developed, partly because of the immaturity of the PS-algol window management technology available [CK87].

#### 4.2.5.3 Napier88 Browser

The first Napier88 browser, developed by Kirby [KD90], implemented the ideas of the refined PS-algol browser in Napier88. Napier88's richer type system required it to be extended to display environments and variants as well as structures and vectors. The browser also supported partitioning of the display by means of self-contained universes, an idea first proposed in the context of the refined PS-algol browser described above.

#### 4.2.5.4 ABERDEEN

The ABERDEEN system, developed by Farkas [Far91], supported interactive program development with compile-time linking to persistent data items. The user specified the linking by attaching a tag identifier to the browser representation of the required data item and using that identifier in a source program. The system also allowed a structured view of the type of a data item, in which each component of the type was represented by a menu similar to those used for values. The user could construct hierarchies of collections of type definitions against which source programs were compiled.

## 4.3 Conclusions

There is much scope for further enhancements to the hyper-programming system described. Possibilities include:

- provision of a graphical display of types, as in ABERDEEN;
- support for performing simple operations, such as assignments to structure fields and creation and deletion of environment bindings, by direct manipulation rather than typing in source code;
- support for controlling access to procedure source code.

The need for facilities to restrict access to procedure source code was described in Chapter 3. Password protection could be implemented easily but the challenge is to develop a mechanism that provides the required control without unduly hindering the developer who does have the right to see the source code.

Another issue is whether the user should be allowed to view the source code of procedures encapsulated within abstract data types. Clearly this would violate the abstraction as the user could discover the implementation of the abstract data type, but again there may be a case for allowing controlled access to the implementor. The need for multiple levels of access to abstracted data is further discussed in [CDM+90].

# 5 Reflective Programming Tools

The preceding chapters have described two ways in which persistent programming systems can be extended, through type-safe linguistic reflection and hyper-programming, and have also described a prototype set of hyper-programming tools. This chapter will describe a further set of tools that have been implemented to support reflective programming in Napier88. These tools allow generators to manipulate hyper-program fragments, giving a new richer style of reflection in which the program representations analysed and synthesised may contain direct links to data in the persistent store. In addition they are designed to make generators easier to program.

## 5.1 Reflection and Hyper-Programming

The concepts of linguistic reflection and hyper-programming are linked, in that linguistic reflection is likely to be used in the implementation of most hyper-programming systems. It might be possible to construct a hyper-programming system in which the source representations did not themselves reside in the persistent store, but it is probably not sensible. The straight-forward implementation strategy is to represent hyper-programs within the language and this then requires linguistic reflection to transform them into executable programs.

Conversely, hyper-programming facilities can be used to widen the applicability of reflection. The central concept of hyper-programming, the ability to embed references to persistent data in source code representations, may be applied to reflective systems to give a flexible and uniform linking mechanism. This facility allows a reference to data created by a generator, or already existing in the persistent store, to be linked directly into the newly generated code. This overcomes the problems in existing run-time reflection systems caused by generators and generated code fragments being evaluated in completely separate environments. In other systems the programmer must use an ad-hoc solution in which data is placed in the persistent store by a generator and later accessed by generated code fragments. This suffers both from a performance overhead, due to the access specification checking required when the generated code is executed, and from a lack of security as there is no guarantee that the data will still be accessible when the generated code is executed.

The ability for generators to reflect over hyper-program source representations opens up new styles of program manipulation. In other systems the representations manipulated are divorced from the persistent data in that they may contain access specifications for data items, but not the data items themselves. Thus the information about a data item that may be determined by analysis of a program fragment that accesses it is limited to its expected access path and type. In a reflective hyper-programming system however, the program fragment may contain a direct link to the data item, in which case the generator can access the data itself.

This has an impact on both the analysis and synthesis of program fragments. When analysing a fragment that contains a direct link to a data item the generator can perform arbitrary computation on that data item in order to discover its properties. For example, a generator performing source level optimisation on a hyper-program might decide whether or not to replace a given loop with an in-line expansion by examining the size, or other properties, of the data linked into the hyper-program within the bounds of the loop.

During program synthesis a generator may construct code representations containing direct links to data items created by the generator or already existing in the persistent store. Thus data operated on by the generated program can be made manifest if it already exists when the program is generated. Where manifest data items are immutable values rather than store locations, the generator may verify that particular properties of the data hold; if so, those properties are guaranteed to continue to hold during subsequent executions of the generated code.  This means that checks can be executed in the generator rather than in the generated code, allowing some varieties of constraint checking which are normally performed dynamically to be performed statically with respect to the generated code.

## 5.2 Ease of Programming Generators

The general structure of a generator was described in Chapter 2.  Each generator contains a prelude and a result expression.  When evaluated the generator first executes the prelude.  This sets up the environment (in the general rather than the Napier88 sense) in which the result expression is evaluated.

The result expression produces the generated program fragment; the code in it lies in the subset $L_L$ of the language $L$.  This subset contains all the language sentences that produce values in $Val_L$ when evaluated.  The result expression may contain components in either or both of $L_{L_{Const}}$ and $L_{L_{Var}}$.  These subsets of $L_L$ contain expressions that give constant and variable results respectively.  Constant expressions represent fixed fragments of source code while variable expressions may contain references to values in the environment populated by the execution of the prelude.  The values in the environment may vary between evaluations of the generator.

Chapter 2 also identified some factors that make generators in reflective systems hard to understand.  One of these was the programmer's difficulty in distinguishing the constant parts of the result definition from the variable parts.  This is combined with a high level of syntactic noise.  The tools described here are designed to make Napier88 generators easier to write and understand.

A window-based generator editor is used to allow the programmer to view a generator at various degrees of detail.  At the most abstract level the programmer sees only the prelude code and the fixed parts of the result definition.  The positions of the variable parts are indicated by buttons embedded in the code.  This level of detail shows the programmer the main structure of the generated result, while abstracting over the variations that depend on the particular specialisation.  To examine the details of the variations the programmer may press a button and view the corresponding code in a separate window.  This use of windows allows much of the noisy syntax involved in combining parts of the result definition to be omitted, making it easier to read.

The usefulness of this ability to separate constant and variable parts of the result definition depends on the style in which generators are written.  It is always possible to write generators in such a way that the entire result definition is variable, but the assumption made here is that programmers will choose to write constant definitions wherever possible.  The separation of $L_{L_{Const}}$ and $L_{L_{Var}}$ code also allows different representations to be used.  A textual form for the fixed code is easy to read while a more structured form for the code produced by the variable parts facilitates the specification of the code manipulation.

## 5.3 Generator Model

The generator model supported by the editor was designed to meet the following criteria:

- to allow generators to manipulate hyper-program source representations;
- to use hyper-program facilities to give a flexible mechanism for communication between generators, generated code and the persistent store;
- to give uniformity between generators and the variable $L_{L_{Var}}$ parts of result definitions, which may themselves be regarded as generators; and

• to allow arbitrary nesting of generators.

In the model each generator has two separate components: a *prelude* and a *result definition*. The prelude is a procedure that processes the parameters input to the generator, while the result definition is a variant that may be either a fragment of hyper-program source code or a procedure that produces one. These source code fragments may contain place-holders corresponding to further generators. Thus each $L_{L_{Var}}$ part of the result definition is represented by a generator, fulfilling the third and fourth criteria.

To evaluate a generator its prelude is executed with the generator parameters passed to it. If the result definition is a procedure then it is executed in turn, with the results produced by the prelude passed to it. The result of this procedure, or the result definition itself in the other case, is a source code fragment which may contain place-holders for other generators. If so these generators are themselves evaluated and the resulting code fragments incorporated into the result. This process is continued until a source code representation without generator place-holders is obtained.

The ability of a generator to produce hyper-program source code containing links to data items means that generated code can refer directly to values constructed by the generator. This is not possible in other generator models, in which generated code is evaluated in a separate environment from the generator. With TRPL's compile-time reflection, for example, a generator can produce code that when executed will construct a new value equivalent to one in scope in the generator, but it cannot be the *same* value, as the generator and generated code are evaluated in different environments. With run-time reflection in standard Napier88 it is possible for generated code to refer to a value in scope in a generator but only indirectly through the persistent store. Thus the generator can link a value into the persistent store from where it is later retrieved by the generated code. The disadvantage of this mechanism is that the link to the value may have been removed in the meantime.

The new model allows direct links from generated code to values in scope in a generator. This also allows generated code to refer directly to values in the persistent store at the time of generator execution. This combination gives the desired communication flexibility.

## 5.4 Napier88 Representation of Generator Model

Figure 5.1 shows the Napier88 type definitions used to implement the model described in the previous section. Looking first at the main definitions, the first component of the structure type *Generator* is a procedure, *prelude*, that takes a Napier88 environment as its parameter and returns another environment. These environments contain the generator parameters and prelude results respectively. The prelude may return a newly created environment, the input environment with new bindings in it, or any other environment.

The second component of a generator, the result definition, is an instance of the variant type *GeneratorResult*. Its value may be an instance of type *GeneratorSource* or a procedure that takes an environment and produces a *GeneratorSource*. In the first case the result definition is a literal code fragment while in the second case it is a procedure that must be executed to produce a code fragment. In both cases the code fragment may contain place-holders both for hyper-program links and for sub-generators. Where sub-generators occur they too are evaluated, at some stage in the evaluation process to be described later, and their results substituted into the result code representation. The sub-generators are represented by an instance of the variant type *Optional*, the *absent* branch indicating that there are none, or the *present* branch containing a vector of generators together with descriptions of where each result is to be substituted into the main result.

The code representations manipulated are hyper-programs. Each representation, of type *HyperSource*, contains a fragment of code, which may be textual or in some parse tree form, and an optional vector of substitutions in the code. A substitution, of type *Binding*, is a reference to a value, store location or type. A store location may be within an environment, a structure, an abstract data type, a vector or a stack frame.

Note the symmetry between generator results and hyper-programs: both consist of a form of source code and a number of substitutions.

```
!********************** Subsidiary Definitions ***********************

type CodeTree is …            ! Parsed form of code representation.

type Code is variant( textual : string ; tree : CodeTree )

type CodeRegion is …          ! Specification of region of source code.

type Optional[ T ] is variant( present : T ; absent : null )

type Substitution[ T ] is structure( val : T ; codeRegion : CodeRegion )

type TypeRep is …             ! Representation of type.

type EnvLocation is structure( pointer : null ; typeRep : TypeRep )

type StructLocation is structure( structValue : any ; field : string )

type VectorLocation is structure( vectorValue : any ; index : int )

type StackPos is structure( Frame, MSoffset, PSoffset : int )

type FrameLocation is structure(  frame : null ; stackPos : StackPos ; typeRep :
                                  TypeRep ; envLoc : bool )

type TypeContainer is structure( typeRep : TypeRep )

type Binding is variant(  value :            any;
                          envLocation :      EnvLocation;
                          structLocation :   StructLocation;
                          abstypeLocation :  StructLocation;
                          vectorLocation :   VectorLocation;
                          frameLocation :    FrameLocation;
                          aType :            TypeContainer )

!*********************** Main Definitions ************************

rec type Generator is structure(  prelude : proc( env → env ) ;
                                  resultDefn : GeneratorResult )

& GeneratorResult is variant(  literal : GeneratorSource ;
                               expression : proc( env → GeneratorSource ) )
```

```
& GeneratorSource is structure(
                                    code : HyperSource ;
                                    generators : Optional[ *Substitution[ Generator ] ] )

& HyperSource is structure(  code : Code ;
                                    bindings : Optional[ *Substitution[ Binding ] ] )
```

**Figure 5.1: Napier88 description of generator model**

Note also that the *literal* branch of *GeneratorResult* is redundant so far as expressiveness is concerned: a literal result definition could be expressed as a procedure which ignored its parameters and always produced the same result. However, the presence of this branch enables the window-based generator editor to display a meaningful representation of the result definition, as will be illustrated later. The generator editor is used by the programmer to construct instances of type *Generator*.

Hyper-program links to data items may occur in various places in the generators and generated code:

- in a prelude—since it is a procedure which may be produced by evaluating a hyper-program containing links to data items;
- in the source code specified by a literal result definition;
- in the body of an expression result definition—since it is a procedure;
- in the source code produced by an expression result definition—this is achieved using one of the pre-defined procedures *mkLink*, *mkEnvLocLink*, *mkStructLocLink*, *mkVecLocLink* or *mkTypeLink* described in Appendix B.

## 5.5 Generator Evaluation

To allow reflection, a pre-defined procedure *expandGenerator* is used to evaluate the generator, producing a code representation that can be passed to the run-time compiler. The procedure evaluates the given generator, passing it its parameters in an environment. Any generators in the resulting code (sub-generators) are expanded in turn until a code representation without place-holders is obtained.

The expansion of each generator is performed by the procedures *dropAndEval* and *resultOf*. The latter returns a structure containing the generator result and the generator environment obtained by executing the prelude. The procedure *dropAndEval* obtains the result of the given generator. It then iterates through the sub-generators if any, and for each one evaluates it and substitutes the resulting source for the sub-generator. The definition of these procedures are shown in Figure 5.2:

```
type SourceAndEnv is structure( source : GeneratorSource ; envir : env )

! Returns the result source code and the environment.
let resultOf = proc( generator : Generator ; initialEnvir : env → SourceAndEnv )
begin
      ! Call the prelude to set up the environment.
      let enrichedEnvir = generator( prelude )( initialEnvir )
```

```
    project generator( result ) as X onto
    expression :
    begin
        ! Call the procedure to evaluate the
        ! expression in the context of the environment.
        SourceAndEnv( X( enrichedEnvir ), enrichedEnvir )
    end
    literal : SourceAndEnv( X, enrichedEnvir ) ! Return the literal result.
    default : dummyValue ! Can't happen.
end

! Evaluates generator and expands one level of sub-generators.
let dropAndEval = proc( generator : Generator ; initialEnvir : env →
                        SourceAndEnv )
begin
    ! Get the result code and environment.
    let result = resultOf( generator, initialEnvir )
    let resultSource := result( source )
    let resultEnvir := result( envir )

    project resultSource( generators ) as generatorVec onto
    present :
    begin
        ! Expand all the sub-generators
        for i = 1 to upb[ Substitution[ Generator ] ]( generatorVec ) do
        begin
            let generatorSubstitution = generatorVec( i )
            ! Expand the sub-generator.
            let expand = resultOf( generatorSubstitution( subs ), resultEnvir )
            resultEnvir := expand( envir )
            ! Substitute the source code into the main result.
            ! Assume substitute defined elsewhere.
            resultSource := substitute( resultSource,
                                        generatorSubstitution( codeRegion ),
                                        expand( source ) )
        end
    end
    default : {} ! No sub-generators.

    SourceAndEnv( resultSource, resultEnvir )
end
```

```
! Fully expands a generator.
let expandGenerator = proc( gen : Generator ; initialEnvir : env → HyperSource )
begin
	! Expand the generator to the first level.
	let result := dropAndEval( gen, initialEnvir )

	! Continue expanding sub-generators until none left.
	while result( source )( generators ) is present do
	begin
		! Make the current source into a generator.
		let nextLevelGenerator = Generator( nullPrelude,
				GeneratorResult( literal : result( source ) ) )
		result := dropAndEval( nextLevelGenerator, result( envir ) )
	end

	! Return only the source code.
	result( source )( code )
end
```

**Figure 5.2: Definition of *expandGenerator***

The evaluation sequence is illustrated in Figure 5.3 which shows the evaluation of a generator with a literal result. The literal source code contains text with one hyper-program link to a value in the persistent store and two sub-generator substitutions. Each sub-generator contains an expression result; these are evaluated in turn, to produce plain source code in the first case and source code with a further sub-generator in the second case. Once all sub-generators have been expanded the generated code fragments are composed to give the resulting hyper-program.

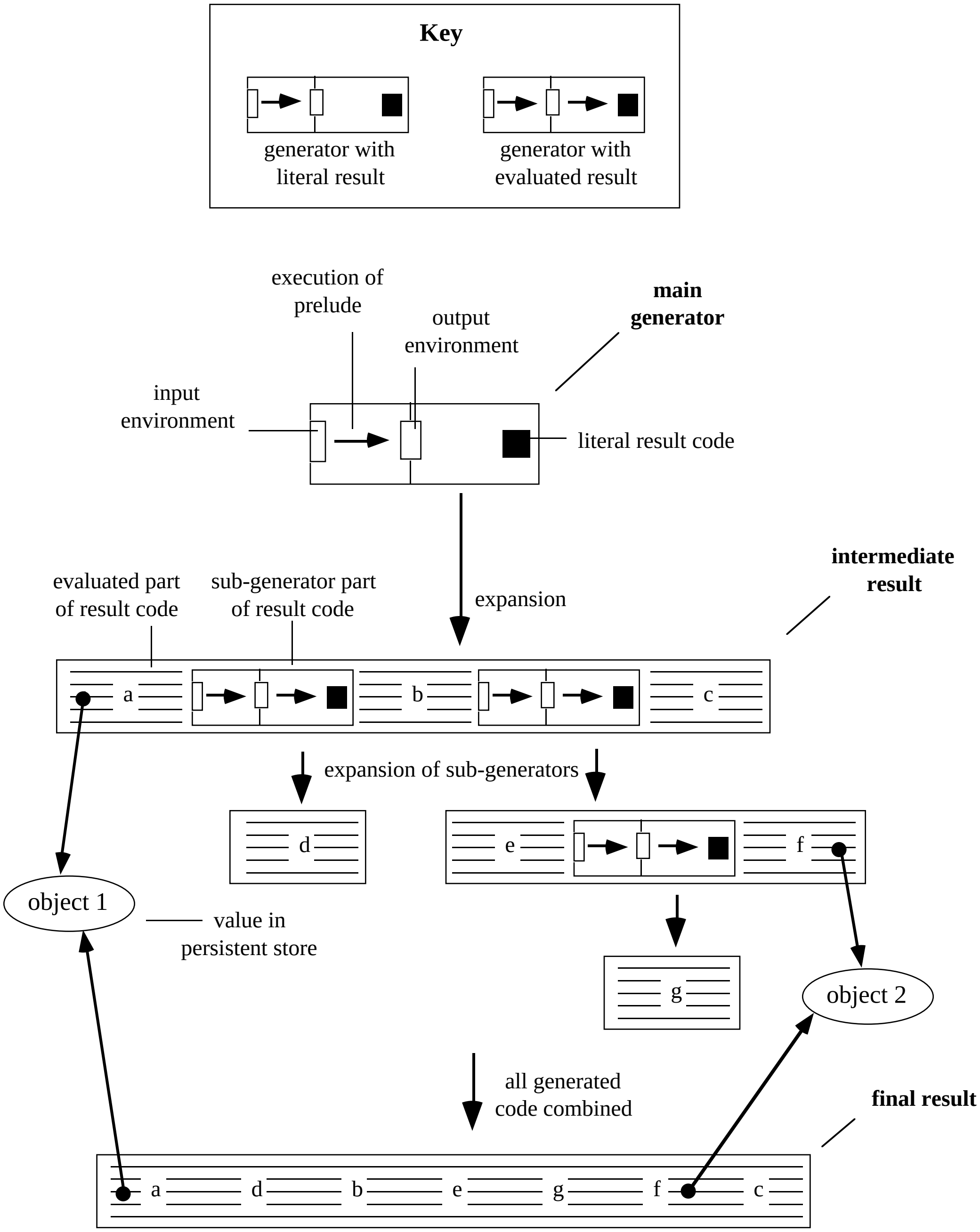

**Figure 5.3: Evaluation of generator by *expandGenerator***

Figure 5.4 shows the pattern of communication between generators: the environment produced by each prelude is passed as input to the prelude of each of the sub-generators immediately below it.

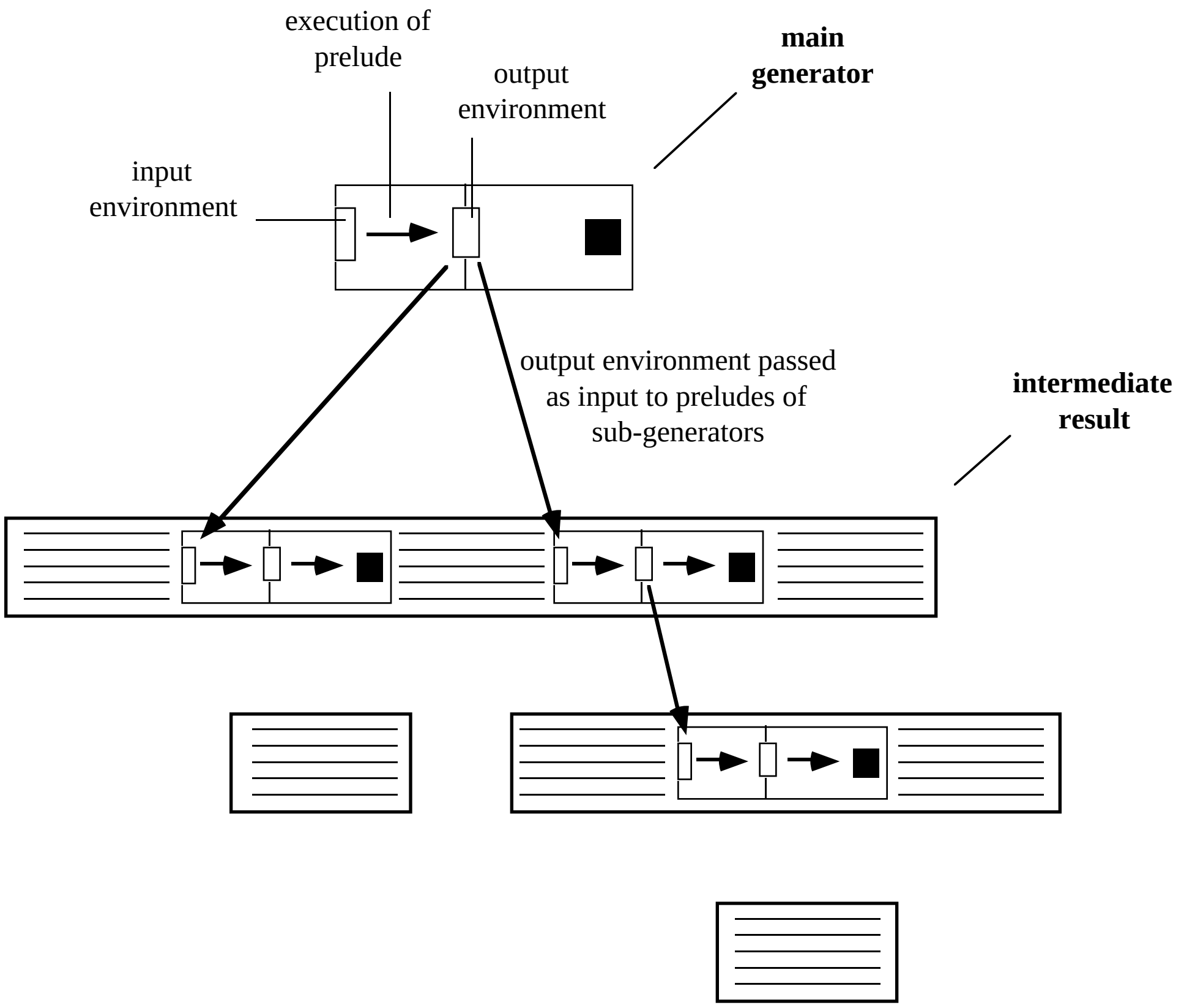

**Figure 5.4: Communication between generator and sub-generators**

## 5.6 Pre-defined Types and Operators

The generator construction system provides a number of pre-defined types and procedures that may be linked into generators and generated code. The types include those shown in Figure 5.1, and a set type. The full list is given in Appendix B. The procedures provide set operations and analysis and synthesis of both type representations and source representations.

To link a procedure or type into a program the programmer selects a representation of it with the browsing tools and creates a link as described in Chapter 4.

## 5.7 Graphical Interface

### 5.7.1 Creating Generators

The graphical generator interface will be introduced with an example used in Chapter 2. Figure 5.5 shows a Napier88 generator which produces the representation of a procedure to calculate a user-specified function:

```
proc( → string )
begin
      writeString( "enter real expression over x" )
      let expr = readString()

      "proc( x : real → real ) ; " ++ expr
end
```

**Figure 5.5: Generator in Napier88**

Using the generator model described in the previous section, this can be represented by the following Napier88 code:

```
! Expanded definition of CodeRegion for string code representation.
type CodeRegion is structure( start, finish : int )

! Dummy values.
let noGenerators = Optional[ *Substitution[ Generator ] ]( absent : nil )
let noBindings = Optional[ *Substitution[ Binding ] ]( absent : nil )
let noPrelude = proc( e : env → env ) ; e

! Procedure to set up environment by reading in function body.
let prelude = proc( e : env → env )
begin
      writeString( "enter real expression over x" )
      in e let expr = Code( textual : readString() )
end

! Define result code with place-holder for function body.
let codeString = "proc( x : real → real ) ; body"
let source = Code( textual : codeString )

! Procedure to generate result code from environment.
let genDefn = proc( e : env → GeneratorSource )
      use e with expr : Code in GeneratorSource( expr, noGenerators )

! Turn it into a full generator.
let bodyGen = Generator( noPrelude, GeneratorResult( expression : genDefn ) )

! Define textual region to be substituted, using character offsets.
let substitutionRegion = CodeRegion( 27, 30 )

! Vector containing single sub-generator.
let generators = Optional[ *Substitution[ Generator ] ]( present :
      vector @1 of [ Substitution[ Generator ]( bodyGen, substitutionRegion )] )

! Make source into hyper-program source.
let hyperSource = HyperSource( source, noBindings )

! Form main generator.
let result = GeneratorResult( literal : GeneratorSource( hyperSource, generators ) )
let mkFun = Generator( prelude, result )
```

**Figure 5.6: Generator in refined model**

Figure 5.7 shows how the example fits with the general structure shown in Figure 5.3. Execution of the prelude populates the evaluation environment with the representation of the user-specified function; this is passed to the sub-generator from which it is returned unchanged and composed with the procedure header to give the result.

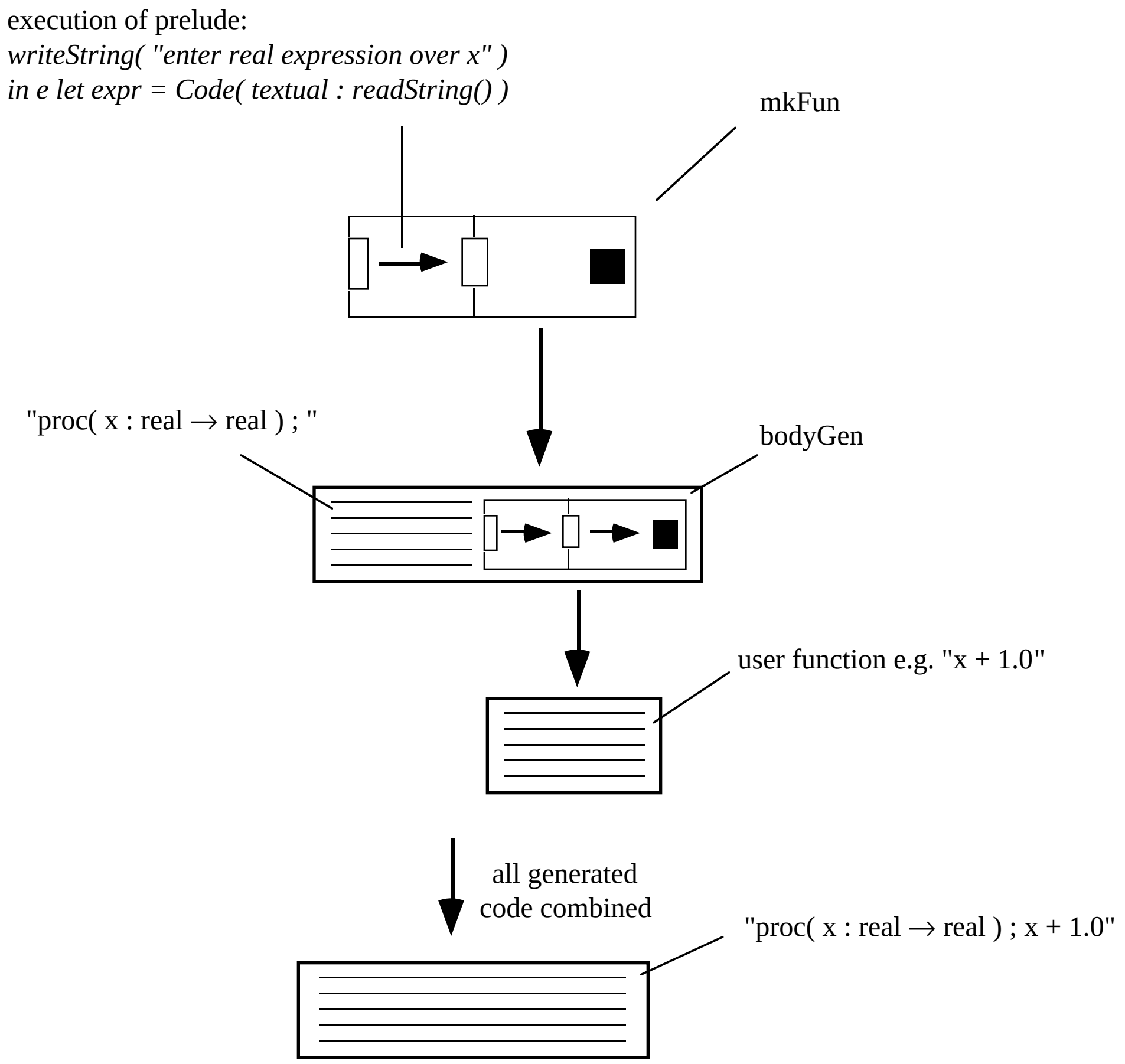

**Figure 5.7: Structure of generator example**

The programmer need not enter the verbose code shown in Figure 5.6 directly, since it is constructed by the interactive generator editor. Figures 5.8 to 5.11 show how the generator is constructed using the editor. For such a simple example it may not be clear that the system provides any improvement in legibility over the original Napier88 encoding, but it will serve to illustrate the way the editor is used. Appendix A gives a non-trivial example, the definition of a generic natural join function, together with corresponding definitions in Napier88 and TRPL.

The window in Figure 5.8 shows the definition of the generator *mkFun* under construction. The top three quarters of the window contains the prelude. The first two sub-windows show the parameters expected by the prelude, the one on the right for type representation parameters and the one on the left for all other values. In this case no parameters are used by the prelude; any present in the input environment will be ignored. The body of the prelude contains a call to a procedure *writeString* linked in from the persistent store: the prefix *L:* on the button label indicates that the button represents a hyper-program link to a location. The radio buttons below the prelude body window have the *unchanged* choice selected, indicating that the prelude returns the input environment, unchanged, as the output environment. The prelude results and result definition have not yet been filled in.

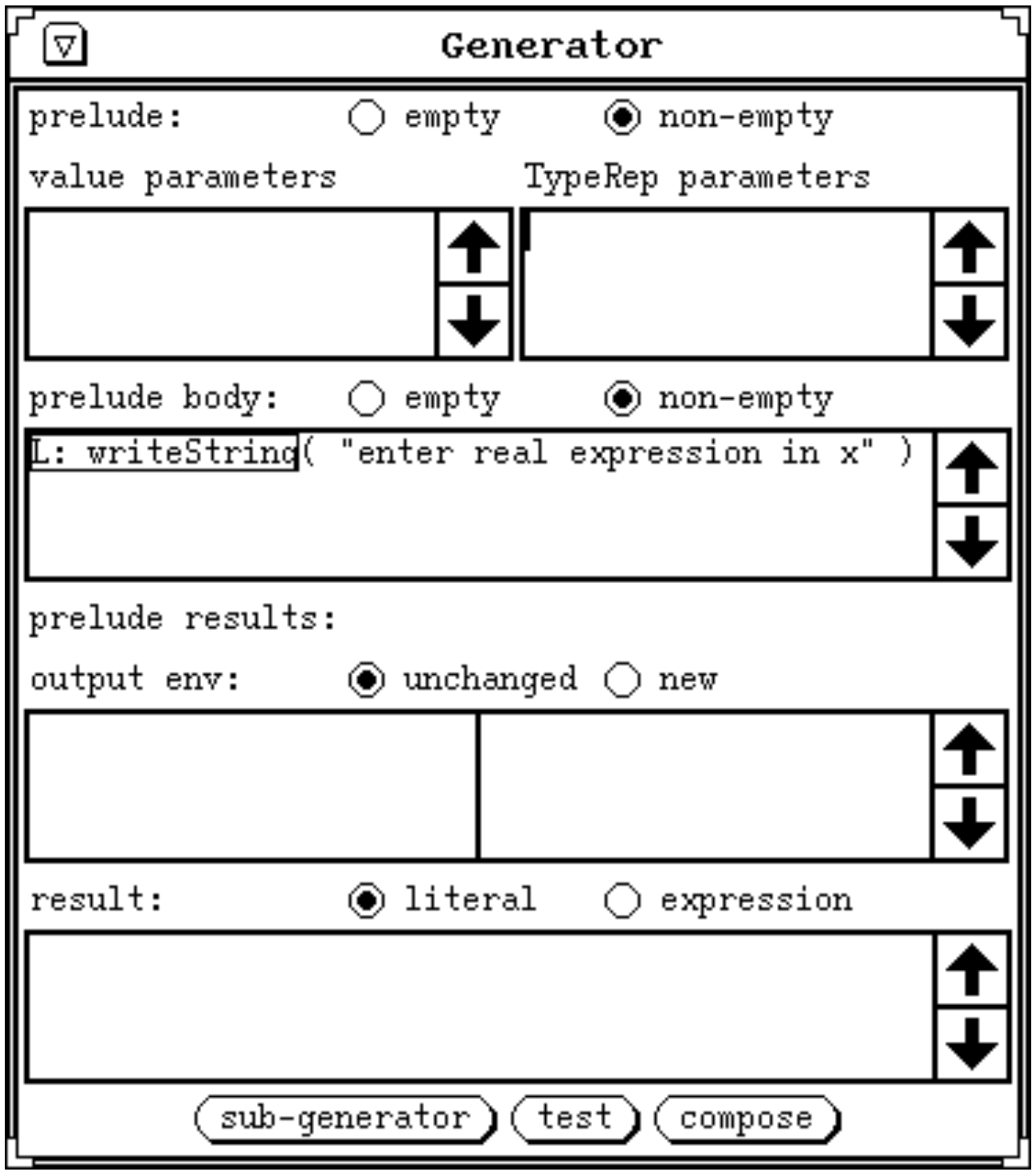

**Figure 5.8: Defining prelude body**

Figure 5.9 shows the generator definition window after the programmer has entered the specification of the prelude outputs, to be placed in the output environment. The first column contains the identifiers to be associated with the outputs and the second column contains expressions for the outputs themselves. In this case there is one output with the identifier *expr*. The output value is obtained by reading in a string and converting it to a source code fragment. This is performed at generator evaluation time, immediately after the prelude body is executed. It involves calling two other procedures, *mkHyperSource* and *readString*, the former being one of the pre-defined procedures provided by the generator system. The code to produce the output value contains hyper-program links to these procedures.

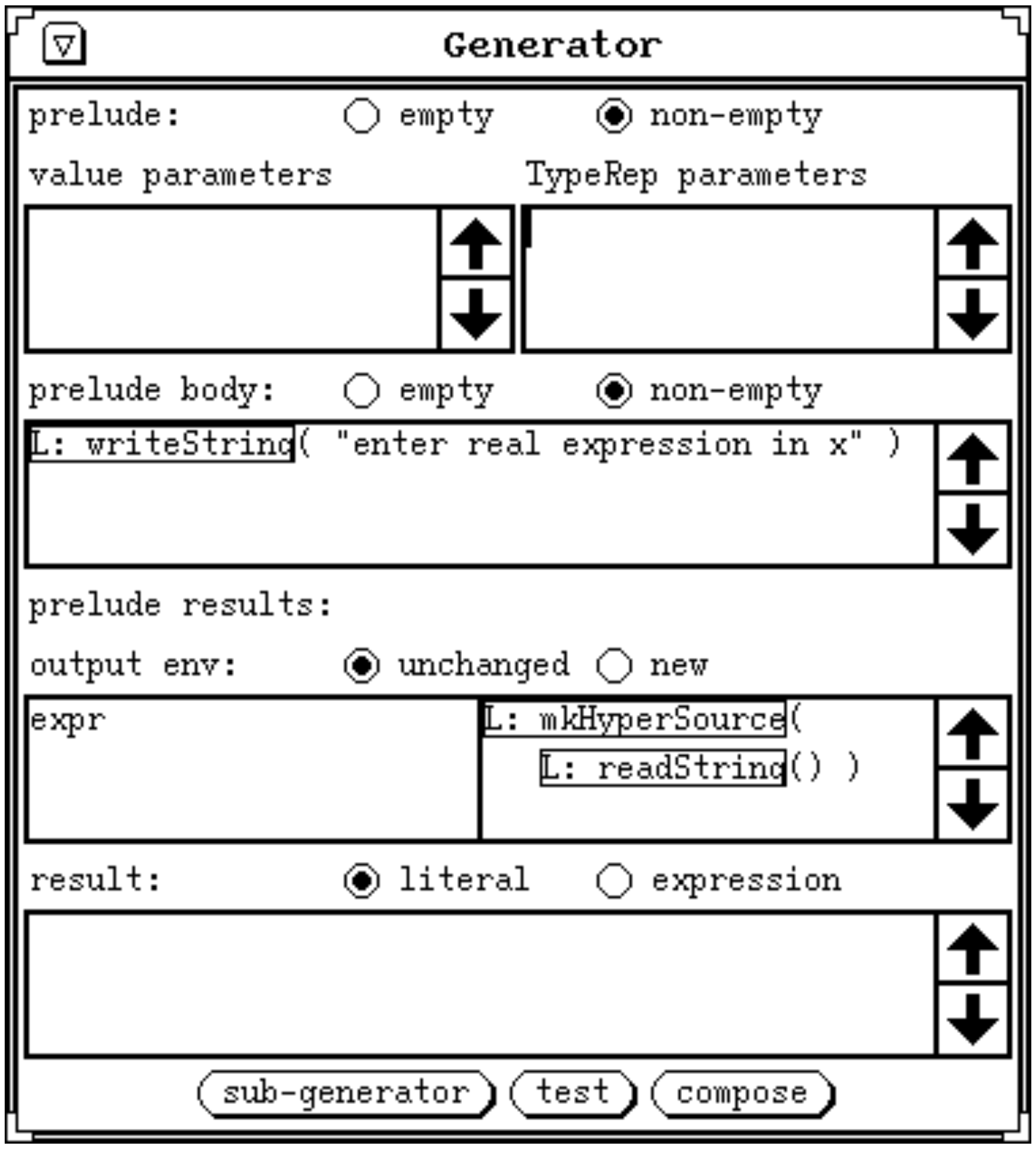

**Figure 5.9: Defining prelude results**

Figure 5.10 shows the window after the result definition has been entered. The *literal* choice is selected to show that the definition is a literal. The result definition window contains the literal source code. The sub-generator that generates the function body has not been entered at this stage.

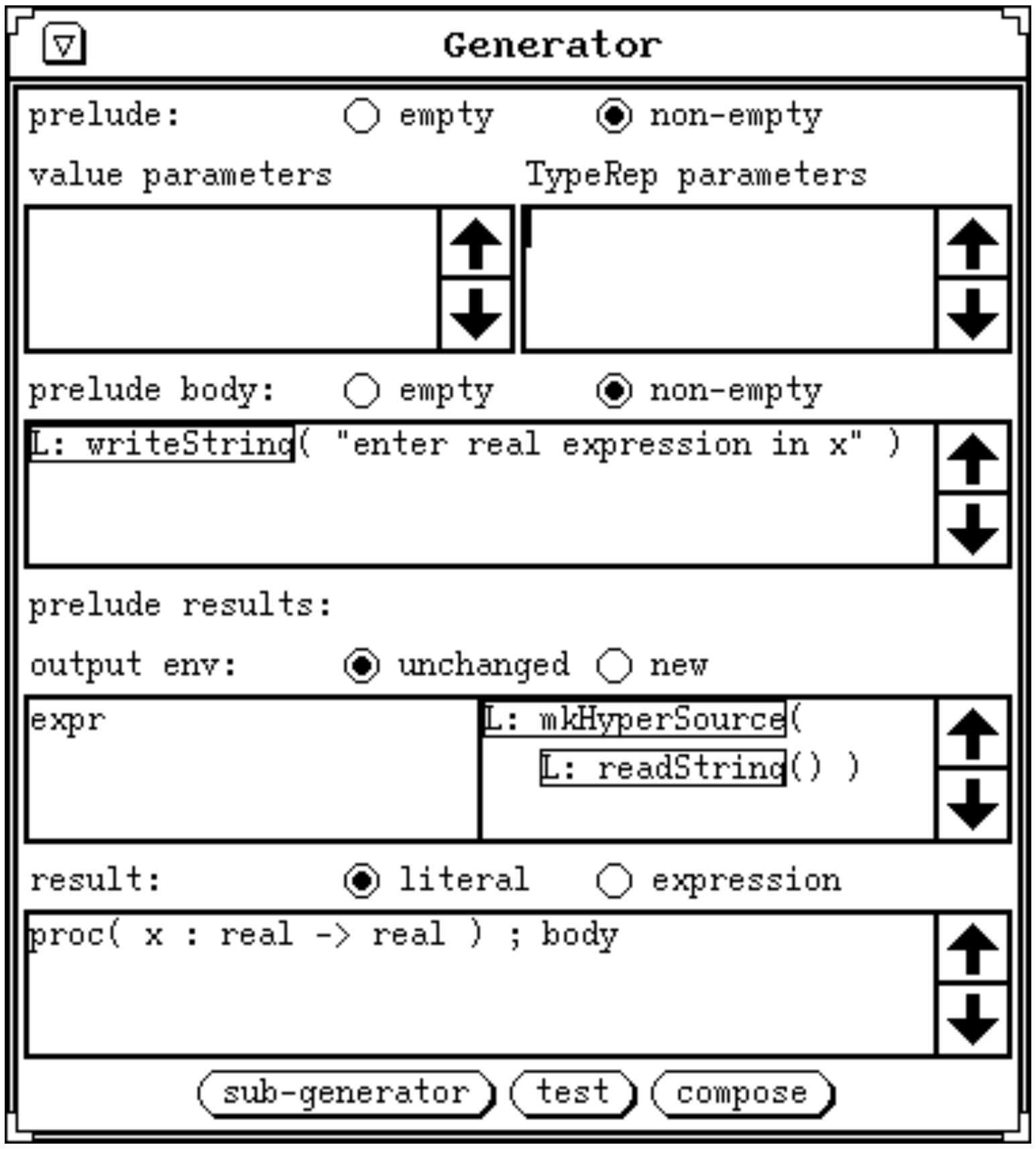

**Figure 5.10: Defining result definition**

Finally, Figure 5.11 shows the editor after the sub-generator has been specified. To do this the programmer selected the text *body* in the result definition and pressed the *sub-generator* button. The editor then replaced the text with a button labelled *G: body*, the prefix indicating a generator, and displayed another window to allow the sub-generator definition to be entered. This sub-generator contains no prelude. The *expression* choice for the result definition specifies that the sub-generator result is formed by evaluating an expression which simply returns the source code passed in. The parameter list has one entry, showing that the parameter *expr* in the input environment is used. Recall that the output environment of the main generator is passed as the input environment to the sub-generator. The parameter has the pre-defined type *HyperSource*; a link to the type is shown by a button with a *T:* prefix.

The point of having a distinguished literal branch in the result definition type, rather than representing all result definitions as procedures, should now be apparent. It means that the fixed parts of the result definition can be displayed in the generator editor without having to evaluate the generator with any particular arguments.

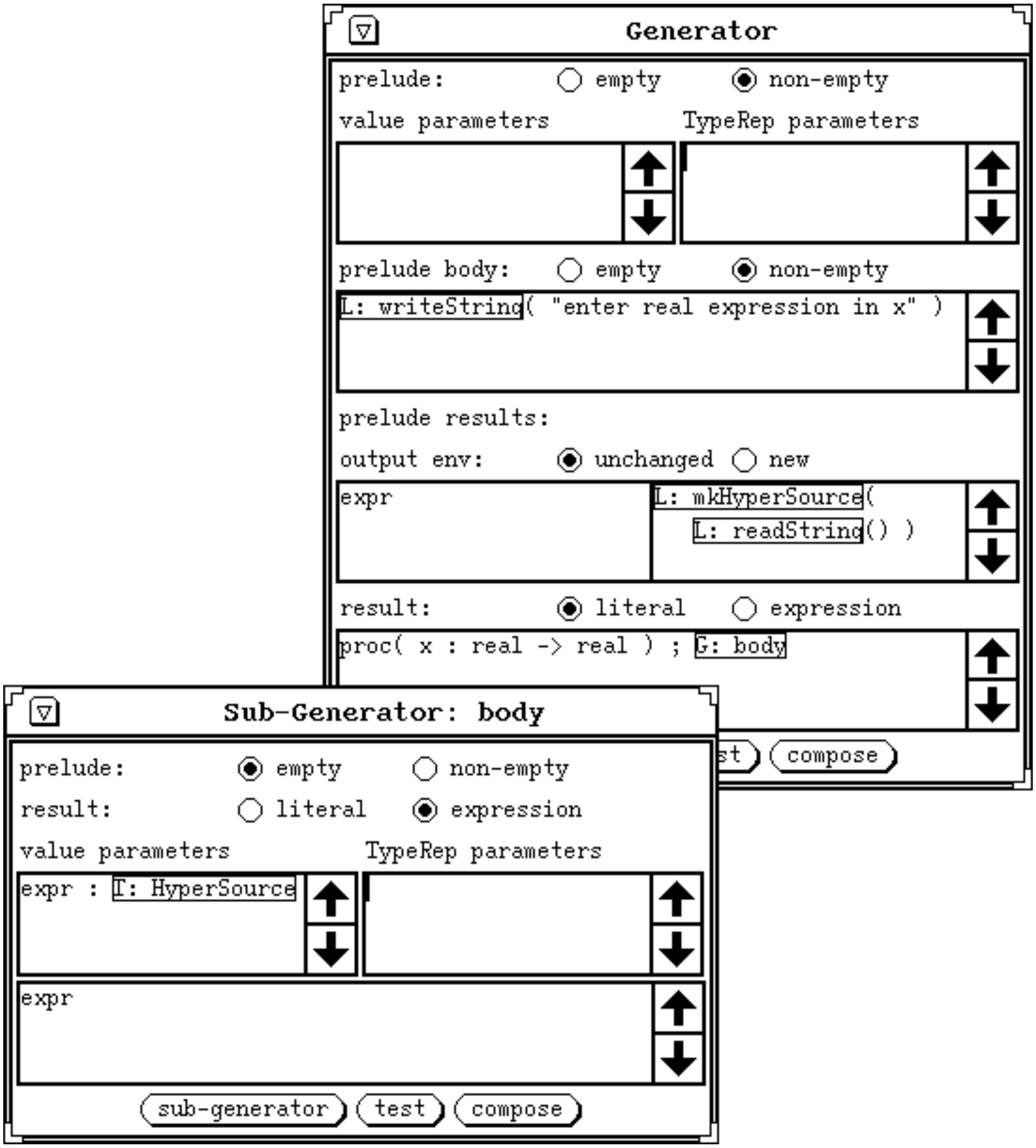

**Figure 5.11: Completed generator definition**

Sub-generator buttons only occur in literal result definitions: if the *sub-generator* button is pressed when the insertion point lies within a literal result definition, a sub-generator button with a *G:* prefix is inserted, as described above. If the insertion point lies within an expression definition, however, the button will denote a hyper-program link to the new generator, and will have a *V:* prefix.

The function of the *test* button is described in the next section. The *compose* button creates an instance of type *Generator* from the current window contents and causes a representation of it to be displayed by the browsing tools so that it may be linked into other programs.

## 5.7.2 Testing Generators

The testing facility allows the programmer to test the generator with various inputs. When the *test* button is pressed a new window is displayed, containing a sub-window in which values for the generator parameters may be entered. The programmer can then press the *generate code* button to evaluate the generator with those parameters. If the generator executes successfully the resulting code representation is displayed in the lower sub-window. One possible reason for failure of the generator is that the parameters supplied are not compatible with those expected by the generator: in this case a message to that effect is displayed. When generated successfully, the code may itself be evaluated by pressing the *evaluate* button. This has the same effect as evaluating code in a hyper-program editor window: if compilation succeeds the code is executed and any resulting value displayed by the browser, otherwise compilation error messages are displayed.

Figure 5.12 shows two examples of test windows for the generator *body*. In the first the parameter *expr* is given the value formed by converting a string literal into source code using the pre-defined procedure *mkHyperSource*. The source code generated is simply the string without any hyper-program links. In the second a fragment of source code from the persistent store is linked in as the value for the parameter. As *body* is only a sub-generator the code fragments generated are not well-formed when taken in isolation. Pressing the *evaluate* button in either case would result in a message that the name *x* had not been declared.

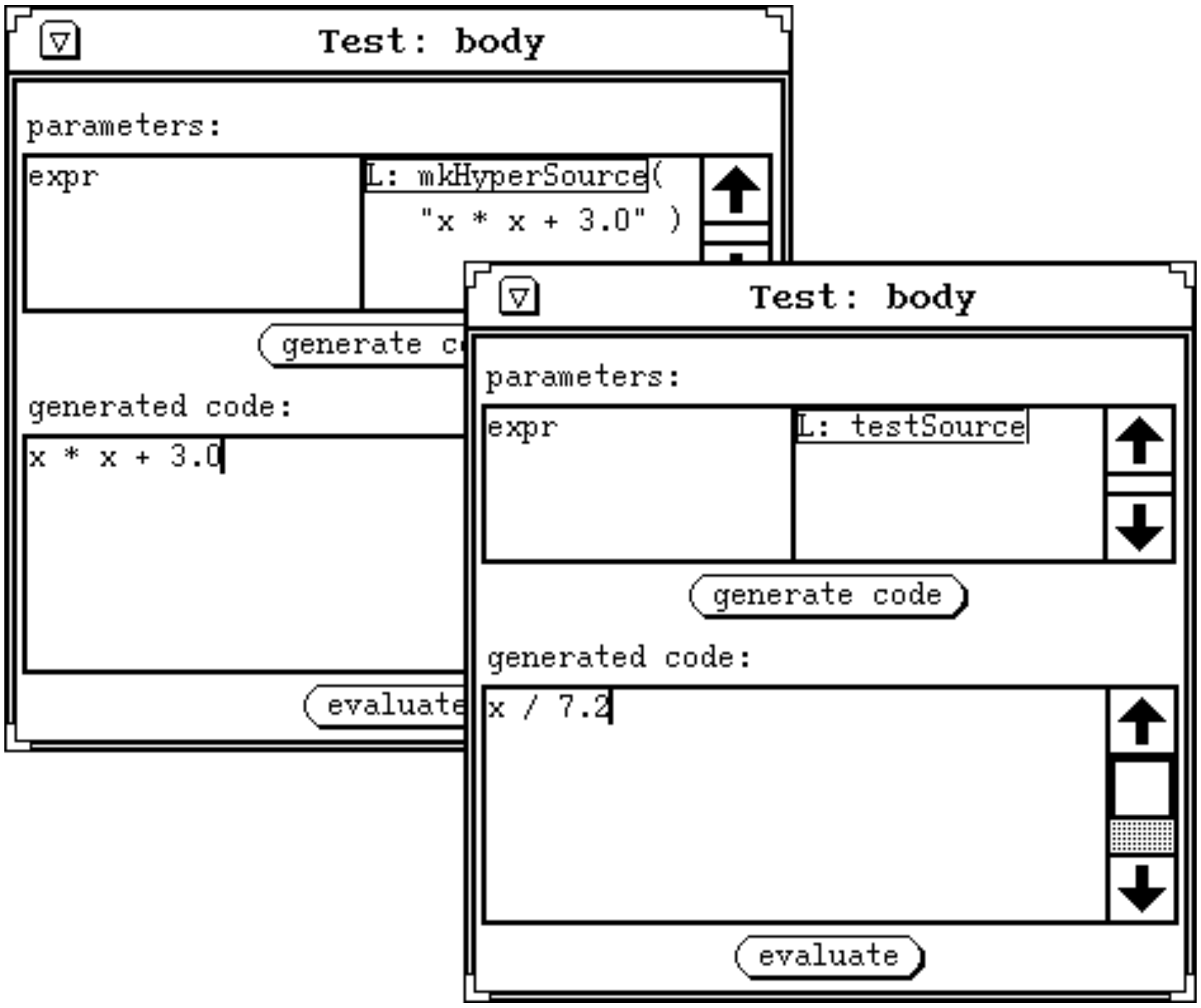

**Figure 5.12: Test windows for sub-generator *body***

Figure 5.13 shows a test window for the main generator *mkFun*, after the *generate code* button has been pressed and the programmer has been prompted for input during the execution of the prelude. In this example the expression input contained hyper-program links to the procedures *sin* and *f* in the persistent store, with the result that the generated code also contains these links.

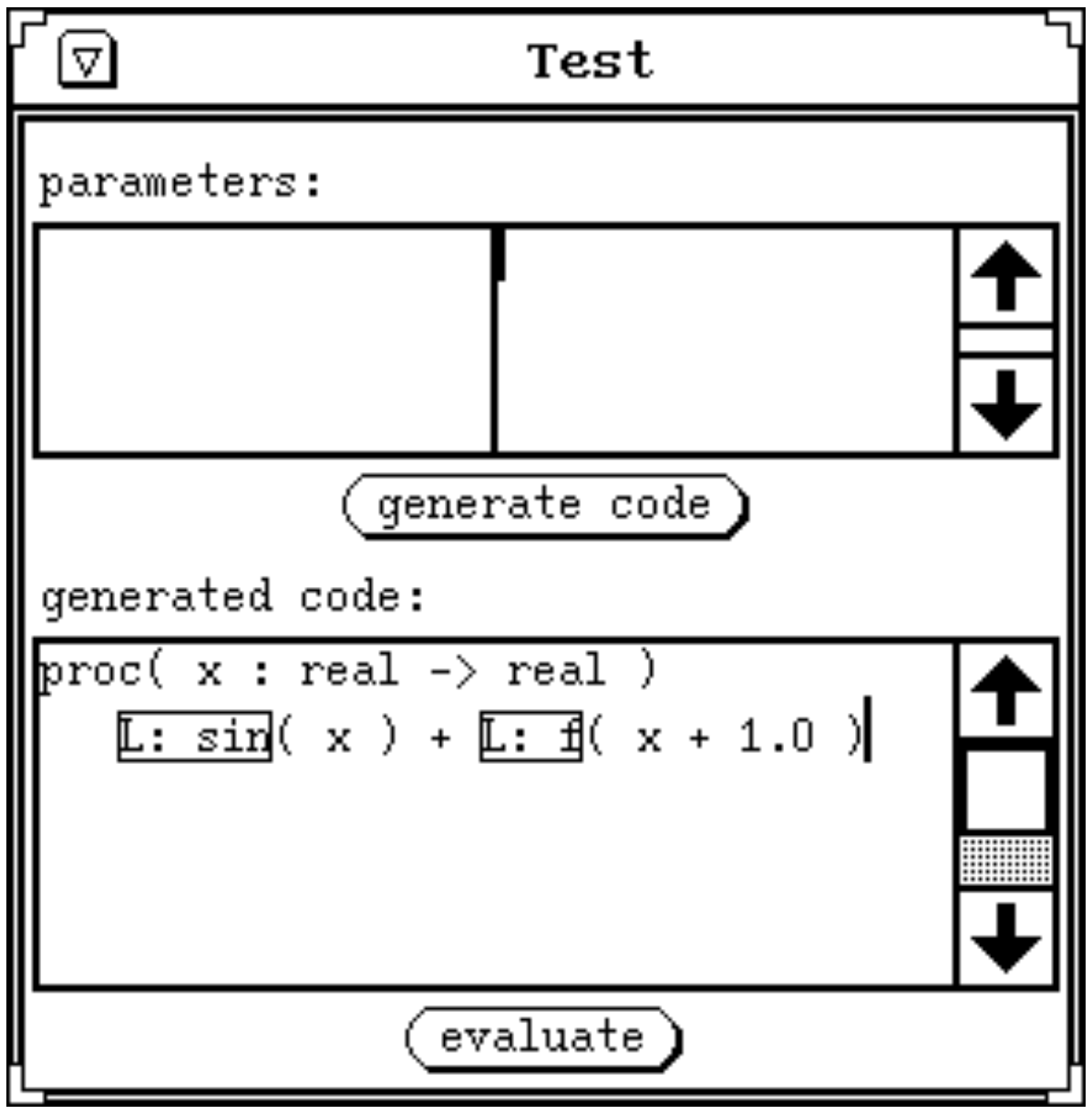

**Figure 5.13: Test window for generator *mkFun***

### 5.7.3 Generating Hyper-Program Links

It has already been described how a generator may produce a source representation that contains embedded direct links to data items in the persistent store. One mechanism for achieving this was illustrated in the example in Figure 5.13, in which source code containing links was passed to a generator as a parameter and formed part of the result. There are two other ways that links may be incorporated in generated code:

- a literal result definition may contain hyper-program links; or

- an expression result definition may contain expressions that evaluate to give hyper-program links.

These will be illustrated in turn. Figure 5.14 shows an example of a generator with a literal result definition that contains both a hyper-program link and a type link, the latter being distinguished by a *T:* prefix on the button. The test window shows that the code generated is simply the result definition code.

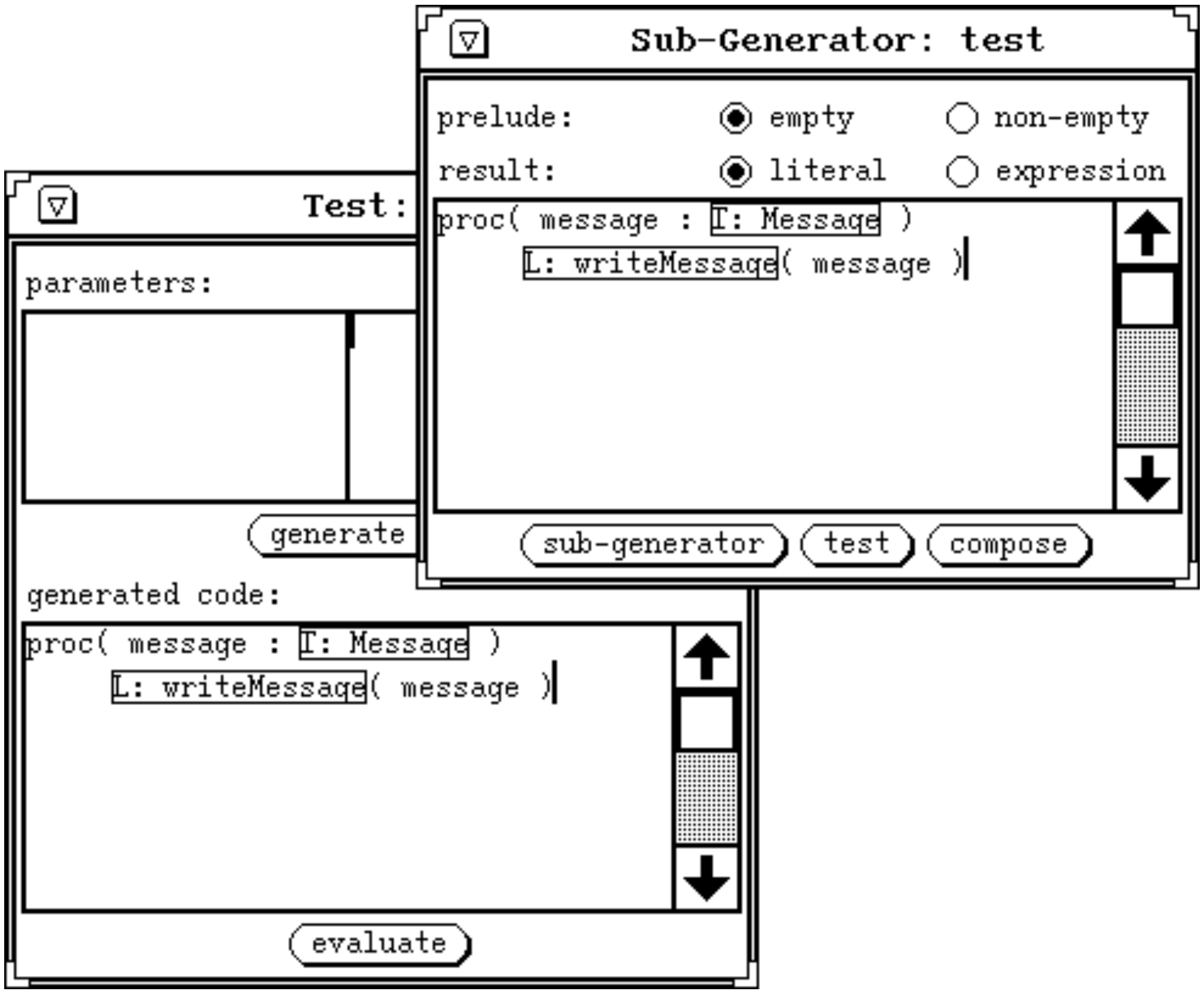

**Figure 5.14: Hyper-program and type links in a literal result definition**

Figure 5.15 shows a generator *test2*, the result definition of which contains embedded sub-generators *newType*, *process* and *vec*. These sub-generators contain references to the pre-defined procedures *mkTypeLink*, *mkEnvLocLink* and *mkLink* which create links to, respectively, a type, a named environment location and a value. The test window at the bottom shows an example of the generated code which contains a link to the type *int* and hyper-program links to a location containing a procedure and to a vector value.

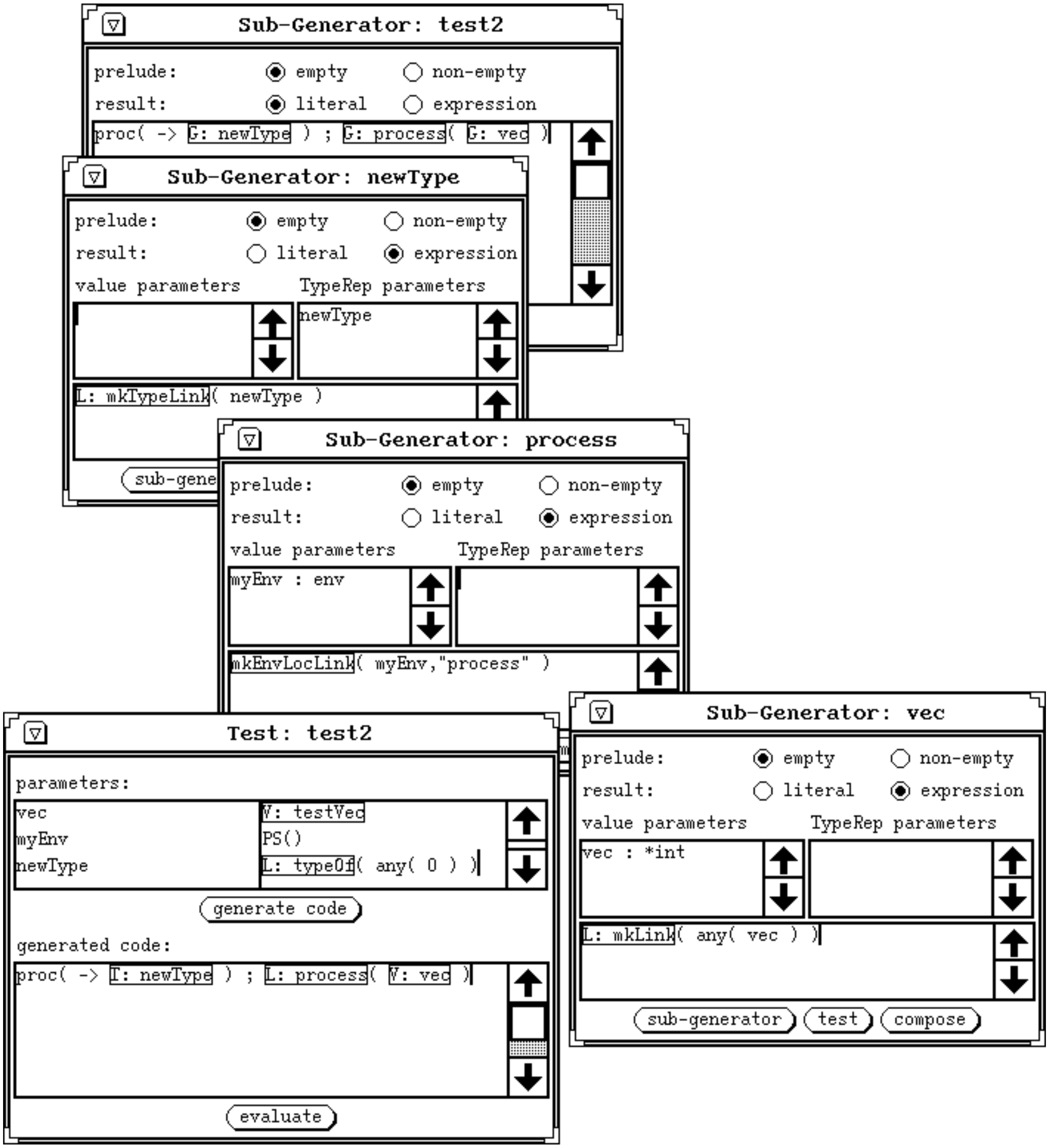

**Figure 5.15: Hyper-program and type links in an expression result definition**

## 5.8 Conclusions

A set of reflective programming tools has been described. They allow the programmer to construct generators that manipulate and produce hyper-program source representations. The tools provide a graphical user interface to a generator model in which generators may produce fixed or variable results, and may contain embedded sub-generators nested to any degree. The hyper-program facilities provide a uniform mechanism by which both generator and generated code may contain links to data in the persistent store. This gives convenient access to the library of pre-defined procedures for source code analysis and manipulation. The graphical interface provides an abstraction mechanism with which the programmer may choose to view a generator at varying degrees of detail.

The use of hyper-program generators has been related to the possibilities for early program checking outlined in the previous chapter. As hyper-programs may contain manifest data, properties of that data can be verified by the generators that produce them. These techniques stretch the spectrum of times available for linking and checking. While allowing very early linking, with associated safety and efficiency benefits, they do not preclude dynamic linking during execution in the cases where it is useful.

# 6 Implementation

## 6.1 Introduction

This chapter describes the principal features of the implementation of the hyper-programming and reflective programming tools. This includes a description of the software systems upon which they are implemented, which are as follows:

- a graphical user interface tool-kit, *WIN*; [KCD+89, CDK90, KCC+92a]
- a set of persistent object browsing tools [KD90, KCC+92a]; and
- the Napier88 compiler [Dea88, Cut92].

Each of these systems is available to the Napier88 programmer and can be used independently of the others. They are implemented in Napier88 although the object browser and the compiler make use of implementation level facilities that are not generally available to the Napier88 programmer.

The user interface aspects of the hyper-programming system are implemented entirely using WIN. Underlying it are the browsing tools and the compiler. The browsing tools allow data linked into hyper-programs to be displayed graphically and data in the persistent store to be selected for linking into new hyper-programs. The compiler allows hyper-programs to be transformed into executable forms. Some changes to the original browsing system and compiler were made; these are described in Section 6.5.

The relationships among the software layers are illustrated in Figure 6.1. This shows that, for example, the browser system is built using WIN, the Napier88 language and some of the facilities of the Napier88 implementation level.

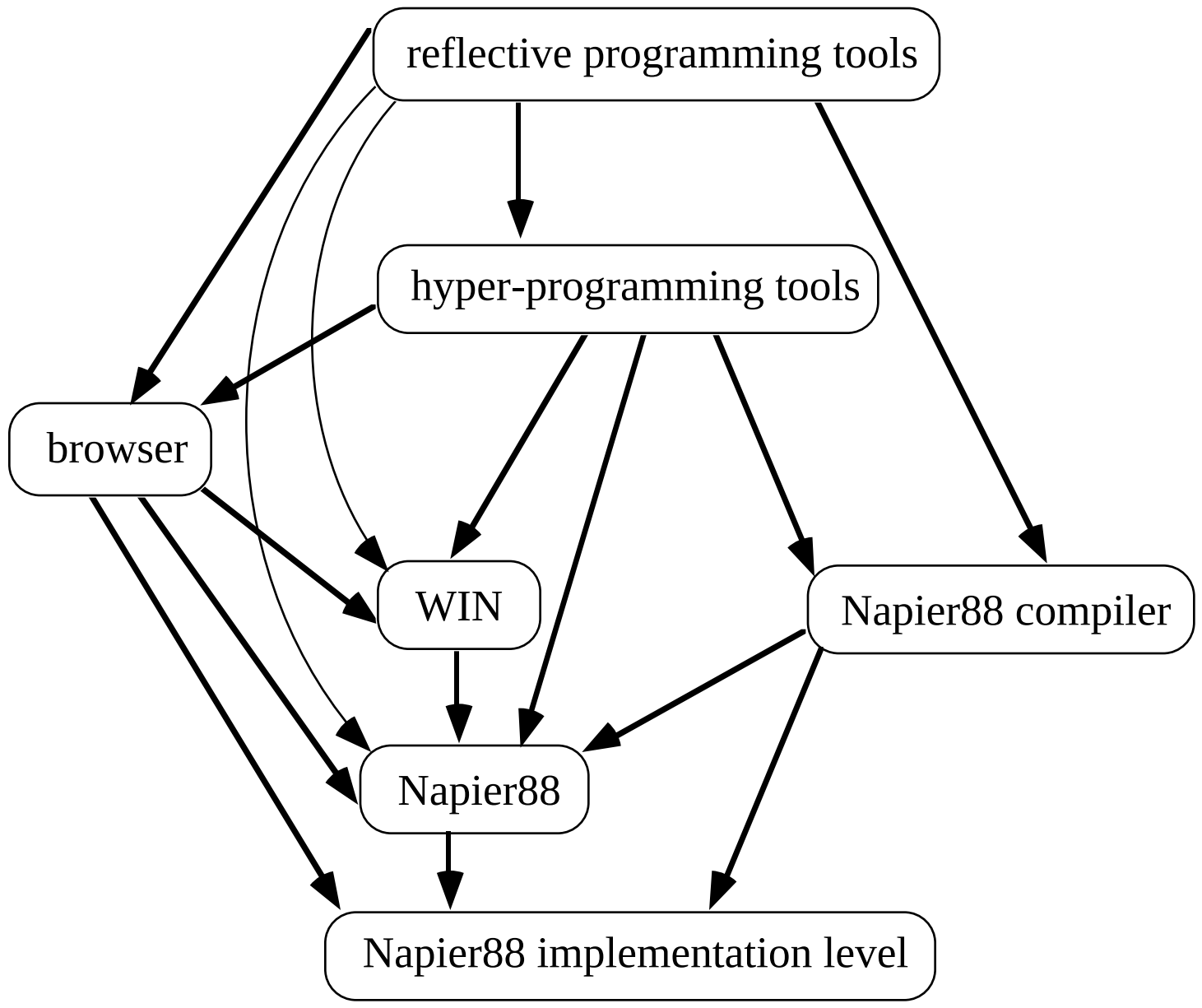

**Figure 6.1: Dependencies among software layers**

## 6.2 User Interface Tool-Kit

The user interface tool-kit, WIN (Windows In Napier88), supports the programming of interactive graphical interfaces for Napier88 applications. Its main features are:

- a user input event distribution system;
- support for creating and displaying overlapping windows;
- facilities for hyper-text editing, a sub-set of which support conventional text editing; and
- a pre-defined library of user interface ‘widgets’.

Design and implementation work was carried out in collaboration with Quintin Cutts, Alan Dearle and Richard Connor. Cutts was heavily involved with the development of the window management system in general, while Dearle and Connor respectively were the principal designers of the virtual window and notifier mechanisms, to be described.

### 6.2.1 History

Work on a persistent window management system began with the development of the PStools system in PS-algol [CK87]. This provided window management and event distribution but planned text editing facilities were never fully implemented. Windows created in the system were permanently associated with particular window managers; this was found to be too inflexible as they could not be stored independently in the persistent store.

Following the implementation of Napier88 the WIN system was developed in a number of stages. The first provided independent windows and window managers, an event distribution mechanism similar to that used in PStools and a limited range of user interface widgets. The next stage added text editing facilities. In the most recent stage hyper-text editing facilities were added, allowing text to contain embedded light-buttons, and a more extensive range of interface widgets provided.

### 6.2.2 Event Distribution

WIN supports multiple applications. Since Napier88 provides a single thread of control only one application can execute at a time and the others are suspended until control is transferred. This transfer of control is event-driven: an application is active only so long as input events are directed to it. Each application is modelled as a procedure that accepts a single input event, performs some action and then returns control to a central event routing system (ERS). At any time the ERS may have several such applications registered with it. When an input event occurs the ERS determines to which application to route it and calls the appropriate application procedure with the event as parameter. As there is only one thread of control the ERS is suspended while the application performs its processing. When the processing terminates the ERS resumes polling for input events. This control structure is illustrated in Figure 6.2:

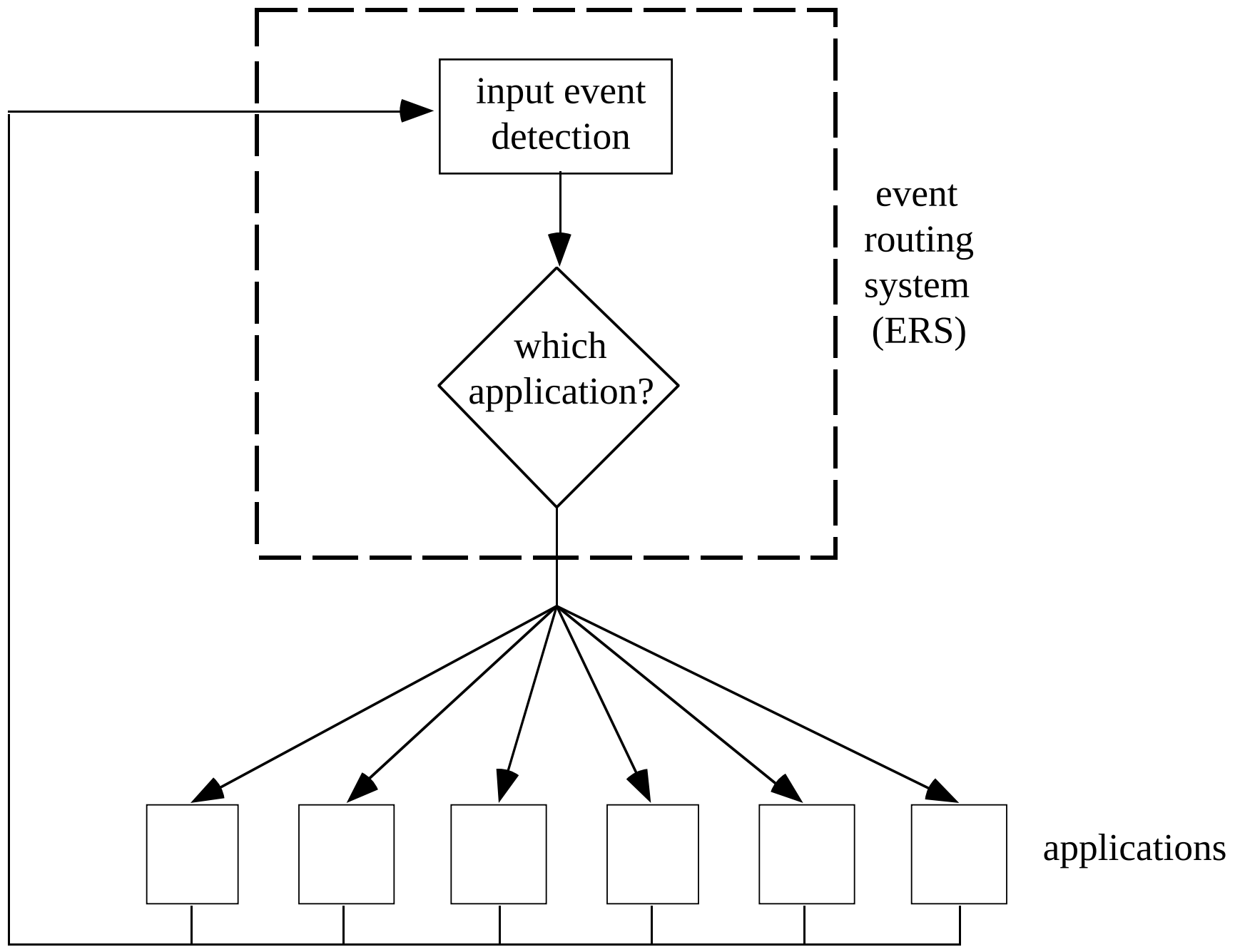

**Figure 6.2: Event distribution in WIN**

The successful operation of this system depends on adherence to a convention that application procedures will return control as soon as possible. In particular it is assumed that applications will not poll for input events themselves but will rely on the ERS to detect and route events.

The ERS deals with fairly low-level input events that describe mouse and keyboard activity. On each event detection/routing cycle the ERS determines whether any keyboard characters have been typed since the last event was routed. If so, all the new characters are concatenated into a single string and routed as an event. Otherwise a mouse event is routed. The representation of the mouse event contains information about the current position of the cursor and whether or not the mouse buttons are currently pressed. Mouse events are routed even if the state of the mouse has not changed since the last event, thus when the system is quiescent with no user input occurring a continuous stream of identical mouse events is routed.

The ERS also generates two kinds of ‘pseudo-event’ which are used to signal to applications when the event distribution path changes. Each time a keyboard or mouse event is routed the ERS first checks whether it is being routed to the same application that received the previous event. If not, before routing the event, the ERS routes a ‘deselect’ event to the application that received the previous event and then a ‘select’ event to the new application. The main purpose of this is to let an application know when it is about to become inactive. For example a light-button might be programmed by an application that highlights and de-highlights an area of the screen depending on whether a mouse button is pressed when the cursor is in the area. The use of deselect events allows the application to de-highlight the light-button if the cursor moves out of the area while the mouse button is pressed down. Without deselect events the application would not be called once the cursor was moved away.

Events are represented by the following Napier88 type:

```
type Event is variant( chars :            string;
                       mouse :            Mouse;
                       select, deselect : null )
```

where

**type** Mouse **is structure**( x, y : **int** ; buttons : ***bool** )

The state of the mouse buttons is represented by the vector of booleans *buttons*, the *i*th element of which is *true* if button *i* is pressed and *false* otherwise. Higher level mouse events such as double clicks and button state transitions are not represented in the ERS but can be detected by computation within applications if required. Keyboard input, however, is treated at a higher level in that events represent characters rather than states of the entire keyboard. Whether or not this approach gives a more suitable level of abstraction for the application programmer, it is forced by the underlying IO facilities of Napier88.

The routing of events is performed by *notifiers*. The main notifier keeps track of the applications registered with the ERS and the criteria that determine how events are routed between them. When each application is registered the programmer supplies a filter procedure that, when invoked, takes an input event as its parameter and returns a boolean. A value of *true* indicates that the application accepts the event for processing while *false* indicates that the event is ignored by the application. The notifier maintains an ordered list of applications and their associated filter procedures. To route an event the notifier scans the list, calling each filter procedure in turn with the new event as parameter, until one of them returns *true*. The event is then passed as a parameter to the corresponding application and deemed to have been consumed. If none of the filter procedures return *true* the event is discarded.

Notifiers can be composed in hierarchies. This is achieved by registering one notifier as an application with another notifier. Thus a top-level notifier might route events between different user applications while sub-notifiers are used to route them to different components within an application.

Notifiers route events given to them; another component is needed to poll for user input, construct events from it, and pass them to the top-level notifier. This role is performed by the *event monitor*, the only system that actively polls the input devices. Figure 6.3 shows the interactions of notifiers, applications and the event monitor:

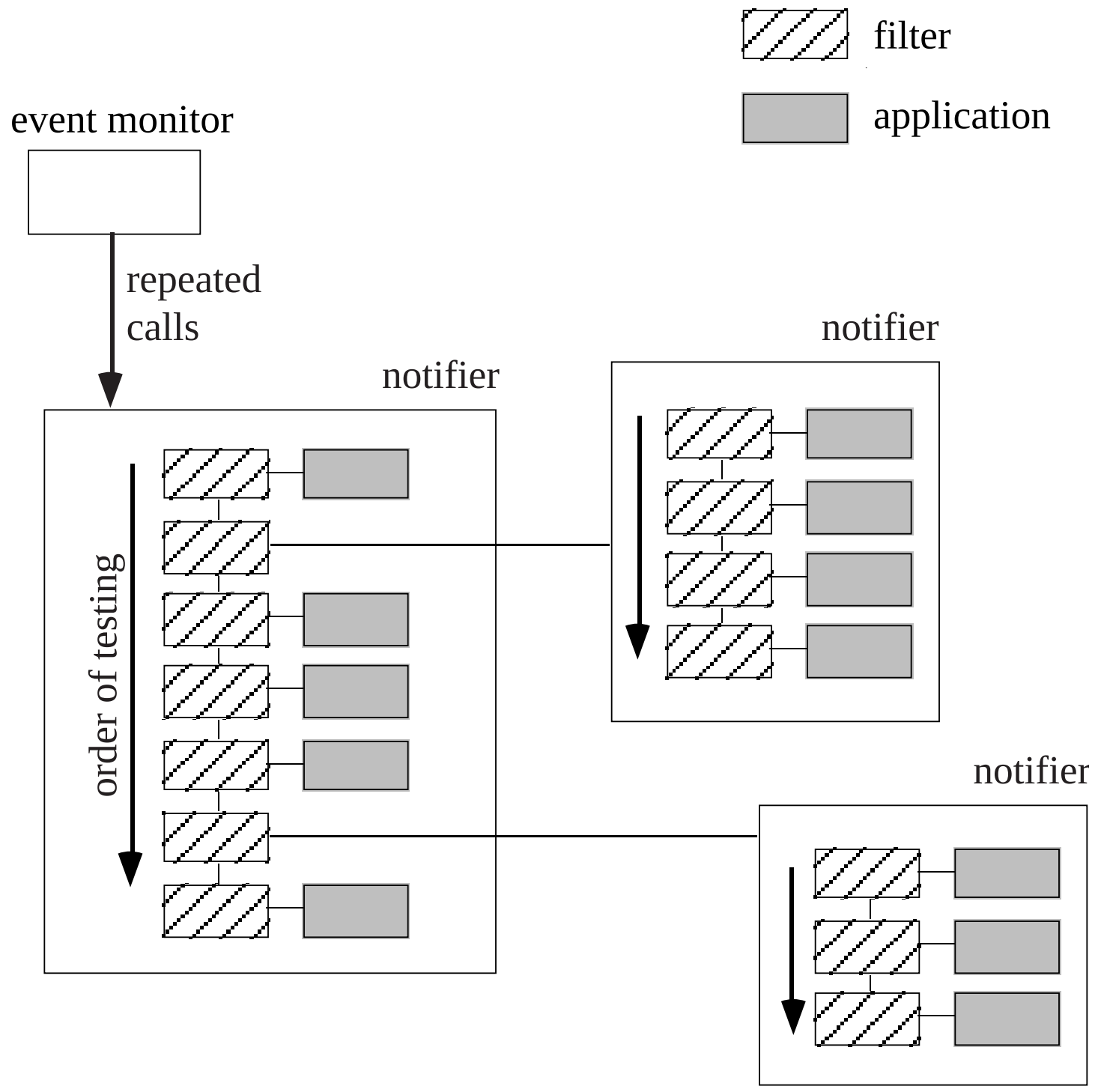

**Figure 6.3: Event monitor and a notifier hierarchy**

Notifiers can be re-configured dynamically, that is, (filter, application) pairs can be removed and new ones added at any point in the list.

### 6.2.3 Windows and Window Managers

Windows are commonly used in user interface systems to partition and organise the screen display area [MM81, App86, Mye86, SG86, WCG87, HP88, Sun89]. When they can overlap the effect is to provide a usable space greater than the physical screen size. WIN supports windows and window managers; the latter are used to organise the display of multiple windows. Each window has an application of the form described in the previous section associated with it. When user input events are directed towards a particular window the ERS routes them to the corresponding application.

Applications manipulate windows. The interactions between windows are handled by the window manager. The full procedural interface of a window is given in Appendix C. The principal operations on a window are raster operations, altering its size and setting the application that handles the events it receives. Raster operations may specify another window or an image as the source or the destination, and a number of raster modes (e.g., *copy*, *xor*, *not*, etc.) may be used. As well as changing the size of a window the programmer can specify the behaviour of the application when the window is resized in future. The programmer of an application does not have deal with matters such as the position of the window or whether it is obscured. The co-ordinates of mouse events are translated relative to the origin of the window, and repainting of the window when it becomes visible is handled automatically.

Window managers are used to display windows on the screen. A window can exist independently of any window manager but it only becomes visible when a window manager is used to display it. The principal operations of a window manager support displaying and un-displaying windows, setting their position and depth, and transforming them to and from their iconic forms. The depth of a window determines which other windows are obscured by it. A window cannot be displayed by more than one window manager simultaneously at present.

Window managers can be nested within one another to arbitrary depth. When a window manager is created the programmer specifies a window within which it operates. The recursion is grounded by a special window manager that operates directly on the physical screen. This is illustrated in Figure 6.4. The dotted region is controlled by the special window manager which is displaying two windows labelled *a* and *d*. Another window manager operates within window *a*, shown by the striped region. That window manager displays a further two windows, *b* and *c*. Window *a* is known as the parent window of *b* and *c*.

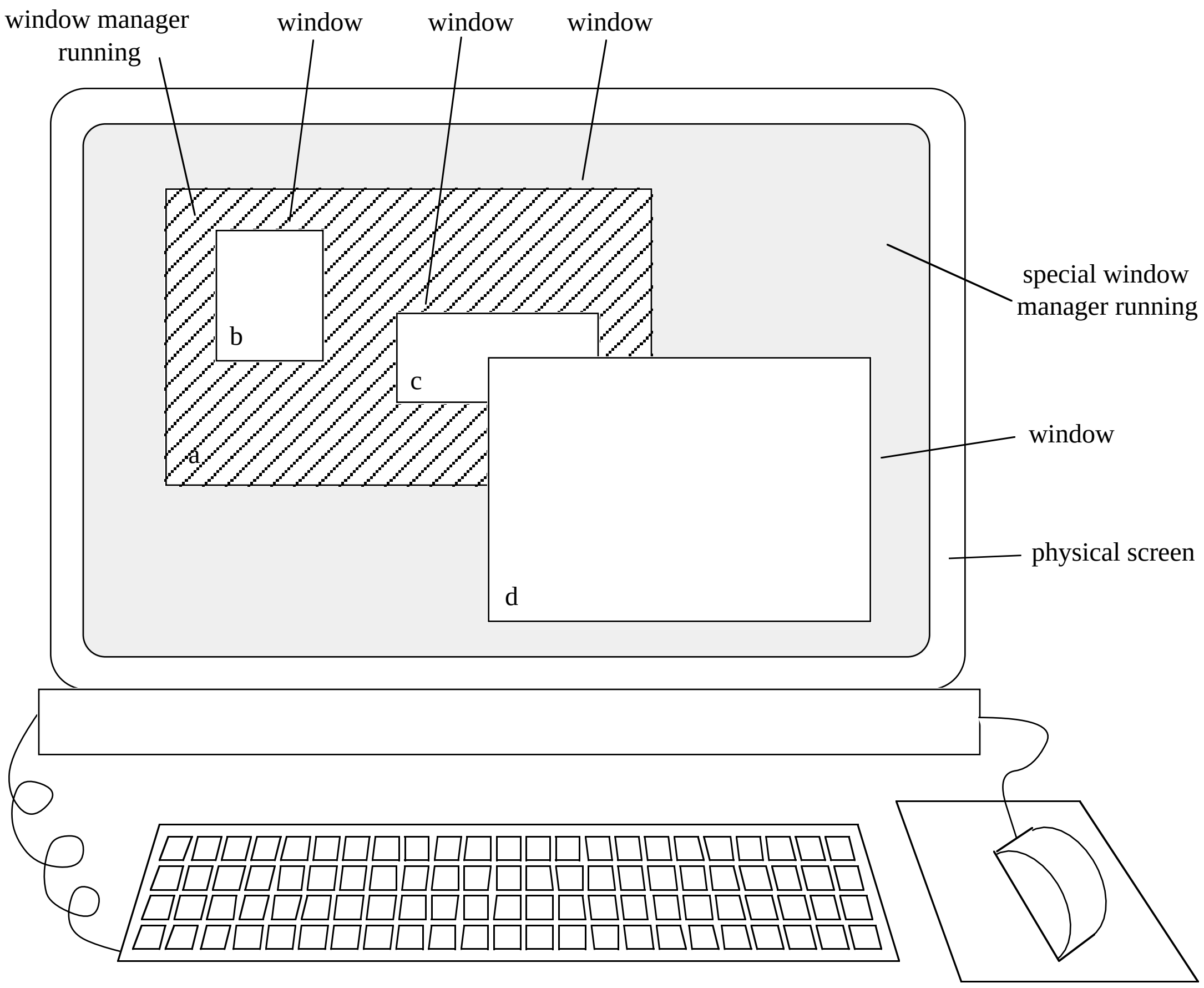

**Figure 6.4: Nested windows and window managers**

### 6.2.4 Hyper-Text Editing

WIN provides hyper-text editors which allow manipulation of text and embedded light-buttons. These editors are used to support hyper-programming facilities. Each hyper-text editor is a window with an application that controls interactive editing of the text. The editor facilities can also be accessed through a procedural interface which is given in Appendix C.

### 6.2.5 Interface Widgets

WIN provides a number of interface widgets which are also implemented as windows with applications. Those available include light-buttons, sliders, independent and mutually exclusive choices, fixed and scrolling menus, and dialogues. Complex applications may be built up by creating widget windows and displaying them with a window manager operating in an application window.

### 6.2.6 Implementation of WIN

#### 6.2.6.1 Window Manager Implementation

Two main approaches to window manager implementation were considered at the design stage. In the first approach each window maintains a complete copy of its display image. This gives simple algorithms for window manipulation as the window manager can access any part of a window image easily whether or not the window is partly or wholly obscured. The costs are that the visible parts of windows are stored both within the windows and on the screen and that drawing on a visible window involves raster operations to both the window's copy and to the screen. This strategy was used to implement the PStools system.

The second approach relies on being able to read the contents of the physical screen as well as write to it. The visible parts of a window's image are stored only on the physical screen and only the obscured parts are stored within the window data structure. This reduces the store overheads and removes the need for double raster operations but the data structures and algorithms to operate over them are correspondingly more complicated. This strategy was chosen for WIN; it will now be described in more detail.

The manner of storage of a window's graphical contents depends on whether it is displayed by a window manager. There are three cases:

- the window has never been displayed by a window manager;
- the window is currently displayed by a window manager;
- the window has been displayed but is currently not displayed.

In the first case the window's contents are stored as a single image in the window's data structure. In the second case the window relinquishes control of its contents to the window manager that displays it. In the third case the window becomes again responsible for its own contents.

The transfer of control is achieved using *virtual windows*. Every window, displayed or not, has encapsulated within it a virtual window with the same procedural interface as a normal window. Whenever one of the window's raster operation procedures is called, the window in turn calls the corresponding procedure of the virtual window.

If a window has never been displayed, its virtual window operates on the single image stored in the window's data structure. When the window is displayed its virtual window is replaced by a new one supplied by the window manager. The new virtual window operates on the window manager's data structure which stores the obscured parts of all the windows displayed by it. The visible parts are, by definition, displayed in visible parts of the parent window, which are displayed in visible parts of its parent window and so on. At the end of the nesting chain the visible parts are stored in the physical screen memory.

When a window is undisplayed the window manager that was displaying it supplies it with another new virtual window that operates on a single image again.

Thus the raster operations are delegated to the virtual window which will result in different procedures being called depending on whether or not the window is displayed. Other operations are dealt with by the window directly; the corresponding procedures in the virtual window are simple stubs as they are never called. This structure is illustrated in Figure 6.5:

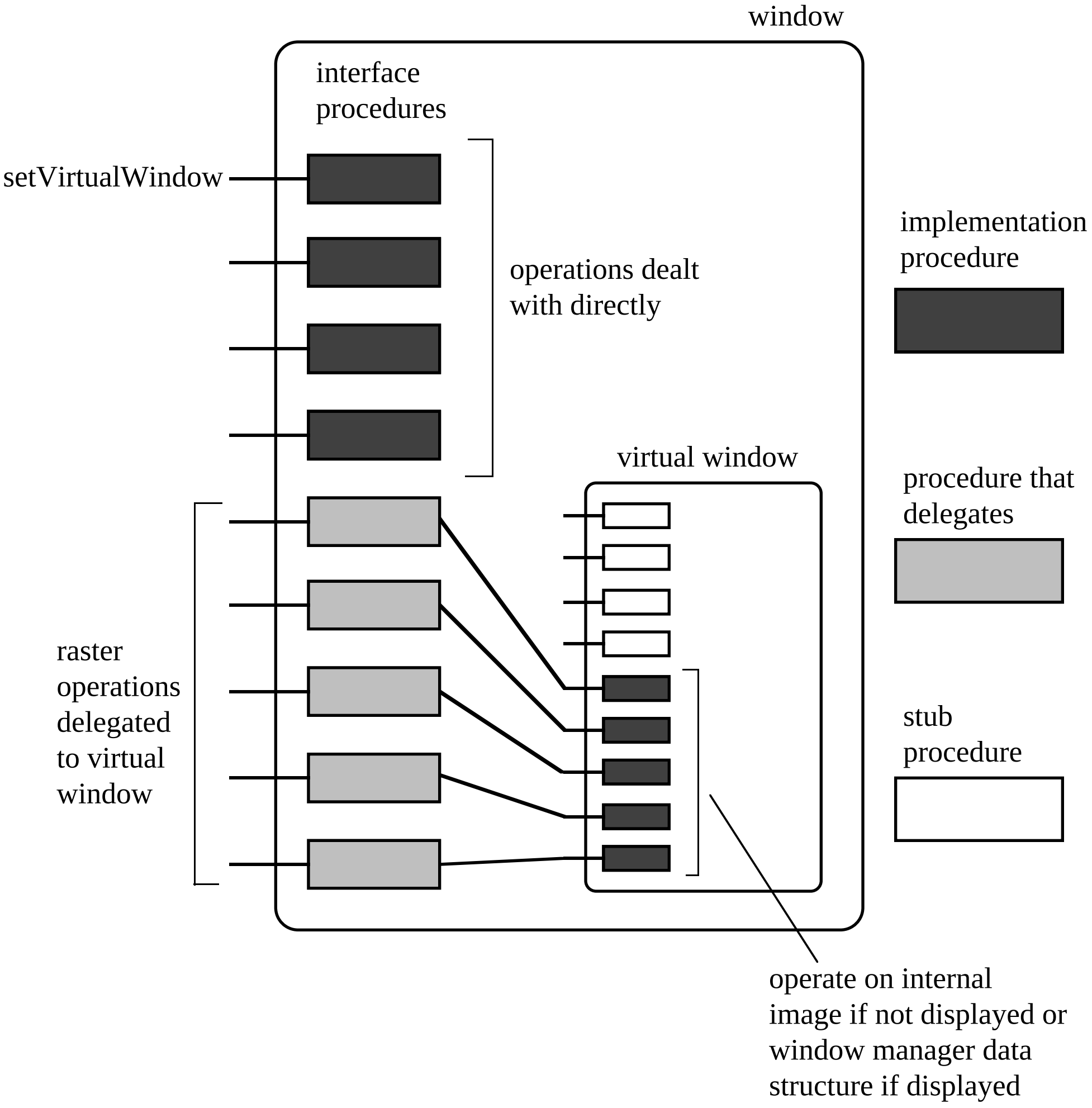

**Figure 6.5: Window with its virtual window**

#### 6.2.6.2 Window Manager Data Structures

The window manager maintains two main data structures. The first of these is a doubly linked list of structures containing information about the positions of the windows displayed and how they obscure one another. The second is a notifier that is used to route input events to the appropriate window applications and border regions. To enable these structures to be described in detail the concepts of *borders* and *current windows* will now be explained.

Windows may be displayed with borders around them to show their outlines and allow the user to manipulate them interactively. A number of border styles are pre-defined and the programmer may also define new styles. Each style consists of a procedure that takes a window as a parameter and returns a list of *Area*s as defined below:

```
type Area is structure( currentImage, nonCurrentImage : image ; pos : Pos ;
                        distributeEvent : Application )
```

A border is defined by a group of rectangular regions displayed around the edge of a window; the list returned by the style procedure contains an instance of type *Area* for each region. Each one contains two images, one displayed when the window is current and the other when it is not current; the position of the region relative to the origin of the window; and an application to process mouse events occurring over the region. This application provides facilities such as interactive moving and resizing of windows. A simple rectangular border can be defined as four regions as illustrated in Figure 6.6:

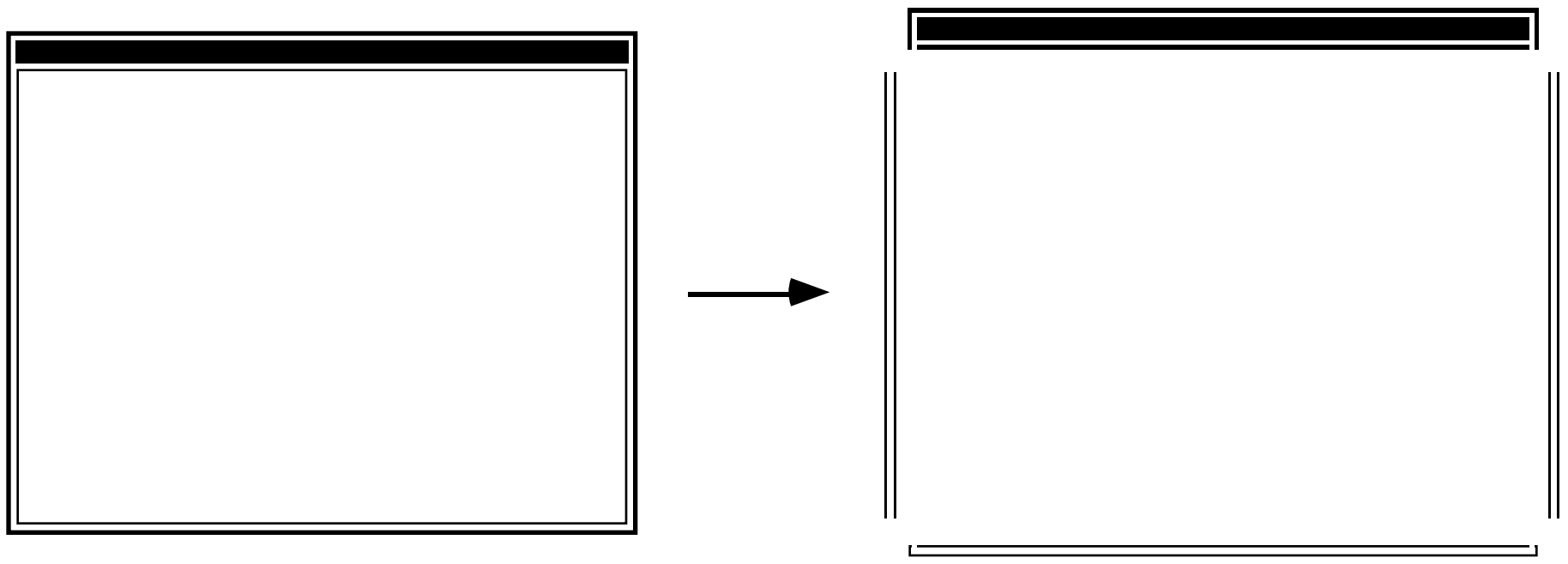

**Figure 6.6: Regions of a border**

At any time either one or none of the windows displayed by a window manager is *current*. This allows the system to determine which window should receive keyboard events: all keyboard events are routed to the application of the current window. The input focus follows the cursor: the current window is always the one immediately below the cursor, or if there is no such window, the one that was below the cursor most recently. Most of the pre-defined border styles identify the current window by drawing its border differently from when it is non-current. This is illustrated in Figure 6.7 in which window 2 is the current window:

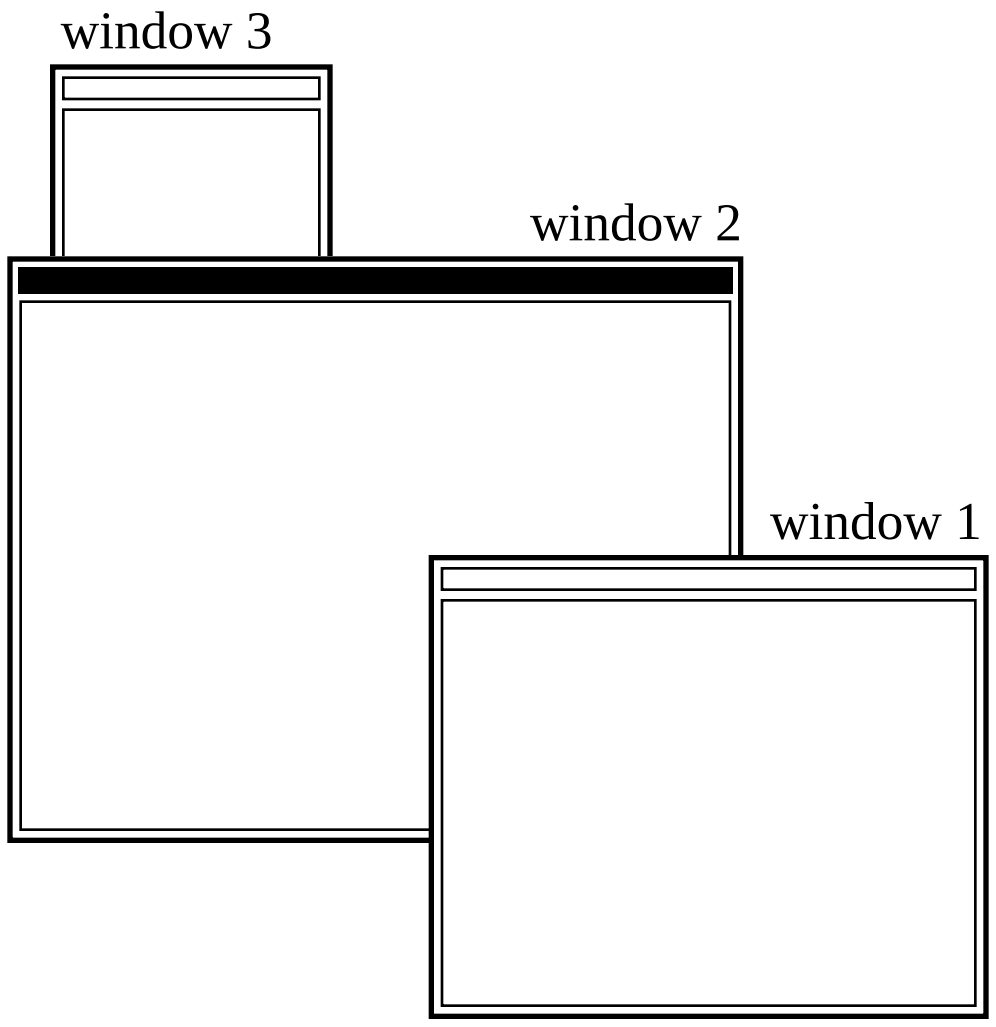

**Figure 6.7: Border styles for current and non-current windows**

When a window that is displayed becomes partly obscured, as with window 2 in Figure 6.7, it is partitioned into rectangular regions each of which is wholly obscured or wholly visible. This is illustrated in Figure 6.8, where window 2 has been partitioned into three regions A, B and C. Regions A and B are wholly visible and region C is wholly obscured.

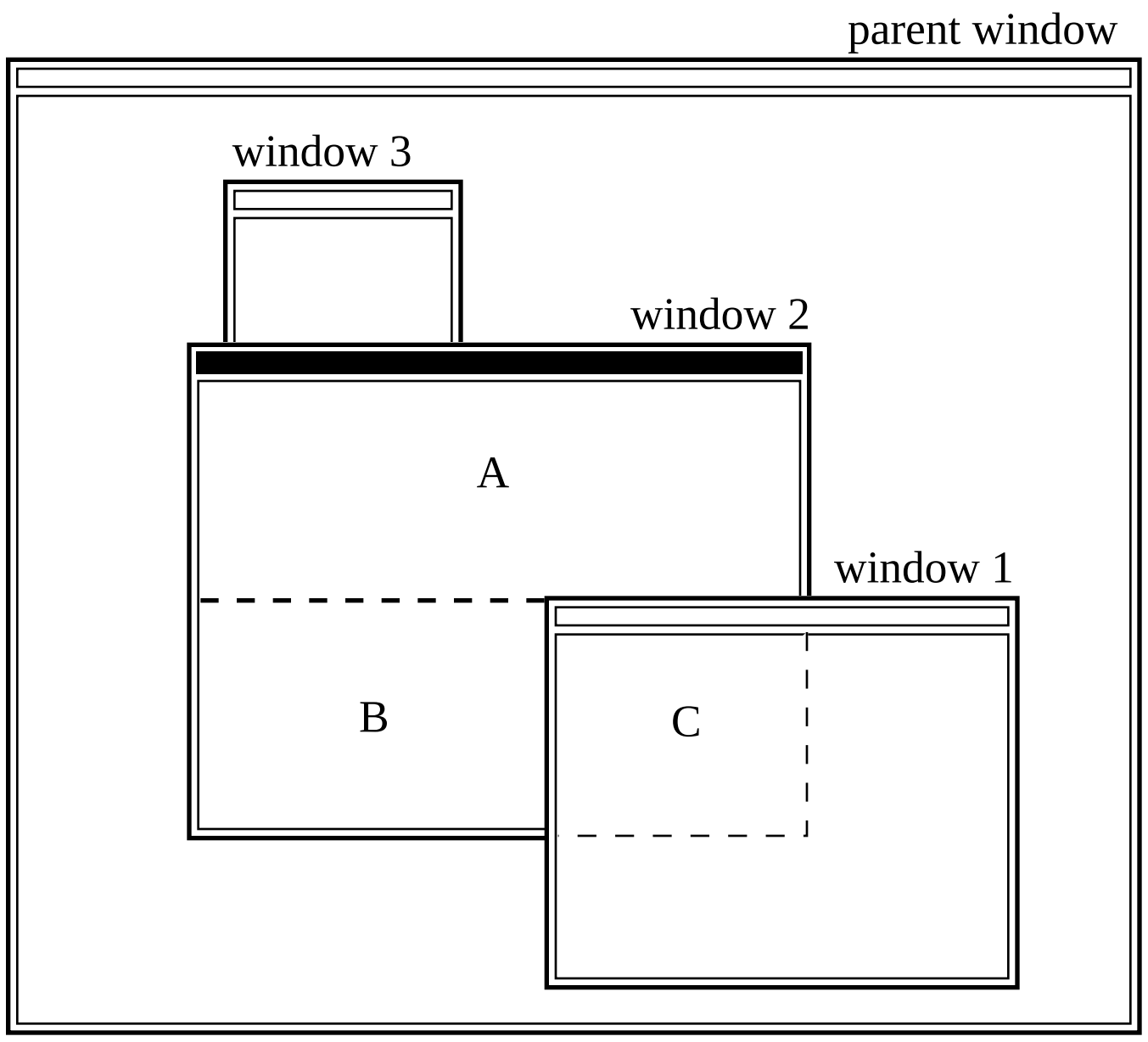

**Figure 6.8: Partitioning of a partly obscured window**

The first window manager data structure contains a list entry for each window. It is ordered by window depth, with the entry for the window nearest the front at the head of the list. Each entry contains a binary tree with leaves corresponding to the regions into which the window has been partitioned. Where a region is visible the leaf records its bounding rectangle in co-ordinates relative to the origin of the window manager's parent window. No image information is stored since visible pixels can be read from the appropriate region of the parent window. For obscured regions the leaf records the bounding rectangle, an image containing the obscured pixels, and a pointer to the obscuring window. Figure 6.9 shows the list for a window manager displaying the three windows of Figure 6.8, with an enlarged view of the tree for window 2. The tree contains an image for region C only.

The bounding rectangles are recorded at internal nodes in the tree, to enable faster searching of the tree. A frequently performed operation is for the window manager to scan a tree to find all leaf nodes whose bounding rectangles enclose a given point or intersect with a given region. An entire sub-tree can be eliminated from a search if the bounding rectangle of the parent node does not meet the intersection criterion. In the example of Figure 6.9, if a given point does not lie within the bounding rectangle stored at the second internal node then it can be safely assumed not to lie within either region B or C without further testing.

Figure 6.9 also shows as an example the details of the notification in the list entry for window 3. It contains a filter procedure which accepts only mouse events that occur within the bounding rectangle of window 3 and its border. The notification's application, executed whenever the notification accepts an event, contains a notifier structure which routes the event either to one of the border regions or to a procedure that deals with events over the main window area. The actions performed by this procedure will be explained shortly.

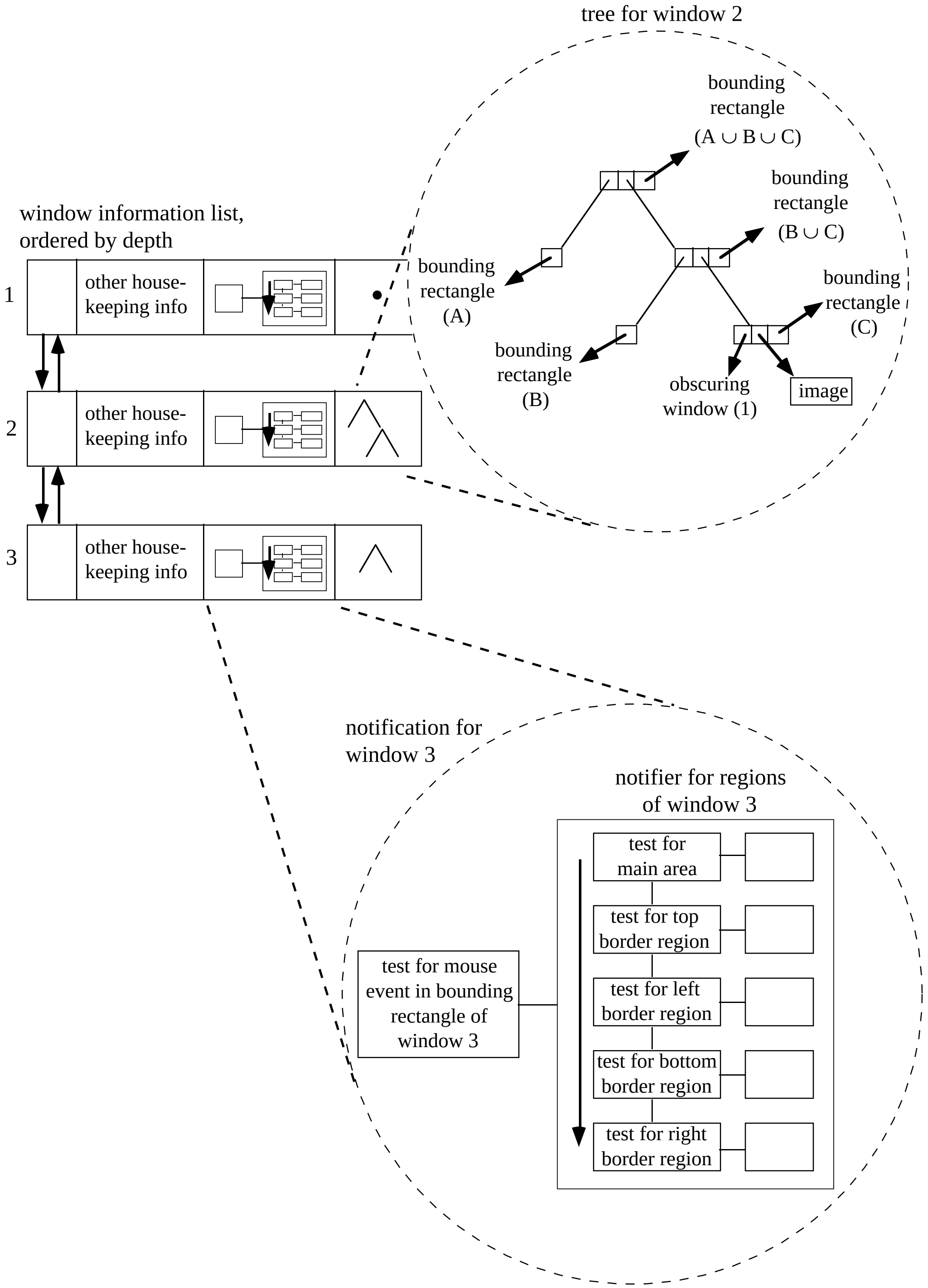

**Figure 6.9: Window manager's list of windows**

Figure 6.10 illustrates the structure of the window manager notifier. The first notification in the notifier list contains a filter procedure that tests for keyboard events or mouse events over a visible region of the current window. To determine whether an event occurs in a visible region it scans the window's tree for a leaf node whose rectangle contains the position of the event. If one is found and it does not contain an image, that region of the window is visible. The corresponding application uses another notifier to route accepted events to either the application of the current window, for keyboard events and mouse events over the main window area, or to the application associated with the region of the border within which the event occurs. As the notification for the current window is always at the head of the notifier list, events directed to the current window are routed without the need to execute multiple filter procedures in the main notifier.

Below the notification for the current window is a notification corresponding to each of the windows displayed. The filter procedures in these notifications test only for mouse events within the bounding rectangles of the corresponding windows: keyboard events are always accepted by the notification for the current window, while the ordering of the notifications by window level ensures that mouse events occurring over an obscured region of a window are not routed to the application of that window. For example, an event occurring over the region of window 3 that is obscured by window 2 will not be routed to window 3 as the event will also be over window 2 and will thus be accepted by its notification first.

The application in each of the notifications corresponding to the non-current windows also routes events using another notifier. Its structure differs from that of the notifier for the current window in that events over the main part of the window are dealt with by a procedure that makes the window current and then discards the event. Making the window current results in the removal of the notification currently at the head of the notifier list and the insertion of a notification for the newly current window in its place. The existing notification for that window is now redundant as any events that its filter procedure would accept will be accepted and consumed by a notification higher up in the list. However the redundant notification is left in place as it will be needed again when some other window becomes current.

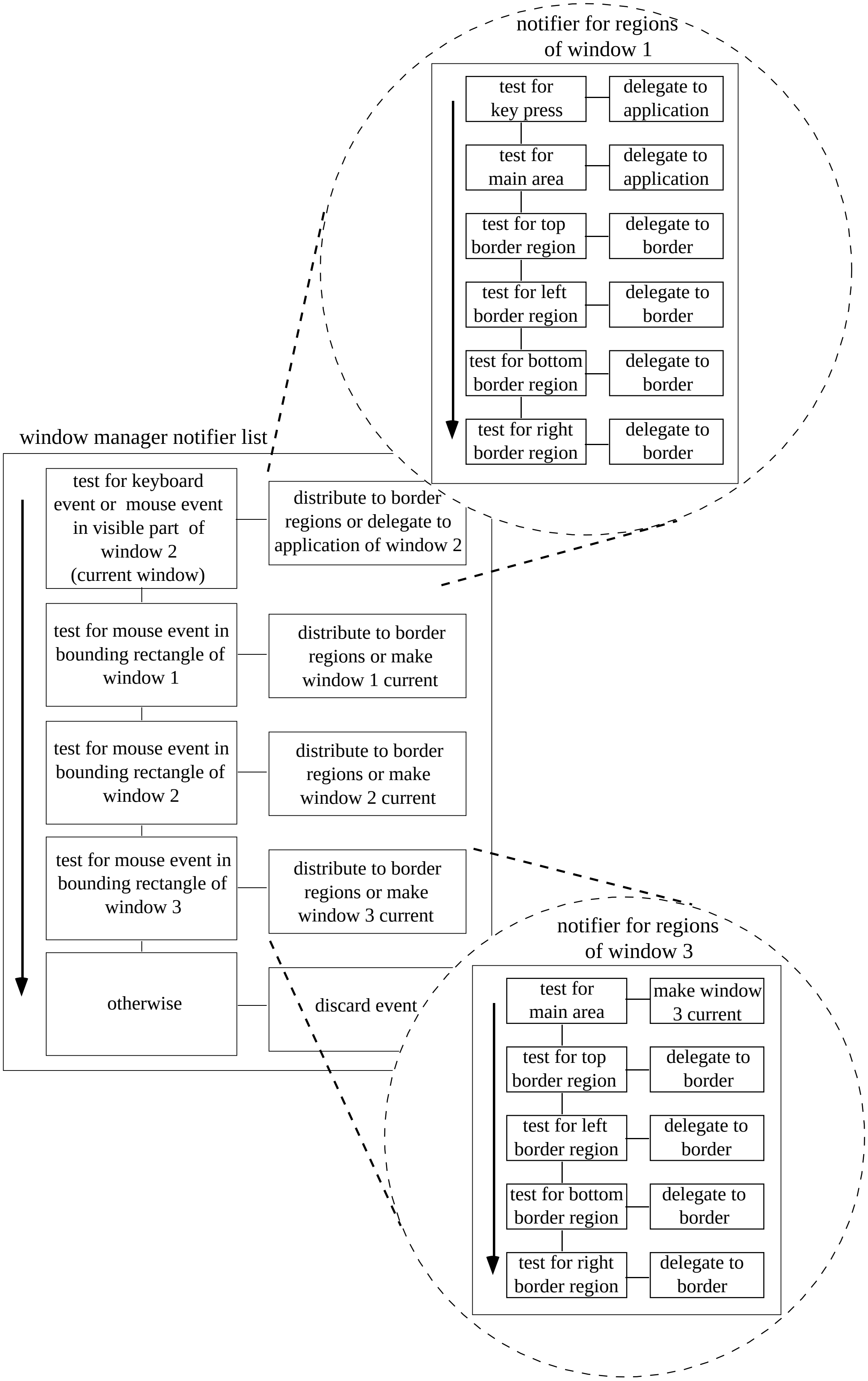

**Figure 6.10: Window manager's notifier list**

The data structures shown in Figures 6.9 and 6.10 are updated whenever a window is displayed or undisplayed or its level is altered.

### 6.2.6.3 Fragmentation

The window manager uses a tree compactor to reduce window fragmentation, which may occur in situations such as that illustrated in Figure 6.11. This shows window 1, which partly obscures window 2, being moved to a new position in four small steps. At each step some of the existing partitions of window 2 are further split into smaller regions.

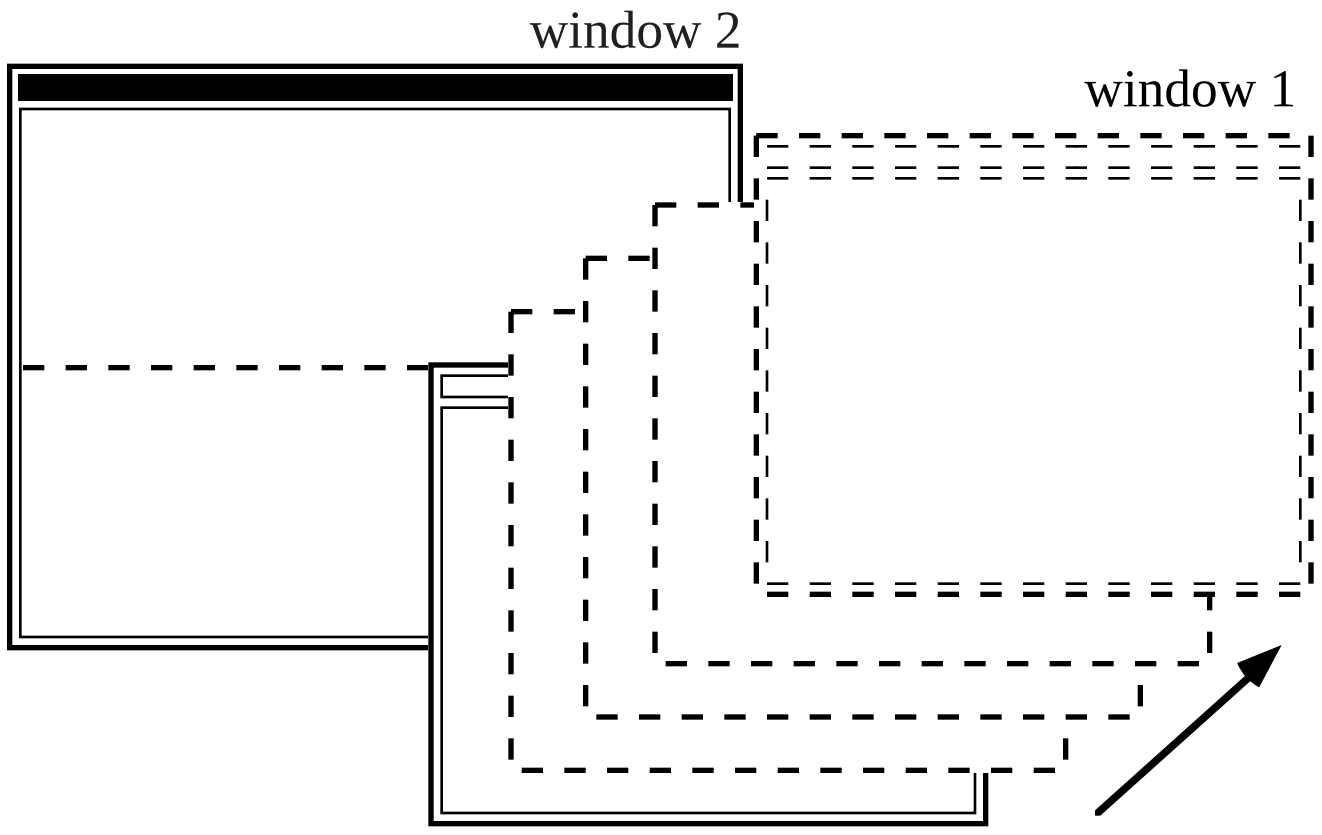

**Figure 6.11: Source of window fragmentation**

Figure 6.12 shows the resulting fragmentation. The tree for window 2 now contains fifteen leaf nodes corresponding to the split-up regions, although the whole window could be represented by a single node since all the regions are visible.

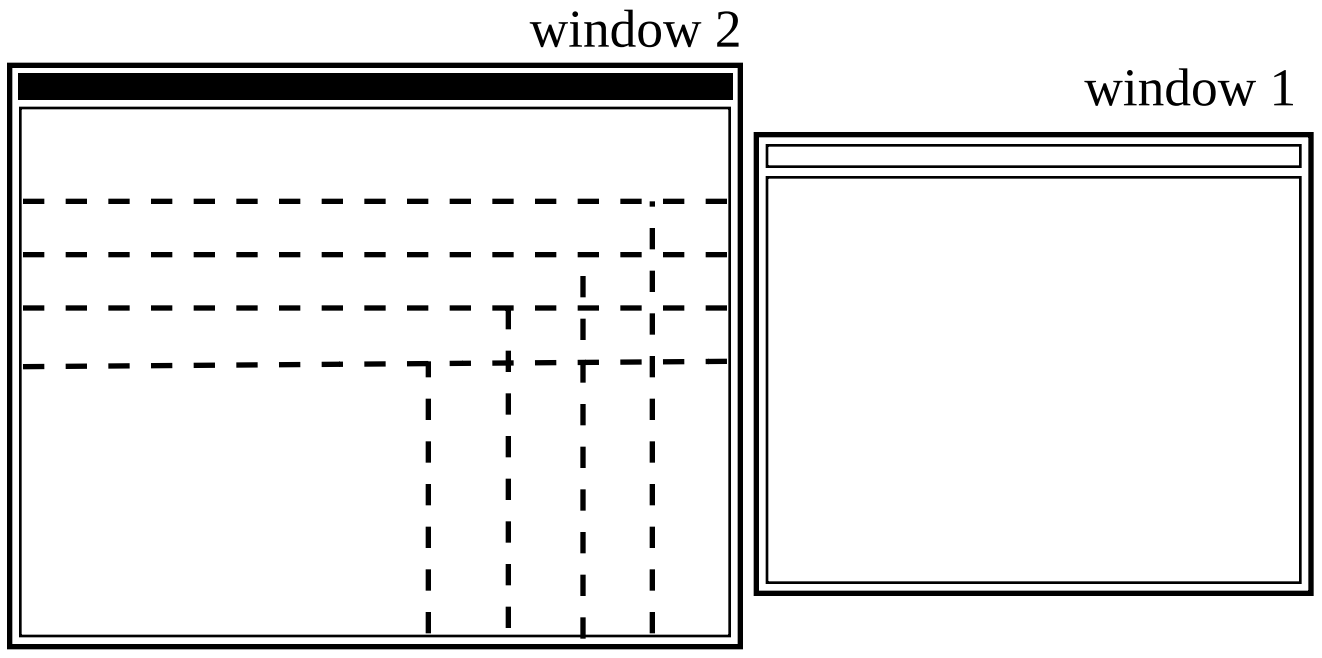

**Figure 6.12: Resulting fragmentation**

There are two problems associated with this fragmentation. The first is that the data structure describing the window's contents becomes complex, so window manipulation algorithms that traverse the structure take longer. The other problem arises if window 2 now becomes completely obscured by another window: its contents are stored as a large number of small images and the memory overheads associated with each image may become significant.

To reduce these problems a buddy-type compactor is used to recombine leaf nodes, by traversing the tree and examining each pair of sibling leaf nodes. If the nodes both represent obscured regions or both represent visible regions the parent node is overwritten by a new leaf node constructed by combining the two siblings. Figure 6.13 shows how the tree of window 2 is reduced to a single node by the compactor:

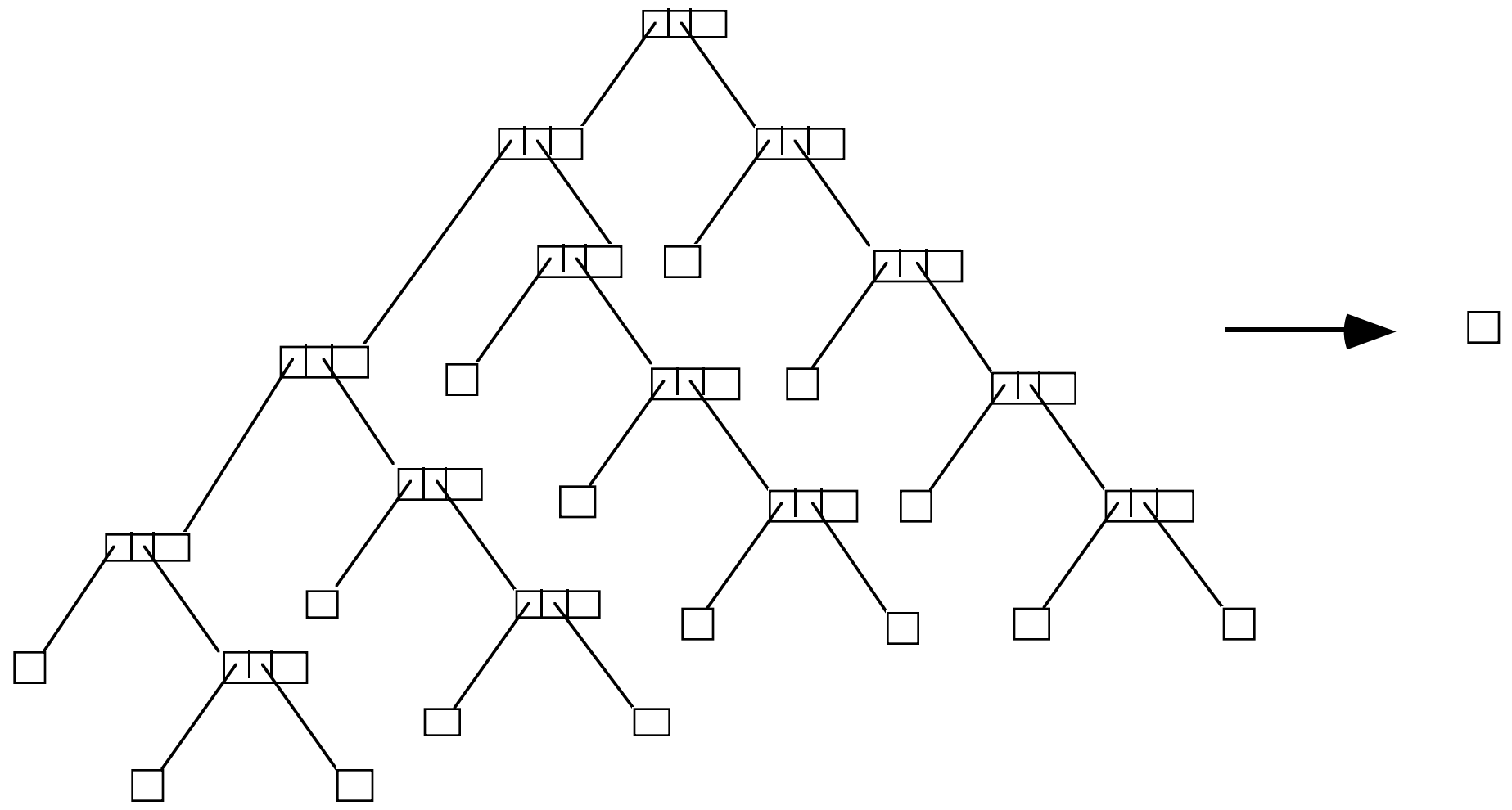

**Figure 6.13: Compaction of a window tree**

Like other buddy algorithms [Kno65], the algorithm does not compact non-sibling pairs even when they are compatible. Its advantage is that little computation is required, making it feasible to invoke it, for every window displayed, whenever any window is moved, displayed or undisplayed.

#### 6.2.6.4 Hyper-Text Editor Data Structures

The hyper-text editor maintains three main data structures. The first structure contains the text itself, the second structure describes which part of the text is visible in the editor window and the third structure describes the positions of the embedded light-buttons and their associated actions.

The text is stored in a doubly linked list of strings, one for each text line. The new-line at the end of each text line is not stored as part of the string but is implicitly present between each consecutive pair of lines. Each list element also contains a line number. This allows the editor to determine the ordering for list elements efficiently.

The user may select regions of the text; the start and finish of the selected text are represented as structures of type *TextPointer*, one field containing the list element for the relevant text line and the other containing an integer offset into the line. The offset refers to a point between two characters, so for example an offset of zero corresponds to the point before the first character and an offset of five to the point between the fifth and sixth characters.

The layout of the text within the window display is recorded in a vector. Each element contains a *TextPointer* specifying the position within the text data structure of the beginning of the corresponding window line. If a text line is longer than the number of characters which fit into a window line, more than one window line points to it. Each element of the window line vector also contains the vertical offset in pixels of the base of the corresponding window line from the base of the window. The offsets could be calculated as required but they are stored as an optimisation since they change only when the number of window lines changes, a relatively infrequent occurrence caused by the window size changing.

Details of embedded light-buttons are stored in a vector. Each element contains:

- an integer index for the button;
- the text displayed on the button;
- the text positions at which it starts and finishes;
- the procedure that will be executed when the button is pressed;
- a value of type *any* that may be set by the programmer.

The light-button vector is ordered by the buttons' positions in the text. This allows the editor to distinguish efficiently between a mouse button press over a light-button, in which case its associated procedure is called, and a press over normal text, in which case the insertion point is set to the new position. The editor calculates the text position corresponding to the (x,y) position of the mouse and then uses a binary split algorithm to determine whether the text position lies between the start and finish of any of the light-buttons.

The data structures will be illustrated with the following example. An editor contains the text shown below:

*And did those feet in ancient time*
*Walk upon England's mountains green?*
*And was the holy lamb of God*
*On England's pleasant pastures seen?*

The text contains two hyper-text buttons, one over the word *mountains* and the other over the word *God*. Part of the text is selected. The text visible in the editor window is shown in Figure 6.14, with the selected text displayed white on black:

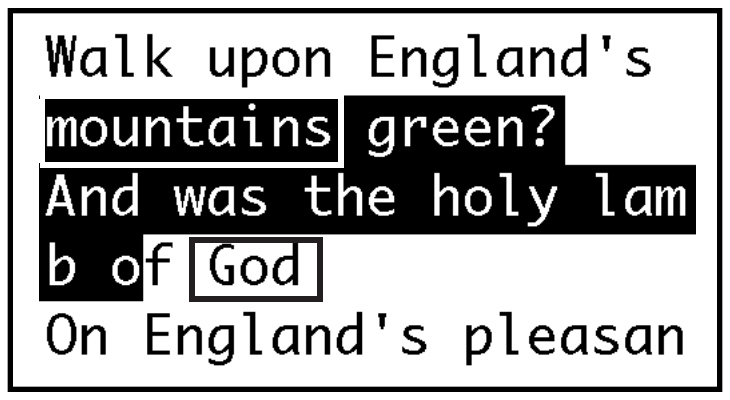

**Figure 6.14: Example of hyper-text editor display**

Figure 6.15 shows the data structures that represent this configuration. The two text pointers at the top record the current selection. The vector on the right records the pixel offsets of each window line from the base of the window: here the height of each line including the inter-line space is 16 pixels. Each element points to the text position lying at the start of the window line. The vector displayed below the doubly linked list contains two elements recording the names and positions of the light-buttons. The button indices are independent of the elements' offsets within the vector; they are used by the programmer to denote particular buttons when calling those editor interface procedures that operate on buttons. The types used are defined in Appendix C.

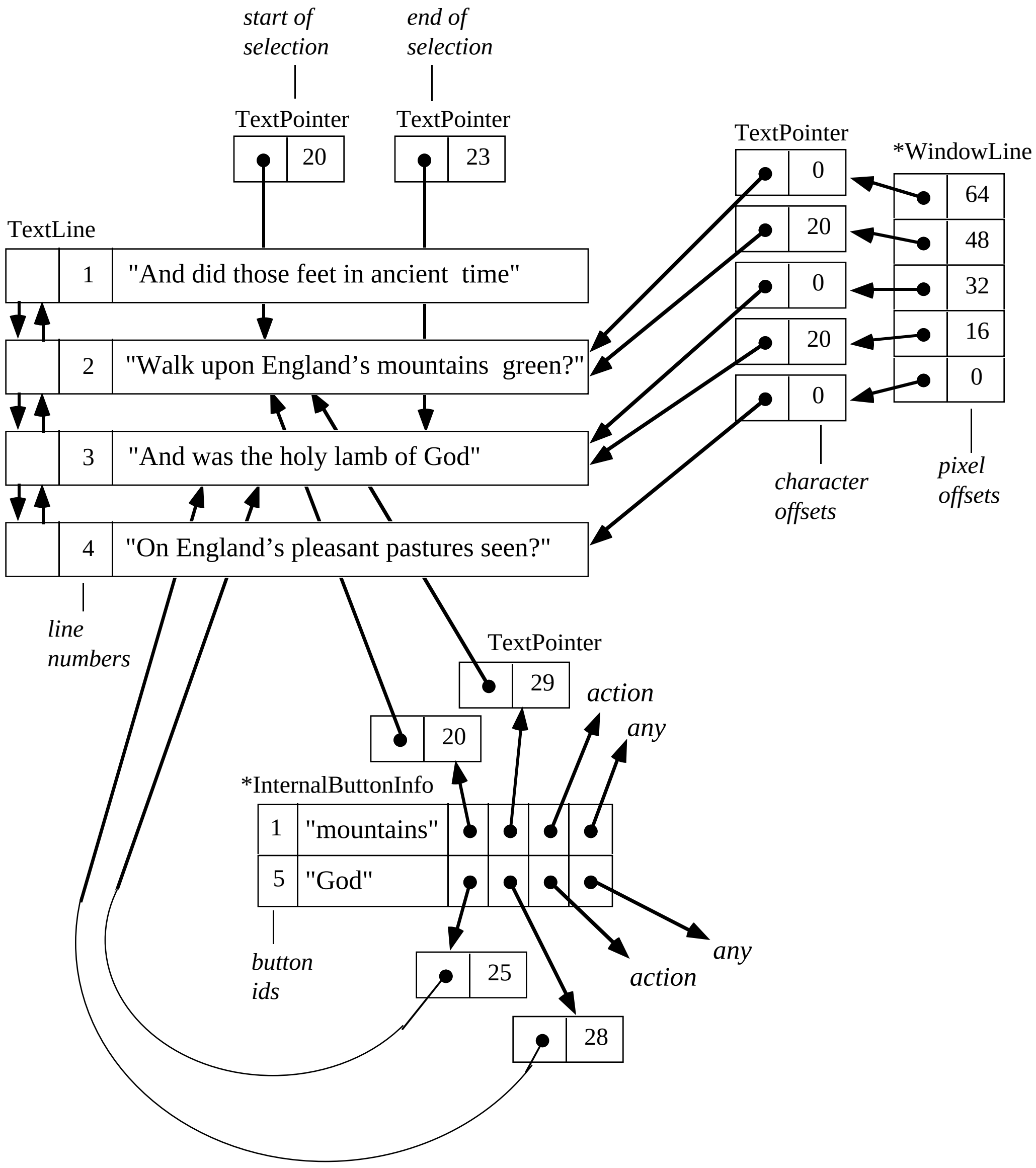

**Figure 6.15: Hyper-text editor data structure**

# 6.3 Browser

The hyper-programming system requires a browsing tool to allow the programmer to locate values in the persistent store for incorporation into hyper-programs. The tool is also needed for examining values bound into existing hyper-programs. The Napier88 browser is capable of displaying a representation of any value passed to it. This representation can be either textual or graphical although the hyper-programming system uses graphical representations only.

## 6.3.1 History

The Napier88 browser is based on a tool written for PS-algol by Dearle and Brown [DB88], and a proposal for its extension [DCK90]. The main differences are that the Napier88 type system is more complex, thus the Napier88 browser has to deal with more type constructors, and in the user interfaces. The PS-algol browser only displays one value at a time, making it difficult for the user to discern the topology of inter-connected structures. With the Napier88 browser multiple values can be displayed and arrows drawn between representations to show references between values.

### 6.3.2 Browser Interface

The programmer accesses the Napier88 browser as a procedure that takes one parameter of the infinite union type *any*. An instance of the browser is generated by calling a generator with an instance of the following type as a parameter:

```
type BrowserType is variant( graphical : WindowManager;
                             textual : proc( string ) )
```

The type *WindowManager* is defined in Appendix C. The value passed to the browser generator determines whether the output of the browser will be in graphical form, in which case the window manager value is used to display value representations, or in textual form, in which case the procedure is called whenever the browser needs to output text. Only in the former case does the browser attempt to display links between values. The graphical user interface was described in Chapter 4.

### 6.3.3 Browser Implementation

Two versions of the browser have been built, with the same user level interface but different underlying implementations. The first version uses type-safe linguistic reflection, while the second is written at a lower level using implementation-level facilities available to the builders of the Napier88 system itself.

#### 6.3.3.1 Reflective Implementation

The action taken by the browser to display a value depends on the value’s type; it executes a different procedure for each type. For example, displaying a variant requires a different kind of window from that required for a structure. As there are an infinite number of Napier88 types, it is not possible to generate all the procedures statically. The browser instead relies on the Napier88 compiler being available as a procedure within the language. When the browser encounters a new type it constructs and compiles a new procedure to display it.

Linguistic reflection is a relatively expensive process; the browser would be more efficient if, for example, it had a generic procedure that could be used to display all structure values without having to use reflection. It would in fact be possible to provide such a generic procedure if all it had to do was to display a menu for a given structure. The labels for the menu entries could be determined by inspection of a representation of the structure’s type. The problem to which reflection provides the solution is dereferencing into the structure to obtain its field values for further browsing.

Structure dereference in Napier88 is expressed by field names which must be specified statically. The reason for this rule is to enable static type checking. There is no way to write a computation that calculates a field name at run-time, as field names are not part of the value space. Because of this a generic structure display procedure could not obtain the structure field values, since the field names could only be determined by dynamic inspection of the type representation.

The solution adopted is use the field names obtained dynamically to generate the representation of a specialised procedure to display values of that particular structure type. With respect to the generated procedure the field names are known statically. The representation is then compiled and executed. A similar process can be used for other type constructors. Figure 6.16 illustrates the procedure representation that might be generated for a particular structure type. For simplification the definition of the procedure *menu3* to display menus with 3 entries is assumed. The generated procedure contains a label for each menu entry and a procedure to be executed when the entry is selected. Recall that apostrophes are used to allow quotes to be included inside a string.

```
"type T is structure( a : proc( … ) ; b : variant( … ) ; c : structure( … ) )
use PS() with browser : proc( any ) in
proc( val : T )
      menu3(  '"a : proc"', proc() ; browser( any( val( a ) ) ),
              '"b : variant"', proc() ; browser( any( val( b ) ) ),
              '"c : structure"', proc() ; browser( any( val( c ) ) ) )"
```

**Figure 6.16: Generated procedure to browse a structure**

The browser maintains a table keyed by representations of the types that have been browsed in the past, containing procedures that will browse values of those types. The table is loaded at the time of its creation with procedures to browse all the base types and procedure values. These procedures can be loaded at the outset because all values of a given base type are browsed in the same way. All procedure values are also browsed in the same way.

When the browser is called it first determines the type of the value in the *any* passed to it. It then searches the table for a representation of that type. If the value being browsed is of one of the base types or the type has been encountered previously, a procedure will be found to browse it.

If, however, the type is not present in the table, the browser constructs a new procedure to browse the value. Browsing involves displaying a menu window representing the value and extracting any other values directly accessible from the value being displayed so that they in turn can be browsed and displayed. To construct the new procedure the browser builds a textual representation of the required procedure and then uses the compiler to compile the text. The code representation is built up by string manipulation, directed by the structure of the type of the value to be displayed. Before executing the procedure which is produced by compilation of the new code, the browser enters it into the table, keyed by a representation of the type of the object which it can browse. In this way the browser learns about new types: the next time an object of that type is encountered there will already be a procedure in the table to browse it. Since the table is persistent, the compilation process is necessary only on the first encounter with the type.

Figure 6.17 illustrates the process of browsing a structure value. If the type of the structure has not been encountered before, the browser analyses a representation of the type to determine the names and types of the structure fields. It uses this information to construct the textual representation of a procedure to display values of the structure type.

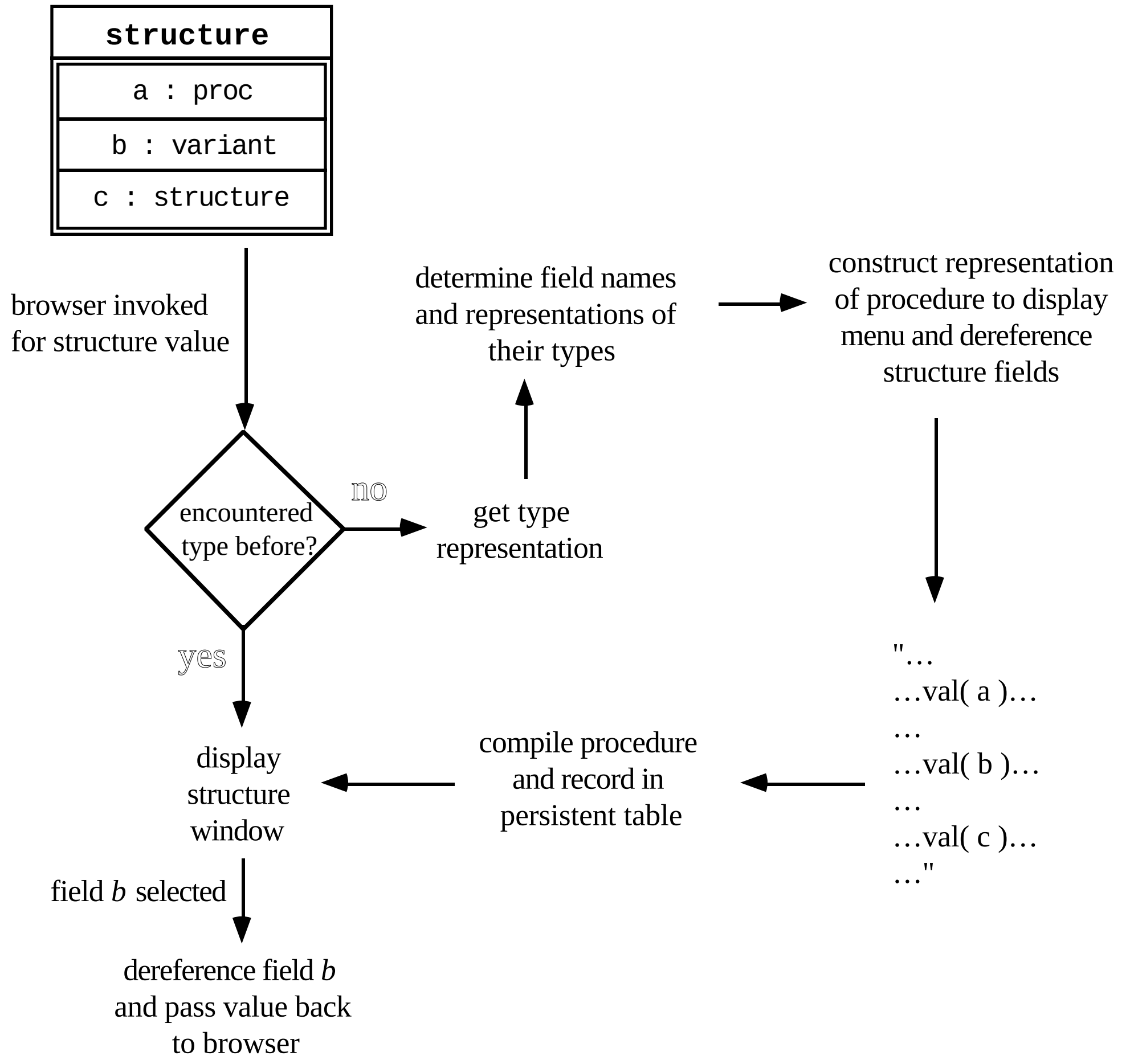

**Figure 6.17: Reflective implementation of structure browsing**

The browsing of environments presents a problem not found with the other base types. As the names and types of the bindings in an environment can only be determined by dynamic inspection, the browser must use reflection to obtain the binding values in the same way as structure field values are obtained. An example of the code generated to browse the value of a binding *x* with type *T* is shown in Figure 6.18. Once compiled the resulting procedure is applied to the environment to display the value.

```
"type T is …  ! Definition of type T
use PS() with browser : proc( any ) in
proc( e : env )
      use e with x : T in browser( any( x ) )"
```

**Figure 6.18: Generated procedure to browse an environment binding**

The overheads associated with performing reflection are more significant with browsing environments than browsing structures because they are incurred every time a different binding is browsed. With structures the names and types of the fields are constant so all the reflection can be performed on the first encounter with the type; since bindings can be added and dropped from environments the reflection is needed each time a new binding is browsed.

### 6.3.3.2 Low-Level Implementation

The browser forms the basis for the hyper-programming environment described in Chapter 4. One of its uses is to allow the programmer to navigate around the persistent store to locate data to be reused. Since environments provide the main store structuring mechanism, this involves much browsing of environments. It was found that the implementation using reflection did not give adequate speed performance to allow serious use. To address this the browser was re-implemented using lower-level technology. The new version uses several procedures that are not available in the standard Napier88 release because they are not type-safe. These procedures operate at the untyped object level that underlies the Napier88 system, and allow words to be read from and written to arbitrary positions in objects.

Implementation at this level is more efficient than using the reflection techniques, because rather than having to construct and compile code to access the components of compound values, the browser can simply read directly from the objects that represent them. This relies on knowledge of the fact that Napier88 values are represented using a small number of object formats, one for each of the type constructors [CBC+90]. The disadvantage of this strategy is that it reduces the portability of the system: its implementation must change if the Napier88 store formats change.

The general structure of the browser is the same in the low-level implementation. It still examines the type of a value passed to it and decides from it how to display the value. If the type has been encountered previously an existing display procedure is located and used, otherwise a new display procedure is generated and stored before use. New procedures are also generated for accessing environment bindings.

The difference is in the way these new procedures are generated. Instead of constructing and compiling new code representations, the system uses one of a small set of existing higher-order procedures, one for each type constructor, that return display procedures. The generator procedures take as parameters information about the type of the value to be displayed. This allows an offset map to be calculated and bound into the returned display procedure. The map contains offsets into the store object implementing the value being displayed. The display procedure uses the offsets with the low-level object access functions to access the values bound into the value being displayed.

All object formats in the current Napier88 system keep pointers and non-pointers separated. This facilitates the location of pointers during garbage collection. Because of this the offset map contains offsets for both the pointer and non-pointer components of each value bound into the value being displayed. Depending on the type of the bound value one of those components may be empty—for example, a real is represented by two non-pointer words, a procedure by two pointer words, and a variant by one pointer and one non-pointer word.

Other low-level functions are used to combine the pointer and non-pointer components back into a typed Napier88 value so that it can be passed in turn to the browser if required. Because the code of every display procedure produced by a given generator is the same, the generation of the new procedures requires only the binding-in of the offsets rather than the construction of new code. This could be viewed as a very limited form of reflection in which the generators can vary the environment of the closures produced but not their code.

The low-level functions used in the implementation of this version of the browser are described below. The functions can only be accessed using a special system-builders' version of the Napier88 compiler.

*makeObject*: This takes as parameters a size and number of pointer fields, and returns a pointer to a new store object.

*formAny*: This takes a pointer to a store object and a representation of the type of the value it implements, and returns the value injected into type *any*.

*splitAny*: This takes an *any* and returns a pointer to the store object implementing it and a representation of the value's type.

*assignPntr, assignInt, lookupPntr, lookupInt*:

These allow pointer and scalar words to be read from and written to store objects.

#### 6.3.3.3 Browsing Structures, Variants and Vectors

To browse a structure the browser needs to build a menu showing the field names and their types and to extract the values stored in the fields of the structure. The field names and types are obtained from the representation of the type of the structure value, while the values in the fields are accessed by reading directly from the store object that implements the structure value.

On encountering a value with an unknown structure type the browser:

- obtains a representation of the structure type;
- extracts the field names and type representations from the type representation;
- uses the field types to work out the store formats of the field values;
- constructs an offset map containing the positions of the field values within the structure object;
- produces a browsing procedure with the offset map bound into it;
- records the procedure in the persistent table and then calls it.

To browse one of the fields of a structure of that type the browser:

- obtains a pointer to the store object implementing the structure value;
- looks up the offsets of the pointer and non-pointer components of the required field value from the offset map;
- reads the pointer and non-pointer components of the field value from the appropriate positions within the structure object;
- converts the components and the field type representation to a typed *any* form;
- calls itself to browse the *any*.

Figure 6.19 illustrates the browsing of a variant field in a structure. In this example the pointer and non-pointer components of the variant are stored non-consecutively in the structure object. The layout of the structure object has been slightly simplified. In reality extra words are stored in the object for housekeeping purposes.

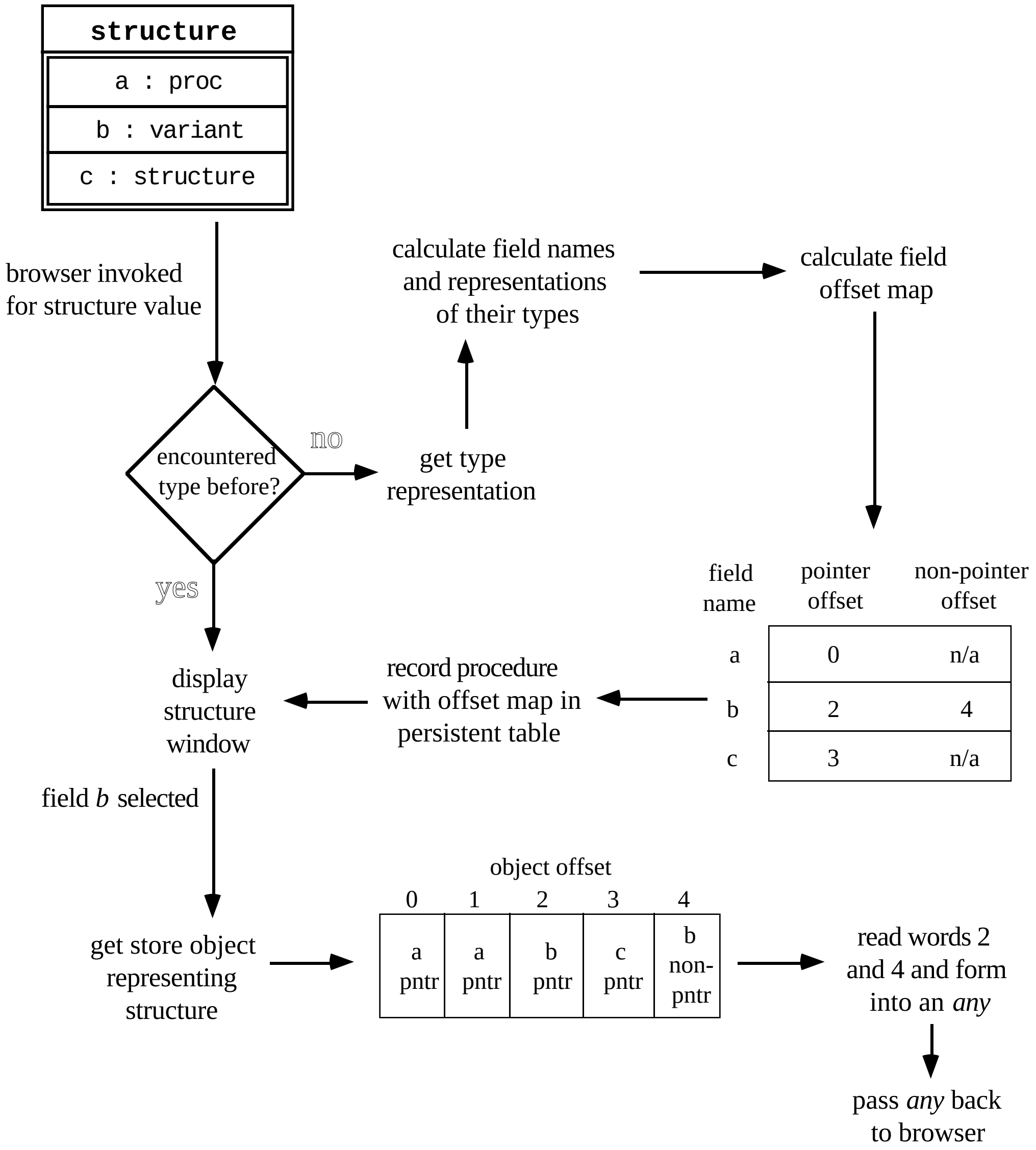

**Figure 6.19: Low-level implementation of structure browsing**

The main improvement in performance over the reflective implementation occurs when the browser first encounters an unfamiliar structure type. The actions in the first list given in this section are all inexpensive compared to invoking the compiler. The amount of work required when a field value is browsed is broadly similar in the two implementations.

The methods for browsing unfamiliar variants and vectors are similar and slightly simpler as the field offset maps are not needed. A variant only holds one value so calculation of its offsets is trivial, while all the elements of a vector must occupy the same number of words, so element offsets can be calculated by multiplication when the elements are accessed.

#### 6.3.3.4 Browsing Environments

As *env* is a base type the browser table contains a single pre-defined procedure for browsing environments. Since the contents of environments can vary there is no point in storing information about environments encountered for future use.

When an environment is encountered the browser uses a standard procedure to scan it, discovering the names of the bindings it contains and representations of their types. To browse a binding from the environment the browser uses a low-level procedure to access the procedures that implement that environment. These procedures are normally hidden from the Napier88 programmer. It then calls one of the procedures to obtain a pointer to the store object that implements the location of the required binding, reads the pointer and non-pointer components of the value, and converts them to an *any* in the same way as for structures. Finally the *any* is passed back to the browser to be displayed.

The amount of work required to display an environment in this implementation is similar to the reflective implementation. The savings occur when bindings in the environment are browsed as the need for compilation is removed.

## 6.4 The Napier88 Compiler

The first Napier88 compiler was implemented by Brown, Connor, Dearle and Morrison [Dea88, Bro89, Con90]. The version used for the experiments described in this thesis is itself implemented in Napier88 and was implemented by Cutts [Cut92]. The compiler is accessible by Napier88 programs through several interfaces. The simplest is shown in Figure 6.20:

```
compileString : proc( string → any )
```

**Figure 6.20: Simple interface to the Napier88 compiler**

This interface to the compiler is a procedure that takes a string parameter and returns a result of type *any*. Projecting the result gives a value that depends on the type of the code represented by the input string, as follows:

- If the input code represents a void program, i.e., a program that does not return a result, the value is a procedure of type *proc()*. Calling the procedure causes the compiled code to be executed.
- If the input code represents a value of type *T*, the value projected from the *any* is a procedure of type *proc( → T )*. Calling the procedure causes the compiled code to be executed, producing the result value.
- If there are compilation errors due to the input code being invalid, the value projected from the *any* is a string that describes the errors.

There is also a more flexible interface to the compiler which abstracts over the nature of the source representation being compiled. It also allows programs to be compiled against existing values, i.e., it supports compilation-time linking. The interface is shown in Figure 6.21:

```
type lValue is …  ! structure containing info about an identifier
type symbolTable is table[ string, lValue ]

compile : proc( env, list[ symbolTable ], *string → any )
```

**Figure 6.21: Flexible interface to the Napier88 compiler**

The procedure providing this version of the compiler interface takes as parameters an environment, a list of symbol tables and a vector of strings. It returns a result of type *any* using the same convention as the simple interface. The environment parameter contains procedures that operate on the source code: one returns the next character from the source code and advances the remembered position, while another returns *true* or *false* depending on whether the end of the source code has been reached. These procedures abstract over the nature of the source code: it could be a string, a file or some other program representation. The list parameter contains symbol tables that form a series of extra ‘outer scopes’ during compilation. Finally the vector parameter contains strings that specify compiler options such as source listing, line numbering, etc. To compile a program against an existing value, the programmer constructs a new symbol table using a procedure available in the persistent store, adds the value to the symbol table and passes it in a list to the compiler. If the compiler encounters an identifier not declared within the source program it searches the extra symbol tables and, if found, plants a reference to the corresponding value or store location in the resulting executable code.

# 6.5 Hyper-Programming Tools

## 6.5.1 Hyper-Program Representations

The hyper-programming environment supports three different representation forms for hyper-programs:

- While being manipulated in a hyper-program editor a hyper-program is represented by a combination of text and embedded light-buttons.

- When exported from an editor a hyper-program has a simpler representation. It is this representation form that is obtained if the contents of an editor are read using the appropriate interface procedure. It is also the form that is manipulated by reflective hyper-program generators and in which source code attached to procedure values is stored. The form is ‘light weight’ thus few storage overheads are incurred.

- A hyper-program is converted to a third form before being passed to the compiler.

The three forms will now be illustrated with reference to the hyper-program shown in Figure 6.22. The hyper-program contains links to a free identifier, a procedure value and an environment location.

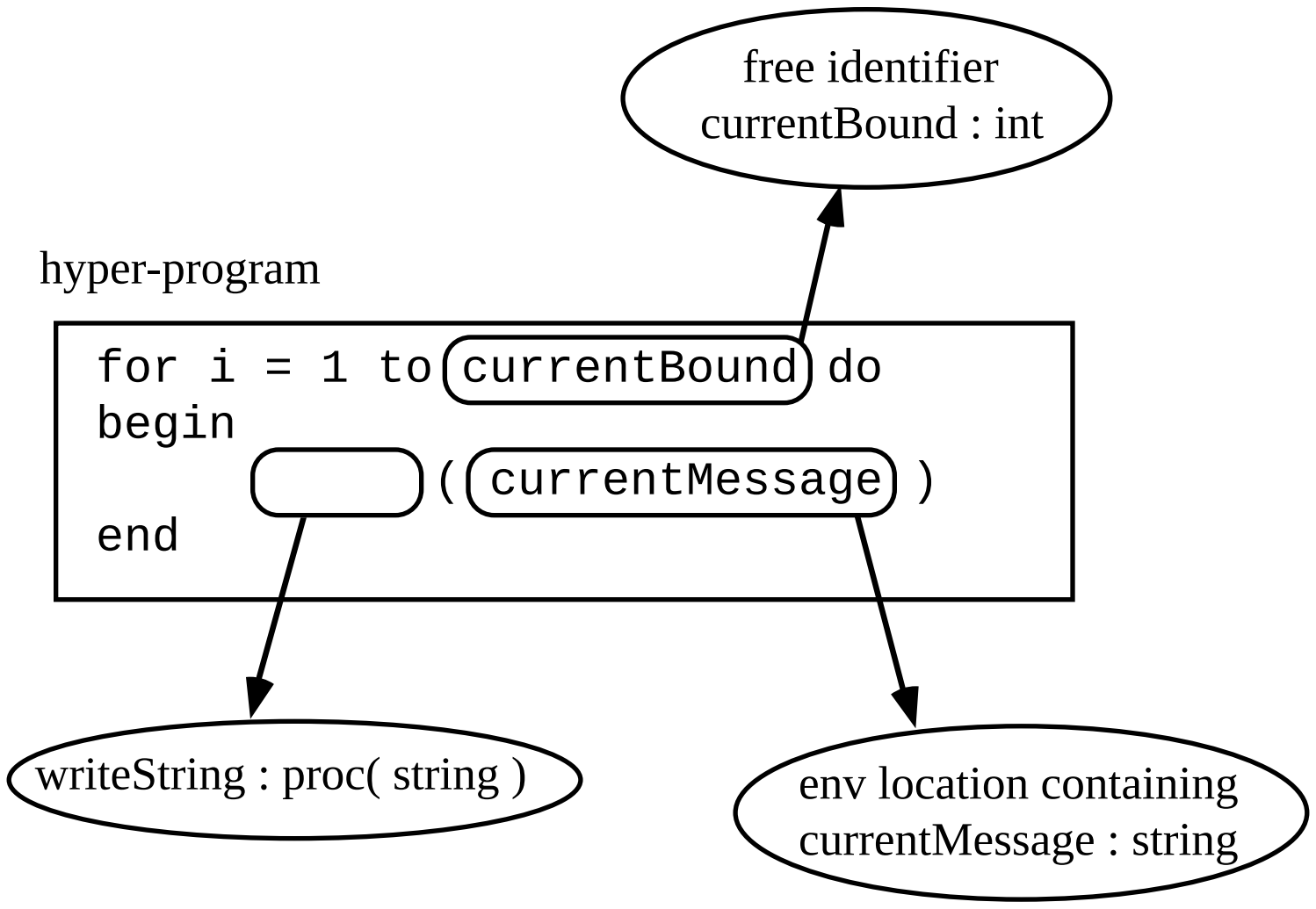

**Figure 6.22: Example hyper-program**

While the hyper-program is stored within a hyper-program editor it is represented by the hyper-text structure shown in Figure 6.23. There is a light-button corresponding to each embedded link. Each element of the light-button vector contains, in addition to the button’s index, text and position:

- a procedure that is called when the light-button is pressed, causing a representation of the linked data to be displayed by the browsing tools;

- a reference to the linked data comprising an instance of type *Binding* injected into type *any*.

In this representation the data linked into the hyper-program is stored in the spare 'user' fields of the light-button representations:

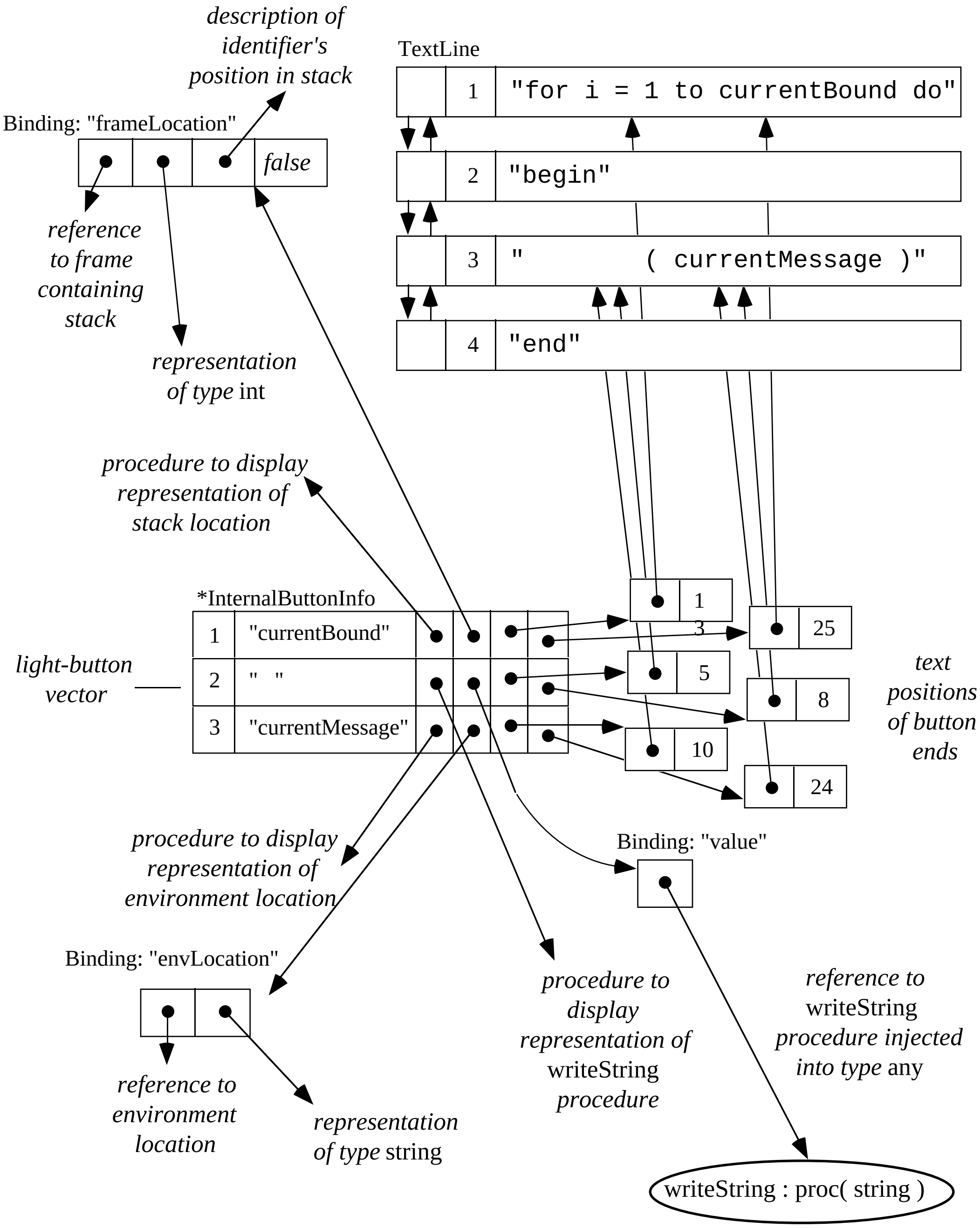

**Figure 6.23: Representation within hyper-program editor**

An exported hyper-program is represented by an instance of type *HyperSource* as defined in Figure 5.1. This contains a program in string or parsed form together with a vector of substitutions. Each substitution specifies a region of the program and the data, an instance of type *Binding*, to be substituted in that region. For a string program representation the regions are specified by a pair of character offsets from the start of the string, the first offset giving the character number for the start of the region and the second giving the number for the end of the region. The example hyper-program is shown in this form in Figure 6.24. It can be seen that this form requires many fewer objects than that in Figure 6.23.

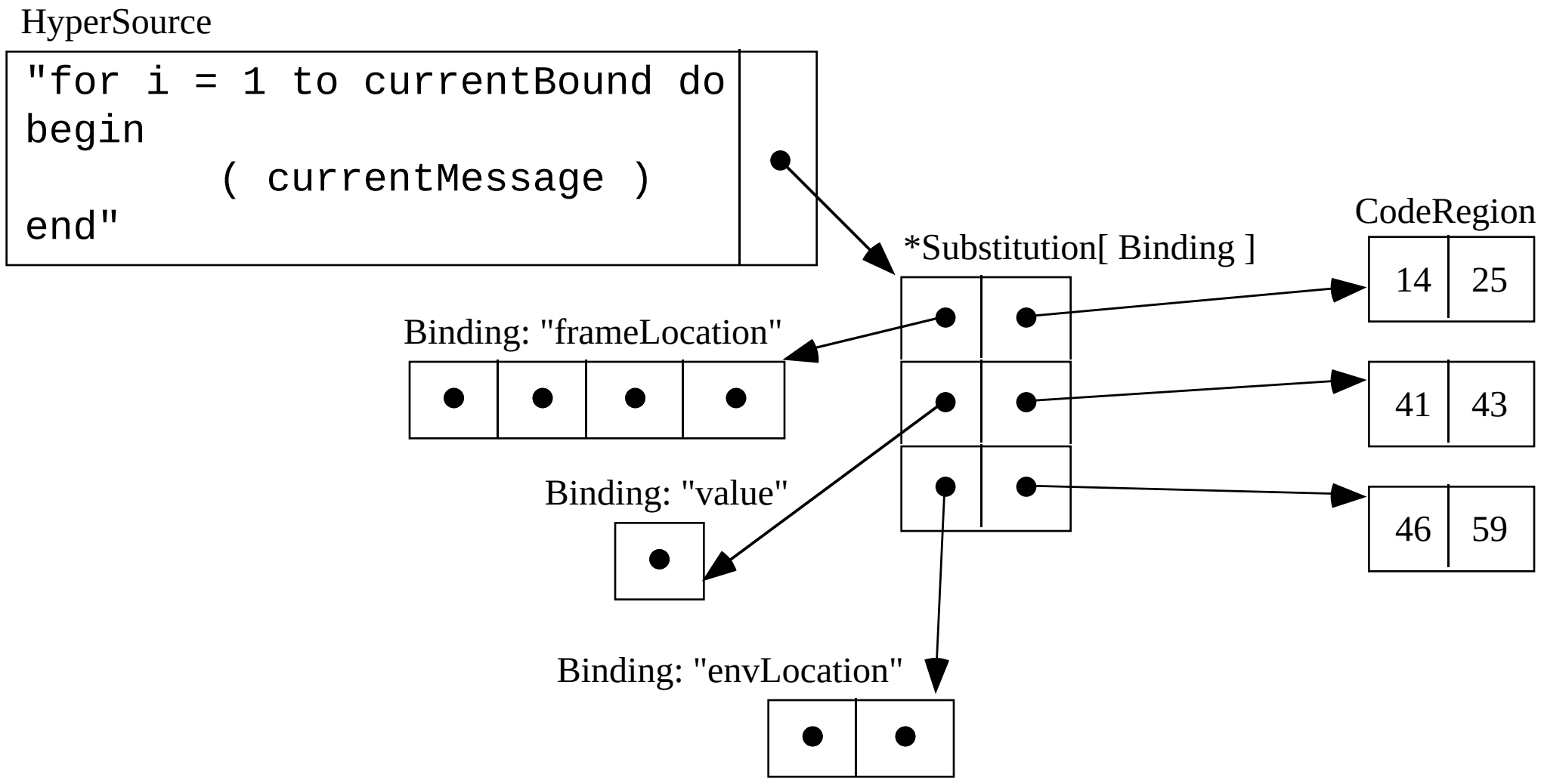

**Figure 6.24: Representation of exported hyper-program**

Figure 6.25 shows the hyper-program form that is processed by the compiler. When the *evaluate* button in a hyper-program editor is pressed the editor converts the hyper-program to this form and passes it to the compiler. Each substitution region in the text string is replaced by a unique identifier of the form *uniqueId*n where *n* is an integer chosen to ensure that the identifier does not occur anywhere else in the processed representation. Associated with the text string is a newly created symbol table which contains an entry for each of the identifiers corresponding to a substitution region. Among other items of information, each entry contains a representation of the type of the linked data and a reference to the data itself. The form of the reference depends on the nature of the data:

- for a value, the reference is to the low-level object representing the value;
- for an environment location, the reference is to the low-level object representing that location;
- for a free identifier, the reference is to the low-level object representing the frame containing the data, along with the position of the data within the frame.

A location in a structure, abstract data type or vector is not represented by a single unique identifier. Instead, the hyper-program contains a unique identifier for the structure, abstract data type or vector value, and code to perform the dereference is inserted after it. This code has the form *( fieldName ), ( fieldName )* or *( index )* respectively.

This hyper-program representation is passed to the compiler using the flexible compiler interface described in Section 6.4. This interface allows external symbol tables to be passed to the compiler along with the text. The compiler then uses the newly created symbol table to resolve uses of the substituted identifiers, which were chosen so that they did not clash with any normal identifiers in the hyper-program.

```
"for i = 1 to uniqueId78 do
begin
     uniqueId317( uniqueId402 )
end"
```

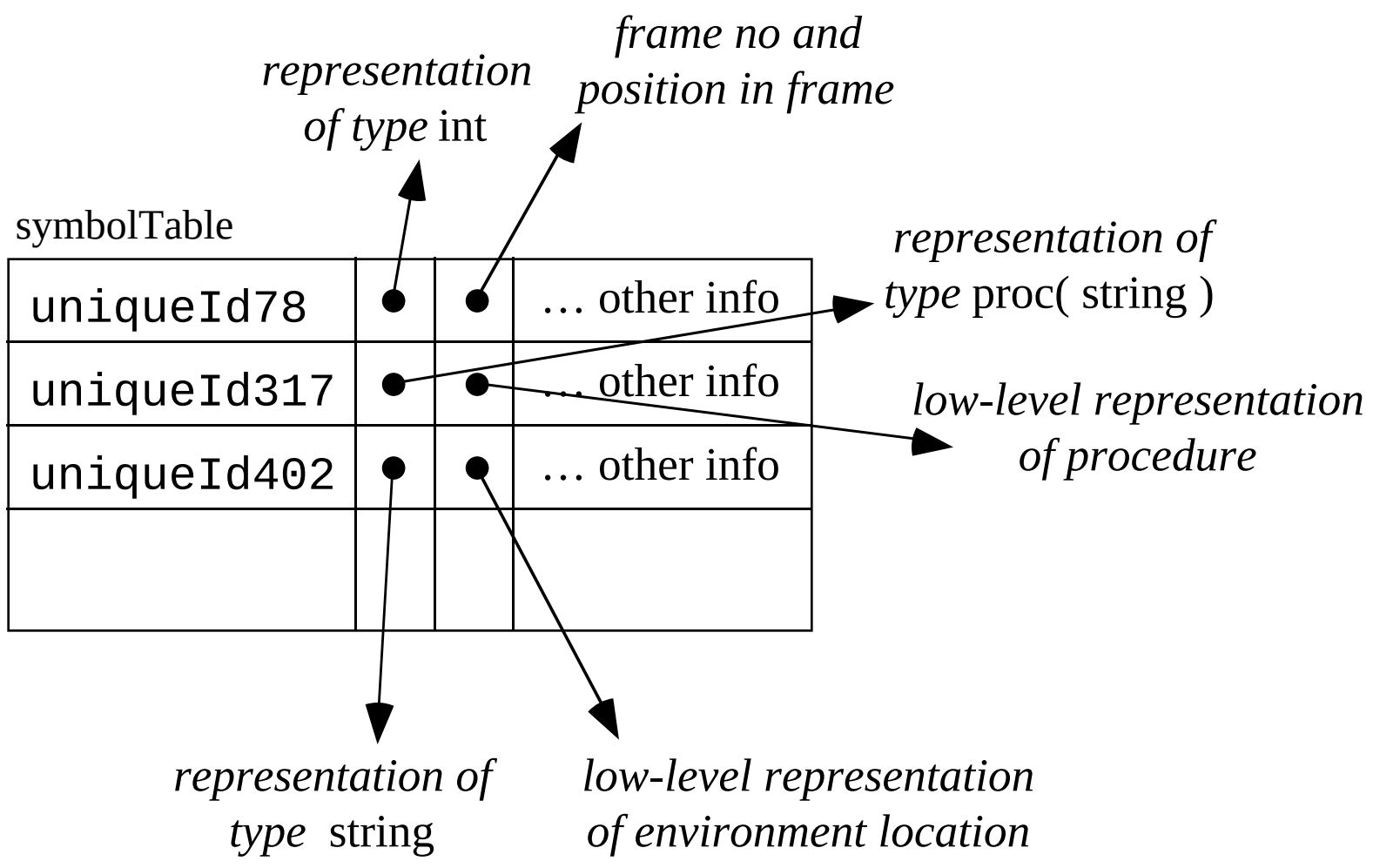

**Figure 6.25: Hyper-program representation passed to the compiler**

### 6.5.2 Constructing Closure Representations

In order to enforce associations from executable programs to the corresponding source programs, the hyper-programming system arranges that whenever a procedure is compiled its source code is retained and stored in the resulting closure. This is achieved by modifying the part of the Napier88 compiler that compiles procedure definitions, the procedure *procLiteral*. When the new version of *procLiteral* reaches the end of a procedure definition it extracts the part of the source code defining the procedure and inserts a reference to it in the newly constructed low-level object representing the code vector. The source code stored is a hyper-program, in the exported form, in which all free identifiers are replaced by hyper-program links.

During execution of the standard Napier88 compiler, the current position within the source text is abstracted over within the lexical analysis procedures. In the hyper-programming system these procedures are modified to make the source text position accessible by other procedures. This enables the modified version of *procLiteral* to note the current text position as it starts to compile a procedure and again at the end, giving the bounds of the procedure definition within the source code. As procedure definitions may be nested, giving nested activations of *procLiteral*, the system maintains a stack of positions of procedure definition starts. An entry is pushed at the start of *procLiteral* and popped at the end.

The system also keeps track of hyper-program links to be inserted into the procedure source code. These occur where the source program itself contains hyper-program links, and also where a free identifier is used within a procedure definition. Free here means that the identifier is declared outside the procedure definition. To determine which identifiers are free, *procLiteral* stores the lexical level of the procedure along with its source start position in each stack element. Whenever the modified lexical analysis procedures encounter an identifier, the symbol table entry for that identifier, if any, is obtained. If an entry exists and it shows that the identifier was declared at a lower lexical level than that of the procedure currently being compiled, then the identifier is free. In that case a new element is added to a list in the element at the top of the procedure stack. The new list element contains the source text position of the identifier and a specification of its corresponding data, of type *Binding*. In the case that the identifier denotes a hyper-program link already present in the source program, then the data already exists and the *Binding* contains a reference to a value or location. Alternatively, where the identifier is defined in the source program outside the procedure definition, the data will not exist until run-time. In this case the *Binding* contains a description of where the data will be at run-time, comprising a frame number and a position within the frame. Each time the end of *procLiteral* is reached the information about the current procedure definition is popped from the stack and used to produce its textual source code together with a vector of substitutions. Each substitution contains the position of an identifier and a *Binding.* The text and the substitutions together form an instance of type *HyperSource*, a hyper-program, and a reference to this is inserted in the newly formed code vector for the procedure.

This process is illustrated in Figure 6.26. The source code contains two procedure definitions *p1* and *p2*, with *p2* nested inside *p1*. The lexical level before the procedure definition is 0; at the start of *p1* it becomes 1; inside *p2* it is 2. The source character offsets of the start and finish of *p1* are denoted by *offset 1* and *offset 4*, while the corresponding offsets for *p2* are *offset 2* and *offset 3*. The identifiers *x* and *y* are declared within the program and *z* represents a hyper-program link to a value in the persistent store. The figure shows the hyper-program source representations recorded for *p1* and *p2*; note that some identifiers appear in both representations. A given identifier may appear normally in one representation and as a hyper-program link in another, as is the case for *y* in this example.

The bottom part of the figure shows the state of the procedure stack at the point that the compiler reaches the end of *p2*. The top element contains information about *p2*: its start offset, its lexical level and a list of the free identifiers used within it, *x* and *y*. Below this on the stack is information about *p1*, the free identifiers being *x*, twice, and *z*. At this point in compilation the contents of the top element are used to form the source for *p2*; the stack is then popped and later the other element is used to form the source for *p1*.

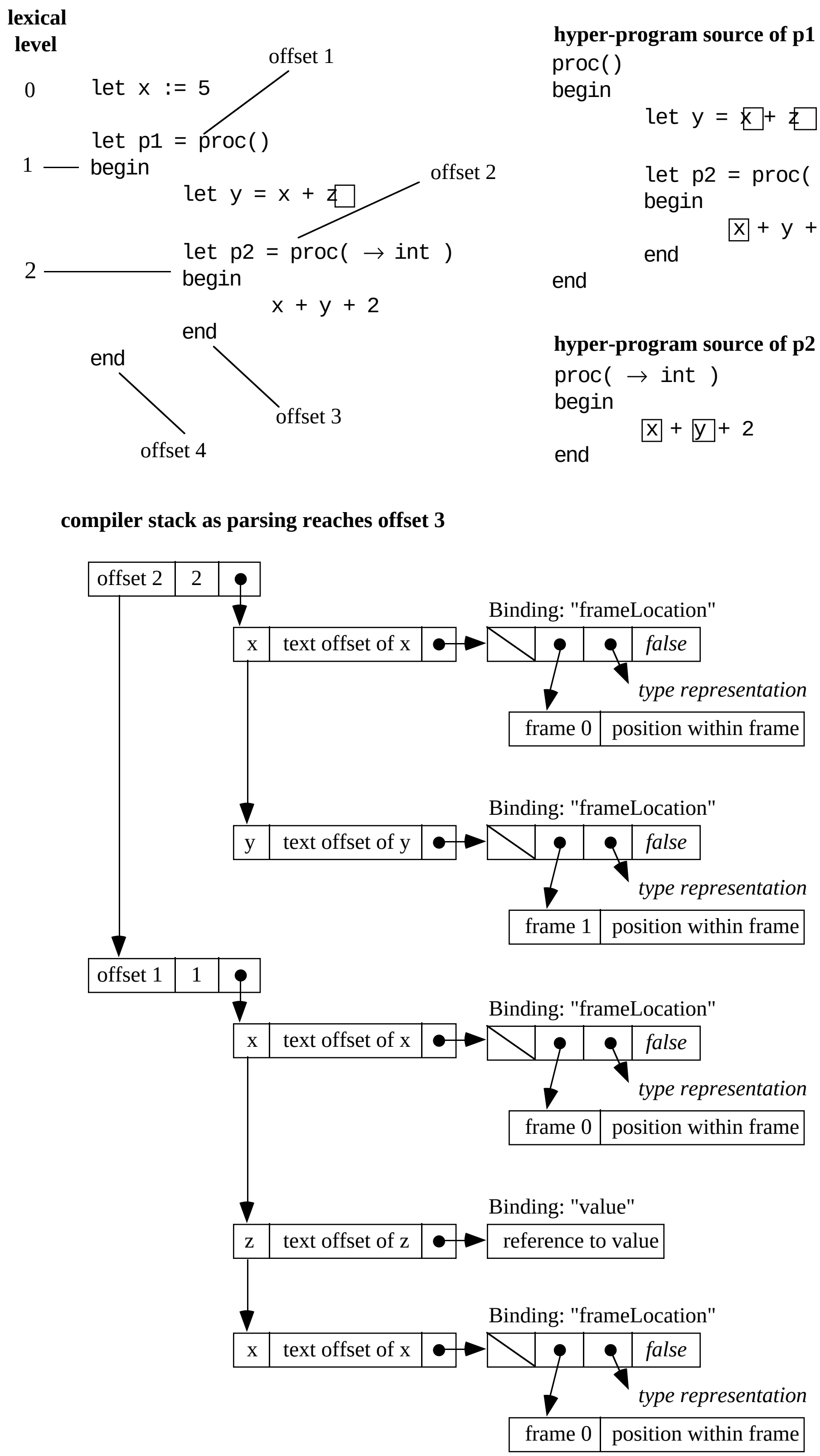

**Figure 6.26: Constructing source representations for nested procedures**

As illustrated in Figure 6.26, the information recorded in the *Binding* for a free identifier during compilation consists of a frame number and a position within that frame. The frame itself cannot be recorded as it does not come into existence until the compiled program is executed, thus the frame pointer field contains a null value. When the source hyper-program of a procedure value is displayed by the browser, the browser scans the hyper-program for null frame pointers and overwrites them with pointers to the appropriate frames. The frames are found by traversing the procedure's static chain to find the appropriate frame numbers. When a light-button corresponding to a free identifier is pressed the associated value is obtained and displayed by reading words from the frame, converting them to a typed value and passing the result to the browser.

The mechanisms described so far allow a source program passed to the compiler to contain hyper-program links to values or locations in the persistent store. Another variation is needed to cater for the possibility that a source program may contain hyper-program links to values or locations within existing frames. This situation arises when the programmer creates a new source program by combining components copied from the source programs of existing procedures with free identifiers, as illustrated in the next two figures. Figure 6.27 shows two source programs that contain references to frames containing free identifiers. Each frame contains a pointer to the next frame in the static chain, eventually terminating in the outer-most frame. Since the two procedures in the example have been produced by executing independently compiled programs, the two static chains are disjoint.

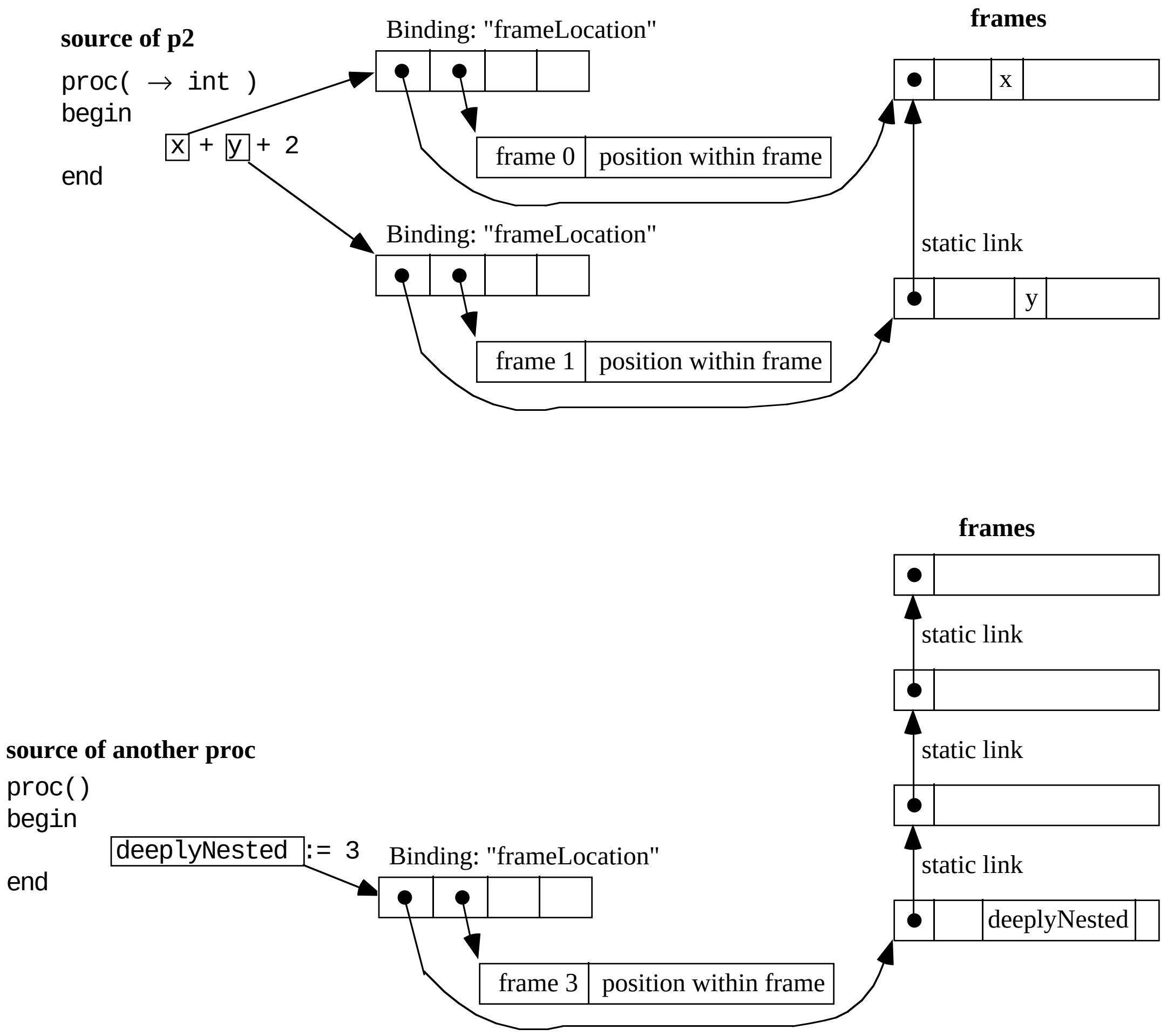

**Figure 6.27: Procedures with disjoint static chains**

Figure 6.28 shows a new source program constructed by copying parts from both existing source programs:

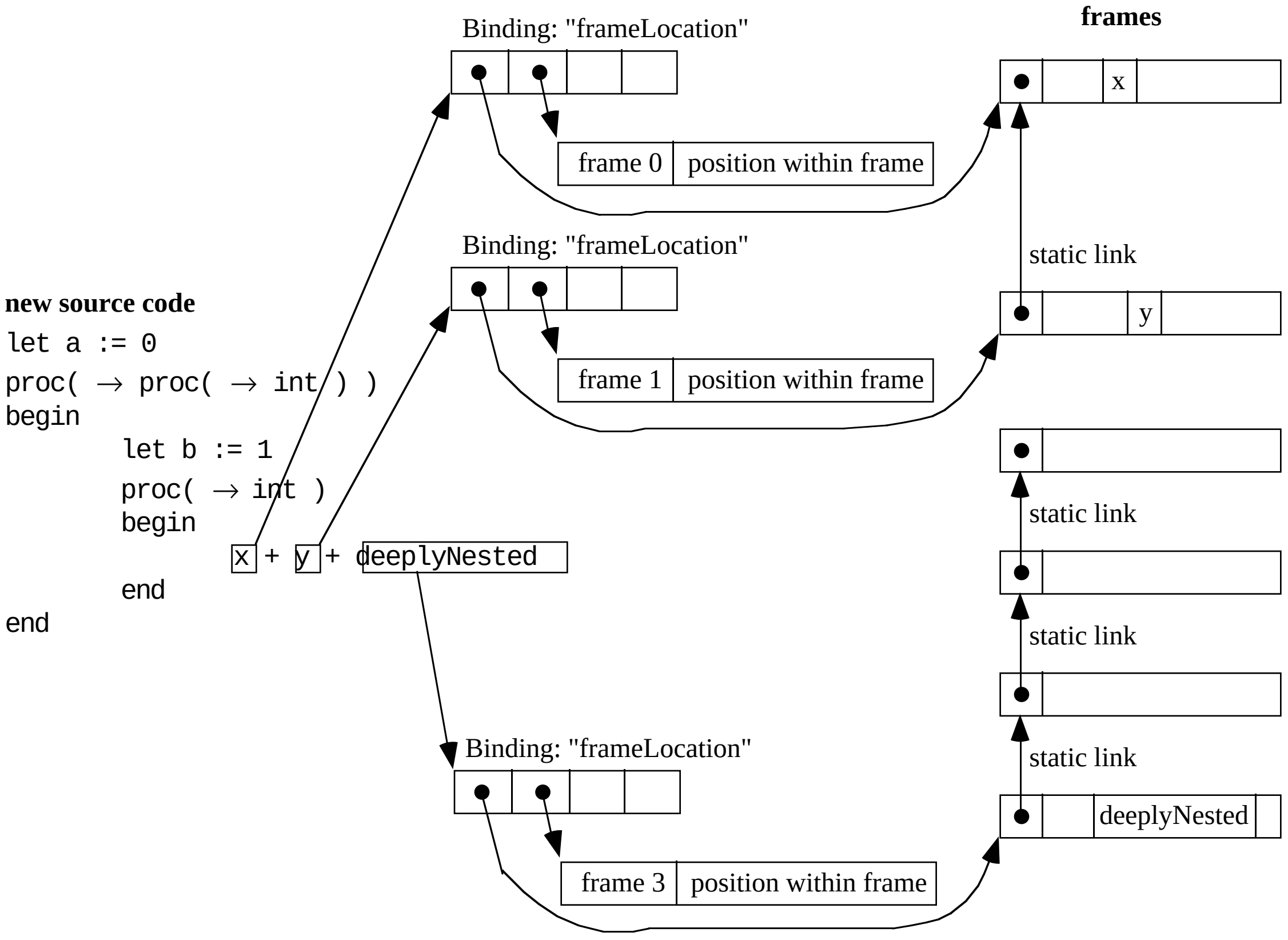

**Figure 6.28: Program with references to existing frames**

When invoked to compile a source program that contains references to external identifiers in existing frames as shown here, the modified compiler first allocates a numbering to each of the frames. The frames are numbered consecutively from 0 and the ordering is unimportant. The compiler then modifies the frame numbers recorded with the external identifiers to reflect the new numbering scheme and sets the lexical level at the beginning of compilation to the number of external frames. In the example shown there are three external frames so the lexical level at the beginning of compilation will be 3. In contrast, the standard compiler always begins compilation with a lexical level of 0.

The final way in which the modified compiler differs from the standard compiler is in the code planted to build the display on entry to a procedure. The standard compiler plants code that is executed on a procedure entry and traverses the procedure's static chain, loading onto the stack a pointer to each frame in the chain. These pointers form the display. The modified compiler also plants additional code that is executed after the standard display has been constructed. The additional code loads a pointer to each of the external frames in decreasing order of frame number. This ensures that references to external identifiers planted in the compiled code will be resolved correctly at run-time.

This mechanism is illustrated in Figure 6.29 which shows the state of the symbol table list at the start of compilation of the body of the inner procedure in the new program. The first two symbol tables contain entries for identifiers declared in the enclosing blocks, in this case the identifiers *a* and *b*. As compilation started at a lexical level of 3, this is the frame number for *a*. Another symbol table contains entries for the unique identifiers assigned to represent the external identifiers. These entries contain the frame numbers assigned to the external frames at the start of compilation.

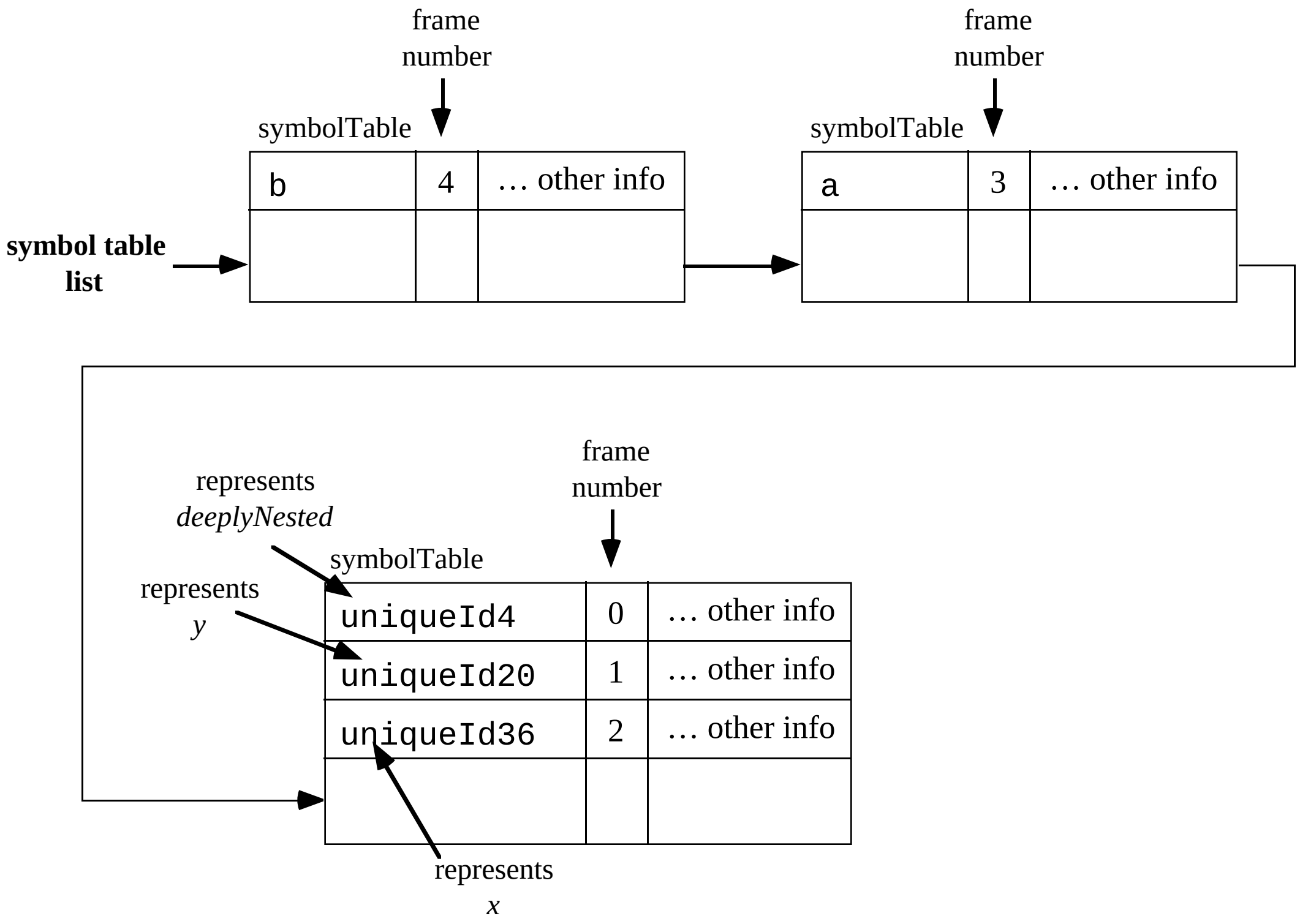

**Figure 6.29: Symbol table list during compilation**

Figure 6.30 shows the current frame at the start of execution of the procedure body. The frame's static link points to the frame for the enclosing block, created at run-time. The display contains pointers to this frame and to each of the external frames.

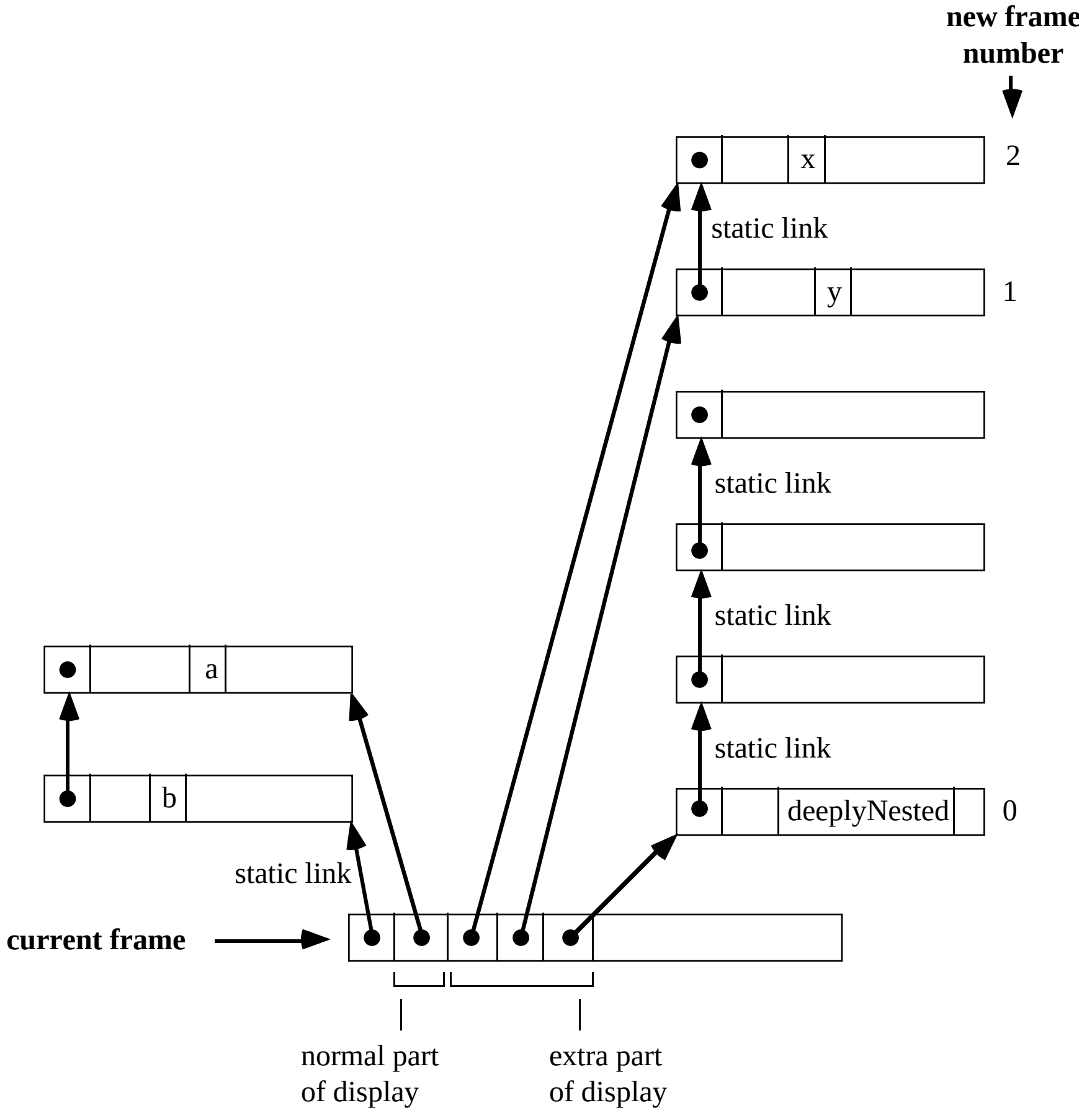

**Figure 6.30: Current frame during execution**

## 6.6 Conclusions

The implementation of the prototype hyper-programming tools has been described. As they are built using the WIN user interface system and the Napier88 browser, these systems have also been described, as have some relevant features of the Napier88 compiler.

The WIN system is implemented entirely in Napier88 and provides overlapping windows, user event distribution and a library of pre-defined interface widgets. Event-driven applications are constructed by writing Napier88 programs that compose selected widgets with additional code to provide application-specific behaviour. The contents of windows are stored in a space-efficient way that avoids duplication of image information in memory or the need for double updates to visible windows. A simple compaction technique is used to limit fragmentation of partially obscured windows.

Two implementation techniques have been used for the adaptive store browser. The original version employed type-safe linguistic reflection to construct browsing code for unfamiliar types, while the current version uses knowledge of the underlying store formats to enable direct access to the components of objects. In both cases, once a procedure for browsing a new type has been created it is stored in a persistent cache so that the work of creating the procedure does not have to be repeated on subsequent encounters with the type.

The hyper-programming tools allow the programmer to construct source programs that contain references to existing data and types. The data may be a value in the persistent store, a location in the persistent store, a value in a frame, or an environment location in a frame. A hyper-program is represented in one of several forms depending on whether it is being edited, exported outside an editor, or processed by the compiler. Modifications to the Napier88 compiler allow it to compile source programs with embedded hyper-program links and to store the relevant part of the source code with the closure produced by the compilation of each procedure definition.

# 7 Conclusions

The motivation for the research described in this thesis is to improve programmer productivity in persistent systems. This has been tackled in three ways:

- by reducing the amount of code that has to be written;
- by increasing the reliability of the code written; and
- by improving the programmer's understanding of the persistent environment.

Uses of type-safe linguistic reflection and hyper-programming techniques to achieve these goals have been investigated. Chapters 2 and 3 describe the techniques and give analyses of their benefits when employed in a persistent programming environment. Chapters 4 and 5 illustrate how the programmer interacts with the tools that support the techniques. Finally Chapter 6 gives details of how the programming tools were implemented.

## 7.1 Type-Safe Linguistic Reflection

Type-safe linguistic reflection allows program representations to be manipulated as data and the results transformed into executable programs. It enables the programmer to write programs that produce new programs. The technique may be used to specify highly generic program generators, extending beyond the genericity available in current polymorphic systems. This provides greater opportunities for software reuse thus reducing the amount of new code to be written. The mode of use of the generators is similar to that of *generics* in Ada [DOD83], in that typically for each specialisation of a generator the specialised form is used many times. Thus the costs of specialisation are amortised over many uses. The difference is that the specialised forms produced by reflection may depend on information available at specialisation time in more interesting ways, such as the structure of the types of the data.

Type-safe linguistic reflection may also be used to enable applications to adapt to changes in the structure of the data on which they operate, while retaining a high degree of static type checking. This reduces the amount of new code required since there is less need to re-implement applications as the data evolves.

Several systems that support type-safe linguistic reflection have been previously implemented. The contributions of this thesis are the following:

- a classification and analysis of the anatomies of reflective systems;
- identification of issues affecting the useability of linguistic reflective systems; and
- investigation of the interaction between linguistic reflection and persistence.

A generator model based on this work has been described, supporting the manipulation of hyper-program representations. Programming tools that aid the programmer in constructing, viewing and editing such generators have also been described.

Type-safe linguistic reflection has been used with persistence in two distinct ways. Firstly, where the reflection takes place in a persistent environment, the program representations manipulated are richer than mere textual forms. As they may contain direct links to persistent data, that data is available for inspection by the generators. This means that the specialised forms produced may depend on properties of the data manipulated other than the type. Similarly the specialised forms may contain links to data, properties of which have been verified by the generators.

Secondly, type-safe linguistic reflection has been used as an implementation technology in constructing an interactive persistent programming environment. Program representations are composed with an editor, itself implemented in the persistent programming language, and transformed using reflection into executable program forms.

## 7.2 Hyper-Programming

A hyper-program is a source program that contains links embedded in the text, in the same way that a fragment of hyper-text contains links to other fragments. The difference is that hyper-program links may to refer to data of any type in the persistent store, rather than being restricted to textual data [FDK+92, KCC+92b].

The provision of hyper-programming facilities assists the three goals of writing less code, writing more reliable code and understanding the persistent environment. The writing of less code is achieved by allowing more succinct programs, as a textual description of how to access a data item may be replaced by a link to the data. Code reliability is improved by enabling certain program checking to be performed statically rather than dynamically. Finally, the use of hyper-programs enables source representations to be supplied for certain programs that may exist in the persistent store but admit no purely textual representation. This assists the programmer in understanding the nature of the software available for reuse.

This thesis describes the first known implementation of hyper-programming. It provides a programming environment in which the programmer may browse the contents of the persistent store and compose hyper-programs linked to data found there. The support technology on which the environment is based is all implemented in Napier88. It includes the Napier88 compiler, a graphical user interface tool-kit, an interactive persistent store browser and a hyper-text editor.

Whereas type-safe linguistic reflection and persistence are orthogonal to one another, hyper-programming and persistence are deeply inter-linked. It is the ability to compose and store program representations within the persistent environment that makes possible the fundamentally different nature of hyper-program representations.

## 7.3 Related Work

### 7.3.1 Reflective Languages

Linguistic reflection is supported by a number of languages including Lisp [MAE+62], POP-2 [BCP71], TRPL [She90], PS-algol [PS88] and Napier88 [MBC+89]. In the last three the reflection is type-safe, that is the new programs that are generated are checked for type correctness before being executed.

Type-safe linguistic reflection has been used in a number of different ways. These include implementation of object browsers [DB88, DCK90], implementation of data models [Coo90a, Coo90b, CQ92], specification of generic program forms [SFS+90], optimisation of implementations [CAD+87, FS91] and validation of specifications [FSS92, SSF92].

It is believed that the generator notations developed in this thesis would enable these uses of type-safe linguistic reflection to be coded in a cleaner and more understandable way.

### 7.3.2 Linking Mechanisms

The hyper-programming environment described allows the links from a program to the data on which it operates to be established during three different phases of the application development process. These are at program composition time, compilation-time and run-time.

No other languages support composition-time linking. Compilation-time linking is available in the ABERDEEN programming environment [Far91]. It also occurs in the interactive languages ML [MTH89], Quest [Car89] and Galileo [ACO85].

Several language systems support a distinct linking phase between compilation-time and run-time, during which unresolved references to data and programs in the compiled program are established to give an executable program. Examples of such languages are Pascal [Wir71], C [KR78] and Ada [DOD83]. It has been shown in [AM84, MAD87, AM88, Con90] how the same effects and benefits of this linking phase may be obtained in a persistent language with first class procedures and no explicit linking phase.

Dynamic systems such as Lisp and Smalltalk-80 [GR83] allow linking at run-time only. While highly flexible this precludes any static checking of the data.

### 7.3.3 User Interface Tool-Kits

A number of user interface tool-kits and interface development systems are commercially available. These include the Apple Macintosh Toolbox [App86]; NeXT's User Interface Builder [Web89]; Sun Microsystems' Graphic User Interface Design Editor [Sun89, Sun90]; the Simple User Interface Toolkit (SUIT) [PYD91]; and IBM's experimental ITS system [WBB+90].

A development trend can be seen in these systems, from the tool-kit approach to the more sophisticated interface development systems. In the earlier systems the composition of interface components is described textually with calls to a program library. The more recent systems allow the interface to be developed interactively, using mouse gesture and reducing the need to write program code. Surveys are given in [Mye89, Shn92].

The user interface tool-kit developed in the course of this thesis is not particularly sophisticated in comparison with the more recent systems listed above. However it has served its purpose as enabling technology. It is also the only known system that can be used with a strongly typed persistent language.

### 7.3.4 Other Languages and Database Systems

This section identifies a number of other programming languages and database systems and for each attempts to indicate whether or not support for run-time linguistic reflection and hyper-programming could be provided. The principal criteria for run-time linguistic reflection are:

- a means of representing programs as data values;
- accessibility of a compiler from within an executing program; and
- a means of binding to a compiled result from within the same executing program.

For hyper-programming the main requirement is for the enforcement of referential integrity i.e. whether once a reference to a value or object is established, it can be guaranteed to refer to the value or object for as long as the reference itself exists.

#### 7.3.4.1 Smalltalk-80

Smalltalk-80 is an object-oriented programming language which supports a 'snap-shot' form of persistence [GR83]. This means that at any point an image of the current state of the system can be dumped to non-volatile storage and later restored. Referential integrity is maintained since there is no explicit deletion of objects and thus dangling references are prevented. Garbage collection is used to remove non-reachable objects automatically. It should be possible to provide reflection and hyper-programming facilities by encapsulating each hyper-program in an object with methods to read and write both characters and hyper-program bindings. A compiler object would provide a method to take a hyper-program and produce either a result object or an error description. The result object, if compilation was successful, would be of class *Object* or a sub-class i.e. any class.

#### 7.3.4.2 GemStone

GemStone [MS87, BOP+89] is an object-oriented database system with a database language, OPAL, based on Smalltalk. It allows the programmer to write queries over objects and their instance variables, and also to specify indices over instance variables. Since the data model is largely the same as Smalltalk, it should be possible to support reflection and hyper-programming in the way described above.

#### 7.3.4.3 Arjuna

Arjuna is a distributed object-oriented programming language [DPS+89, SDP91]. It does not support orthogonal persistence; the programmer must write code to flatten and reconstruct objects at class definition time. Explicit deletion of objects is allowed and object identity is not preserved over flattening and reconstruction. Because of this the language is not suited to hyper-programming, since a hyper-program link in a hyper-program could not be guaranteed to always a refer to the same object. No facilities for linguistic reflection are currently provided although it appears that run-time linguistic reflection could be supported by making the compiler available as an object.

#### 7.3.4.4 OSS

OSS [SM90] is an object storage system for the SOS operating system [Sha86]. It is based on C++ [Str86], and the programmer manipulates OSS objects as though they were C++ objects. As with Arjuna, explicit object deletion is permitted, allowing dangling references and thus making the system unsuitable for hyper-programming. Again, a compiler could be made available within the system in order to provide linguistic reflection.

#### 7.3.4.5 Iris

Iris [LDF+87, FBC+90] provides an object-oriented data model based on DAPLEX [Shi81] and Taxis [MBW80]. Queries over the data are translated into a relational algebra and the database itself is implemented above an underlying relational storage system. Several interfaces to the database are provided, including OSQL, an object-oriented extension of SQL, and an extended version of a Lisp structure browser. Explicit deletion of objects is permitted but only in cases where there exist no references from other objects to the object to be deleted. This implies that, as with Smalltalk, both hyper-programming and run-time reflection could be supported.

#### 7.3.4.6 VBASE

VBASE [AH87] attempts to integrate an object based database system with an object-oriented programming language. Although strong typing has been presented as one of the design goals, the language COP (C Object Processor) which is used by the programmer to implement applications in the system is a strict superset of C. Thus there is no way to prevent arbitrary address arithmetic being performed below the level of the type system. The database system supports automatic maintenance of inverse relationships, thus all references to a particular object can be found simply. Various clustering strategies can be specified on a per-object basis. Objects can be deleted explicitly, thus the referential integrity problem occurs. As with Arjuna and OSS this makes VBASE unsuitable for hyper-programming, while again run-time reflection could be supported. The language provides a dynamic type checking mechanism that could be used to assert the expected type of the result of a reflective computation.

#### 7.3.4.7 $O_2$

$O_2$ is another object-oriented database system [BBB+88, LRV90]. A number of language interfaces to the database are provided, including C ($O_2C$) and BASIC ($BO_2$). Application programs written in these languages are compiled to C, during which type violations such as arbitrary address arithmetic may be detected. Persistence is defined by reachability from one or more roots specified in a schema, and there is no explicit object deletion. $O_2$ could thus support hyper-programming and run-time reflection.

#### 7.3.4.8 ML

ML [Mil78, MTH89] is different from the previous languages and database systems in that it is (largely) functional, and statically typed. Although each compilation unit is statically type checked, the incremental nature of the system makes it possible to perform a variety of linguistic reflection in which a program representation is compiled and the current environment enriched with the results. These results can then be referred to in subsequent programs. It is not possible however to bind to the result of a particular instance of reflection within the compilation unit in which it is created, since this would require a means of dynamic type checking. Nevertheless, hyper-programming could be implemented in this way, with persistent versions of ML [Mat89] deriving the greatest benefits.

## 7.4 Future Research

### 7.4.1 Programming Support

Existing programming environments offer rich sets of tools to assist the programmer in the various phases of software development. Some examples of such environments are Unix [RT78]; Turbo Pascal [Bor89]; Think C™[BM89]; the Cornell Program Synthesizer [TR81]; Interlisp [TM84]; Cedar [Tei84]; PECAN [Rei84]; Trellis [OHK87]; Mesa [Swe85] and GANDALF [Not85]. Tools provide such functions as source code editing; comparing, linking and analysing source programs; checking for inconsistencies; configuring a system according to component dependencies; debugging; maintaining associations between executable and source programs; and many others.

Future research will investigate the tool support needed by the persistent hyper-programmer. Undoubtedly the functions listed above will be required but new needs will also arise as the increasing sophistication of persistent programming environments changes the nature of program construction. Tool integration has been a theme of much work on programming environments; the rich type systems of persistent systems will provide a high band width channel for communication between future programming tools.

### 7.4.2 Hyper-Worlds

As the size of persistent stores increases the problems of change management will grow. The very advantage provided by the hyper-programming paradigm, that of allowing collections of data to be tightly coupled together, creates a tension with the need for flexibility to accommodate change. If unrestricted linking throughout the persistent store is permitted, inter-application dependencies may build up to the extent that it become difficult or costly to make the related changes necessary to restore consistency after a change to an application component. The persistent store can be compared to a large pot containing spaghetti and cheese sauce [Mai90]. After the sauce has set it becomes difficult to extract a single strand of spaghetti without disturbing the others.

If however the mixture is poured into small bowls before the sauce sets, the task of removing a strand is easier. Only the strands in the same bowl as the one to be removed stick to it; those in the other bowls are unaffected. The hyper-world model outlined in Chapter 3 proposes a similar strategy to manage the persistent store. The store is partitioned into many small hyper-worlds or application spaces; a given component may only be linked to by other components in the same hyper-world. This limits the propagation of a change to a component, such as its replacement with another of a different type, to a relatively small region of the persistent store. Research is required to further investigate the suitability of this model and to implement tools to support it.

### 7.4.3 Linking Control

A consequence of the tightly coupled nature of hyper-program representations is that it becomes difficult to use them outside of the persistent store. For example, how can a hyper-program listing be published in an article? How can a hyper-program be sent to a co-worker at a different site for installation in a different persistent store? Possible mechanisms include making a deep copy of a hyper-program and its closure—with the danger of copying the whole persistent store—or cutting direct links in some manner and re-establishing them in another store.

These are facets of a more general problem of relinking. Given a source program or executable program created in a particular environment, how can it be reused in a different environment? Where the program establishes links dynamically, it is only loosely coupled to its environment and is easy to transplant. Future research will seek mechanisms to combine the benefits of tight coupling with the ability for reuse in new environments.

Reflective hyper-programming provides one approach to the particular problem of cutting links in a hyper-program and re-establishing them in another store. This involves implementing a generator generator which takes as input a hyper-program and produces a generator. The result generator contains within it the textual part of the hyper-program and descriptions of the access paths within the persistent store of the data to which the hyper-program is linked. The generator is self-contained thus it admits a purely textual source representation which can be exported to a foreign store. Once installed there the generator is evaluated. It uses the information about the access paths of the data in the original store to access the corresponding data in the foreign store and produce a fully linked hyper-program which is isomorphic to the original.

### 7.4.4 Type-Safe Linguistic Reflection

Even with a graphical user interface, the generators used in type-safe linguistic reflection are still hard to write. Typically they will be written by the systems programmer and made available in the persistent store for general use. It is an open question whether future improvements in generator models and user interfaces will ever reduce the difficulty to the level of writing polymorphic programs in current systems. This seems a worthwhile goal to aim for.

Further refinement of the generator model and the supporting tools may involve the development of a model in which the typing of the generators is more tightly controlled. In the current model each generator takes a single environment as a parameter; this gives a flexible mechanism and has the advantage that the model can be implemented in the type system of Napier88. However it does mean that generator calls with incompatible arguments are not detected until generator execution time. One approach is to define *generator* as a type constructor and allow the parameters to become part of the type in the same way that the types of a procedure’s parameters are part of its type. It would be possible to implement a version of Napier with such a type constructor using a text pre-processor that converted programs into standard Napier88. Indeed it would make an appropriate test of the existing generator system to implement the pre-processor as a generator.

Other challenges in the field of reflection include the development of an integrating model for compile-time and run-time reflection, and second order type checking issues. Although the effects of compile-time reflection can be obtained in a persistent run-time reflective system, different processes occur during the two kinds of reflection as the action of the compiler is more complex in the first. It is not clear as yet whether any advantages would be obtained in a combined system. The other research area addresses the question of whether a generator definition language can be sufficiently restricted to allow the generator results to be type checked statically, while retaining enough flexibility to be useful.

## 7.5 Conclusions

The main theme of this thesis is that the use of a persistent environment as a base for supporting programming activities can lead to major productivity gains. The nature of programs developed within a persistent environment may be fundamentally different from the traditional view of programs as static descriptions of manipulations on data. Linguistic reflection has been used both as an implementation technology for this kind of programming and as a programming tool that increases productivity in its own right. The persistent environment also impacts on the reflection process and the implications of generators being able to access that environment are only beginning to be explored.

Whether or not the hyper-program ever becomes as widely accepted as its poor relative the hyper-text document, it has been exciting to build and to use. Let us hope that some of these ideas survive in persistent systems of the future.

# Appendix A. Generator Tool Example: Natural Join

This appendix illustrates the use of the generator tools to provide a reflective solution to the problem of specifying a generic natural join function. This is contrasted with a similar solution in standard Napier88. Both cases involve analysis of the input types and the construction of tailored code that performs natural join on arguments of those types.

For brevity, in both cases the existence of various pre-defined procedures is assumed. With the generator tools these procedures are indeed pre-defined and are available for general use; with standard Napier88 the programmer would have to define the procedures also. The procedures available with the generator tools are listed in Appendix B.

Relations are modelled here as sets of Napier88 structures. The definition of a generic set type and a procedure for creating empty sets are assumed.

The general solution in both cases is as follows:

- Construct a representation of the result type of the join from the input type.

- Construct the representation of a procedure to perform the join. The procedure is recursive and for the base case, when the first set in the join is empty, it returns an empty set. Otherwise it forms a set containing the tuples obtained by joining the first tuple of the first set with the second set. The result is obtained by performing a union operation with this and the result of joining the remainder of the first set with the second set.

Following these two examples a solution in TRPL is given.

## Example Using Generator Tool

Figures A.1 to A.4 show windows containing generator and procedure definitions to implement natural join. The first window in Figure A.1 shows the main generator, *join*. It takes two parameters of type *TypeRep*, representing the tuple types of the input relations. For brevity the checks to ensure that they represent structure types are not shown. The prelude enriches the input environment with three new values: *resultType*, which represents the tuple type of the result relation, and *type1Fields* and *type2Fields* which are sets containing the field information for the two input types. The value *resultType* is obtained by calling a procedure *joinResultType*, a direct link to which is contained in the prelude code. The structure field information is obtained using the pre-defined procedure *getStructureFields* which returns a set of (name,type) pairs. The result definition is a literal and contains the definition of the resulting join procedure. The code contains a number of buttons representing calls to sub-generators. These are used to define, in order of appearance, the first input type, the second input type, the result type (twice), a procedure to compare instances of the result type and a procedure to perform a join between a single tuple and a relation.

The result definition also contains a direct link to the pre-defined procedure *mkEmptySet*. As will be illustrated later this contrasts with the Napier88 solution in which the result definition contains code to link to the procedure in the persistent store. The direct link notation is both more concise and more secure, as there is no danger of access to the procedure being removed between the times of evaluation of the generator and execution of the generated result.

Figure A.1 also shows the definition of the procedure *joinResultType* which computes a representation of the result type. This is achieved by constructing the union of the two sets containing the names and types of the fields of the input types and using the pre-defined procedure *mkStructureType* to create a type representation. The other windows show the sub-generators *type1* and *type2*. These generators use the pre-defined procedure *mkTypeLink* to obtain links to the types represented by the input type representations.

The first window in Figure A.2 shows the generator *resultType* which produces a link to the join result type. The next, *compareResult*, produces code to construct an instance of type *Comparison* specialised to the result type, using the sub-generator *matchBody* which is explained below. The code contains a direct link to the pre-defined procedure *mkComparison*. The next window shows the generator *onejoin* which produces a definition of the procedure to join a single tuple and a relation. If the set representing the relation is empty the procedure returns an empty set. Otherwise the procedure picks a tuple from the relation and recursively obtains the result of joining the original tuple with the remainder of the relation. If the selected tuple does not match with the original tuple then the result of the join is returned. If it does match then a new tuple is formed by concatenating the two tuples and inserted in the result relation. The last window in Figure A.2 shows the generator that produces the procedure to determine whether two tuples match according to the rules for natural join. Another sub-generator *matchBody* is used to define the body of this procedure.

Figure A.3 shows *matchBody* which produces an expression that will evaluate to true when two tuples match. The expression consists of the ‘and’ing together of a number of boolean expressions, each of which tests for equality of the tuples over a particular attribute. To produce this the generator first computes the intersection of the sets of field information for the two input types, using the pre-defined procedure *intersection*. In computing this intersection, two fields are considered equal if their names are equal and their types are equivalent. The generator then iterates over the intersection set and builds up a new set containing fragments of source code, each of which tests for equality over one attribute/field. Each fragment is constructed by passing the field name to the pre-defined procedure *evalWithString*, along with the sub-generator *mkFieldNameTest*. This procedure evaluates the generator, passing the string value to it in its input environment. Finally the result definition of *matchBody* consists of a call to the pre-defined procedure *andCompose* which produces an expression in which the elements of the set passed to it are ‘and’ed together. The other windows in Figure A.3 contain the sub-generators *mkFieldNameTest* and *fieldName*. The former produces a literal expression comparing two tuples over a field/attribute name determined by the latter, which simply converts a string into a source code fragment.

Figure A.4 shows the generator *concat* which produces the representation of a procedure to concatenate two tuples together. The body of the procedure is constructed by the sub-generator *concatBody* shown below. The structure of *concatBody* is similar to that of *matchBody*. First it calculates the union of the sets of field information. It then iterates over the resulting set and builds up a new set containing expressions that select the attribute value from the first or second tuple as appropriate. These expressions are formed by the sub-generators *takeFrom1* and *takeFrom2*. The result definition of *concatBody* consists of a call to the pre-defined procedure *mkStruct* to form the representation of a code fragment to create a new structure/tuple from the field names and expressions.

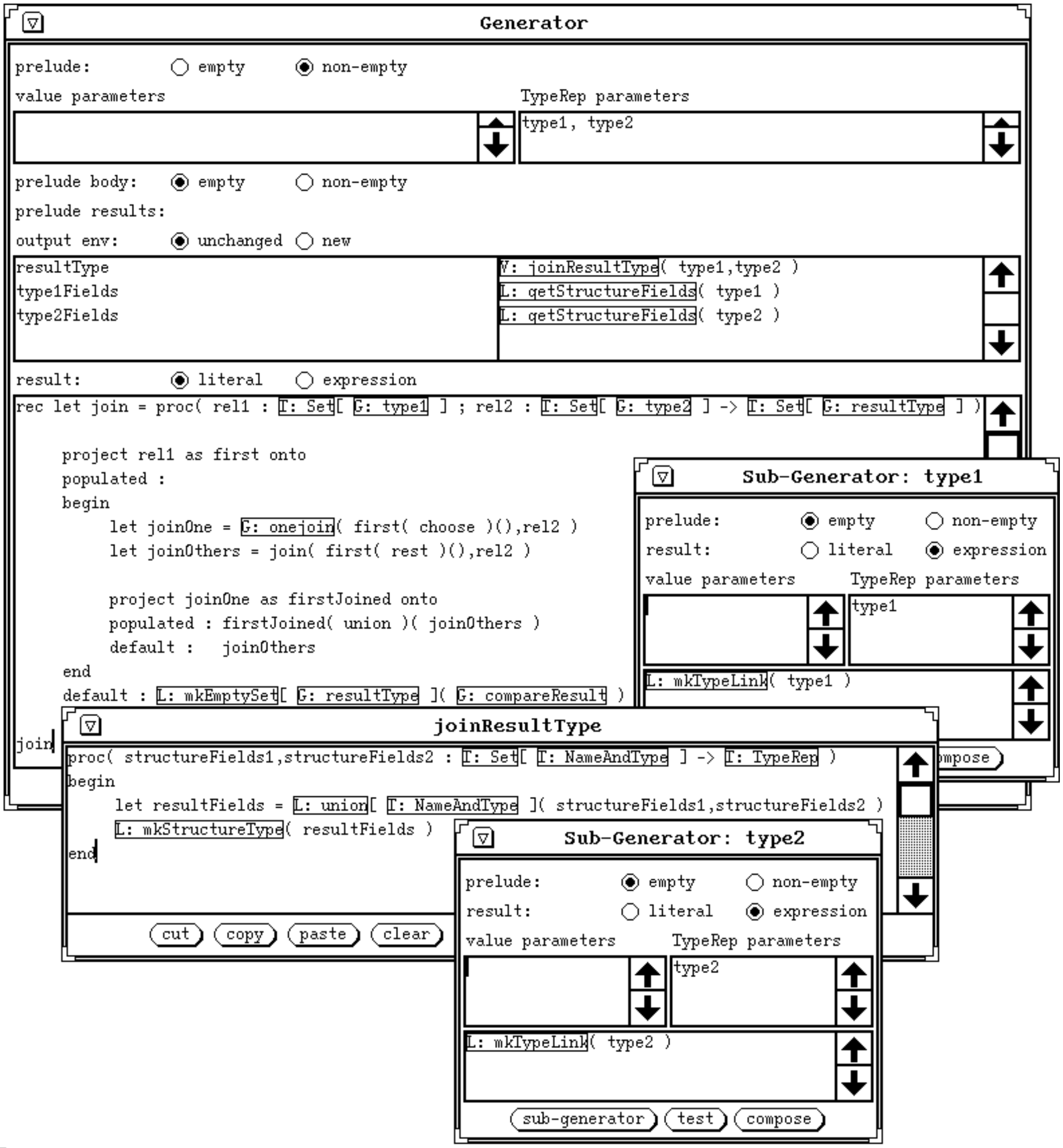

**Figure A.1: Generators *join, type1* and *type2***

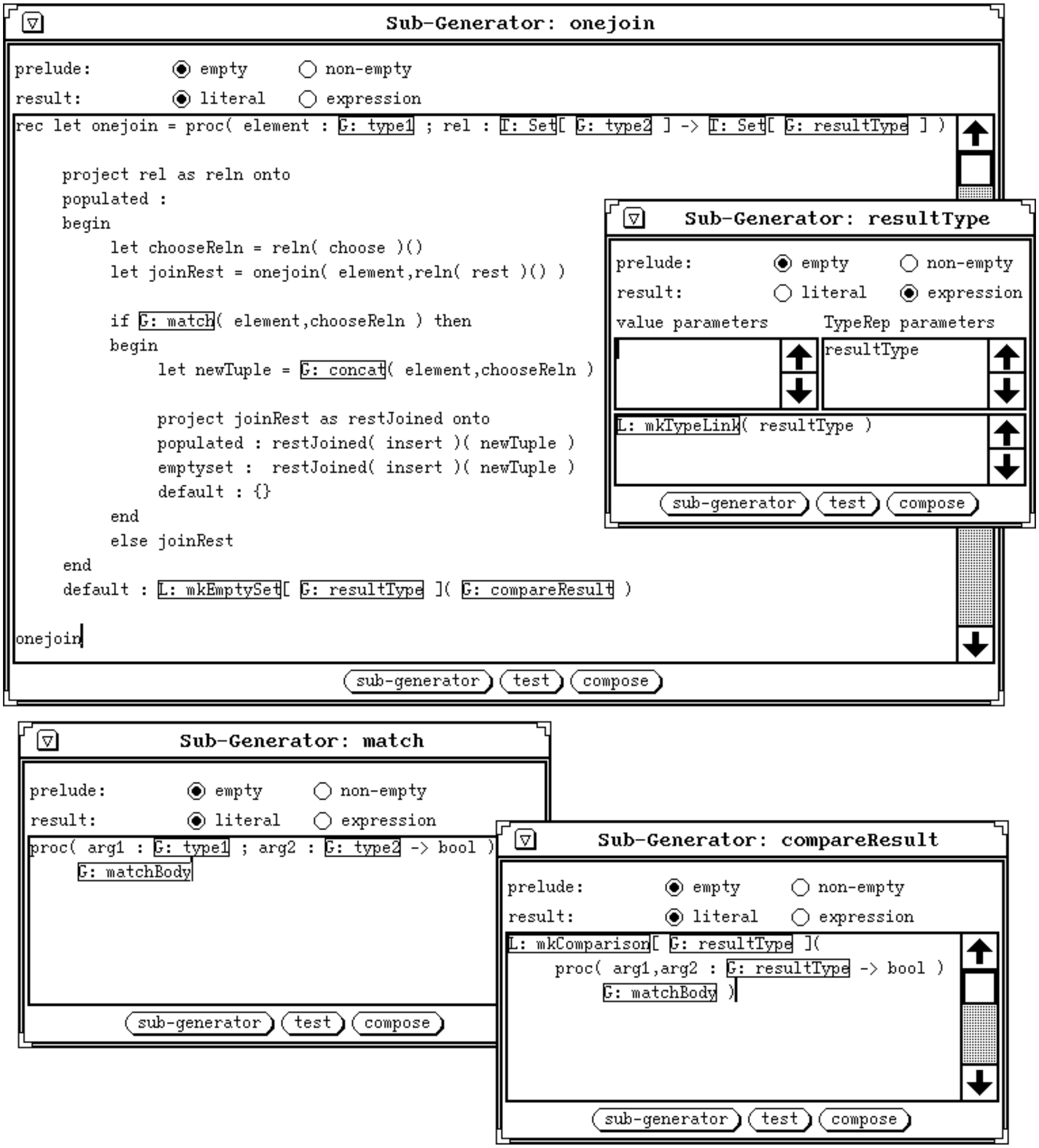

**Figure A.2: Generators *resultType*, *compareResult*, *onejoin* & *match***

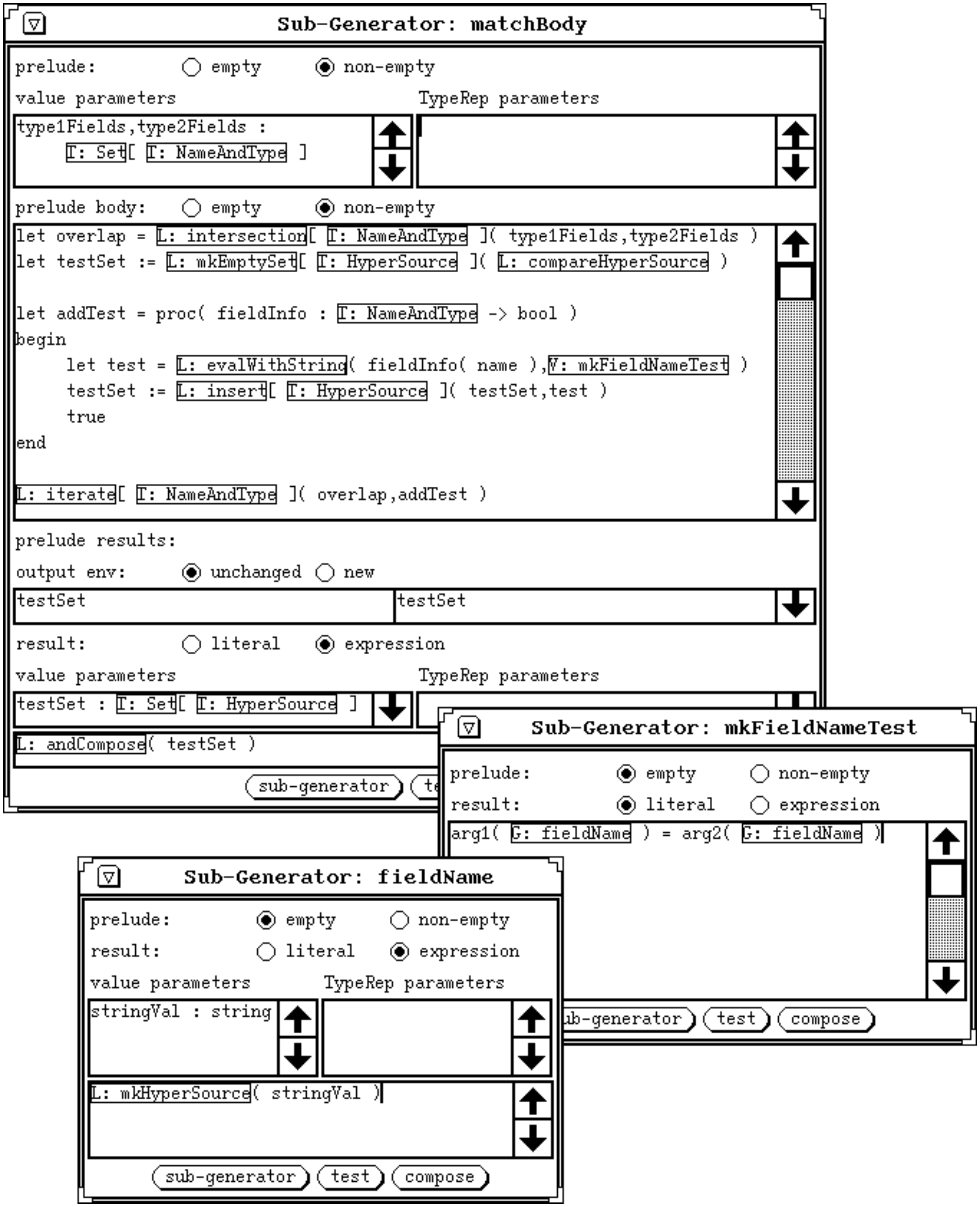

**Figure A.3: Generators *matchBody*, *mkFieldNameTest* & *fieldName***

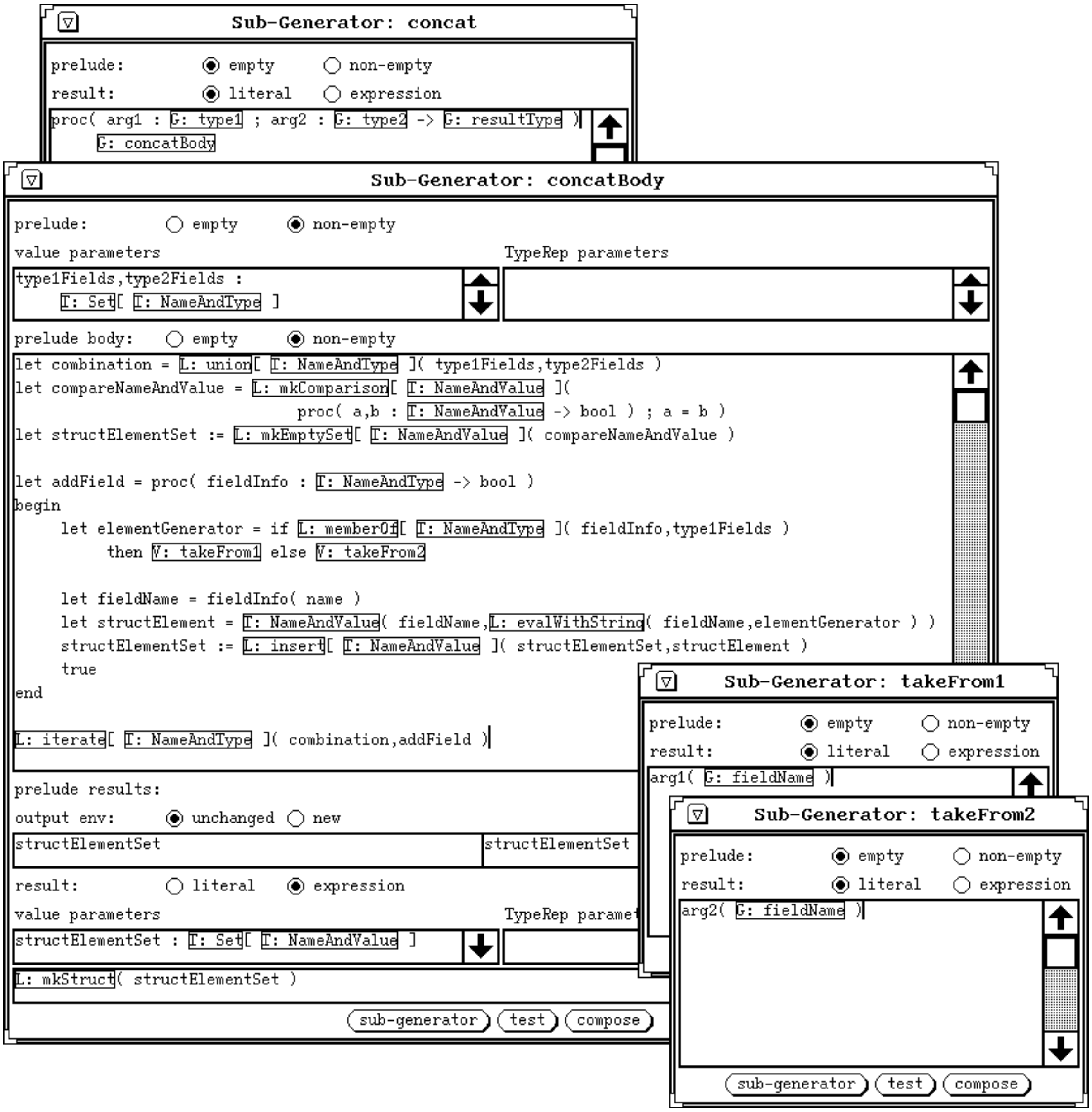

**Figure A.4: Generators *concat, concatBody, takeFrom1* & *takeFrom2***

## Example in Napier88

Figure A.6 shows an implementation of generic natural join in standard Napier88, using as far as possible the same algorithm as the previous solution. The principal differences from the previous solution are as follows:

- The source code being manipulated is in string form rather than hyper-program form.
- There is more syntactic noise in the generator result definitions in the Napier88 version, due to the many string concatenation and quote symbols. Quotes and apostrophes appearing within strings are difficult to read as they are preceded by extra apostrophes as escape symbols.
- The program includes textual type definitions. These are required both in the main program and in the generated code.
- The program contains code to link to the pre-defined procedures in the persistent store. This is also required both in the main program and in the generated code. Thus it is possible that the generated code will fail when it is executed due to pre-defined procedures no longer being present.
- Although the definitions of the pre-defined procedures have not been shown here, they must be defined by the programmer as they are not part of the Napier88 standard user environment [MBC+89].
- The existence of an additional pre-defined procedure *writeType* is assumed. It produces a textual definition of a type from a type representation.

```
rec type  TypeRep is structure(  label, misc, random : int ;
                                 name : string ; others : Var )
&         Var is variant( none : null ; one, unique : TypeRep ; many : *TypeRep )

type NameAndType is structure( name : string ; typeRep : TypeRep )

type NameAndValue is structure( name, value : string )

type Comparison[ T ] is variant(
      ordered :     structure( equal, lessThan : proc( T, T → bool ) );
      unordered :   structure( equal : proc( T, T → bool ) ) )

rec type Set[ T ] is variant(
      emptyset :   structure(  insert :        proc( T → Set[ T ] );
                               union :         proc( Set[ T ] → Set[ T ] );
                               intersection :  proc( Set[ T ] → Set[ T ] );
                               difference :    proc( Set[ T ] → Set[ T ] ) );

      populated :  structure(  insert :        proc( T → Set[ T ] );
                               union :         proc( Set[ T ] → Set[ T ] );
                               intersection :  proc( Set[ T ] → Set[ T ] );
                               difference :    proc( Set[ T ] → Set[ T ] );
                               delete :        proc( T → Set[ T ] );
                               choose :        proc( → T );
                               rest :          proc( → Set[ T ] );
                               includes :      proc( T → bool );
```

```
                              scan :          proc( proc( T → bool ) );
                              size :          proc( → int ) )

use PS() with
    mkComparison :      proc[ T ]( proc( T,T → bool ) → Comparison[ T ] );
    mkEmptySet :        proc[ T ]( Comparison[ T ] → Set[ T ] );
    intersection :      proc[ T ]( Set[ T ], Set[ T ] → Set[ T ] );
    union :             proc[ T ]( Set[ T ], Set[ T ] → Set[ T ] );
    memberOf :          proc[ T ]( T, Set[ T ] → bool );
    insert :            proc[ T ]( T, Set[ T ] → Set[ T ] );
    iterate :           proc[ T ]( Set[ T ], proc( T → bool ) );
    getStructureFields : proc( TypeRep → Set[ NameAndType ] );
    andCompose :        proc( Set[ string ] → string );
    mkStruct :          proc( Set[ NameAndValue ] → string );
    mkStructureType :   proc( Set[ NameAndType ] → TypeRep );
    compareString :     Comparison[ string ];
    writeType :         proc( TypeRep → string ) in
begin

let defineMatch = proc( type1, type2, resultType : TypeRep ;
                        type1Fields, type2Fields : Set[ NameAndType ] → string )
begin
    let overlap = intersection[ NameAndType ]( type1Fields, type2Fields )
    let testSet := mkEmptySet[ string ]( compareString )

    let addTest = proc( fieldInfo : NameAndType → bool )
    begin
        let test =  "arg1( " ++ fieldInfo( name ) ++
                    " ) = arg2( " ++ fieldName ++ " )"
        testSet := insert[ string ]( testSet, test )
        true
    end

    iterate[ NameAndType ]( overlap, addTest )

    "proc( arg1 : " ++ writeType( type1 ) ++ " ; arg2 : " ++ writeType( type2 ) ++
    " → bool ) ; " ++ andCompose( testSet )
end

let defineConcat = proc(  type1, type2, resultType : TypeRep ;
                          type1Fields, type2Fields : Set[ NameAndType ] → string )
begin
    let combination = union[ NameAndType ]( type1Fields, type2Fields )
    let compareNameAndValue = mkComparison[ NameAndValue ](
                        proc( a,b : NameAndValue → bool ) ; a = b )
    let structElementSet := mkEmptySet[ NameAndValue ](
                                                compareNameAndValue )

    let addField = proc( fieldInfo : NameAndType → bool )
    begin
        let fieldName = fieldInfo( name )
        let takeFrom = if memberOf[ NameAndType ]( fieldInfo, type1Fields )
                       then "arg1" else "arg2"
        let structElement =  NameAndValue( fieldName,
                                  takeFrom ++ "( " ++ fieldInfo( name ) ++ " )" )
```

```
        structElementSet := insert[ NameAndValue ](  structElementSet,
                                                     structElement )
        true
    end

    iterate[ NameAndType ]( combination, addField )

    "proc( arg1 : " ++ writeType( type1 ) ++ " ; arg2 : " ++ writeType( type2 ) ++
    " → " ++ writeType( resultType ) ++ " ) ; " ++
    mkStruct( structElementSet )
end

let defineOneJoin = proc( type1, type2, resultType : TypeRep → string )
begin
    "begin
        rec let onejoin = proc(  element : " ++ writeType( type1 ) ++ " ;
            rel : Set[ " ++ writeType( type2 ) ++ " ] →
            Set[ " ++ writeType( resultType ) ++ " ] )

        project rel as reln onto
        populated :
        begin
            let chooseReln = reln( choose )()
            let joinRest = onejoin( element, reln( rest )() )

            if " ++ defineMatch() ++ "( element, chooseReln ) then
            begin
                let newTuple = " ++ defineConcat( type1, type2, resultType ) ++ "
                ( element, chooseReln )

                project joinRest as restJoined onto
                populated :   restJoined( insert )( newTuple )
                emptyset :    restJoined( insert ) ( newTuple )
                default : {}
            end
            else joinRest
       end
       default : mkEmptySet[ " ++ writeType( resultType ) ++ " ]( " ++
                 defineCompareResult( resultType ) ++ " )

       oneJoin
    end"
end

let defineCompareResult = proc( resultType : TypeRep → string )
    "mkComparison[ " ++ writeType( resultType ) ++ " ]( proc( a, b : " ++
    writeType( resultType ) ++ " → bool ) ; a = b )"

let defineTypes =
    "type Comparison[ T ] is variant(
        ordered :      structure( equal, lessThan : proc( T, T → bool ) );
        unordered :    structure( equal : proc( T, T → bool ) ) )

    rec type Set[ T ] is variant(
        emptyset :    structure(  insert :       proc( T → Set[ T ] );
                                  union :        proc( Set[ T ] → Set[ T ] );
```

```
                                intersection : proc( Set[ T ] → Set[ T ] );
                                difference :   proc( Set[ T ] → Set[ T ] ) );

        populated :   structure(  insert :       proc( T → Set[ T ] );
                                  union :        proc( Set[ T ] → Set[ T ] );
                                  intersection : proc( Set[ T ] → Set[ T ] );
                                  difference :   proc( Set[ T ] → Set[ T ] );
                                  delete :       proc( T → Set[ T ] );
                                  choose :       proc( → T );
                                  rest :         proc( → Set[ T ] );
                                  includes :     proc( T → bool );
                                  scan :         proc( proc( T → bool ) );
                                  size :         proc( → int ) )"

let getPredefined =
    "use PS() with
        mkEmptySet :      proc[ T ]( Comparison[ T ] → Set[ T ] );
        mkComparison :    proc[ T ]( proc( T, T → bool ) → Comparison[ T ] ) in"

let defineJoin = proc( type1, type2, resultType : TypeRep → string )
begin
    defineTypes ++ getPredefined ++

    "begin
        rec let join = proc( rel1 : Set[ " ++ writeType( type1 ) ++" ] ;
                             rel2 : Set[ " ++ writeType( type2 ) ++ " ] →
                             Set[ " ++ writeType( resultType ) ++ " ] )

        project rel1 as first onto
        populated :
        begin
            let joinOne = " ++ defineOneJoin( type1, type2, resultType ) ++ "
            ( first( choose )(), rel2 )
            let joinOthers = join( first( rest )(), rel2 )

            project joinOne as firstJoined onto
            populated :  firstJoined( union )( joinOthers )
            default :    joinOthers
        end
        default : mkEmptySet[ " ++ writeType( resultType ) ++ " ]( " ++
                  defineCompareResult( resultType ) ++ " )

        join
    end"
end

let joinResultType = proc(  type1Fields, type2Fields : Set[ NameAndType ] →
                            TypeRep )
begin
    let resultFields = union[ NameAndType ]( type1Fields, type2Fields )
    mkStructureType( resultFields )
end

let join = proc( type1, type2 : TypeRep → string )
begin
    let type1Fields = getStructureFields( type1 )
```

```
    let type2Fields = getStructureFields( type2 )
    let resultType = joinResultType( type1Fields, type2Fields )

    defineJoin( type1, type2, resultType )
end

end
```

**Figure A.6: Implementation of generic natural join in standard Napier88**

## Example in TRPL

Figure A.7 shows an implementation of generic natural join in TRPL. The principal differences from the previous solutions are as follows:

- The solution involves the definition of a context sensitive macro that is evaluated at compilation-time.

- The source code being manipulated is in parsed rather than in string or hyper-program form.

- The solution assumes the existence of a polymorphic join function, *join*, that takes the *match* and *concat* functions as arguments.

- The input types are set types rather then tuple types.

The macro definition begins with the extraction of the types of the input values *r* and *s* from the current compilation environment *e* using a built-in function *type_of*. This uses an environment variable defined in the header as the current compiler environment. These types are expanded using another built-in function *expandtype*. This expands all type variables contained in a type representation into their structural forms. The next two equations extract the list of component names by using pattern recognition on the representation of the input types. A representation of a legal type for this macro call is of the form *parametric_rep ("set", cons (struct_rep ("constrName", componentList), nil))*. The case statement either matches this for each type representation or returns an error. When a match is made the variables in the pattern are bound to their matched components and the case body is evaluated. Question marks stand for parts of the value to be matched by anything and ignored. The case bodies here are just the extracted list of component name and type pairs. New names are then generated for the output type and a constructor function for its tuples.

The macro then computes the unique and overlapping components of the two input relations and generates the output type definition. This code uses pattern matching lambda expressions, the expressions starting with *[x & ?, y & ?]*. In these functions the input arguments are first matched with the patterns in the brackets. The patterns here are pairs since *&* is the infix pair construction operator. As before, successful pattern matching causes the variables in the patterns to be bound to the matching components of the values. In this case *x* and *y* are bound to the names of the components.

The unique and overlapping components are computed by *set_difference* and *set_intersection* using the lambda functions over the component lists. Pattern matching lambda expressions capture the criterion that components are equal when their names represented as strings are equal. If the names are equal but the types are not, the *match* function will produce a type error when it is passed to the compiler. The *units* section contains only the output type definition using another built-in function *define_type*. The first parameter gives the computed type name and the second supplies the representation of the type expression including the tuple constructor function name, bound to *constr*. Note the use of the constructor functions, *parametric_rep* and *struct_rep*, to construct the typed representation of the new type.

Next is the code for generating the representations of the *match* and *concat* function bodies. The *match* body is an expression of the form *rt.a=st.a && rt.b=st.b && ... && true*, where && denotes logical *and*. It is to be used in the inline expansion as the body of a lambda function having *rt* and *st* as variables standing for the tuples of the input relations, *r* and *s*.

This portion of the definition uses a macro, *EREP*, to facilitate the generation of expression representations. *EREP* takes as its first argument an expression which gives a pattern for the representation it generates. Optional arguments may follow which give values to be substituted in the representation of the first argument. This allows computed representations to be inserted into constant expressions. A simple example of this is *EREP (f (x), x := s2id ("y"))*, where *s2id* is a function that converts a string value into the representation of an identifier. This evaluates to the representation of *f (y)*. The *match* body is produced by mapping the *eqterm* function over the overlapping component name and type pairs. The *eqterm* function takes a component pair, extracts the component name and constructs an equality expression that compares the named projection of *rt* and *st* tuples. The list of these terms is used to construct a boolean expression ‘and’ing all the equality terms with *true*. This uses a reduction function over the mapped list. The reduction uses a binary lambda function and *EREP* to build the representation of the *and* expression. Starting the reduction with *true* defines the base case of no common component names to be the cartesian product.

The *concat* body is generated by using *EREP* and *listmap*, together with a feature that allows variable length constructs in the pattern used in *EREP*. The ellipsis before *args* marks it as a parameter that accepts a list for its substitution. The list of representations of component names is produced by the *append3* and *listmap* functions, the former a function that appends three lists. An example of a *concat* body is *make_a_b_c_d (rt.a, rt.b, rt.c, st.d)*. The inline expansion uses *EREP* and the computed bodies of *match* and *concat* to generate the representation of a call to *join*.

```
macro NATJOIN (r, s) ; env e;
let   ertype :=  type_of (r, e),              @ get the types of r and s
      estype :=  type_of (s, e),
      rtype :=   expandtype (ertype, e),      @ expand set types to
      stype :=   expandtype (estype, e),      @ remove any type variables

@ build component lists for r and s
      rcomps :=  case rtype
            {parametric_rep ("set", cons (struct_rep (?, rcompslist), nil))
                        → rcompslist,     @ ? indicates tuple constructor name unimportant
            others  → warning ("first argument not a set of tuple", nil)},
      scomps := case stype
            {parametric_rep ("set", cons (struct_rep (?, scompslist), nil))
                        → scompslist,
            others  → warning ("second argument not a set of tuple", nil)},

@ generate symbols for new type
      tn := genstring ("type$"),

@ and constructor function for output tuples
      constr := genstring ("constr$"),

      runique := set_difference (rcomps, scomps, [x & ?, y & ?] → string_eq (x, y)),
      sunique := set_difference (scomps, rcomps, [x & ?, y & ?] → string_eq (x, y)),
      overlap := set_intersection (scomps, rcomps, [x & ?, y & ?] → string_eq (x, y))
```

```
in
units
      LIST (   @ the new type definition
               define_type (   tn,
                               parametric_rep ("set", LIST (struct_rep (constr,
                                      append3 (overlap, runique, sunique))))))

@ build bodies of match and concat
@ first a representation for the body of the match lambda
@ expression which looks like rt.a=st.a && rt.b=st.b && ... && true

let   eqterm := [x & ?] → EREP ((rt.field) = (st.field), field := s2id (x)),
      match := listreduce (  listmap( overlap, eqterm),
                             [term, exp] → EREP (t && e, t := term, e := exp),
                             EREP (true)),

@ build a representation for the body of the concat
@ lambda expression which looks like
@ construct (rt.common1, … rt.unique1, … st.unique1, …)

concat := EREP (  con ( …args),
                  con := s2id (constr),
                  args := append3 (
                             listmap (overlap, [x & ?] → EREP (rt.f, f := s2id (x))),
                             listmap (runique, [x & ?] → EREP (rt.f, f := s2id (x))),
                             listmap (sunique, [x & ?] → EREP (st.f, f := s2id (x)))))

@ the inline expansion is a call to join with lambda functions for match and concat
in
      EREP (  join (r, s, [rt, st] → mtch, [rt, st] → cnct),
              mtch := match,
              cnct := concat)
```

**Figure A.7: Implementation of generic natural join in TRPL**

# Appendix B. Generator Interfaces

## Pre-defined Types

```
type CodeTree is …   ! Parsed form of code representation

type Code is string

type CodeRegion is structure( start,finish : int )

type Optional[ T ] is variant( present : T ; absent : null )

type Substitution[ T ] is structure( value : T ; region : CodeRegion )

rec type  TypeRep is structure(  label, misc, random : int ;
                                 name : string ; others : Var )
&         Var is variant( none : null ; one, unique : TypeRep ; many : *TypeRep )

type EnvLocation is structure( pointer : null ; typeRep : TypeRep )

type StructLocation is structure( structValue : any ; field : string )

type VectorLocation is structure( vectorValue : any ; index : int )

type StackPos is structure( Frame,MSoffset,PSoffset : int )

type FrameLocation is structure(  frame : null ; stackPos : StackPos ;
                                  typeRep : TypeRep ; envLoc : bool )

type TypeContainer is structure( typeRep : TypeRep )

type Binding is variant(  value :             any;
                          envLocation :       EnvLocation;
                          structLocation :    StructLocation;
                          abstypeLocation :   StructLocation;
                          vectorLocation :    VectorLocation;
                          frameLocation :     FrameLocation;
                          aType :             TypeContainer )

rec type Generator is structure(  prelude : proc( env → env ) ;
                                  resultDefn : GeneratorResult )

& GeneratorResult is variant(  literal : GeneratorSource ;
                               expression : proc( env → GeneratorSource ) )

& GeneratorSource is structure(  code : HyperSource ;
                                 generators : Optional[ *Substitution[ Generator ] ] )

& HyperSource is structure(  code : Code ;
                             bindings : Optional[ *Substitution[ Binding ] ] )

type NameAndType is structure( name : string ; typeRep : TypeRep )

type NameAndValue is structure( name : string ; value : HyperSource )
```

```
type Comparison[ T ] is variant(
        ordered :      structure( equal, lessThan : proc( T, T → bool ) );
        unordered :    structure( equal : proc( T, T → bool ) ) )

rec type Set[ T ] is variant(
        emptyset :     structure(  insert :         proc( T → Set[ T ] );
                                   union :          proc( Set[ T ] → Set[ T ] );
                                   intersection :   proc( Set[ T ] → Set[ T ] );
                                   difference :     proc( Set[ T ] → Set[ T ] ) );

        populated :    structure(  insert :         proc( T → Set[ T ] );
                                   union :          proc( Set[ T ] → Set[ T ] );
                                   intersection :   proc( Set[ T ] → Set[ T ] );
                                   difference :     proc( Set[ T ] → Set[ T ] );
                                   delete :         proc( T → Set[ T ] );
                                   choose :         proc( → T );
                                   rest :           proc( → Set[ T ] );
                                   includes :       proc( T → bool );
                                   scan :           proc( proc( T → bool ) );
                                   size :           proc( → int ) )
```

## Pre-defined Procedures

intersection : **proc**[ T ]( Set[ T ], Set[ T ] → Set[ T ] )
Returns the intersection of two sets.

union : **proc**[ T ]( Set[ T ], Set[ T ] → Set[ T ] )
Returns the union of two sets.

memberOf : **proc**[ T ]( T, Set[ T ] → **bool** )
Determines whether the given element belongs to the given set.

insert : **proc**[ T ]( Set[ T ], T → Set[ T ] )
Returns the set obtained by inserting the element in the given set.

getStructureFields : **proc**( TypeRep → Set[ NameAndType ] )
Takes a structure type representation and returns a set of (name,type) pairs, empty if the representation isn't of a structure type.

map : **proc**[ S,T ]( Set[ S ], **proc**( S → T ), Comparison[ T ] → Set[ T ] )
Takes a set, a procedure operating on the element type and an equality function, and returns the set obtained by applying the procedure to all elements of the set.

orCompose : **proc**( Set[ HyperSource ] → HyperSource )
Returns the source consisting of the boolean 'or'ing of the elements of the set.

andCompose : **proc**( Set[ HyperSource ] → HyperSource )
Returns the source consisting of the boolean 'and'ing of the elements of the set.

mkStruct : **proc**( Set[ NameAndValue ] → HyperSource )
Returns the source for a structure creation with the given field names and elements.

evalWithString : **proc**( Generator, **string** → HyperSource )
Evaluates the given generator, passing the given string to it as a parameter with the name *stringVal*.

mkStructureType : **proc**( Set[ NameAndType ] → TypeRep )
Creates a structure type representation from the given field information. A fail value is returned if field names are duplicated.

mkEmptySet : **proc**[ T ]( Comparison[ T ] → Set[ T ] )
Returns an empty set with elements of the given type.

mkLink : **proc**( **any** → HyperSource )
Returns the source for a link to the given value.

mkEnvLocLink : **proc**( **env**, **string** → HyperSource )
Returns the source for a link to the location with the given name in the given environment.

mkStructLocLink : **proc**( **any**, **string** → HyperSource )
Returns the source for a link to the location with the given field name in the given structure or abstype.

mkVecLocLink : **proc**( **any**, **int** → HyperSource )
Returns the source for a link to the location with the given index in the given vector.

mkTypeLink : **proc**( TypeRep → HyperSource )
Returns the source for a link to the type represented by the given type representation.

mkHyperSource : **proc**( **string** → HyperSource )
Converts the given string to source code.

mkGeneratorSource : **proc**( HyperSource → GeneratorSource )
Converts the given source code to generator source code with no generator place-holders.

concatHyperSource : **proc**( HyperSource,HyperSource → HyperSource );
Concatenates the two fragments of hyper-program source code.

concatGeneratorSource : **proc**( GeneratorSource,GeneratorSource → GeneratorSource )
Concatenates the two fragments of generator source code.

extractHyperSource : **proc**( HyperSource,**int**,**int** → HyperSource )
Extracts the hyper-program source code between and including the two character offsets.

extractGeneratorSource : **proc**( GeneratorSource,**int**,**int** → GeneratorSource )
Extracts the generator source code between and including the two character offsets.

mkComparison : **proc**[ T ]( **proc**( T, T → **bool** ) → Comparison[ T ] )
Converts the given equality testing procedure to an instance of *Comparison* for that type.

compareHyperSource : Comparison[ HyperSource ]
Compares two instances of *HyperSource* for equality.

iterate : **proc**[ T ]( Set[ T ], **proc**( T → **bool** ) )
Iterates through the given set calling the given procedure with each element as argument until it returns false or the set is exhausted.

expandGenerator : **proc**( Generator, **env** → HyperSource )
Expands the given generator with the parameters in the given environment to produce a hyper-program source representation.

compileAndProcess : **proc**( HyperSource,**proc**( **any** ) )
Compiles the given hyper-program source code and passes the result to the given procedure.

equalType : **proc**( TypeRep,TypeRep → **bool** );
Tests the two type representations for equivalence.

typeOf : **proc**( **any** → TypeRep )
Returns a representation of the type of the value injected into the given *any*.

# Appendix C. WIN Interfaces

## User Types

```
type Pos is              structure( x,y : int )
type Size is             structure( x,y : int )
type Limit is            structure( pos : Pos ; size : Size )
type Rect is             structure( origin,corner : Pos )
type Level is            structure( fromFront : bool ; position : int )
type InputOption is      variant( all,none,normal : null )

type Optional[ T ] is variant( present : T ; absent : null )

rec type List[ T ] is variant( cons : structure( hd : T ; tl : List[ T ] ) ; tip : null )

rec type DoubleList[ T ] is variant(  cons : structure(  hd : T ;
                                                         before,after : DoubleList[ T ] ) ;
                                      tip : null )

type Pair[ S,T ] is structure( fst : S ; snd : T )

type Mouse is structure( x,y : int ; buttons : *bool )
type Event is variant(  chars :            string;
                        mouse :            Mouse;
                        select,deselect :  null )
type EventType is variant( up,down,enter,leave,click,doubleClick : null )
type MouseEvent is structure( button : int ; event : EventType )

type Application is proc( Event )

type EventTest is proc( Event → bool )

type Notification is structure(  examineEvent : EventTest ;
                                 processEvent : Application )

type Notifier is structure(  distributeEvent :   Application;
                             addNotification :   proc( Notification,Level → proc() ) )

type ResizeControl is structure( before : proc( Rect → Rect ) ; after : proc( Rect ) )

rec type DisplayInfo is structure(  window : Window ; pos : Pos ;
                                    level : Level ; style : BorderStyle )

& Window is structure(
        windowRaster :        proc( Limit,Limit,Window,int,bool );
        imageRaster :         proc( Limit,image,int,bool );
        drawLine :            proc( Pos,Pos,int );
        resize :              proc( Rect );
        takeInput :           proc( InputOption );
        getSize :             proc( → Size );
        setApplication :      proc( Application );
        getApplication :      proc( → Application );
        setTitle :            proc( string );
        getTitle :            proc( → string );
        setResizeControl :    proc( ResizeControl );
        getResizeControl :    proc( → ResizeControl );
```

```
        setMinSize :            proc( Size );
        setMaxSize :            proc( Size );
        getWindowManager :      proc( → WindowManager );
        setVirtualWindow :      proc( Window ) )

& WindowManager is structure(
        display :               proc( DisplayInfo,bool );
        undisplay :             proc( Window );
        makeCurrent :           proc( Window );
        setPos :                proc( Window,Pos );
        getPos :                proc( Window → Pos );
        setLevel :              proc( Window,Level );
        getLevel :              proc( Window,bool → Level );
        setCursor :             proc( Window,image );
        getCursor :             proc( Window → image );
        getWindows :            proc( → *Window );
        getBorderExtent :       proc( Window → Rect );
        getNotifier :           proc( → Notifier );
        getDisplayWindow :      proc( → Window );
        getIconManager :        proc( → IconManager );
        setBackgroundApp :      proc( Application );
        getBackgroundApp :      proc( → Application ) )

& IconManager is structure(  close :              proc( Window );
                             open :               proc( Window );
                             getIconState :       proc( Window → DisplayInfo );
                             getWindowState :     proc( Window → DisplayInfo ) )

& BorderStyle is proc( Window → List[ Area ] )
& Area is structure(  currentImage,nonCurrentImage : image ;
                      pos : Pos ; distributeEvent : Application )

type AreaList is List[ Area ]

type Font is structure(  characters : *image ;
                         fontHeight,descender : int ; info : string )
type FontPack is structure(  font : Font ;
                             stringToTile,charToTile : proc( string → image ) )

type Result[ Data ] is variant( ok : Data ; fail : null )

type Table[ Key,Data ] is structure(
        enter :     proc( Key,Data );
        lookup :    proc( Key → Result[ Data ] );
        remove :    proc( Key );
        scan :      proc( proc( Key,Data → bool ) );
        firstKey :  proc( → Result[ Key ] ) )

type Comparison[ Key ] is variant(
        ordered :    structure( equal,lessThan : proc( Key,Key → bool ) );
        unordered :  structure( equal : proc( Key,Key → bool ) ) )

type Index is variant(  characters :  int;
                        lines :       structure( line,char : int ) )
```

```
type ButtonInfo[ TextPointer ] is structure( name :    string;
                                            start,
                                            finish :  TextPointer;
                                            action :  proc( int );
                                            extra :   any )

type Text is structure( characters : string ; buttons : *int )

rec type SimpleEditor is abstype[ TextPointer ](
        copyText :        proc( SimpleEditor );
        cutText :         proc( SimpleEditor );
        pasteText :       proc( SimpleEditor );
        insertText :      proc( string );
        readFromFile :    proc( file );
        writeToFile :     proc( file );
        select :          proc( TextPointer,TextPointer );
        firstSelection :  proc( → TextPointer );
        lastSelection :   proc( → TextPointer );
        firstLine :       proc( → TextPointer );
        lastLine :        proc( → TextPointer );
        frontOfLine :     proc( TextPointer → TextPointer );
        endOfLine :       proc( TextPointer → TextPointer );
        nextLine :        proc( TextPointer → TextPointer );
        previousLine :    proc( TextPointer → TextPointer );
        peek :            proc( → Text );
        read :            proc( → Text );
        readLine :        proc( → Text );
        selectedText :    proc( → Text );
        before :          proc( TextPointer,TextPointer → bool );
        endOfText :       proc( → bool );
        lineCount :       proc( → int );
        new :             proc();
        offset :          proc( TextPointer,bool → Index );
        search :          proc( string,bool → bool );
        seek :            proc( Index → TextPointer );
        insertButton :    proc( string,proc( int ) → int );
        setButtonInfo :   proc( int,string,proc( int ),any );
        lookupButton :    proc( int → Result[ ButtonInfo[ TextPointer ] ] );
        scanButtons :     proc( proc( int,ButtonInfo[ TextPointer ] ) ) )

type HyperEditor is  abstype[ TextPointer ](
        copyText :         proc( SimpleEditor );
        cutText :          proc( SimpleEditor );
        pasteText :        proc( SimpleEditor );
        insertText :       proc( string,bool );
        readFromFile :     proc( file );
        writeToFile :      proc( file );
        select :           proc( TextPointer,TextPointer );
        firstSelection :   proc( → TextPointer );
        lastSelection :    proc( → TextPointer );
        firstLine :        proc( → TextPointer );
        lastLine :         proc( → TextPointer );
        topLine :          proc( → TextPointer );
        bottomLine :       proc( → TextPointer );
        frontOfLine :      proc( TextPointer → TextPointer );
        endOfLine :        proc( TextPointer → TextPointer );
        nextLine :         proc( TextPointer → TextPointer );
        previousLine :     proc( TextPointer → TextPointer );
```

```
        peek :            proc( → Text );
        read :            proc( → Text );
        readLine :        proc( → Text );
        selectedText :    proc( → Text );
        before :          proc( TextPointer,TextPointer → bool );
        endOfText :       proc( → bool );
        getFont :         proc( → FontPack );
        getHighlight :    proc( → bool );
        getWindow :       proc( → Window );
        interactiveEdit : proc(  SimpleEditor,EventTest,EventTest,EventTest →
                                 Application );
        invert :          proc( TextPointer,TextPointer );
        lineCount :       proc( → int );
        new :             proc();
        offset :          proc( TextPointer,bool → Index );
        position :        proc( Pos → TextPointer );
        redisplay :       proc( TextPointer );
        scroll :          proc( int,bool );
        search :          proc( string,bool → bool );
        seek :            proc( Index → TextPointer );
        setFont :         proc( FontPack );
        setHighlight :    proc( bool );
        setWindow :       proc( Window );
        unbindWindow :    proc();
        insertButton :    proc( string,proc( int ) → int );
        setButtonInfo :   proc( int,string,proc( int ),any );
        lookupButton :    proc( int → Result[ ButtonInfo[ TextPointer ] ] );
        scanButtons :     proc( proc( int,ButtonInfo[ TextPointer ] ) ) )

type Appearance is variant( graphical : image ; textual : string )

type HyperEditorPack is structure( window :       Window;
                                   editor :       HyperEditor;
                                   append :       proc( string );
                                   getText,
                                   getFileName :  proc( → string ) )

type HyperProgramPack is structure(  window :  Window;
                                     editor :  HyperEditor;
                                     getTitle : proc( → string );
                                     insert :   proc( HyperSource );
                                     getText :  proc( → HyperSource ) )

type MenuPack is structure(
        window :        Window;
        setTop :        proc( int );
        getTop :        proc( → int );
        setNoVisible :  proc( int );
        getNoVisible :  proc( → int );
        setHighlight :  proc( int,bool );
        getHighlight :  proc( int → bool );
        getNoEntries :  proc( → int );
        locate :        proc( Appearance → Pair[ int,bool ] );
        doAction :      proc( int,MouseEvent );
        addEntry :      proc( Appearance,proc( int,MouseEvent ),int );
        removeEntry :   proc( int ) )

type ButtonPack is structure( window : Window ; flash : proc() )
```

```
type SliderPack is structure( window :      Window;
                              set :         proc( real );
                              setBounds :   proc( real,real,real ) )

type CheckBoxPack is structure( window : Window ; set : proc( bool ) )

type ChoicePack is structure( window : Window ; set : proc( int,int,bool ) )

type DialoguePack is structure( window : Window ; set : proc( string ) )

type BrowserType is variant( graphical : WindowManager ; textual : proc( string ) )

type WindowState is structure( window : Window ; pos : Pos ;
                               level : Level ; open,displayed : bool )
```

## Implementation Types

```
type Association is structure( windowDisplayInfo,iconDisplayInfo : DisplayInfo )
type AssociationList is List[ Association ]

rec type WindowInfo is structure( rect,borderRect :        Rect;
                                  notification :           Notification;
                                  tree :                   VisTree;
                                  areaList :               AreaList;
                                  removeNotification :     proc();
                                  background :             bool;
                                  cursor :                 image;
                                  style :                  BorderStyle;
                                  inputOption :            InputOption;
                                  window :                 Window )

& VisTree is  variant( node : structure( rect : Rect ; parent : Parent ;
                                         order : int ; left,right : VisTree );
                       leaf : structure( rect : Rect ; parent : Parent ;
                                         content : Content ; covering : Covering ) )
& Parent is  variant( tree : VisTree ; root : WindowInfo ; none : null )
& Content is  Optional[ image ]
& Covering is variant( present : WindowInfo ; absent,offScreen : null )

type WinList is DoubleList[ WindowInfo ]

rec type TextLine is variant( cons : structure( hd : string ; index : int ;
                                                before,after : TextLine );
                              tip :  null )
type TextPointer is structure( line : TextLine ; offset : int )
type TextRecord is  structure( firstSelection,lastSelection : TextPointer ;
                               firstLine : TextLine )

type WindowLine is  structure( lineStart : TextPointer ; lineBase : int )
type WindowRecord is structure(
        window :                                  Window ;
        textRecord :                              TextRecord ;
        lineArray :                               *WindowLine ;
        font :                                    FontPack;
        firstSelectionIndex,lastSelectionIndex :  int )
```

```
type InternalButtonInfo is structure( id : int ; name : string ;
                                      start,finish : TextPointer ;
                                      action : proc( int ) ; extra : any )

type Destination is variant( w : Window ; i : image )

rec type Keeper is structure( binding :      Binding;
                              window :       Window ;
                              refersTo :     List[ ReferenceTo ];
                              referedToBy :  List[ ReferenceFrom ] )
& ReferenceTo is structure( keeper : Keeper ; references : int )
& ReferenceFrom is structure( keeper :                      Keeper;
                              arrowImage,obscuredImage :    image;
                              menuOffset,fieldNo,objectX,
                              objectY,parentX,parentY :     int;
                              lineSet :                     bool )

type Traverser is proc( Binding,Binding,env,int,int )

type BrowserImplementation is structure( browse : proc( Binding ) ;
                                         localEnv : env )

type BindingInfo is structure( binding :     Binding;
                               name :        string;
                               menuOffset :  int;
                               fieldNo :     int )

type Selection is Optional[ Binding ]

type ProcInfo is structure( freeIds : List[ Substitution[ Binding ] ] ;
                            lexLevel,startOffset : int )
```

# References

[AAC+91] Albano, A., Atkinson, M.P., Connor, R.C.H., Delobel, C., Ghelli, G., Lécluse, C., Mancini, L., Matthes, F., Morrison, R., Orsini, R., Philbrow, P., Rabitti, F., Richard, P., Schmidt, J. & Watt, D. **FIDE Course on Database Programming Languages and Persistent Systems** (1991).

[AB87] Atkinson, M.P. & Buneman, O.P. "Types and Persistence in Database Programming Languages". ACM Computing Surveys 19, 2 (1987) pp 105-190.

[ABC+83] Atkinson, M.P., Bailey, P.J., Chisholm, K.J., Cockshott, W.P. & Morrison, R. "An Approach to Persistent Programming". Computer Journal 26, 4 (1983) pp 360-365.

[ABC+84] Atkinson, M.P., Bailey, P.J., Cockshott, W.P., Chisholm, K.J. & Morrison, R. "Progress with Persistent Programming". Universities of Glasgow and St Andrews Technical Report PPRR-8-84 (1984).

[ACC82] Atkinson, M.P., Chisholm, K.J. & Cockshott, W.P. "PS-algol: An Algol with a Persistent Heap". ACM SIGPLAN Notices 17, 7 (1982) pp 24-31.

[ACO85] Albano, A., Cardelli, L. & Orsini, R. "Galileo: a Strongly Typed, Interactive Conceptual Language". ACM Transactions on Database Systems 10, 2 (1985) pp 230-260.

[AGO88] Albano, A., Ghelli, G. & Orsini, R. "The Implementation of Galileo's Values Persistence". In **Data Types and Persistence**, Atkinson, M.P., Buneman, O.P. & Morrison, R. (ed), Springer-Verlag (1988) pp 253-263.

[AH87] Andrews, T. & Harris, C. "Combining Language and Database Advances in an Object-Oriented Development Environment". In Proc. OOPSLA'87, Orlando, Florida (1987).

[AHM88] Altmann, R.A., Hawke, A.N. & Marlin, C.D. "An Integrated Programming Environment Based on Multiple Concurrent Views". Australian Computer Journal 20, 2 (1988) pp 65-72.

[Ala90] Alagic, S. "Persistent Metaobjects". In **Implementing Persistent Object Bases**, Dearle, A., Shaw, G.M. & Zdonik, S.B. (ed), Morgan Kaufmann (1990) pp 27-38.

[ALP+91] Atkinson, M.P., Lécluse, C., Philbrow, P. & Richard, P. "Design Issues in a Map Language". In **Bulk Types & Persistent Data**, Kanellakis, P. & Schmidt, J.W. (ed), Morgan Kaufmann (1991) pp 20-32.

[AM84] Atkinson, M.P. & Morrison, R. "Persistent First Class Procedures are Enough". In **Lecture Notes in Computer Science 181**, Joseph, M. & Shyamasundar, R. (ed), Springer-Verlag (1984) pp 223-240.

[AM85] Atkinson, M.P. & Morrison, R. "Procedures as Persistent Data Objects". ACM Transactions on Programming Languages and Systems 7, 4 (1985) pp 539-559.

[AM86] Atkinson, M.P. & Morrison, R. “Integrated Persistent Programming Systems”. In Proc. 19th International Conference on Systems Sciences, Hawaii (1986) pp 842-854.

[AM88] Atkinson, M.P. & Morrison, R. “Types, Bindings and Parameters in a Persistent Environment”. In **Data Types and Persistence**, Atkinson, M.P., Buneman, O.P. & Morrison, R. (ed), Springer-Verlag (1988) pp 3-20.

[AMP86] Atkinson, M.P., Morrison, R. & Pratten, G.D. “A Persistent Information Space Architecture”. In Proc. 9th Australian Computing Science Conference, Australia (1986).

[App86] Apple Computer. **Inside Macintosh**. Addison-Wesley, Reading, Massachusetts (1986).

[Atk91] Atkinson, M.P. “POP-2 Example”. Personal communication (1991).

[BBB+88] Bancilhon, F., Barbedette, G., Benzaken, V., Delobel, C., Gamerman, S., Lécluse, C., Pfeffer, P., Richard, P. & Valez, F. “The Design and Implementation of $O_2$, an Object-Oriented Database System”. In **Lecture Notes in Computer Science 334**, Dittrich, K.R. (ed), Springer-Verlag (1988) pp 1-22.

[BCP71] Burstall, R.M., Collins, J.S. & Popplestone, R.J. **Programming in POP-2**. Edinburgh University Press, Edinburgh, Scotland (1971).

[BM89] Borenstein, P. & Mattson, J. **Think C™ User Manual**. Symantec Corporation, Cupertino, California (1989).

[BMM+92] Brown, A.L., Mainetto, G., Matthes, F., Müller, R. & McNally, D.J. “An Open System Architecture for a Persistent Object Store”. In Proc. 25th International Conference on Systems Sciences, Hawaii (1992) pp 766-776.

[BOP+89] Bretl, B., Otis, A., Penney, J., Schuchardt, B., Stein, J., Williams, E.H., Williams, M. & Maier, D. “The GemStone Data Management System”. In **Object-Oriented Concepts, Applications, and Databases**, Kim, W. & Lochovsky, F. (ed), Morgan-Kaufman (1989).

[Bor89] Borland International. **Turbo Pascal**. Borland International, Scotts Valley, California (1989).

[Bow86] Bowen, K. “Meta-level Techniques in Logic Programming”. In Proc. International Conference on Artificial Intelligence and its Applications, Singapore (1986).

[BPR91] Bruynooghe, R.F., Parker, J.M. & Rowles, J.S. “PSS: A System for Process Enactment”. In Proc. 1st International Conference on the Software Process: Manufacturing Complex Systems (1991).

[Bro89] Brown, A.L. “Persistent Object Stores”. Ph.D. Thesis, University of St Andrews (1989).

[Bru91] Bruynooghe, R.F. “PML Reference Manual”. ICL Technical Report ICL/4R2F/00070 (1991).

[CAD+87] Cooper, R.L., Atkinson, M.P., Dearle, A. & Abderrahmane, D. "Constructing Database Systems in a Persistent Environment". In Proc. 13th International Conference on Very Large Data Bases (1987) pp 117-125.

[Car85] Cardelli, L. "Amber". AT&T Bell Labs, Murray Hill, Technical Report AT7T (1985).

[Car89] Cardelli, L. "Typeful Programming". DEC Technical Report 45 (1989).

[CBC+90] Connor, R.C.H., Brown, A.L., Carrick, R., Dearle, A. & Morrison, R. "The Persistent Abstract Machine". In **Persistent Object Systems**, Rosenberg, J. & Koch, D.M. (ed), Springer-Verlag (1990) pp 353-366.

[CDK90] Cutts, Q.I., Dearle, A. & Kirby, G.N.C. "WIN Programmers' Manual". University of St Andrews Technical Report CS/90/17 (1990).

[CDM+90] Connor, R.C.H., Dearle, A., Morrison, R. & Brown, A.L. "Existentially Quantified Types as a Database Viewing Mechanism". In **Lecture Notes in Computer Science 416**, Bancilhon, F., Thanos, C. & Tsichritzis, D. (ed), Springer-Verlag (1990) pp 301-315.

[CK87] Cutts, Q.I. & Kirby, G.N.C. "An Event-Driven Software Architecture". Universities of Glasgow and St Andrews Technical Report PPRR-48-87 (1987).

[CM84] Clocksin, W.F. & Mellish, C.S. **Programming in PROLOG (2nd Edition)**. Springer-Verlag, New York (1984).

[Con90] Connor, R.C.H. "Types and Polymorphism in Persistent Programming Systems". Ph.D. Thesis, University of St Andrews (1990).

[Con91] Connor, R.C.H. "A language which can manipulate its own syntactic entities". Personal communication (1991).

[Con92] Connor, R.C.H. "Panel on Persistent Type Systems". In Proc. 5th International Workshop on Persistent Object Systems, San Miniato, Italy (1992).

[Coo90a] Cooper, R.L. "On The Utilisation of Persistent Programming Environments". Ph.D. Thesis, University of Glasgow (1990).

[Coo90b] Cooper, R.L. "Configurable Data Modelling Systems". In Proc. 9th International Conference on the Entity Relationship Approach, Lausanne, Switzerland (1990) pp 35-52.

[CQ92] Cooper, R.L. & Qin, Z. "A Graphical Data Modelling Program With Constraint Specification and Management". In Proc. 10th British National Conference on Databases, Aberdeen (1992).

[Cut92] Cutts, Q.I. "Delivering the Benefits of Persistence to System Construction and Execution". Ph.D. Thesis, University of St Andrews (1992).

[DB88] Dearle, A. & Brown, A.L. "Safe Browsing in a Strongly Typed Persistent Environment". Computer Journal 31, 6 (1988) pp 540-544.

[DCC92] Dearle, A., Cutts, Q.I. & Connor, R.C.H. "An Application Architecture Using Type-Safe Incremental Linking". University of St Andrews Technical Report CS/92/13 (1992).

[DCK90] Dearle, A., Cutts, Q.I. & Kirby, G.N.C. "Browsing, Grazing and Nibbling Persistent Data Structures". In **Persistent Object Systems**, Rosenberg, J. & Koch, D.M. (ed), Springer-Verlag (1990) pp 56-69.

[Dea87] Dearle, A. "Constructing Compilers in a Persistent Environment". In Proc. 2nd International Workshop on Persistent Object Systems, Appin, Scotland (1987).

[Dea88] Dearle, A. "On the Construction of Persistent Programming Environments". Ph.D. Thesis, University of St Andrews (1988).

[Dea89] Dearle, A. "Environments: A flexible binding mechanism to support system evolution". In Proc. 22nd International Conference on Systems Sciences, Hawaii (1989) pp 46-55.

[DM90] Davie, A.J.T. & McNally, D.J. "Statically Typed Applicative Persistent Language Environment (STAPLE) Reference Manual". University of St Andrews Technical Report CS/90/14 (1990).

[DMD92] Dearle, A., Marlin, C.D. & Dart, P. "A Hyperlinked Persistent Software Development Environment". In Proc. Hyper-Oz '92: A Workshop on Hypertext Activities in Australia, Adelaide, Australia (1992).

[DOD83] "Reference Manual for the Ada Programming Language". U.S. Department of Defense Technical Report ANSI/MIL-STD-1815A (1983).

[DPS+89] Dixon, G.N., Parrington, G.D., Shrivastava, S.K. & Wheater, S.M. "The Treatment of Persistent Objects in Arjuna". Computer Journal 32, 4 (1989) pp 323-332.

[Eve85] Evered, M. "LEIBNIZ - A Language to Support Software Engineering". Dr. Ing. Thesis, University of Darmstadt (1985).

[Far91] Farkas, A.M. "ABERDEEN: A Browser allowing intERactive DEclarations and Expressions in Napier88". University of Adelaide Honours Project (1991).

[FBC+90] Fishman, D.H., Beech, D., Cate, H.P., Chow, E.C., Connors, T., Davis, J.W., Derrett, N., Hoch, C.G., Kent, W., Lyngbaek, P., Mahbod, B., Neimat, M.A., Ryan, T.A. & Shan, M.C. "Iris: An Object-Oriented Database Management System". In **Readings in Object-Oriented Database Systems**, Zdonik, S.B. & Maier, D. (ed), Morgan Kaufman (1990) pp 216-226.

[FDK+92] Farkas, A.M., Dearle, A., Kirby, G.N.C., Cutts, Q.I., Morrison, R. & Connor, R.C.H. "Persistent Program Construction through Browsing and User Gesture with some Typing". In Proc. 5th International Workshop on Persistent Object Systems, San Miniato, Italy (1992) pp 375-394.

[FID90] "The FIDE Project". Esprit II Basic Research Action 3070 (1990).

[FS91] Fegaras, L. & Stemple, D. "Using Type Transformation in Database System Implementation". In Proc. 3rd International Conference on Database Programming Languages, Nafplion, Greece (1991) pp 289-305.

[FSS92] Fegaras, L., Sheard, T. & Stemple, D. "Uniform Traversal Combinators: Definition, Use and Properties". In Proc. 11th International Conference on Automated Deduction (CADE-11), Saratoga Springs, New York (1992).

[FW84] Friedman, D. & Wand, M. “Reification: Reflection Without Meta-physics”. In Proc. ACM Symposium on Lisp and Functional Programming (1984) pp 348-355.

[GMD85] Gray, P.M.D., Moffat, D.S. & Du Boulay, J.B.H. “Persistent Prolog: A Searching Storage Manager for Prolog”. In Proc. 1st International Workshop on Persistent Object Systems, Appin, Scotland (1985) pp 353-368.

[GR83] Goldberg, A. & Robson, D. **Smalltalk-80: The Language and its Implementation**. Addison Wesley, Reading, Massachusetts (1983).

[HKS92] Hook, J., Kieburtz, R.B. & Sheard, T. “Generating Programs by Reflection”. Oregon Graduate Institute of Science & Technology Technical Report CS/E 92-015 (1992).

[HP88] Hewlett Packard. **NewWave Environment General Information Manual**. Hewlett-Packard (1988).

[HS90] Hurst, A.J. & Sajeev, A.S.M. “A Capability Based Language for Persistent Programming”. In **Persistent Object Systems**, Rosenberg, J. & Koch, D.M. (ed), Springer-Verlag (1990) pp 186-201.

[HWA+90] Hudak, P., Wadler, P., Arvind, Boutel, B., Fairbairn, J., Fasel, J., Hughes, J., Johnsson, T., Kieburtz, D., Peyton-Jones, S., Nikhil, R., Reeve, M., Wise, D. & Young, J. “Report on the Functional Programming Language Haskell”. University of Glasgow (1990).

[IBM78] “IBM Report on the Contents of a Sample of Programs Surveyed”. IBM, San Jose, California (1978).

[KCC+92a] Kirby, G.N.C., Cutts, Q.I., Connor, R.C.H., Dearle, A. & Morrison, R. “Programmers’ Guide to the Napier88 Standard Library, Edition 2.1”. University of St Andrews (1992).

[KCC+92b] Kirby, G.N.C., Connor, R.C.H., Cutts, Q.I., Dearle, A., Farkas, A.M. & Morrison, R. “Persistent Hyper-Programs”. In Proc. 5th International Workshop on Persistent Object Systems, San Miniato, Italy (1992) pp 73-95.

[KCD+89] Kirby, G.N.C., Cutts, Q.I., Dearle, A. & Marlin, C.D. “WIN: A Persistent Window Management System”. Universities of Glasgow and St Andrews Technical Report PPRR-73-89 (1989).

[KD90] Kirby, G.N.C. & Dearle, A. “An Adaptive Graphical Browser for Napier88”. University of St Andrews Technical Report CS/90/16 (1990).

[Kir92] Kirby, G.N.C. “Persistent Programming with Strongly Typed Linguistic Reflection”. In Proc. 25th International Conference on Systems Sciences, Hawaii (1992) pp 820-831.

[Kno65] Knowlton, K.C. “A Fast Storage Allocator”. Communications of the ACM 8, 10 (1965) pp 623-625.

[Kow79] Kowalski, R. “Algorithm = Logic + Control”. Communications of the ACM 22, (1979) pp 424-436.

[KR78] Kernighan, B.W. & Ritchie, D.M. **The C programming language**. Prentice-Hall (1978).

[LDF+87] Lyngbaek, P., Derrett, N.P., Fishman, D.H., Kent, W. & Ryan, T.A. "Design and Implementation of the Iris Object Manager". In Proc. 2nd International Workshop on Persistent Object Systems, Appin, Scotland (1987) pp 25-51.

[LRN86] Laird, J., Rosenbloom, P. & Newell, A. "Chunking in SOAR: The Anatomy of a General Learning Mechanism". Machine Intelligence 1, 1 (1986).

[LRV90] Lécluse, C., Richard, P. & Velez, F. "$O_2$, an Object-Oriented Data Model". In **Readings in Object-Oriented Database Systems**, Zdonik, S.B. & Maier, D. (ed), Morgan Kaufman (1990) pp 227-236.

[MAD87] Morrison, R., Atkinson, M.P. & Dearle, A. "Flexible Incremental Bindings in a Persistent Object Store". Universities of Glasgow and St Andrews Technical Report PPRR-38-87 (1987).

[MAE+62] McCarthy, J., Abrahams, P.W., Edwards, D.J., Hart, T.P. & Levin, M.I. **The Lisp Programmers' Manual**. M.I.T. Press, Cambridge, Massachusetts (1962).

[Mae87] Maes, P. "Concepts and Experiments in Computational Reflection". In Proc. OOPSLA'87, Orlando, Florida (1987) pp 147-155.

[Mai90] Mainetto, G. "Italian Cookery". Personal communication (1990).

[Mat85] Matthews, D.C.J. "Poly Manual". University of Cambridge Technical Report 65 (1985).

[Mat89] Matthews, D.C.J. "Papers on Poly/ML". University of Cambridge Technical Report 161 (1989).

[MBC+87] Morrison, R., Brown, A.L., Connor, R.C.H. & Dearle, A. "Polymorphism, Persistence and Software Reuse in a Strongly Typed Object Oriented Environment". Universities of Glasgow and St Andrews Technical Report PPRR-32-87 (1987).

[MBC+89] Morrison, R., Brown, A.L., Connor, R.C.H. & Dearle, A. "The Napier88 Reference Manual". University of St Andrews Technical Report PPRR-77-89 (1989).

[MBC+90] Morrison, R., Brown, A.L., Connor, R.C.H., Cutts, Q.I., Kirby, G.N.C., Dearle, A., Rosenberg, J. & Stemple, D. "Protection in Persistent Object Systems". In **Security and Persistence**, Rosenberg, J. & Keedy, J.L. (ed), Springer-Verlag (1990) pp 48-66.

[MBW80] Mylopoulos, J., Bernstein, P.A. & Wong, H.K.T. "A Language Facility for Designing Database-Intensive Applications". ACM Transactions on Database Systems 5, 2 (1980) pp 185-207.

[Mil78] Milner, R. "A Theory of Type Polymorphism in Programming". Journal of Computer and System Sciences 17, 3 (1978) pp 348-375.

[MM81] Meyrowitz, N. & Moser, M. "BRUWIN: An Adaptable Design Strategy for Window Manager/Virtual Terminal Systems". Journal of the ACM 28 (1981) pp 180-189.

[MMS92] Matthes, F., Müller, R. & Schmidt, J.W. "Object Stores as Servers in Persistent Programming Environments—The P-Quest Experience". ESPRIT BRA Project 3070 FIDE Technical Report FIDE/92/48 (1992).

[MS87] Maier, D. & Stein, J. “Development and Implementation of an Object-Oriented DBMS”. In **Research Directions in Object-Oriented Programming**, Shriver, B. & Wegner, P. (ed), MIT Press (1987) pp 355-392.

[MS89] Matthes, F. & Schmidt, J.W. “The Type System of DBPL”. In Proc. 2nd International Workshop on Database Programming Languages, Salishan, Oregon (1989) pp 219-225.

[MTH89] Milner, R., Tofte, M. & Harper, R. **The Definition of Standard ML**. MIT Press, Cambridge, Massachusetts (1989).

[Mye86] Myers, B.A. “A Complete and Efficient Implementation of Covered Windows”. IEEE Computer September (1986) pp 57-67.

[Mye89] Myers, B.A. “User Interface Tools: Introduction and Survey”. IEEE Software 6, 1 (1989) pp 15-23.

[Not85] Notkin, D. “The GANDALF Project”. Journal of Systems and Software 5, (1985) pp 91.

[OHK87] O’Brien, P.D., Halbert, D.C. & Kilian, M.F. “The Trellis Programming Environment”. In Proc. OOPSLA’87, Orlando, Florida (1987) pp 91-102.

[Per87] Perry, N. “Hope+”. Flagship project, Imperial College London Technical Report IC/FPR/LANG/2.5.1/7 (1987).

[Phi90] Philbrow, P.C. “Indexing Strongly Typed Heterogeneous Collections Using Reflection and Persistence”. In Proc. ECOOP/OOPSLA Workshop on Reflection and Metalevel Architectures in Object-Oriented Programming, Ottawa, Canada (1990).

[PS88] “PS-algol Reference Manual, 4th edition”. Universities of Glasgow and St Andrews Technical Report PPRR-12-88 (1988).

[PYD91] Pausch, R., Young, N. & DeLine, R. “Simple User Interface Toolkit (SUIT): The Pascal of User Interface Toolkits”. In Proc. ACM Symposium on User Interface Software and Technology (1991) pp 117-125.

[RC86] Rees, J. & Clinger, W. “Revised Report on the Algorithmic Language Scheme”. ACM SIGPLAN Notices 21, 12 (1986) pp 37-43.

[RC90] Richardson, J.E. & Carey, M.J. “Implementing Persistence in E”. In **Persistent Object Systems**, Rosenberg, J. & Koch, D.M. (ed), Springer-Verlag (1990) pp 175-199.

[Rei84] Reiss, S.P. “Graphical Program Development with PECAN Program Development Systems”. ACM SIGPLAN Notices 19, 5 (1984) pp 30-41.

[RT78] Ritchie, D.M. & Thompson, K. “The UNIX Time-Sharing System”. The Bell System Technical Journal 63, 6 (1978) pp 1905-1930.

[SDP91] Shrivastava, S.K., Dixon, G.N. & Parrington, G.D. “An Overview of the Arjuna Distributed Programming System”. IEEE Software 8, 1 (1991) pp 66-73.

[SFS+90] Stemple, D., Fegaras, L., Sheard, T. & Socorro, A. "Exceeding the Limits of Polymorphism in Database Programming Languages". In **Lecture Notes in Computer Science 416**, Bancilhon, F., Thanos, C. & Tsichritzis, D. (ed), Springer-Verlag (1990) pp 269-285.

[SG86] Scheifler, R.W. & Gettys, J. "The X Window System – An Overview". ACM Transactions on Graphics 5, 2 (1986).

[Sha86] Shapiro, M. "SOS: a Distributed Object-Oriented Operating System". In Proc. 2nd ACM SIGOPS European Workshop on "Making Distributed Systems Work", Amsterdam, Netherlands (1986).

[She90] Sheard, T. "A user's Guide to TRPL: A Compile-time Reflective Programming Language". COINS, University of Massachusetts Technical Report 90-109 (1990).

[She91] Sheard, T. "Automatic Generation and Use of Abstract Structure Operators". ACM Transactions on Programming Languages and Systems 19, 4 (1991) pp 531-557.

[Shi81] Shipman, D. "The Functional Data Model and the Data Language DAPLEX". ACM Transactions on Database Systems 6, 1 (1981) pp 140-173.

[Shn92] Shneiderman, B. **Designing the User Interface**. Addison-Wesley, Reading, Massachusetts (1992).

[Sjø92] Sjøberg, D. "Measuring Name and Identifier Usage in Napier88 Applications". ESPRIT BRA Project 3070 FIDE Technical Report FIDE/92/37 (1992).

[SM90] Shapiro, M. & Mosseri, L. "A Simple Object Storage System". In **Persistent Object Systems**, Rosenberg, J. & Koch, D.M. (ed), Springer-Verlag (1990) pp 272-276.

[SS89] Sheard, T. & Stemple, D. "Automatic Verification of Database Transaction Safety". ACM Transactions on Database Systems 12, 3 (1989) pp 322-368.

[SS91] Sheard, T. & Stemple, D. "Examples in TRPL". COINS, University of Massachusetts (1991).

[SSF92] Stemple, D., Sheard, T. & Fegaras, L. "Linguistic Reflection: A Bridge from Programming to Database Languages". In Proc. 25th International Conference on Systems Sciences, Hawaii (1992) pp 844-855.

[SSS+92] Stemple, D., Stanton, R.B., Sheard, T., Philbrow, P., Morrison, R., Kirby, G.N.C., Fegaras, L., Cooper, R.L., Connor, R.C.H., Atkinson, M.P. & Alagic, S. "Type-Safe Linguistic Reflection: A Generator Technology". ESPRIT BRA Project 3070 FIDE Technical Report FIDE/92/49 (1992).

[Str67] Strachey, C. **Fundamental Concepts in Programming Languages**. Oxford University Press, Oxford (1967).

[Str86] Stroustrup, B. **The C++ Programming Language**. Addison-Wesley (1986).

[Sun89] Sun Microsystems. **Open Look™ Graphical User Interface Functional Specification**. Addison-Wesley, Mountain View, California (1989).

[Sun90] Sun Microsystems. **Open Windows Developer's Guide 1.1: User's Guide**. Addison-Wesley, Mountain View, California (1990).

[Swe85] Sweet, R.E. "The Mesa Programming Environment". In Proc. ACM SIGPLAN Symposium on Programming Languages and Programming Environments (1985) pp 216-229.

[Tei84] Teitelman, W. "A Tour Through Cedar". IEEE Software April (1984) pp 44-73.

[TM84] Teitelman, W. & Masinter, L. "The Interlisp Programming Environment". In **Interactive Programming Environments**, Barstow, D.R., Shrobe, H.E. & Sandewall, E. (ed), McGraw-Hill (1984).

[TR81] Teitelbaum, T. & Reps, T. "The Cornell Program Synthesizer: A Syntax-Directed Programming Environment". Communications of the ACM 24, 9 (1981) pp 563.

[Tur85] Turner, D.A. "Miranda: A non-strict functional language with polymorphic types". In **Lecture Notes in Computer Science 201**, Jouannaud, J. (ed), Springer-Verlag (1985) pp 1-16.

[WA86] Wile, D.S. & Allard, D.G. "Worlds: An Organizing Structure for Object-Bases". In Proc. 2nd ACM SIGSOFT/SIGPLAN Symposium on Practical Software Development Environments, Palo Alto, California (1986).

[Wai87] Wai, F. "Distribution and Persistence". In Proc. 2nd International Workshop on Persistent Object Systems, Appin, Scotland (1987) pp 207-225.

[War89] Warboys, B. "The IPSE 2.5 Project: Process Modelling as the Basis for a Support Environment". In Proc. 1st International Conference on System Development Environments and Factories, Berlin, Germany (1989).

[WBB+90] Wiecha, C., Bennett, W., Boies, S., Gould, J. & Greene, S. "ITS: A Tool for Rapidly Developing Interactive Applications". ACM Transactions on Information Systems 8, 3 (1990) pp 204-236.

[WCG87] Williams, A., Crampton, C. & Goswell, C. "Unix Window Management Systems Client-Server Interface Specification". Rutherford Appleton Laboratory Technical Report 11/3/87 (1987).

[Web89] Webster, B.F. **The NeXT Book**. Addison-Wesley, Reading, Massachussetts (1989).

[Wir71] Wirth, N. "The Programming Language Pascal". Acta Informatica 1 (1971) pp 35-63.

[Wir83] Wirth, N. **Programming in Modula-2**. Springer-Verlag (1983).